\documentclass[12pt,twoside,notitlepage]{book}

\RequirePackage{lineno}
\usepackage{enumitem}
\usepackage{amssymb, amsmath}
\usepackage{mathpazo}
\usepackage{float,graphicx}
\usepackage[figuresright]{rotating}
\usepackage[usenames,dvipsnames]{color}
\usepackage{microtype}
\usepackage[small,bf]{caption}
\usepackage{tabulary}
\usepackage{booktabs,multirow,dcolumn,bigdelim}

\usepackage[pdftex,plainpages=false,pdfpagelabels,backref,pdfborder={0 0 0}]{hyperref}
\usepackage{url}
\usepackage[toc,page,titletoc]{appendix}
\usepackage{afterpage}
\usepackage[small,bf]{caption}
\usepackage{moresize}
\usepackage[scaled]{helvet}
\usepackage{sectsty}
\allsectionsfont{\normalfont\sffamily}
\usepackage{titletoc}
\titlecontents{chapter}
  [1.5em]
  {\linespread{0.9}\normalfont\sffamily\bfseries}
  {\contentslabel{1em}} 
  {\hspace*{-2.3em}} 
  {\mdseries\titlerule*[1pc]{.}\contentspage} 
\titlecontents{section}
  [3.5em]
  {\linespread{0.9}\normalfont\sffamily}
  {\contentslabel{2.3em}} 
  {\hspace*{-2.3em}} 
  {\titlerule*[1pc]{.}\contentspage} 
\titlecontents{subsection}
  [4.5em]
  {\linespread{0.9}\normalfont\sffamily}
  {\contentslabel{2.3em}} 
  {\hspace*{-2.3em}} 
  {\titlerule*[1pc]{.}\contentspage} 
\titlecontents{subsubsection}
  [5.5em]
  {\linespread{0.9}\normalfont\sffamily}
  {\contentslabel{2.3em}} 
  {\hspace*{-2.3em}} 
  {\titlerule*[1pc]{.}\contentspage} 
\usepackage[text={6.5in,8.75in},headheight=15pt,centering]{geometry}
\usepackage{xspace}

\DeclareGraphicsExtensions{.pdf,.png}
\usepackage{fancyhdr}
\renewcommand{\chaptermark}[1]{\markboth{#1}{}}

\fancypagestyle{plain}{%
  \fancyhf{}
  \fancyfoot[RO,LE]{\sffamily \thepage}
}

\newcommand {\AuAu}{\mbox{Au$+$Au}\xspace}
\newcommand {\auau}{\mbox{Au$+$Au}\xspace}
\newcommand {\pbpb}{\mbox{Pb$+$Pb}\xspace}

\newcommand {\pp}{\mbox{$p$$+$$p$}\xspace}
\newcommand {\ppbar}{\mbox{$p$$+$$\overline{p}$}\xspace}
\newcommand{\pT}{\mbox{${p_T}$}\xspace}
\newcommand{\jpsi}{\mbox{$J/\psi$}}

\newcommand{\qgp}{\mbox{quark-gluon plasma}\xspace}

\newcommand{\PbPb}{\mbox{Pb$+$Pb}}

\newcommand{\ET}{\mbox{$E_T$}}

\newcommand{\pt}{\mbox{${p_T}$}\xspace}

\newcommand{\dAu}{\mbox{$d$$+$Au}\xspace}

\newcommand{\fastjet}{\mbox{\sc FastJet}\xspace}
\newcommand{\geant}{\mbox{\sc Geant4}\xspace}
\newcommand{\antikt}{\mbox{anti-$k_T$}\xspace}
\newcommand{\pythia}{\mbox{\sc Pythia}\xspace}
\newcommand{\pyquen}{\mbox{\sc Pyquen}\xspace}
\newcommand{\hijing}{\mbox{\sc Hijing}\xspace}

\newcommand{\roounfold}{\mbox{\sc RooUnfold}\xspace}
\newcommand{\beetle}{\mbox{\sc Beetle}\xspace}
\newcommand{\gj}{\mbox{$\gamma$+jet}\xspace}
\newcommand{\gh}{\mbox{$\gamma$+hadron}\xspace}
\newcommand{\martinimusic}{\mbox{\sc Martini+Music}\xspace}
\newcommand{\martini}{\mbox{\sc Martini}\xspace}
\newcommand{\music}{\mbox{\sc Music}\xspace}

\newcommand{\dijet}{\mbox{dijet}\xspace}
\newcommand{\fake}{\mbox{fake}\xspace}
\newcommand{\fast}{\mbox{fast}\xspace}
\newcommand{\veryfast}{\mbox{very fast}\xspace}

\newcommand{\onewidth}{0.6\linewidth}
\newcommand{\twowidth}{0.48\linewidth}

\begin{document} 


\frontmatter

\pagestyle{empty}

\renewcommand*\familydefault{\sfdefault}
{\sffamily
\vfill
\vspace{4cm}
\begin{figure}[H]
  \begin{center}
  \includegraphics[width=0.7\linewidth]{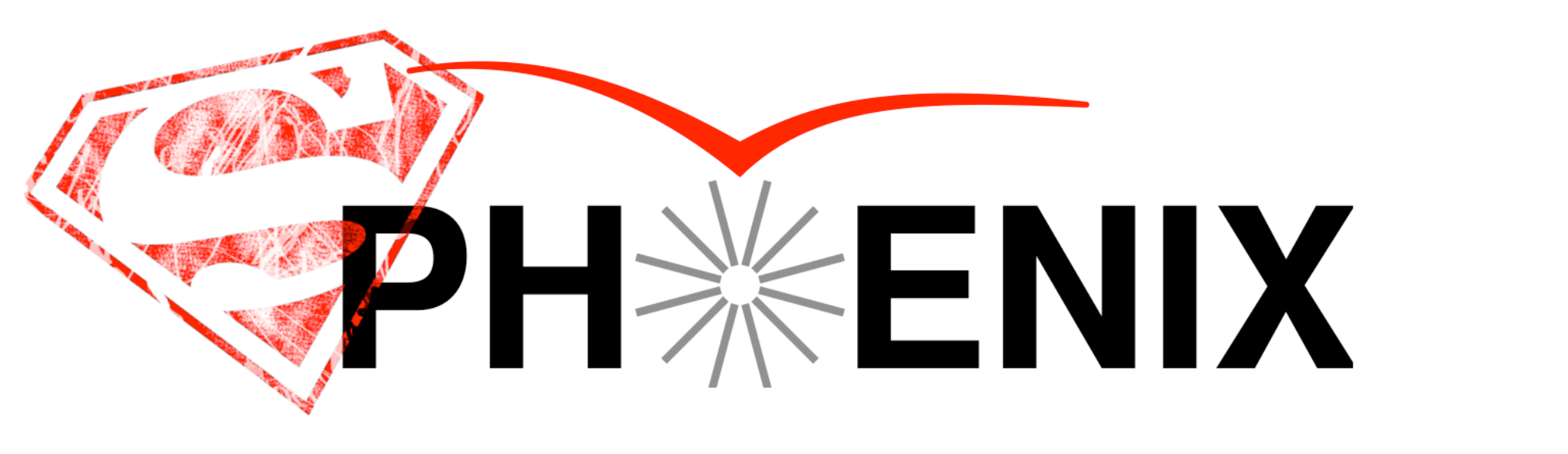}
  \end{center}
\end{figure}

\begin{center}
  \large
  An Upgrade Concept from the PHENIX Collaboration  

  July 1, 2012
\end{center}

\vspace{2cm}

\begin{figure}[H]
  \begin{center}
    \includegraphics[width=\linewidth]{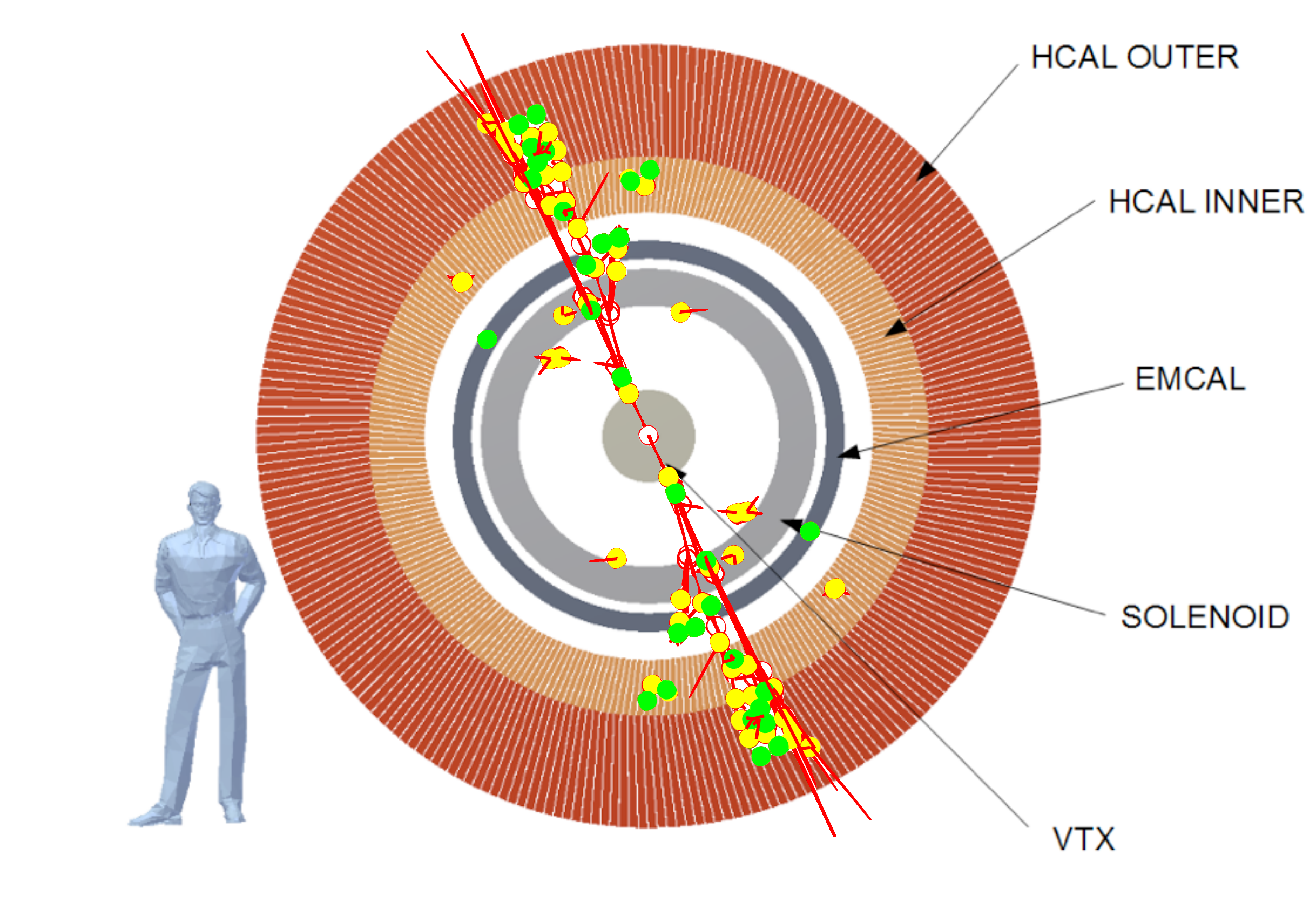}
  \end{center}
\end{figure}

%
}

\vfill
\renewcommand*\familydefault{\rmdefault}

\cleardoublepage
\pagestyle{fancy}

\chapter*{Executive Summary}
\label{executive_summary}
\setcounter{page}{1}

The PHENIX collaboration presents in this document a proposal for a
major upgrade to the PHENIX detector at the Relativistic Heavy Ion
Collider.  This upgrade, referred to as sPHENIX, brings exciting new
capability to the RHIC program by opening new and important channels
for experimental investigation and utilizing fully the luminosity of
the recently upgraded RHIC facility.  sPHENIX enables a compelling jet
physics program that will address fundamental questions about the
nature of the strongly coupled \qgp discovered experimentally at RHIC
to be a perfect fluid. Fundamental questions such as how and why the \qgp 
behaves as a perfect fluid in the vicinity of strongest coupling, near 1--2 $T_{c}$,
can only be fully addressed with world-class jet observables at RHIC
energies.  Comparing these measurements to higher temperature \qgp
measurements at the Large Hadron Collider will provide invaluable
insight into the thermodynamics of QCD.

The proposed upgrade addresses specific questions whose answers are
necessary to advance our understanding of the \qgp:
\begin{enumerate}[topsep=2pt, partopsep=2pt, parsep=4pt, itemsep=4pt]
\item How to reconcile the observed strongly coupled \qgp with the
asymptotically free theory of quarks and gluons?
%
%
\item What are the dynamical changes to the \qgp in terms of
  quasiparticles and excitations as a function of temperature?
\item How sharp is the transition of the \qgp from the most strongly
  coupled regime near $T_c$ to a weakly coupled system of partons
  known to emerge at asymptotically high temperatures?
\end{enumerate}

To pursue this physics we are proposing an upgrade consisting of a
2\,T magnetic solenoid of radius 70\,cm surrounded by electromagnetic
and hadronic calorimetry with uniform coverage over $|\eta|<1.0$.
With the now completed RHIC luminosity upgrade, a 20 week run will
deliver over 50 billion \auau collisions, and sPHENIX will thus sample
over 10 million dijet events with $\ET > 20$\,GeV, along with a
correspondingly large $\gamma$$+$jet sample. The newly developed
flexibility of RHIC enabled by the Electron Beam Ion Source and the
high rate capability of sPHENIX will provide critical precision
control data sets in \pp, $p(d)$$+$A, and a full range of collision species.

\begin{figure}[p]
 \begin{center}
    \includegraphics[width=0.75\linewidth]{OverviewCutaway051112}
    \\
    \vskip 0.8in
    \includegraphics[width=0.95\linewidth]{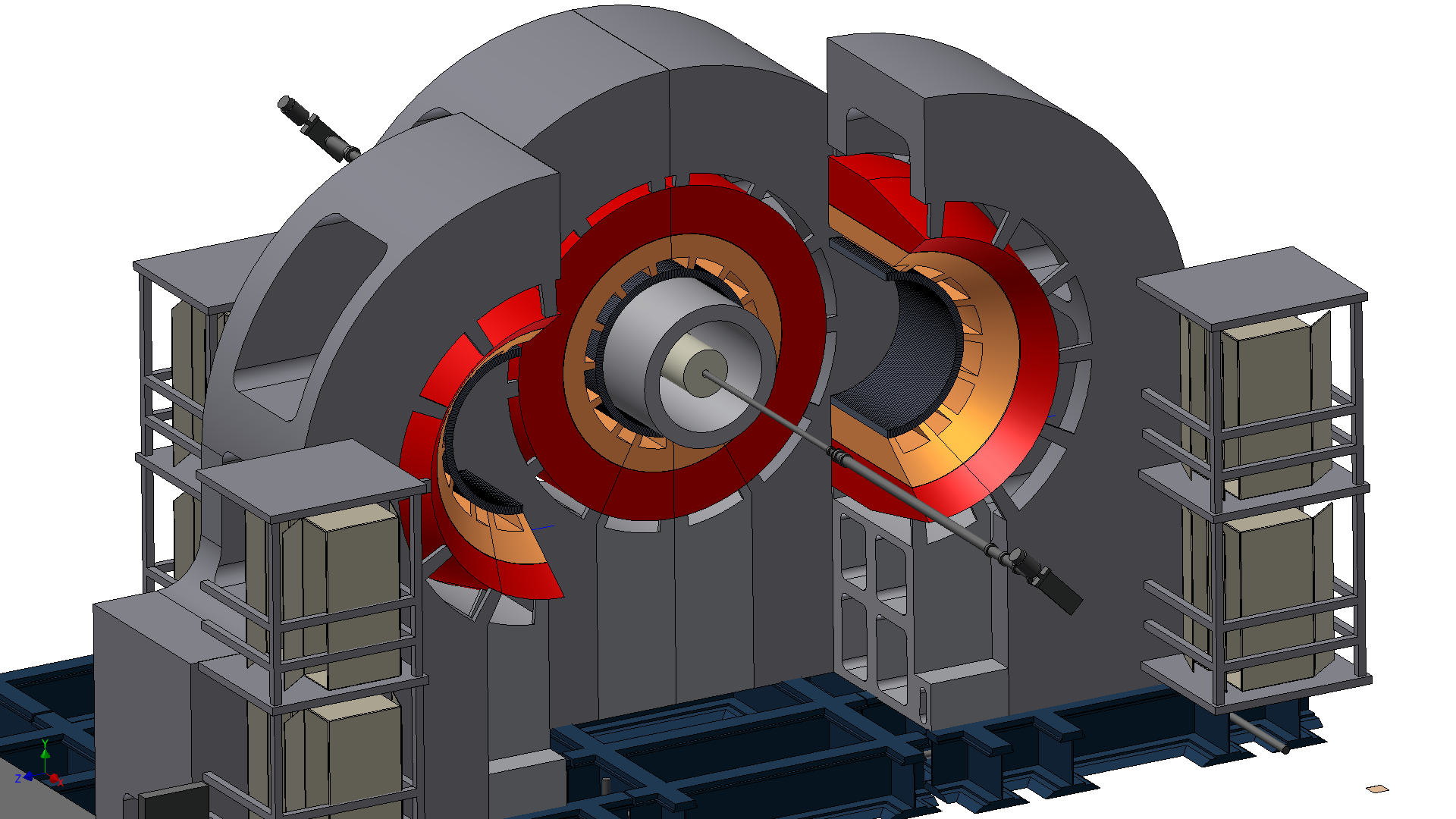}
    \caption{(top) An engineering rendering of the
      sPHENIX upgrade showing the inner silicon tracker (VTX), the solenoid, and the calorimeters.
      (bottom) A view showing how sPHENIX would fit into and be supported
      and serviced in the current PHENIX interaction region.}
    
    \label{fig:superdrawing}

 \end{center}
\end{figure}

An engineering rendering of the upgraded detector and its
incorporation into the PHENIX interaction region are shown in
Figure~\ref{fig:superdrawing}.  The design of sPHENIX takes advantage
of technological developments enabling the detector to be very
compact, which allows for a significantly lower cost per unit solid
angle coverage. Further cost savings are achieved by reusing
significant elements of the existing PHENIX mechanical and electrical
infrastructure. We have obtained budgetary guidance from well-regarded
vendors for the major components of sPHENIX.  We have estimated the
cost of engineering, management, and construction, and applied
standard guidance for overhead and contingency.  From this we
conclude that the cost of sPHENIX is on the order of \$25M, and that
designing, building and installing the detector could be done within
five years.

The sPHENIX upgrade proposed in this document represents a major
scientific instrument. Its physics capabilities can be augmented in
the future through modest incremental upgrades that have been an
integral part of the design considerations from the outset.  Note that
these additional upgrades are not included in the scope of this
proposal and are described separately in
Appendices~\ref{chap:barrel_upgrade} and \ref{chap:fsPHENIX}.

The specific future options considered for installation inside the solenoid
magnet are additional charged particle tracking outside the existing
PHENIX silicon vertex detector (VTX) and a preshower with fine
segmentation in front of the electromagnetic calorimeter.  The possibility of
extending the sPHENIX capabilities has attracted international interest.  For
example, RIKEN has expressed very strong interest in providing
additional charged particle tracking outside of the existing PHENIX
silicon tracker.  These future additions will expand the sPHENIX
physics program to include: (a) heavy quarkonia suppression via the
three upsilon states, (b) tagging of charm and beauty jets, (c) jet
fragmentation function modifications, (d) nuclear suppression of
$\pi^{0}$ yields up to $p_{T} = 40$\,GeV/c, and (e) a possible low
mass dilepton program.  The open geometry of the magnetic solenoid
also allows for a forward angle spectrometer upgrade option aimed at
measuring photon, jet, and lepton observables relevant to answering
questions in $p(d)+A$ collisions about cold nuclear matter and in
transversely polarized \pp collisions about transversity.

The design for a future Electron Ion Collider (EIC) at RHIC consists
of adding a 5--30\,GeV electron beam to the current hadron and nuclear
beam capabilities.  The proposed initial construction would consist of
a 5--10\,GeV electron beam, referred to as Phase 1 of eRHIC.  We have
designed sPHENIX so that it would also serve as the foundation for a
future EIC detector, referred to as ePHENIX.  The sPHENIX proposal,
covering $|\eta|<1.0$, when combined with future upgrades in the
backward ($\eta < -1.0$) and forward ($\eta > 1.0$) regions is
compatible with a full suite of EIC physics measurements.

The document is organized as follows.  In
Chapter~\ref{chap:physics_case}, we detail the physics accessible via
jet, dijet, and $\gamma$$+$jet measurements at RHIC to demonstrate the
mission need.  In Chapter~\ref{chap:detector_requirements}, we detail
the sPHENIX detector upgrade and the subsystem requirements to achieve
the physics goals.  In Chapter~\ref{chap:detector_concept}, we detail
the specific detector design and \geant simulation results.  In
Chapter~\ref{chap:jet_performance}, we detail the physics performance
with full detector simulations.  In the Appendices we detail the additional physics
capabilities gained through further upgrades.
Appendix~\ref{chap:barrel_upgrade} describes two midrapidity detector
additions, Appendix~\ref{chap:fsPHENIX} details a forward rapidity
upgrade, and Appendix~\ref{chap:ePHENIX} shows an evolution to an
ePHENIX detector suitable for a future Electron Ion Collider at RHIC.

\cleardoublepage

\resetlinenumber

\tableofcontents

\cleardoublepage

\mainmatter

\renewcommand{\thepage}{\arabic{page}}
\setcounter{chapter}{0}
\setcounter{page}{1}



\chapter{The Physics Case for sPHENIX}
\label{chap:physics_case}

\begin{figure}[!hbt]
 \begin{center}
   \raisebox{4pt}{\includegraphics[trim = 2 2 2 2, clip, width=0.46\linewidth]{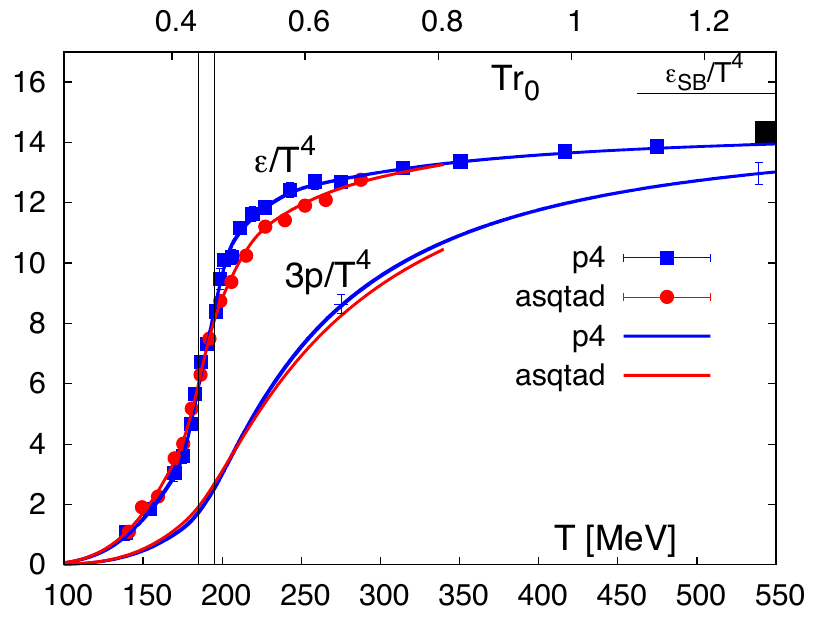}}
    \hfill
    \includegraphics[trim = 2 2 2 2, clip, width=0.52\linewidth]{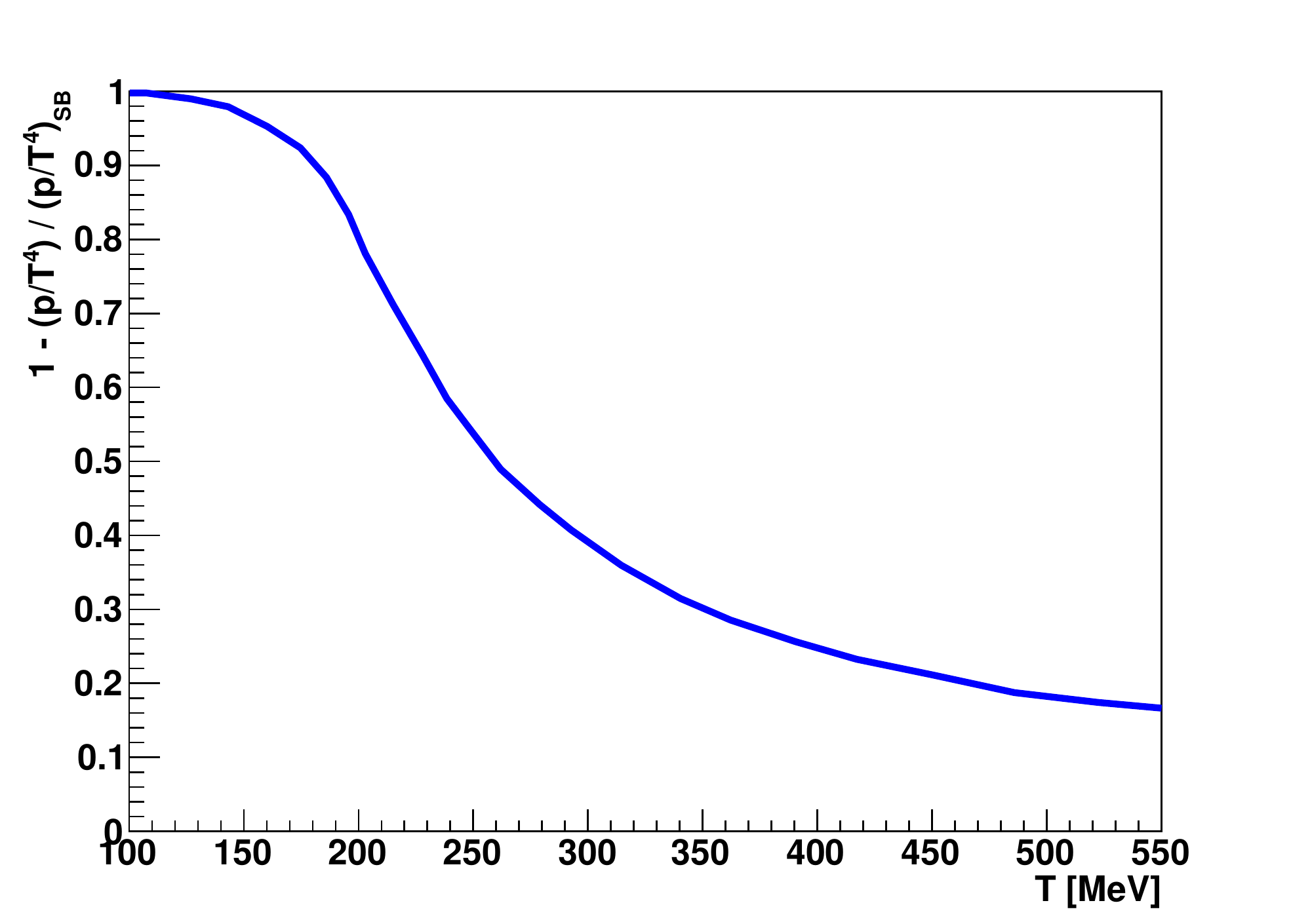}
    \caption{(Left) The energy density and three times the pressure
      normalized by $1/T^{4}$ as a function of
      temperature~\protect\cite{PhysRevD.80.014504}{}. (Right) Deviation in 
      $p/T^{4}$ relative to the Stefan-Boltzmann value as a function
      of temperature.  The deviation from the Stefan-Boltzmann value is 23\%, 39\%, 53\%, and 80\% at
      temperatures of 420, 300, 250, and 200\,MeV, respectively.
    \label{fig:lattice}}
 \end{center}
\end{figure}

Hadronic matter under conditions of extreme temperature or net baryon
density transitions to a new state of matter called the \qgp.  Lattice
QCD calculations at zero net baryon density indicate a smooth
crossover transition at $T_{c} \approx 170$\,MeV, though with a rapid
change in properties at that temperature as shown in the left panel of
Figure~\ref{fig:lattice}~\cite{PhysRevD.80.014504}.  This \qgp
dominated the early universe for the first six microseconds of its
existence.  Collisions of heavy nuclei at the Relativistic Heavy Ion
Collider (RHIC) have sufficient initial kinetic energy that is then
converted into heat to create \qgp with an initial
temperature---measured via the spectrum of directly emitted
photons---of greater than 300\,MeV~\cite{Adare:2008ab}. The higher
energy collisions at the Large Hadron Collider (LHC) produce an even
higher initial temperature $T > 420$\,MeV~\cite{Luzum:2009sb}.

In materials where the dominant forces are electromagnetic, the
coupling $\alpha_{\mathrm{em}}$ is always much less than one.  Even so,
many-body collective effects can render perturbative calculations
non-convergent and result in systems with very strong
coupling~\cite{Adams:2012th}.  In cases where the nuclear force is
dominant, and at temperature scales of order 1--3\,$T_{c}$, the
coupling constant $\alpha_{s}$ is not much less than one and the
system is intrinsically non-perturbative.  In addition, the many-body
collective effects in the \qgp and their temperature dependence near
$T_{c}$ are not well understood.

Lattice QCD results for the deviation of the pressure, normalized by
$1/T^{4}$, from the Stefan-Boltzmann limit are shown in
Figure~\ref{fig:lattice}.  The Stefan-Boltzmann limit holds for a
non-interacting gas of massless particles (i.e., the extreme of the weakly
coupled limit), and as attractive inter-particle interactions grow stronger the
pressure decreases.  Thus, one might expect that the \qgp would
transition from a weakly coupled system at high temperature to a more
strongly coupled system near $T_{c}$.  However, a direct quantitative
extraction of the coupling strength warrants caution as string theory
calculations provide an example where the coupling is very strong and
yet the deviation from the Stefan-Boltzmann limit is only
25\%~\cite{Gubser:2009fc,Gubser:1996de}.  The change in initial
temperature between RHIC and LHC collisions is thus expected to be
associated with important changes in the nature of the \qgp~\cite{Wiedemann:2009sa}.  
If not, the question is why not.

The collisions at RHIC and the LHC involve a time evolution during
which the temperature drops as the \qgp expands.  The real constraint
on the temperature dependence of the \qgp properties will come from
calculations which simultaneously describe observables measured at both energies.
Since we are studying a phase transition, it is crucial to do
experiments near the phase transition and compare them with
experiments done further above $T_c$.  Typically, all the non-scaling
behavior is found near the transition.

For many systems the change in coupling strength is related to
quasiparticle excitations, strong coherent fields, etc., and to study
these phenomena one needs to probe the medium at a variety of length
scales.  For example, in a superconductor probed at long length
scales, one scatters from Cooper pairs; in a superconductor probed at
short distance scales one observes the individual electrons.  Hard
scattered partons generated in heavy ion collisions that traverse the
\qgp serve as the probes of the medium.  Utilizing these partonic
probes, measured as reconstructed jets, over the broadest possible
energy scale is a key part of unraveling the quasiparticle puzzle in
the \qgp.  Jet probes at the LHC reach the highest energies and with
large total energy loss probe the shortest distance scales; the lower
backgrounds at RHIC will push the jet probes to lower energies thus
probing the important longer distance scales in the medium.

This Chapter is organized into Sections as follows.  We first describe
the key ways of 'pushing' and 'probing' the \qgp to understand its
properties.  We then discuss three different aspects in which the RHIC
jet results are crucial in terms of (1) the temperature dependence of
the QGP, (2) the microscopic inner workings of the QGP, and (3) the
QGP time evolution along with the parton shower evolution.  We then
discuss the current state of jet probe measurements from RHIC and LHC
experiments, followed by a review of theoretical calculations for RHIC
jet observables.  Finally, we review the jet, dijet, and $\gamma$-jet
rates relevant for measurements at RHIC.

\section{Pushing and probing the QGP}

Results from RHIC and LHC heavy ion experiments have provided a wealth
of data for understanding the physics of the \qgp.  One very
surprising result discovered at RHIC was the fluid-like flow of the
\qgp~\cite{Adcox:2004mh}, in stark contrast to some expectations that
the \qgp would behave as a weakly coupled gas of quarks and gluons.
It was originally thought that even at temperatures as low as
2--5\,$T_{c}$, the \qgp could be described with a weakly coupled
perturbative approach despite being quite far from energy scales
typically associated with asymptotic freedom.  The \qgp created in
heavy ion collisions expands and cools, eventually passing through the
phase transition to a state of hadrons, which are then measured by
experiment.  Extensive measurements of the radial and elliptic flow of
hadrons, when compared to hydrodynamics calculations, imply a very
small ratio of shear viscosity to entropy density,
$\eta/s$~\cite{Luzum:2008cw}.  In the limit of very weak coupling
(i.e., a non-interacting gas), the shear viscosity is quite large as
particles can easily diffuse across a velocity gradient in the medium.
Stronger inter-particle interactions inhibit diffusion to the limit
where the strongest interactions result in a very short mean free path
and thus almost no momentum transfer across a velocity gradient,
resulting in almost no shear viscosity.  The shortest possible mean
free path is of order the de~Broglie wavelength, which sets a lower
limit on $\eta/s$~\cite{Danielewicz:1984ww}.  A more rigorous
derivation of this limit of $\eta/s \ge 1/4\pi$ has been calculated
within string theory for a broad class of strongly coupled gauge
theories by Kovtun, Son, and Starinets (KSS)~\cite{Kovtun:2004de}.
Viscous hydrodynamic calculations assuming $\eta/s$ as temperature
independent through the heavy ion collision time evolution are
consistent with the experimental data where $\eta/s$ is within 50\% of
this lower bound for strongly coupled
matter~\cite{Luzum:2008cw,Song:2007ux,Alver:2010dn,Teaney:2009qa,Schenke:2011zz,Adare:2011tg}.
Even heavy quarks (i.e., charm and beauty) are swept up in the fluid
flow and theoretical extractions of the implied $\eta/s$ are equally
small~\cite{Adare:2006nq}.

\begin{figure}[t]
 \begin{center}
   \includegraphics[width=\onewidth]{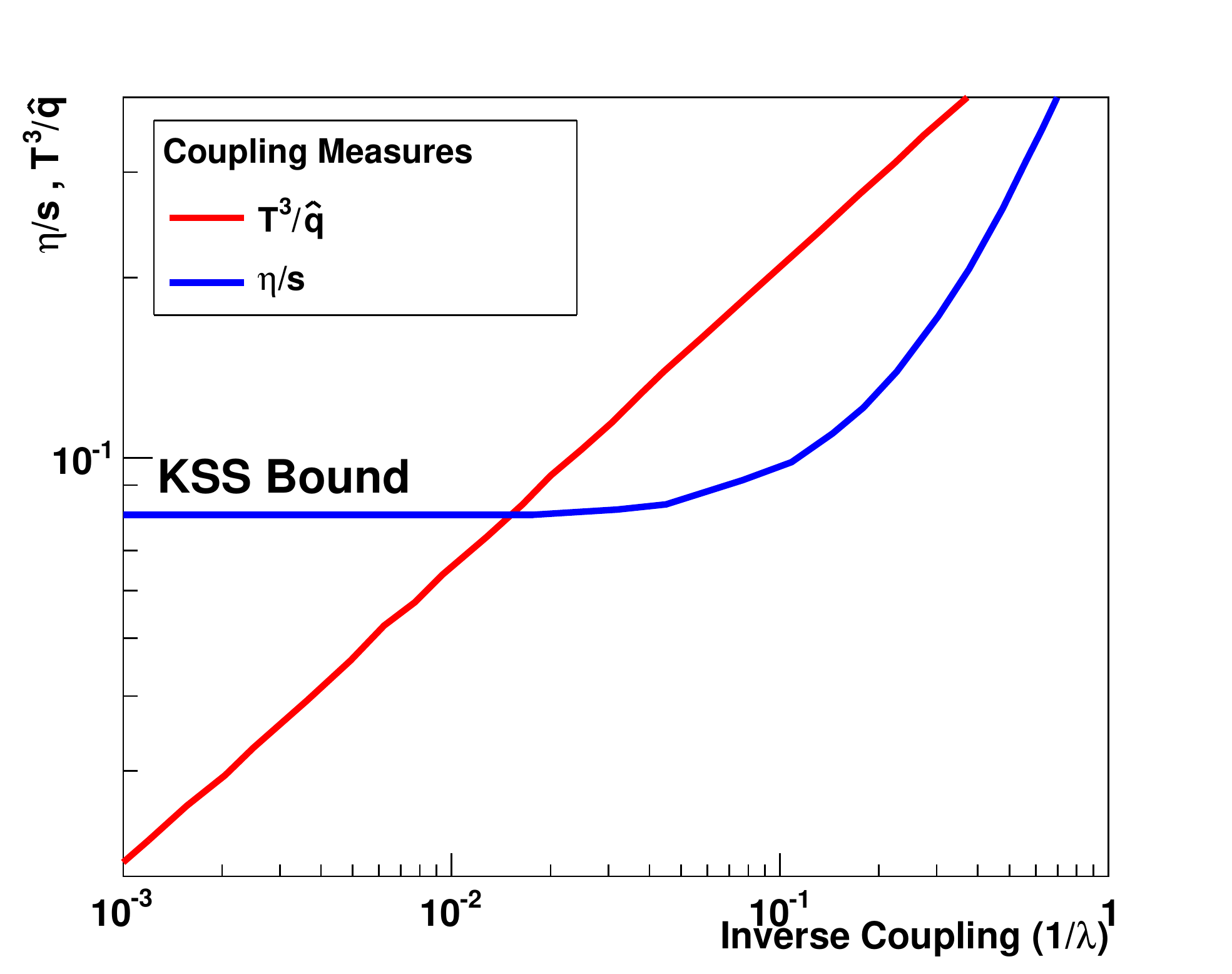} 
   \caption{ $\eta/s$ (blue) and $T^3/\hat{q}$ (red) as a function of
     the inverse of the 't Hooft coupling\protect\cite{Majumder:2007zh}{}.  For
     large $\lambda$ (i.e., small $1/\lambda$), $\eta/s$ approaches the
     quantum lower bound asymptotically, losing its sensitivity to
     further changes in the coupling strength.}
    \label{fig:etas_versus_qhat} 
 \end{center}
\end{figure}

Other key measures of the coupling strength to the medium are found in
the passage of a hard scattered parton through the \qgp.  As the
parton traverses the medium it accumulates transverse momentum as
characterized by $\hat{q} = d(\Delta p_{T}^{2})/dt$ and transfers
energy to the medium via collisions as characterized by $\hat{e} =
dE/dt$.  Ref.~\cite{Liu:2006ug} has calculated $\hat{q}/T^3$ in
${\cal N} = 4$ supersymmetric Yang-Mills theory to be proportional to
the square root of the coupling strength whereas $\eta/s$
asymptotically approaches the quantum lower bound as the coupling
increases.  Both of these ratios are shown as a function of the
inverse coupling in Figure~\ref{fig:etas_versus_qhat}.  For large 't
Hooft coupling ($\lambda$), $\eta/s$ is already quite close to
$1/4\pi$, whereas $T^3/\hat{q}$ is still changing.  This behavior has
caused the authors of Ref.~\cite{Majumder:2007zh} to comment: ``The
ratio $T^{3}/\hat{q}$ is a more broadly valid measure of the coupling
strength of the medium than $\eta/s$.''

In vacuum, the hard scattered parton creates a shower of particles
that eventually form a cone of hadrons, referred to as a jet.  In the
\qgp, the lower energy portion of the shower may eventually be
equilibrated into the medium, thus giving a window on the rapid
thermalization process in heavy ion collisions.  This highlights part
of the reason for needing to measure the fully reconstructed jet
energy and the correlated particle emission with respect to the jet at
all energy scales.  In particular, coupling parameters such as
$\hat{q}$ and $\hat{e}$ are scale dependent and must take on weak
coupling values at high enough energies and very strongly
coupled values at thermal energies.

\begin{figure}[ht]
 \begin{center}
   \includegraphics[trim = 2 2 2 2, clip, width=\onewidth]{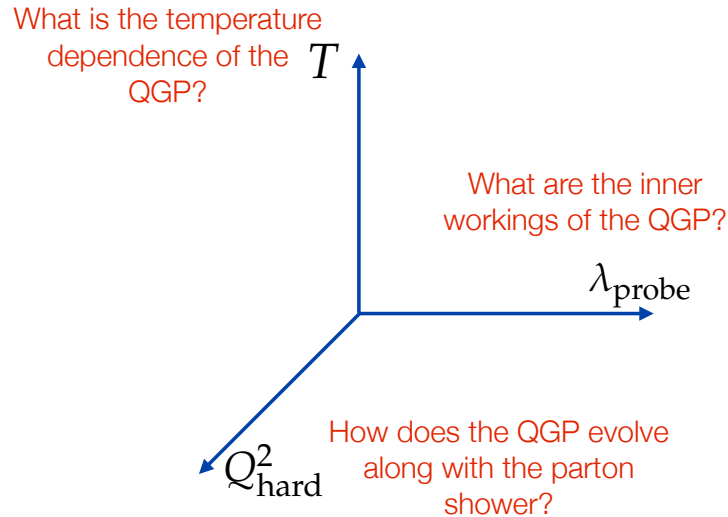}
   \caption{Three illustrative axes along which the \qgp may be pushed and probed.  The axes are
the temperature of the \qgp, the $Q^2_{\mathrm{hard}}$ of the hard process that sets of the scale for the virtuality
evolution of the probe, and the wavelength with which the parton probes the medium $\lambda_{\mathrm{probe}}$.}
   \label{fig:threeaxes}
 \end{center}
\end{figure}

The focus of this proposal is the measurement of jet probes of the
medium as a way of understanding the coupling of the medium, the
origin of this coupling, and the mechanism of rapid equilibration.
Some of these jet probe measurements are already being carried out by
the LHC experiments.  The \qgp is one form of the ``condensed matter''
of QCD and in any rigorous investigation of condensed matter of any
type, it is critical to make measurements as one pushes the system
closer to and further from a phase transition and with probes at
different length scales.  
Substantially extending these scales with measurements at RHIC, particularly closer 
to the transition temperature and at longer distance scales, is the unique ability provided by this proposal.


The critical variables to manipulate for this program are the temperature of
the \qgp, the length scale probed in the medium, and the virtuality of
the hard process as shown schematically in Figure~\ref{fig:threeaxes}.
In the following three sections we detail the physics of each axis.

\section{What is the temperature dependence of the QGP?}

\begin{figure}[t]
 \begin{center}
    \includegraphics[trim = 2 2 2 2, clip, width=\twowidth]{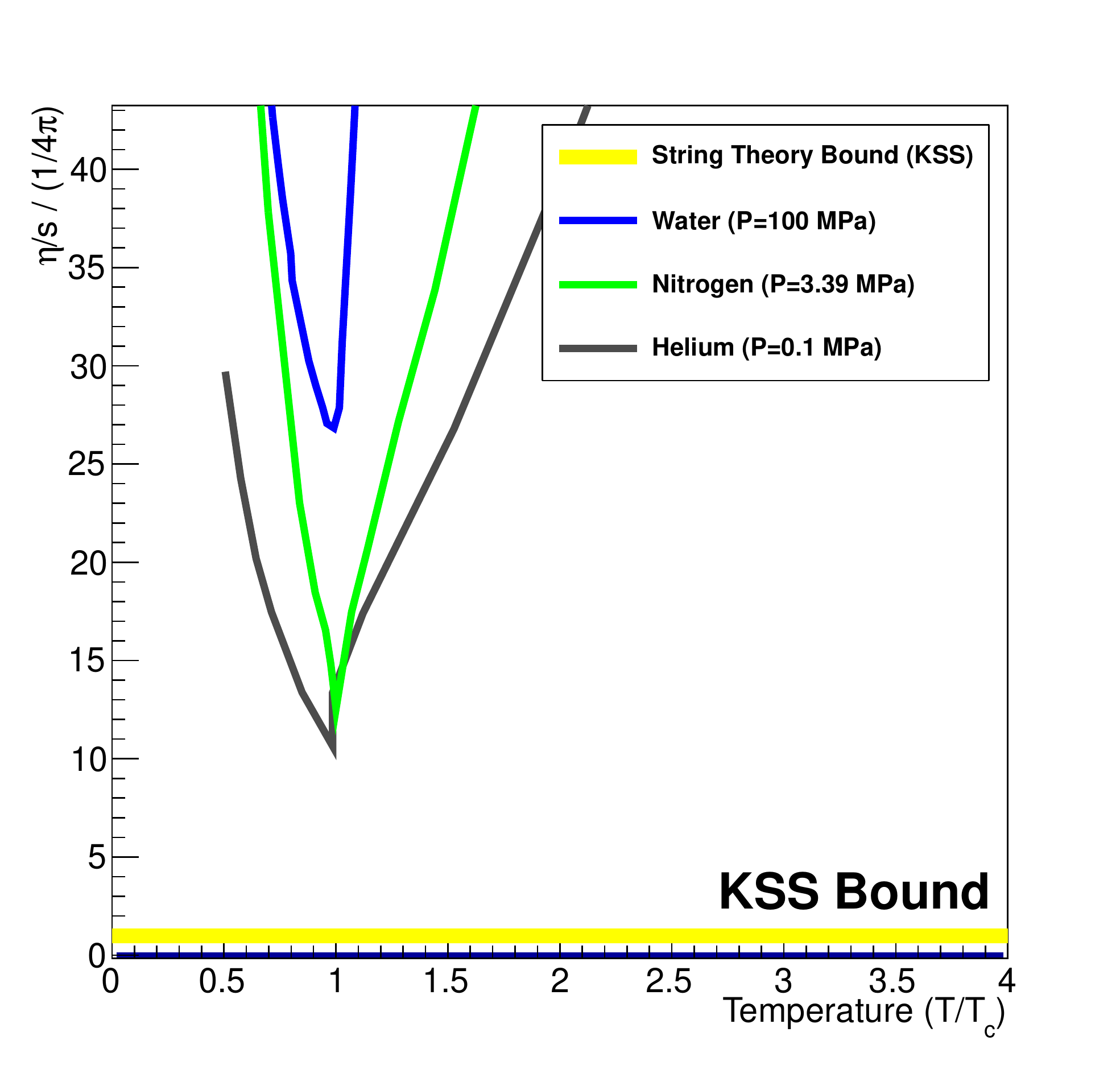}
    \hfill
    \includegraphics[trim = 2 2 2 2, clip, width=\twowidth]{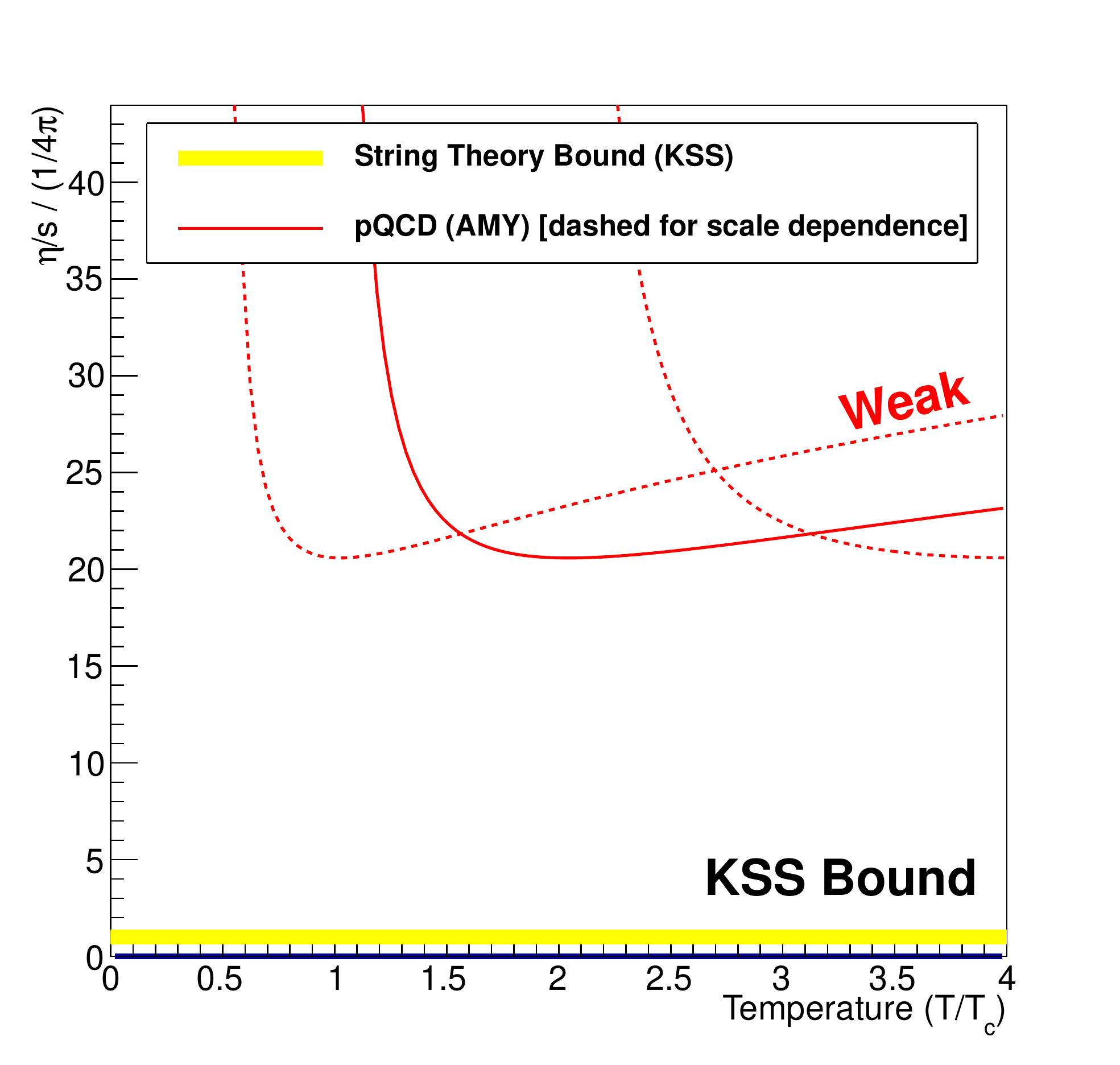}
    \caption{(Left) The ratio of shear viscosity to entropy density,
      $\eta/s$, normalized by the conjectured KSS bound as a function
      of the reduced temperature, $T/T_{c}$, for water, Nitrogen, and
      Helium. The cusp for Helium as shown corresponds to the case at
      the critical pressure.  (Right) Calculation of hot QCD matter
      (quark-gluon plasma) for a weakly coupled system.  Dashed lines
      show the scale dependence of the perturbative calculation.}
    \label{fig:etaovers_1}
 \end{center}
\end{figure}

The internal dynamics of more familiar substances---the subjects of
study in conventional condensed matter and material physics---are
governed by quantum electrodynamics.  It is well known that near a
phase boundary they demonstrate interesting behaviors, such as the
rapid change in the shear viscosity to entropy density ratio,
$\eta/s$, near the critical temperature, $T_c$. This is shown in
Figure~\ref{fig:etaovers_1} for water, nitrogen, and
helium~\cite{Csernai:2006zz}.  Despite the eventual transition to
superfluidity at temperatures below $T_{c}$, $\eta/s$ for these
materials remains an order of magnitude above the conjectured quantum
bound of Kovtun, Son, and Starinets (KSS) derived from string
theory~\cite{Kovtun:2004de}.  These observations provide a deeper
understanding of the nature of these materials: for example the
coupling between the fundamental constituents, the degree to which a
description in terms of quasiparticles is important, and the
description in terms of normal and superfluid components.

The dynamics of the QGP is dominated by quantum chromodynamics and any
experimental characterization of the dependence of $\eta/s$ on
temperature will lead to a deeper understanding of strongly coupled
QCD near this fundamental phase transition.  Theoretically,
perturbative calculations in the weakly coupled limit indicate that
$\eta/s$ decreases slowly as one approaches $T_{c}$ from above, but
with a minimum still a factor of 20 above the KSS
bound~\cite{Arnold:2003zc} (as shown in the right panel of
Figure~\ref{fig:etaovers_1}).  However, as indicated by the dashed
lines in the figure, the perturbative calculation has a large
renormalization scale dependence and results for different values of
the scale parameter ($\mu, \mu/2, 2\mu$) diverge from each other near
$T_{c}$.

\begin{figure}[tp]
 \begin{center}
    \includegraphics[trim = 2 2 2 2, clip, width=\twowidth]{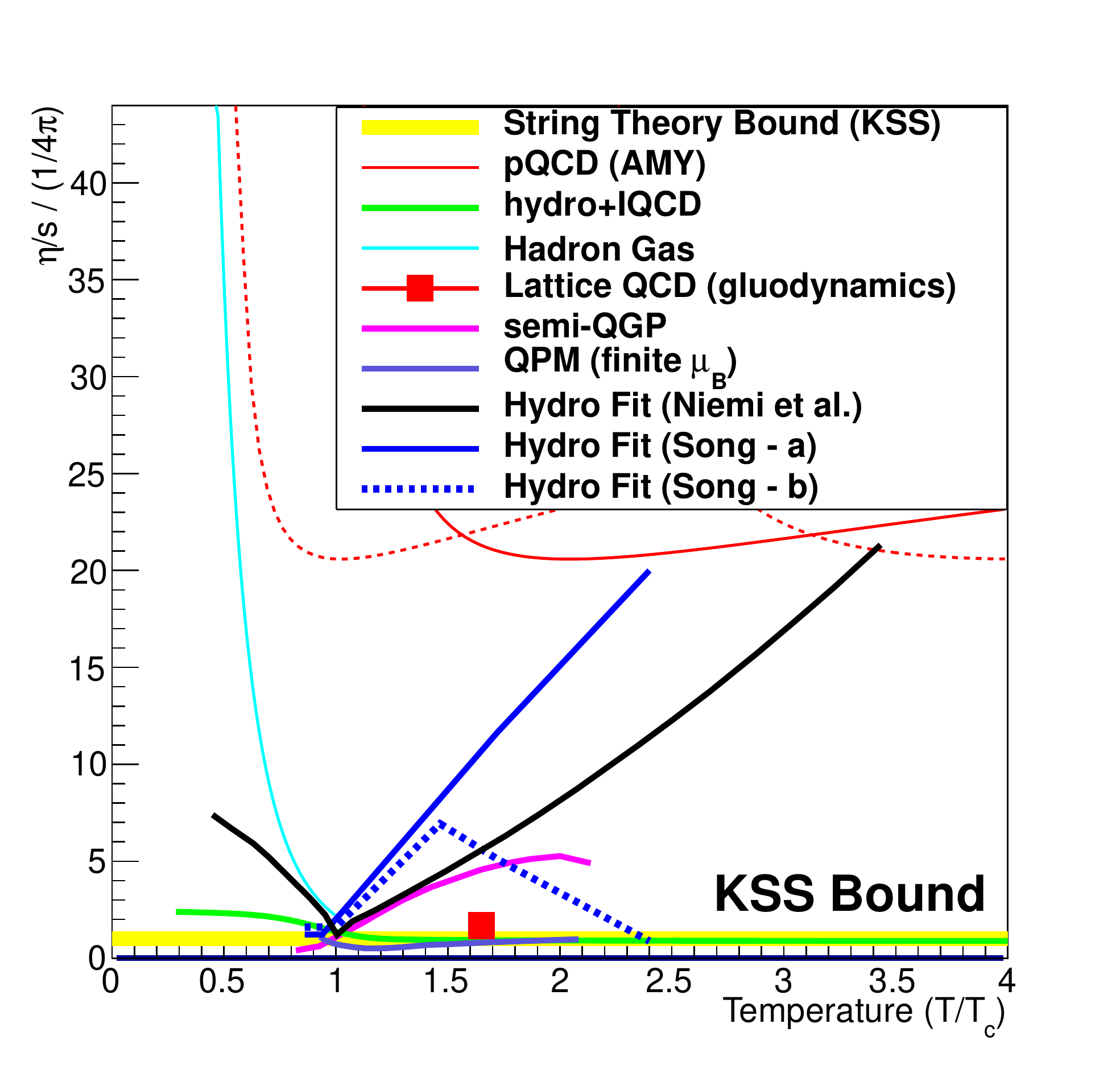}
    \hfill
    \includegraphics[trim = 2 2 2 2, clip, width=\twowidth]{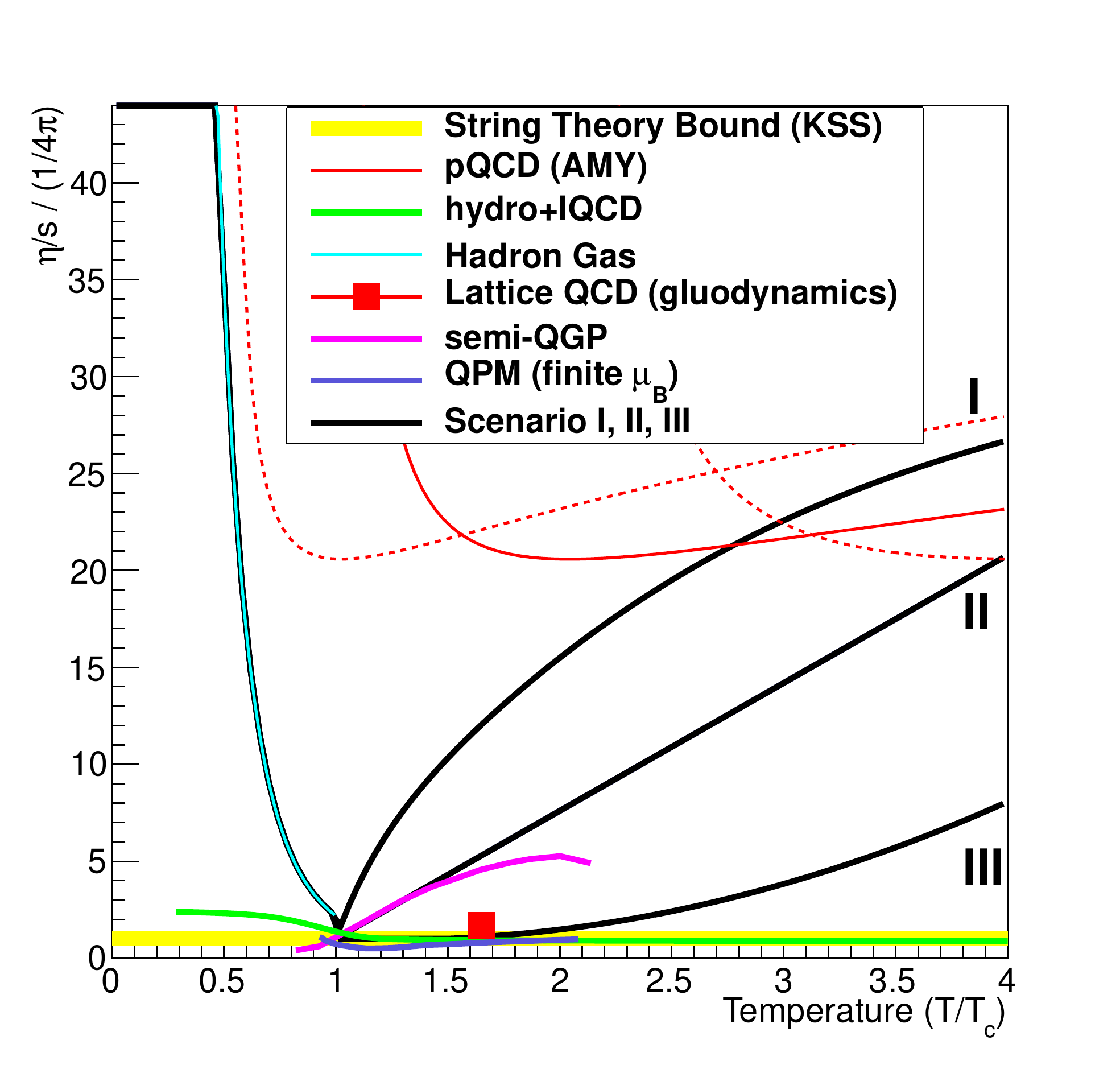}
    \caption{\label{fig:etaovers_2}(Left) Shear Viscosity divided by
      entropy density, $\eta/s$, renormalized by the conjectured KSS
      bound as a function of the reduced temperature, $T/T_{c}$, with 
      various calculations for the quark-gluon plasma case. See text
      for discussion.  (Right) Figure
      with three conjectured scenarios for the quark-gluon plasma
      transitioning from the strongly coupled bound (as a near perfect
      fluid) to the weakly coupled case.}
 \end{center}
\end{figure}

Figure~\ref{fig:etaovers_2} (left panel) shows several
state-of-the-art calculations for $\eta/s$ as a function of
temperature. Hadron gas calculations show a steep increase in $\eta/s$
below $T_c$~\cite{Prakash:1993bt}, and similar results using the UrQMD
model have also been obtained~\cite{Demir:2008tr}.  Above $T_c$ there
is a lattice calculation in the SU(3) pure gauge
theory~\cite{Meyer:2007ic} resulting in a value near the KSS bound at
$T=1.65\,T_c$.  Calculations in the semi-QGP
model~\cite{Hidaka:2009ma}, in which color is not completely ionized,
have a factor of five increase in $\eta/s$ in the region of
1--2\,$T_c$.  Also shown are calculations from a quasiparticle model
(QPM) with finite $\mu_B$~\cite{Srivastava:2012sb} indicating little
change in $\eta/s$ up to 2\,$T_c$.  There is also an update on the
lower limit on $\eta/s$ from second order relativistic viscous
hydrodynamics~\cite{Kovtun:2011np}, with values remaining near
$1/4\pi$.  It is safe to say that little is known in a theoretically
reliable way about the nature of this transition or the approach to
weak-coupling.


Hydrodynamic modeling of the bulk medium does provide constraints on
$\eta/s$, and recent work has been done to understand the combined
constraints on $\eta/s$ as a function of temperature utilizing both
RHIC and LHC flow data sets~\cite{Song:2011qa,Nagle:2011uz,Niemi:2011ix}.  
The results from~\cite{Niemi:2011ix} as
constrained by RHIC and LHC data on hadron transverse momentum spectra
and elliptic flow are shown in Figure~\ref{fig:etaovers_2} (left
panel). These reach the pQCD weak coupled value at $20 \times 1/4\pi$
for $T = 3.4 T_{c}$.  Also shown are two scenarios, labeled ``Song-a''
and ``Song-b'', for $\eta/s(T)$ in ~\cite{Song:2011qa} from which the 
authors conclude that ``one cannot unambiguously determine the
functional form of $\eta/s(T)$ and whether the QGP fluid is more
viscous or more perfect at LHC energy.''

Shown in Figure~\ref{fig:etaovers_2} (right panel) are three possible
scenarios for a more or less rapid modification of the medium from the
strong to the weak coupling limit.  Scenario I has the most rapid
change in $\eta/s(T)$ following the ``Song-a'' parametrization and
Scenario III has the least rapid change going through the lattice QCD
pure glue result~\cite{Meyer:2007ic}.  It is imperative to map out
this region in the `condensed matter' physics of QCD and extract the
underlying reason for the change.

The above discussion has focused on $\eta/s$ as the measure of the
coupling strength of the \qgp.  However, both $\eta/s$ and jet probe
parameters such as $\hat{q}$ and $\hat{e}$ are sensitive to the
underlying coupling of the matter, but in distinct ways. Establishing
for example the behavior of $\hat{q}$ around the critical temperature
is therefore essential to a deep understanding of the quark-gluon
plasma.  Hydrodynamic modeling may eventually constrain $\eta/s(T)$
very precisely, though it will not provide an answer to the question
of the microscopic origin of the strong coupling (something naturally
available with jet probes).

\begin{figure}[tp]
 \begin{center}
    \includegraphics[trim = 2 2 2 2, clip, width=\twowidth]{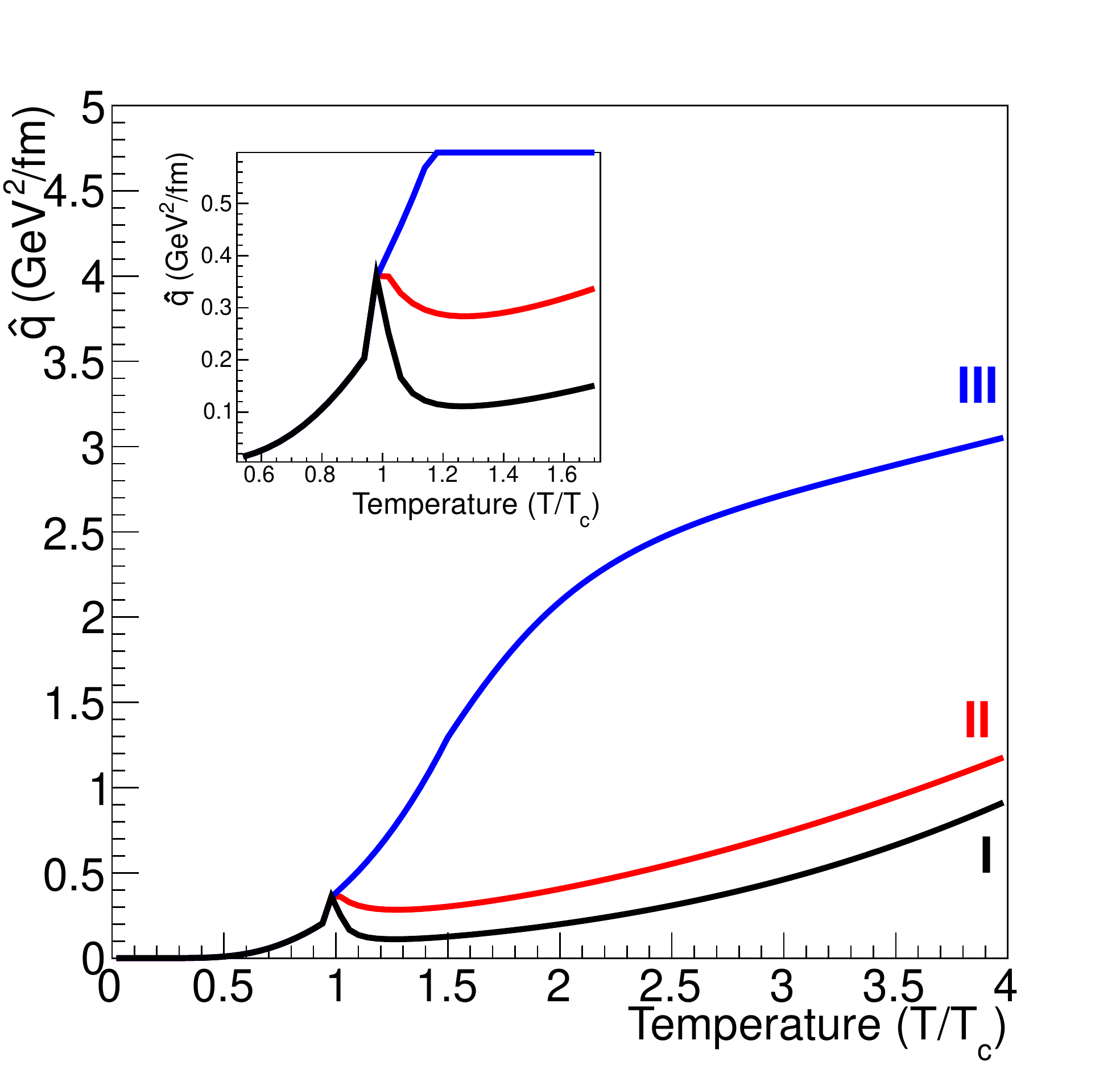}
    \hfill
    \includegraphics[trim = 2 2 2 2, clip, width=\twowidth]{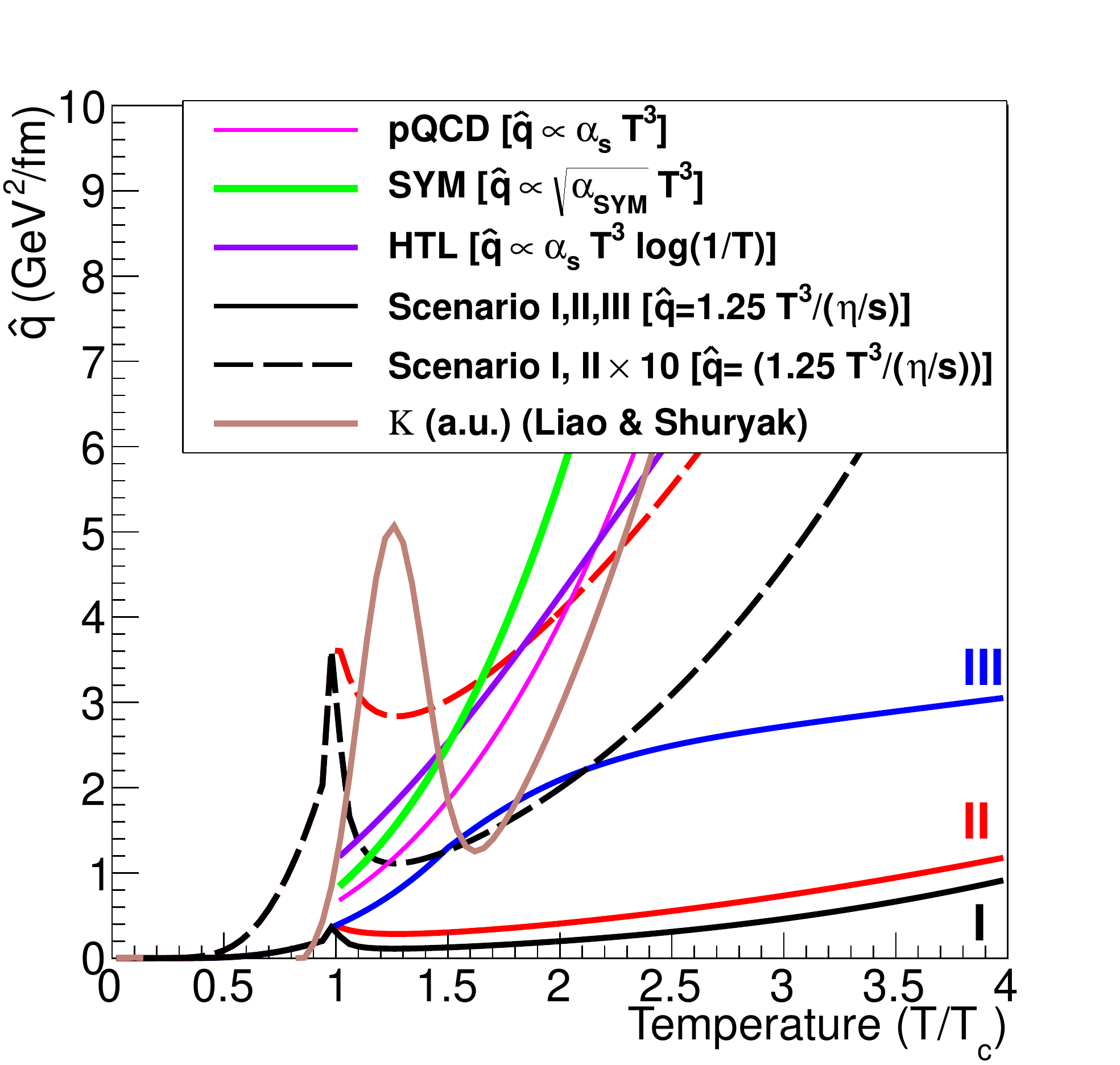}
    \caption{(Left) $\hat{q}$ as a function of
      $T/T_{c}$ in the three scenarios as related with the
      weak-coupling calculation. (Right) Different calculations for
      the scaling of $\hat{q}$ under weak and strong coupling
      assumptions.}
    \label{fig:qhatmap}
 \end{center}
\end{figure}

The authors of Ref~\cite{Majumder:2007zh} propose a test of the strong
coupling hypothesis by measuring both $\eta/s$ and $\hat{q}$.  They
derive a relation between the two quantities expected to hold in the
weak coupling limit.
\begin{equation}
\hat{q} \stackrel{?}{=} \frac{1.25 T^{3}}{\eta/s}
\label{eq:qhat2etas}
\end{equation}
The authors conclude that ``an unambiguous determination of both sides
of [the equation] from experimental data would thus permit a model
independent, quantitative assessment of the strongly coupled nature of
the quark-gluon plasma produced in heavy ion collisions.''  For the
three scenarios of $\eta/s(T)$ shown in Figure~\ref{fig:etaovers_2}
(right panel), we calculate $\hat{q}$ as a function of temperature
assuming the equivalence case in Eqn.~\ref{eq:qhat2etas} and the
result is shown in Figure~\ref{fig:qhatmap} (left panel).  The inset in
Figure~\ref{fig:qhatmap} shows a magnified view of the region around
$T_c$ and a significant local maximum in $\hat{q}$ is observed in
scenarios I and II.



Figure~\ref{fig:qhatmap} (right panel) shows that for the equivalence
relation of Eqn.~\ref{eq:qhat2etas}, all three scenarios have a result
that differs significantly from the simple perturbative expectation of
$\alpha_{s} T^{3}$~\cite{Armesto:2011ht}.  Also shown in
Figure~\ref{fig:qhatmap} are the predicted temperature dependence of
$\hat{q}$ in the strongly coupled AdS/CFT (supersymmetric Yang-Mills)
case~\cite{Liu:2006ug} and the Hard Thermal Loop (HTL)
case~\cite{arnold:2000dr}.

Since the expected scaling of $\hat{q}$ with temperature is such a strong
function of temperature, jet quenching measurements should be
dominated by the earliest times and highest temperatures.  In order to
get sensitivity to the temperatures around 1--2 $T_c$, measurements at
RHIC are needed as opposed to the LHC where larger initial temperatures are produced.

In a recent paper~\cite{Liao:2008dk}, Liao and Shuryak use 
RHIC measurements of single hadron suppression and azimuthal
anisotropy to infer that ``the jet quenching is a few times stronger
near $T_c$ relative to the quark-gluon plasma at $T > T_c$.''  This
enhancement of $\hat{q}$ is shown in Figure~\ref{fig:qhatmap} (right panel)
and is the result of color magnetic monopole excitations in the plasma
near $T_{c}$.   A more detailed discussion of constraints from current
experimental measurements is given in Section~\ref{currentjetdata}. We
note that enhancements in $\hat{q}$ above the critical temperature may
be a generic feature of many models, as illustrated by the three
conjectured evolutions, and so underscore the need for detailed
measurements of \qgp properties near the transition temperature.

All measurements in heavy ion collisions are the result of emitted
particles integrated over the entire time evolution of the reaction,
covering a range of temperatures.  Similar to the hydrodynamic model constraints, the
theory modeling requires a consistent temperature and scale dependent
model of the \qgp and is only well constrained by precision data through
different temperature evolutions, as measured at RHIC and the LHC.

\begin{figure}[!hbt]
 \begin{center}
   \raisebox{0.05in}{\includegraphics[trim = 2 2 2 2, clip,width=0.57\linewidth]{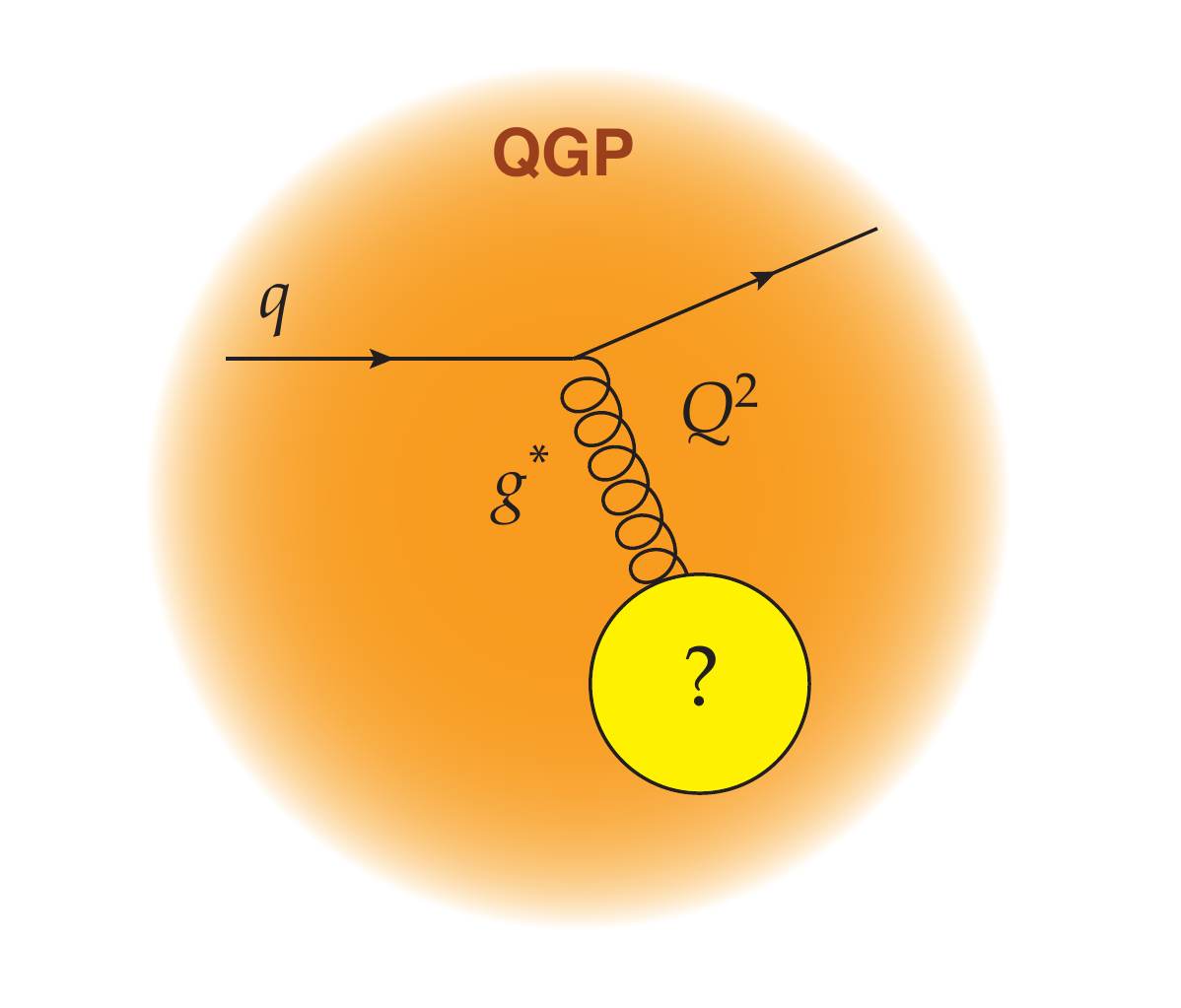}}
    \hfill
    \includegraphics[trim = 2 2 2 50, clip,width=0.4\linewidth]{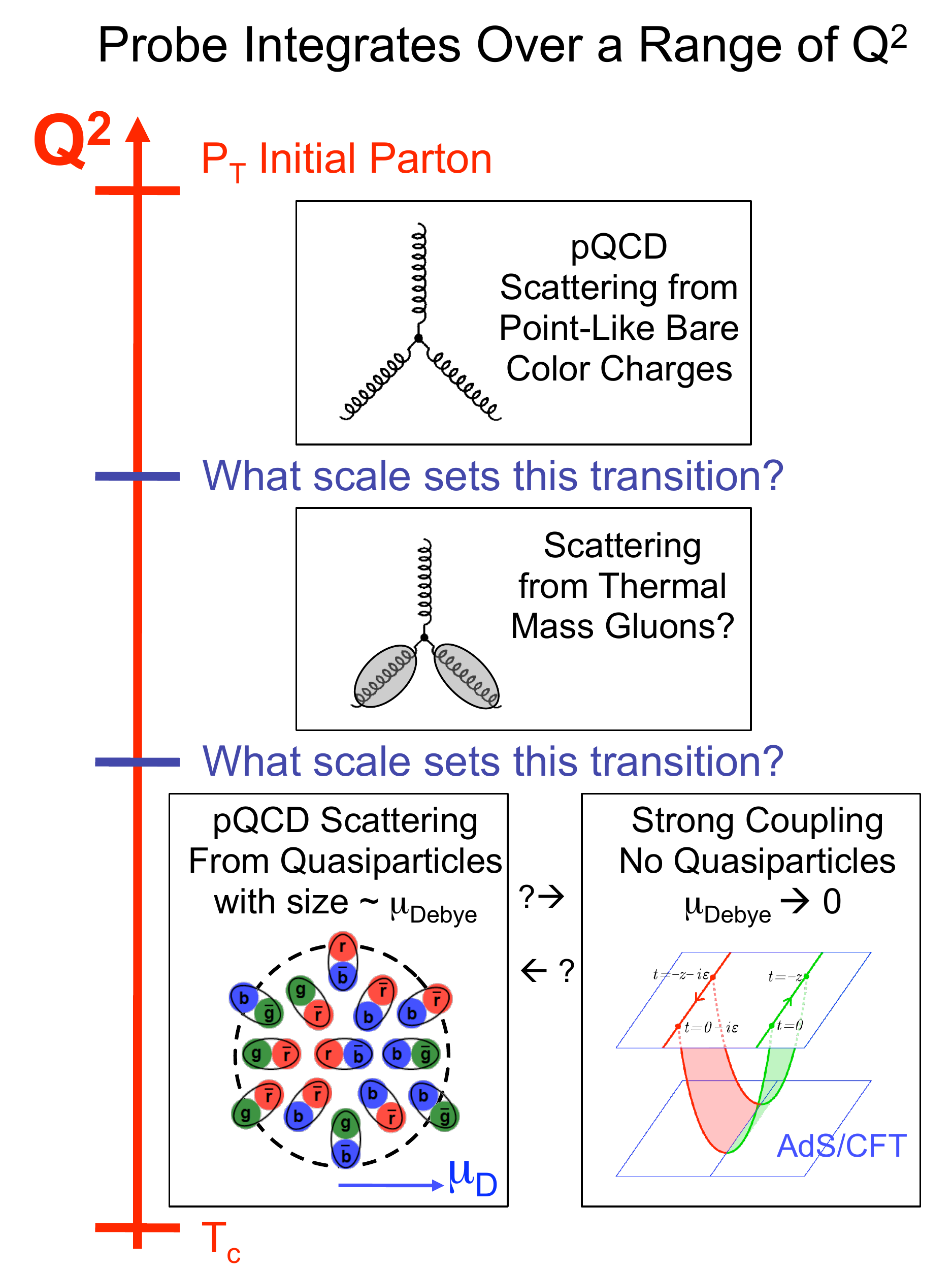}
    \caption{(Left) Diagram of a quark exchanging a virtual gluon with an unknown object in the QGP.  This highlights
the uncertainty for what sets the scale of the interaction and what objects or quasiparticles are recoiling.  (Right) Diagram
as a function of the $Q^{2}$ for the net interaction of the parton with the medium and the range of possibilities
for the recoil objects.\label{fig:probescale}}
 \end{center}
\end{figure}

\section{What are the inner workings of the QGP?}

A second axis along which one can investigate the underlying
structure of the \qgp concerns the question of what length scale of
the medium is being probed by jet quenching processes.  In
electron scattering, the scale is set by the virtuality of the
exchanged photon, $Q^{2}$. By varying this virtuality one can
obtain information over an enormous range of scales: from
pictures of viruses at length scales of $10^{-5}$ meters, to the
partonic make-up of the proton in deep inelastic electron scattering
at length scales of less than $10^{-18}$ meters. For the case of hard
scattered partons in the quark-gluon plasma, the length scale probed
is related not to the virtuality of the initial hard process discussed
above, but rather to the virtuality of the gluon exchanged with the
color charges in the medium, as shown in the left panel of
Figure~\ref{fig:probescale}.  However, it is theoretically unclear
whether the length scale is simply set by the individual exchange
gluon virtuality or instead by the total coherent energy loss through
the medium.

Figure~\ref{fig:probescale} (right panel) shows that if the length
scale probed is very small then one expects scattering directly from
point-like bare color charges, most likely without any influence from
quasiparticles or deconfinement.  As one probes longer length
scales, the scattering may be from thermal mass gluons and eventually
from possible quasiparticles with size of order the Debye screening
length.  Rajagopal states that ``at some length scale, a
quasiparticulate picture of the QGP must be valid, even though on its
natural length scale it is a strongly coupled fluid.  It will be a
challenge to see and understand how the liquid QGP emerges from
short-distance quark and gluon quasiparticles.~\cite{krishna}''

The extension of jet measurements over a wide range of energies and
with different medium temperatures again gives one the largest span
along this axis.  What the parton is scattering from in the medium is
tied directly to the balance between radiative energy loss and
inelastic collisional energy loss in the medium. In the limit that
the scattering centers in the medium are infinitely massive, one only
has radiative energy loss---as was assumed for nearly 10 years to be
the dominant parton energy loss effect.  In the model of Liao and
Shuryak~\cite{Liao:2008dk}, the strong coupling near the quark-gluon
plasma transition is due to the excitation of color magnetic
monopoles, and this should have a significant influence on the
collisional energy loss and equilibration of soft partons into the
medium.

As a parton traverses the medium if it scatters from infinitely massive
scattering centers, then the energy loss can only be through radiative processes.
As the mass of the objects being scattered from lowers, the contributions
of elastic energy loss become more significant.  That is why measurements
of jet observables that help disentangle these different energy loss processes
are needed at both RHIC and LHC where the length scale probed and the possible
coupling strength of the QGP are different.

\section{How does the QGP evolve along with the parton shower?}

\begin{figure}[!hbt]
 \begin{center}
    \includegraphics[trim = 2 2 2 2, clip, width=0.95\linewidth]{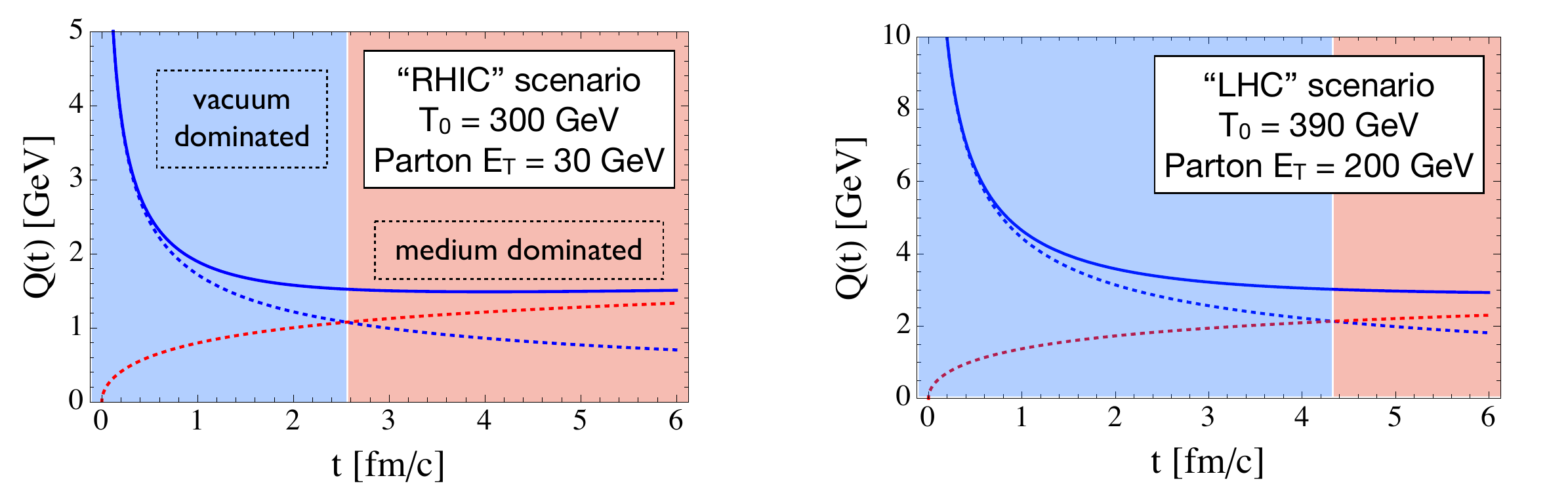}
    \caption{\label{fig:virtualityevolution}Jet virtuality evolution
      at RHIC (left) and LHC (right).  Vacuum contributions to
      virtuality (blue dashed lines) decrease with time and medium
      induced contributions (red dashed lines) increase as the parton
      scatters in the medium.  The total virtuality (blue solid lines)
      is the quadrature sum of the two contributions.  At RHIC the
      medium induced virtuality dominates by 2.5\,fm/c while at the LHC
      the medium term does not dominate until 4.5\,fm/c.  From
      Ref.~\protect\cite{Muller:usersmeeting}{}.}
 \end{center}
\end{figure}

The initial hard scattered parton starts out very far off-shell and in
$e^{+}e^{-}$, \pp or \ppbar~collisions the virtuality evolves in
vacuum through gluon splitting down to the scale of hadronization.  In
heavy ion collisions, the vacuum virtuality evolution is interrupted
at some scale by scattering with the medium partons which increase the
virtuality with respect to the vacuum evolution.
Figure~\ref{fig:virtualityevolution} shows the expected evolution of
virtuality in vacuum, from medium contributions, and combined for a
quark-gluon plasma at $T_0 = 300$\,MeV with the traversal of a 30\,GeV
parton (left) and at $T_{0}=390$\,MeV with the traversal of a 200\,GeV
parton (right)~\cite{Muller:usersmeeting,Muller:2010pm}.  If this
picture is borne out, it ``means that the very energetic parton [in
the right picture] hardly notices the medium for the first 3--4 fm of
its path length~\cite{Muller:2010pm}.'' Spanning the largest possible
range of virtuality (initial hard process $Q^{2}$) is very important,
but complementary measurements at both RHIC and LHC of produced jets
at the same virtuality (around 50\,GeV) will test the interplay between
the vacuum shower and medium scattering contributions.

\section{Current jet probe measurements}
 \label{currentjetdata}

\begin{figure}[!hbt]
 \begin{center}
   \includegraphics[trim = 117 578 275 57, clip, width=\onewidth]{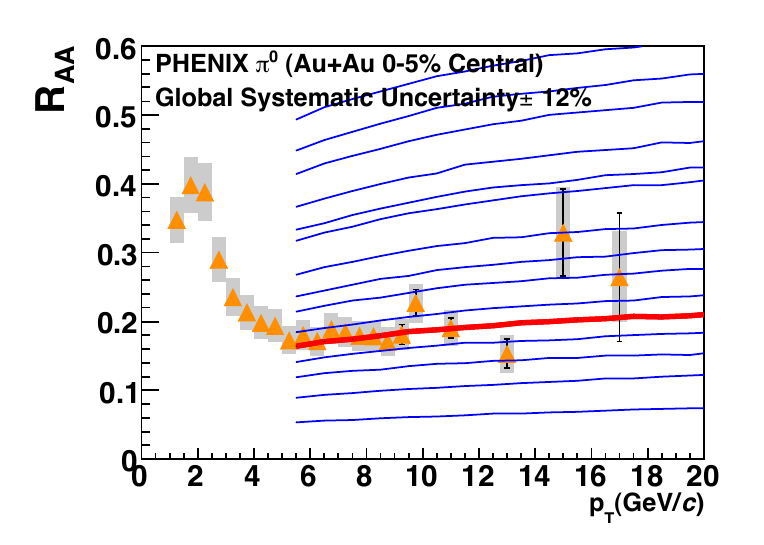}
   \caption{\label{fig:phenixconstraint}$\pi^0$ $R_{AA}$ for central \AuAu~ collisions compared
     to PQM Model calculations~\cite{Dainese:2004te,Loizides:2006cs} for various values 
     of $\langle\hat{q}\rangle$~\cite{Adare:2008cg}.  The red line corresponds to 
     $\langle\hat{q}\rangle=13.2$\,GeV$^2$/fm and is the best fit to the data.  }
 \end{center}
\end{figure}

\begin{figure}[!hbt]
 \begin{center}
    \includegraphics[trim = 2 2 2 2, clip, width=0.6\linewidth]{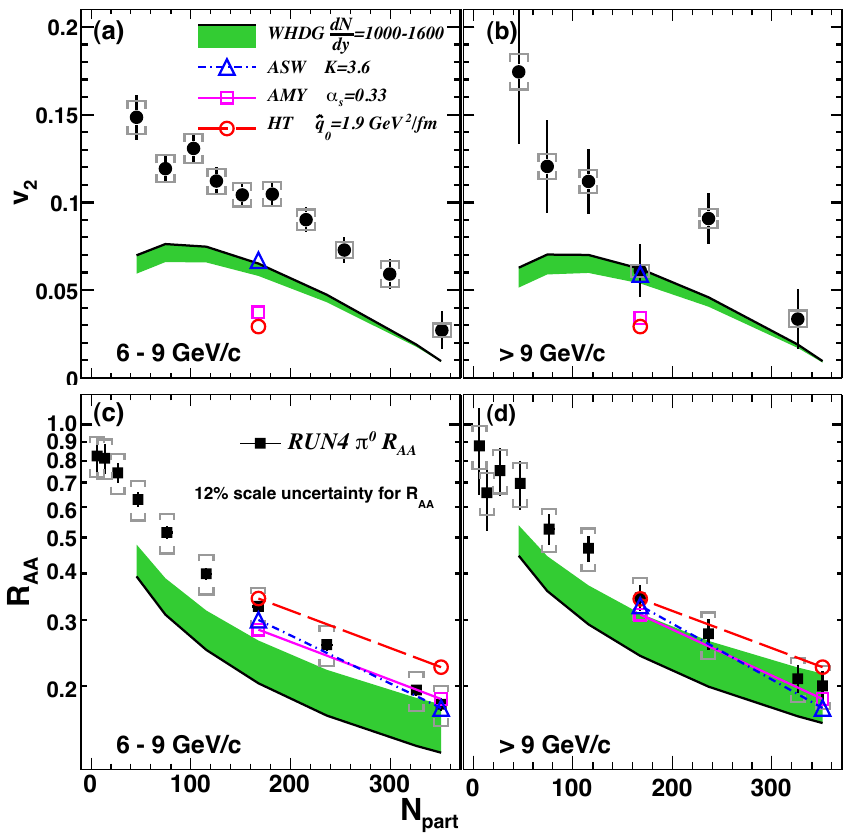}
    \caption{\label{fig:phenixpi0v2} $\pi^0$ $v_2$ (top panels) and $R_{AA}$ (bottom panels)
      for $6<p_T<9$\,GeV/c (left panels) and $p_T>9$\,GeV/c (right panels).  Calculations from
      four weakly coupled energy loss models are shown as well~\cite{Bass:2008rv,Wicks:2005gt}.
      From Ref.~\protect\cite{Adare:2010sp}{}.}
  \end{center}
\end{figure}

Jet quenching (i.e., the significant loss of energy for partons
traversing the QGP) was discovered via measurements at RHIC of the
suppression of single hadron yields compared to expectations from \pp
collisions~\cite{Adcox:2001jp,Adler:2002xw}.
Figure~\ref{fig:phenixconstraint}~\cite{Adare:2008cg} shows a
comparison between the PHENIX $\pi^{0}$ $R_{AA}$ data and the Parton
Quenching Model (PQM)~\cite{Dainese:2004te,Loizides:2006cs} with
various values of $\hat{q}$.  This calculation assumes radiative
energy loss only in a weakly coupled picture and with no recoil
collisional energy loss with partons or quasiparticles in the medium.
The coupling parameter value $\hat{q}=13.2$\,GeV$^{2}$/fm implies a
very strong coupling and violates the weak coupled assumption of the
model formalism.

However, as detailed in Ref.~\cite{Adare:2008cg,Bass:2008rv}, other
formalisms also assuming weak coupling are able to achieve an equally
good description of the data and with substantially smaller values of
$\hat{q}$.  Thus, the single high $p_T$ hadron suppression constrains
the $\hat{q}$ value within a model, but is not able to discriminate
between different energy loss mechanisms and formalisms for the
calculation.  Two-hadron correlations measure the correlated
fragmentation between hadrons from within the shower of one parton and
also between the hadrons from opposing scattered partons.  These
measurements, often quantified in terms of a nuclear modification
$I_{AA}$~\cite{Adare:2010ry,Adare:2010mq,Adams:2005ph}, are a
challenge for models to describe simultaneously~\cite{Nagle:2009wr}.

One observable that has been particularly challenging for energy loss
models to reproduce is the azimuthal anisotropy of $\pi^0$ production
with respect to the reaction plane.  A weak dependence on the path
length in the medium is expected from radiative energy loss.  This
translates into a small $v_2$ for high $p_T$ particles (i.e., only a
modest difference in parton energy loss when going through a short
versus long path through the QGP).  Results of $\pi^0$ $v_2$ are shown
in Figure~\ref{fig:phenixpi0v2}~\cite{Adare:2010sp}.  Weakly coupled
radiative energy loss models are compared to the $R_{AA}$ (bottom
panels) and $v_2$ (top panels) data.  These models reproduce the
$R_{AA}$, but they fall far short of the $v_2$ data in both $p_T$
ranges measured (6--9\,GeV/c and $>9$\,GeV/c).  
This large path length dependence is naturally described by 
strongly coupled energy loss models~\cite{Marquet:2009eq,Adare:2010sp}.  Note that one can match
the $v_2$ by using a stronger coupling, larger $\hat{q}$, but at the
expense of over-predicting the average level of suppression.

The total energy loss of the leading parton provides information on
one part of the parton-medium interaction.  Key information on the
nature of the particles in the medium being scattered from is
contained in how the soft (lower momentum) part of the parton shower
approaches equilibrium in the \qgp.  This information is only
accessible through full jet reconstruction, jet-hadron correlation,
and $\gamma$-jet correlation observables.

There are preliminary results on fully reconstructed jets from both
STAR~\cite{Caines:2011ew,Putschke:2011zz,Putschke:2008wn,Jacobs:2010wq}
and PHENIX experiments~\cite{Lai:2011zzb,Lai:2009zq}.  However, these
have not proceeded to publication in part due to limitations in the
measurement capabilities.  In this proposal we demonstrate that a
comprehensive jet detector (sPHENIX) with large, uniform acceptance
and high rate capability, combined with the now completed RHIC
luminosity upgrade can perform these measurements to access this key
physics.

\begin{figure}[t]
 \begin{center}
   \includegraphics[trim = 50 410 300 150,clip,width=\onewidth]{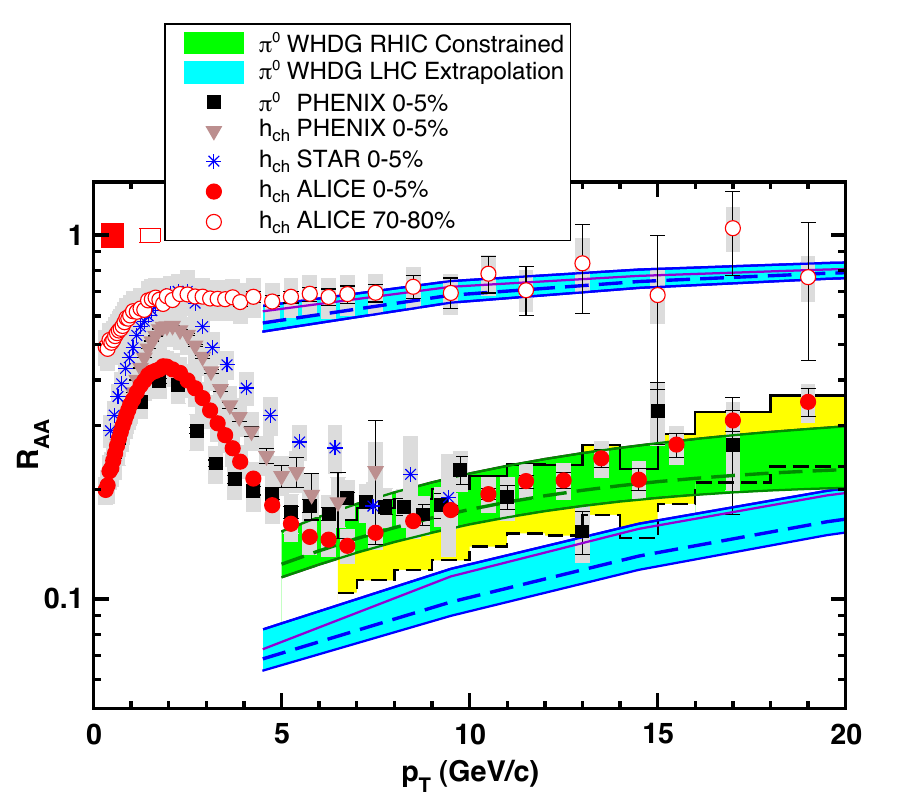}
    \caption{$R_{AA}$ measurements from RHIC and the LHC compared to
      WHDG calculations.  The parameters are constrained by the RHIC
      data and extrapolated to 2.76\,TeV.  The prediction for the LHC
      is shown (blue band) and lies below the ALICE data for central
      collisions (red circles).  From Ref.~\protect\cite{Horowitz:2011gd}{}.}
    \label{fig:WHDG} 
 \end{center}
\end{figure}

New data from the LHC experiments has significantly expanded the
information on jet probes of the QGP.  Figure~\ref{fig:WHDG} shows
the aforementioned $\pi^{0}$ nuclear modification factor from the
PHENIX experiment together with an energy loss calculation wherein the
value of $\hat{q}$ is constrained to match the
data~\cite{Horowitz:2011gd}.  Also shown are recent results from the
ALICE experiment at the LHC~\cite{Aamodt:2010jd} compared to the same
energy loss calculation scaled by the expected increase in the color
charge density created in the higher energy LHC collisions, shown as
the light blue band.  The over-prediction based on the assumption of an
unchanging probe-medium coupling strength led to title of
Ref.~\cite{Horowitz:2011gd}: ``The surprisingly transparent sQGP at the LHC.''
They state that ``one possibility is the sQGP produced at the LHC is
in fact more transparent than predicted.''  Similar conclusions have been reached by other
authors~\cite{Chen:2011vt,Zakharov:2011ws,Buzzatti:2011vt}.  

The measurements of fully reconstructed jets and the particles correlated
with the jet (both inside the jet and outside) are crucial to testing
this hypothesis.  Not only does the strong coupling influence the
induced radiation from the hard parton (gluon bremsstrahlung)
and its inelastic collisions with the medium, but it also influences the way
soft partons are transported by the medium outside of the jet cone
as they fall into equilibrium with the medium.  Thus, the jet
observables combined with correlations get directly at the coupling of
the hard parton to the medium and the parton-parton coupling for the
medium partons themselves.

\begin{figure}[t]
 \begin{center}
    \includegraphics[trim = 2 2 2 2, clip, width=0.9\linewidth]{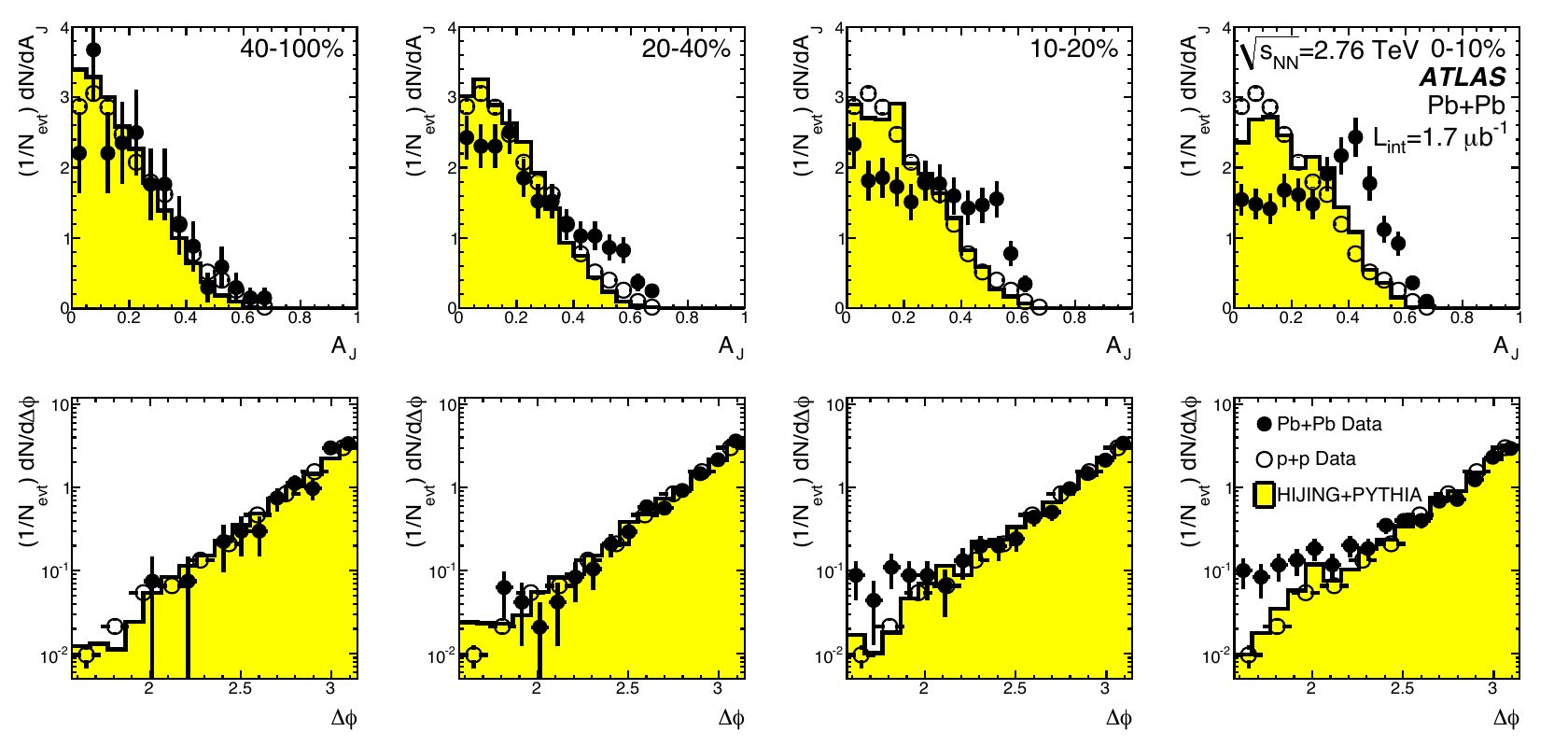}
    \caption{\label{fig:atlasdijet}$A_J$ (top row) and dijet $\Delta\phi$ distribution
from ATLAS~\protect\cite{Aad:2010bu}{}.  Jets are reconstructed with the anti-$k_T$ algorithm with
$R=0.4$.  The leading jet has $E_T>100$\,GeV and the associated jet has $E_T>25$\,GeV.  \PbPb~
data (solid points), \pp~data at 7\,TeV (open points) and \pythia embedded in \hijing events
and run through the ATLAS Monte Carlo (yellow histograms) are shown.  From Ref.~\protect\cite{Aad:2010bu}{}.}
 \end{center}
\end{figure}

\begin{figure}[t]
 \begin{center}
    \includegraphics[trim = 2 2 2 2, clip, width=\onewidth]{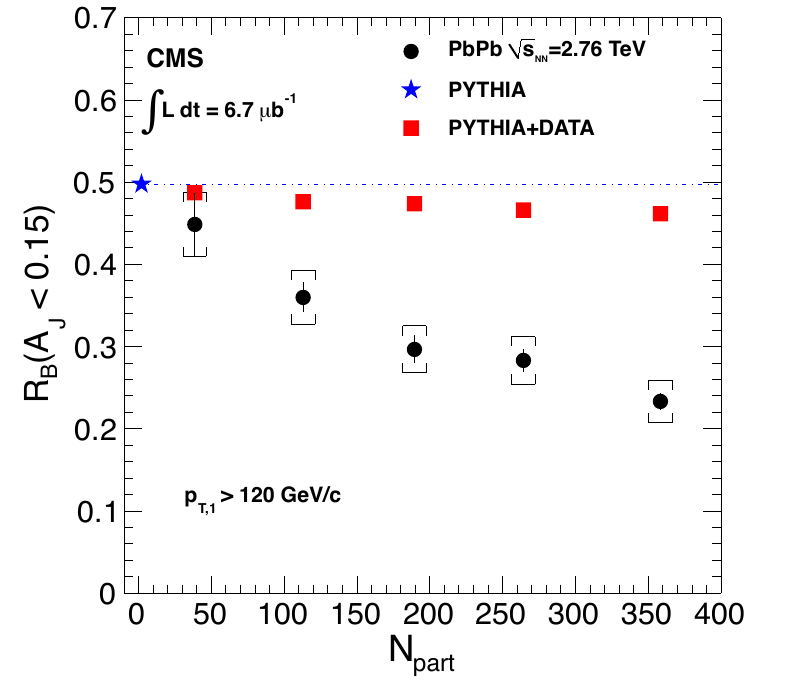}
    \caption{\label{fig:cmsdijetsupp} Fraction of dijets which
      have $A_J<0.15$ in \PbPb~collisions as a function of centrality.
      Jets are reconstructed with an iterative cone algorithm with a
      radius of 0.5.  The leading jet is required to have an
      $E_T>120$\,GeV and the associated jet has $E_T>50$\,GeV.  Results
      are shown for \PbPb~data (circles), \pythia (star) and \pythia
      jets embedded into real data (squares).  From
      Ref.~\protect\cite{Chatrchyan:2011sx}{}.}
 \end{center}
\end{figure}

These jet observables are now becoming available at the LHC.
The first results based on reconstructed jets in heavy ion collisions were the centrality
dependent dijet asymmetries measured by ATLAS~\cite{Aad:2010bu}. These
results, shown in Figure~\ref{fig:atlasdijet}, indicate a substantial broadening of
dijet asymmetry $A_{J} = (E_{1}-E_{2})/(E_{1}+E_{2})$
distribution for increasingly central \PbPb~collisions and the lack of modification 
to the dijet azimuthal correlations.
The broadening of the $A_J$ distribution points to substantial energy loss for
jets and the unmodified azimuthal distribution shows that the opposing jet $\Delta\phi$
distribution is not broadened as it traverses the matter.  
Figure~\ref{fig:cmsdijetsupp} shows CMS results~\cite{Chatrchyan:2011sx} quantifying the fraction
of dijets which are balanced (with $A_J<0.15$) decreases with increasing centrality.


Direct photon-jet measurements are a powerful tool to study jet
quenching. Unlike dijet measurements the photon passes through the
matter without losing energy, providing a much cleaner handle on the
expected jet $p_T$~\cite{Wang:1996yh}.  CMS has first results for photons
with $p_T>60$\,GeV/c correlated with jets with $p_T>30$\,GeV/c~\cite{Chatrchyan:2012}.  
Though with modest statistical precision, the measurements indicate energy transported outside
the $R=0.3$ jet cone through medium interactions.


These and other reconstructed jet measurements have been complementary
to one and two particle measurements at the LHC.  Reconstructed jets have
significantly extended the kinematic range for jet quenching studies
at the LHC, and quenching effects are observed up to the highest
reconstructed jet energies ($>300$\,GeV)~\cite{Chatrchyan:2012ni}.
They also provide constraints on the jet modification that are not
possible with particle based measurements.  For example, measurements from ATLAS
constrain jet fragmentation modification from vacuum fragmentation to
be small~\cite{Steinberg:2011qq} and  CMS results on jet-hadron
correlations have shown that the lost energy is recovered in low $p_T$
particles far from the jet cone~\cite{Chatrchyan:2011sx}.  
At the LHC the lost energy is transported to very large angles
and the remaining jet fragments as it would in the vacuum.


Detector upgrades to PHENIX and STAR at RHIC with micro-vertex detectors will
allow the separate study of $c$ and $b$ quark probes of the medium, as tagged
via displaced vertex single electrons and reconstructed $D$ and $\Lambda_{c}$ hadrons.   Similar 
measurements at the LHC provide tagging of heavy flavor probes as well.  These
measurements will also provide insight on the different energy loss mechanisms,
in particular because initial measurements of non-photonic electrons from RHIC
challenge the radiative energy loss models.

It is clear that in addition to extending the RHIC observables to include fully
reconstructed jets and $\gamma$-jet correlations, theoretical development work is
required for converging to a coherent 'standard model' of the medium coupling strength
and the nature of the probe-medium interaction.  In the next section, we detail positive
steps in this direction.

\section{Theoretical calculations of jets at RHIC}
\label{sec:jetcalculations}

Motivated in part by the new information provided by early LHC jet results
and the comparison of RHIC and LHC single and di-hadron results, 
the theoretical community is actively working to understand the detailed
probe-medium interactions.  The challenge is to understand not only the
lost energy of the leading parton, but how the parton shower evolves in medium
and how much of the lost energy is re-distributed in the \qgp.  Theoretical
calculations attempting to describe the wealth of new data from RHIC and the LHC
currently have not reconciled some of the basic features, with some models
including large energy transfer to the medium as heat (for example~\cite{CasalderreySolana:2011rq}) and
others with mostly radiative energy loss (for example~\cite{Renk:2012cb,Renk:2011wb}).  None of
the current calculations available has been confronted with the full set of jet probe
observables from RHIC and the LHC.
Measurements of jets at RHIC energies and with jets over a different kinematic range allow
for specific tests of these varying pictures.  In this section, we give a brief review
of a subset of calculations for jet observables at RHIC enabled by the sPHENIX upgrade and
highlight the sensitivity of these observables to the underlying physics.

Much of this work has been carried out under the auspices of the Department of Energy
Topical Collaboration on Jet and Electromagnetic Tomography of Extreme Phases of Matter in Heavy-ion Collisions
~\cite{jetcollaboration}.  A workshop was held by the JET
Collaboration at Duke University in March 2012 dedicated to the topic 
of jet measurements at RHIC which was attended by theorists as well 
as experimentalists from both RHIC and the LHC.  There was active participation by a 
number of theory groups and there has been significant continued effort, including follow up
video conferences connecting theorists and experimentalists.

In order to overcome specific theoretical hurdles regarding analytic parton energy loss
calculations and to couple these calculations with realistic models of the QGP space-time evolution, 
Monte Carlo approaches have been developed (as examples~\cite{Zapp:2009pu,Renk:2010zx,Young:2011va,ColemanSmith:2011wd,Lokhtin:2011qq,Armesto:2009zc}).  
Here we describe RHIC energy jet probe results from four specific theory groups utilizing different
techniques for calculating the jet-medium interactions.
These efforts indicate a strong theoretical interest and the potential constraining power of a comprehensive
jet physics program at RHIC.  



\begin{figure}[t]
 \begin{center}
    \includegraphics[trim = 2 2 2 2, clip, width=\twowidth]{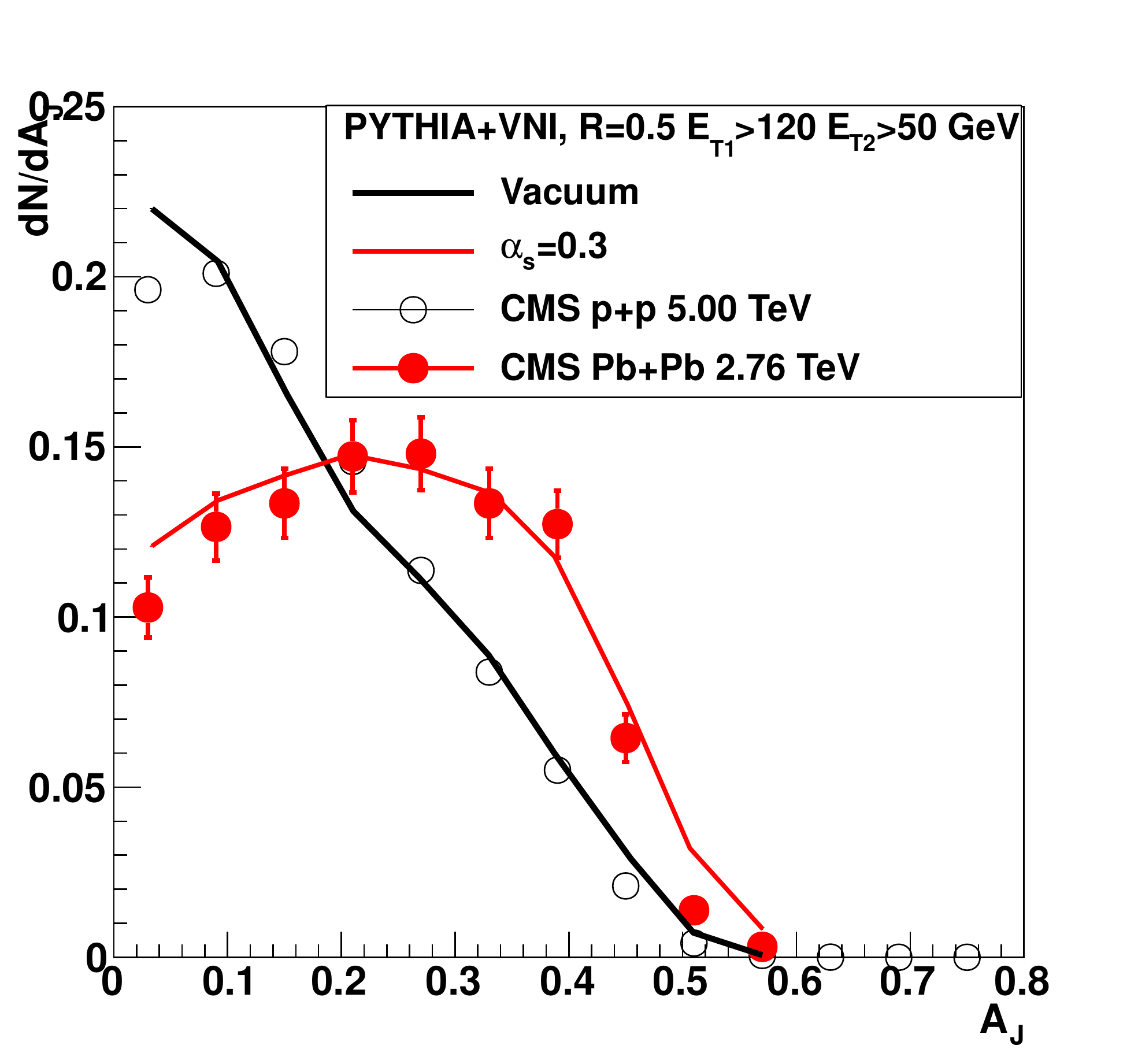}
    \hfill
    \includegraphics[trim = 2 2 2 2, clip,width=\twowidth]{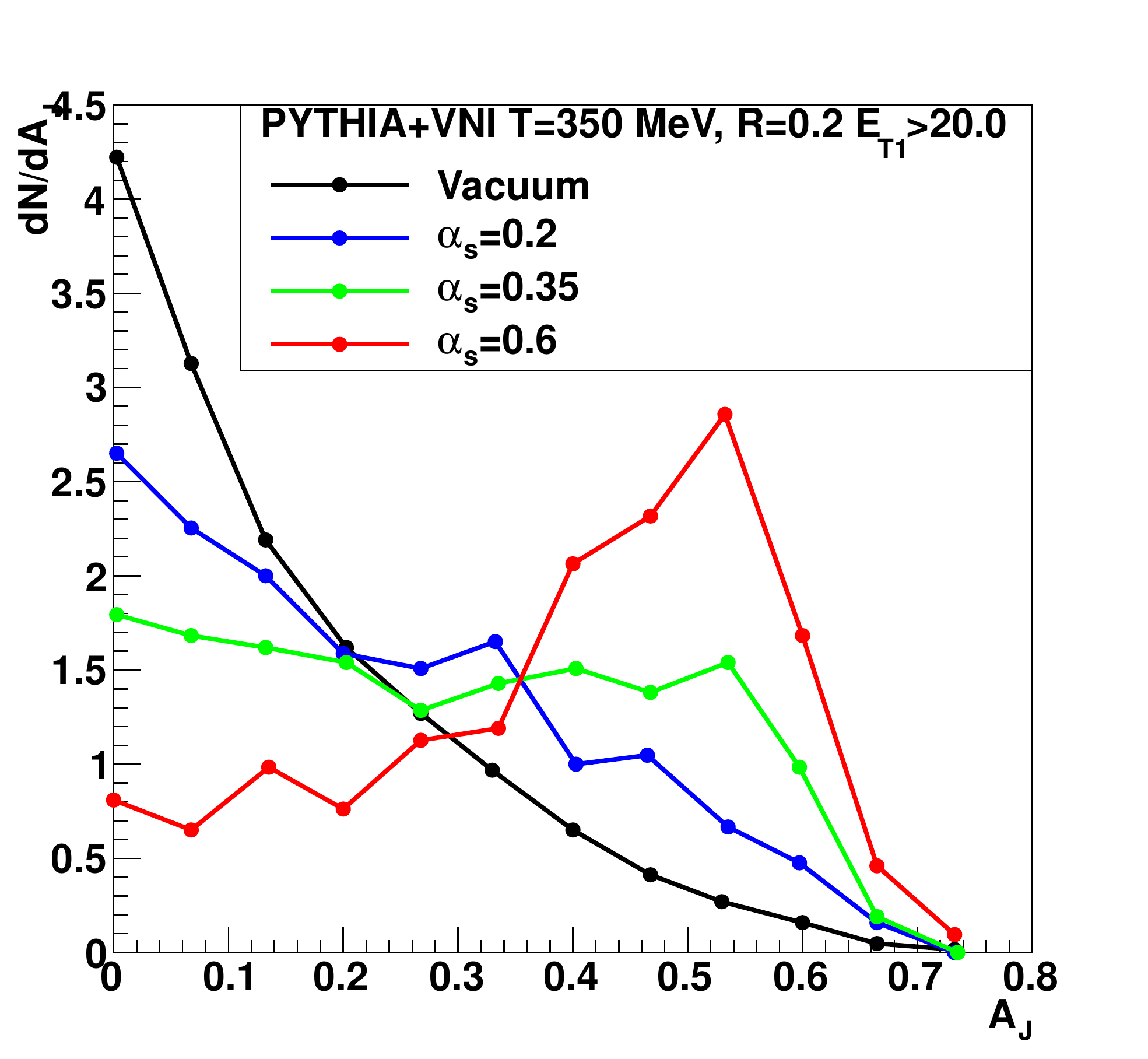}
        \caption{\label{fig:colemansmith1}(Left) Calculation in VNI parton cascade of
dijet $A_J$ with $T=0.35$\,GeV and $\alpha_s=0.3$ compared to the CMS data~\protect\cite{ColemanSmith:2011rw}{}.
(Right) Calculation for RHIC jet energies, $E_{T,1}>20$\,GeV, for a circular geometry of radius 5\,fm
of $A_J$ for different values of $\alpha_s$ increasing to $\alpha_s=0.6$ 
(red line)~\protect\cite{ColemanSmith:2012vr}{}.}
 \end{center}
\end{figure}


The first results are from Coleman-Smith and collaborators~\cite{ColemanSmith:2011rw,ColemanSmith:2011wd}
 where they extract jet parton showers from \pythia (turning off hadronization) and then
embed the partons into a deconfined \qgp, modeled with the VNI parton
cascade~\cite{Geiger:1991nj}.  For the
calculations shown here, the background medium consists of a cylinder
of deconfined quarks and gluons at a uniform temperature.  One
excellent feature of the calculation is that it provides the ability
to track each individual parton and thus not only look at the full
time evolution of scattered partons from the shower, but also medium
partons that are kicked up and can contribute particles to the
reconstructed jets.

Calculation results for the dijet asymmetry $A_{J} = (E_{1}-E_{2})/(E_{1}+E_{2})$
in a QGP with a temperature appropriate for LHC collisions and fixed $\alpha_{s}=0.3$
are shown in Figure~\ref{fig:colemansmith1} (left panel)~\cite{ColemanSmith:2011rw}.   The jets in the calculation are
reconstructed with the anti-$k_T$ algorithm with radius parameter $R = 0.5$ and then smeared by a 
simulated jet resolution of 100\%/$\sqrt{E}$, and with requirements
of $E_{T1}>120$\,GeV and $E_{T2}>50$\,GeV on the leading and sub-leading jet, respectively.  
The calculated $A_J$ distributions reproduce the CMS experimental data~\cite{Chatrchyan:2011sx}.

In Figure~\ref{fig:colemansmith1} (right panel) the calculation is repeated with a medium temperature
appropriate for RHIC collisions and with RHIC observable jet energies, $E_{T1} >
20$\,GeV and $R = 0.2$.  The calculation is carried out for different coupling strengths $\alpha_{s}$ between
partons in the medium themselves and the parton probe and medium partons.  The variation in the value of $\alpha_{s}$ should
be viewed as changing the effective coupling in the many-body environment of the QGP.  It is interesting to note
that in the parton cascade BAMPS, the authors find a coupling of $\alpha_{s} \approx 0.6$ is required to describe the
bulk medium flow~\cite{Wesp:2011yy}.  These results indicate sizable modification to the dijet asymmetry and thus 
excellent sensitivity to the effective coupling to the medium at RHIC energies.  


\begin{figure}[!hbt]
 \begin{center}
    \includegraphics[trim = 2 7 2 2, clip, width=\twowidth]{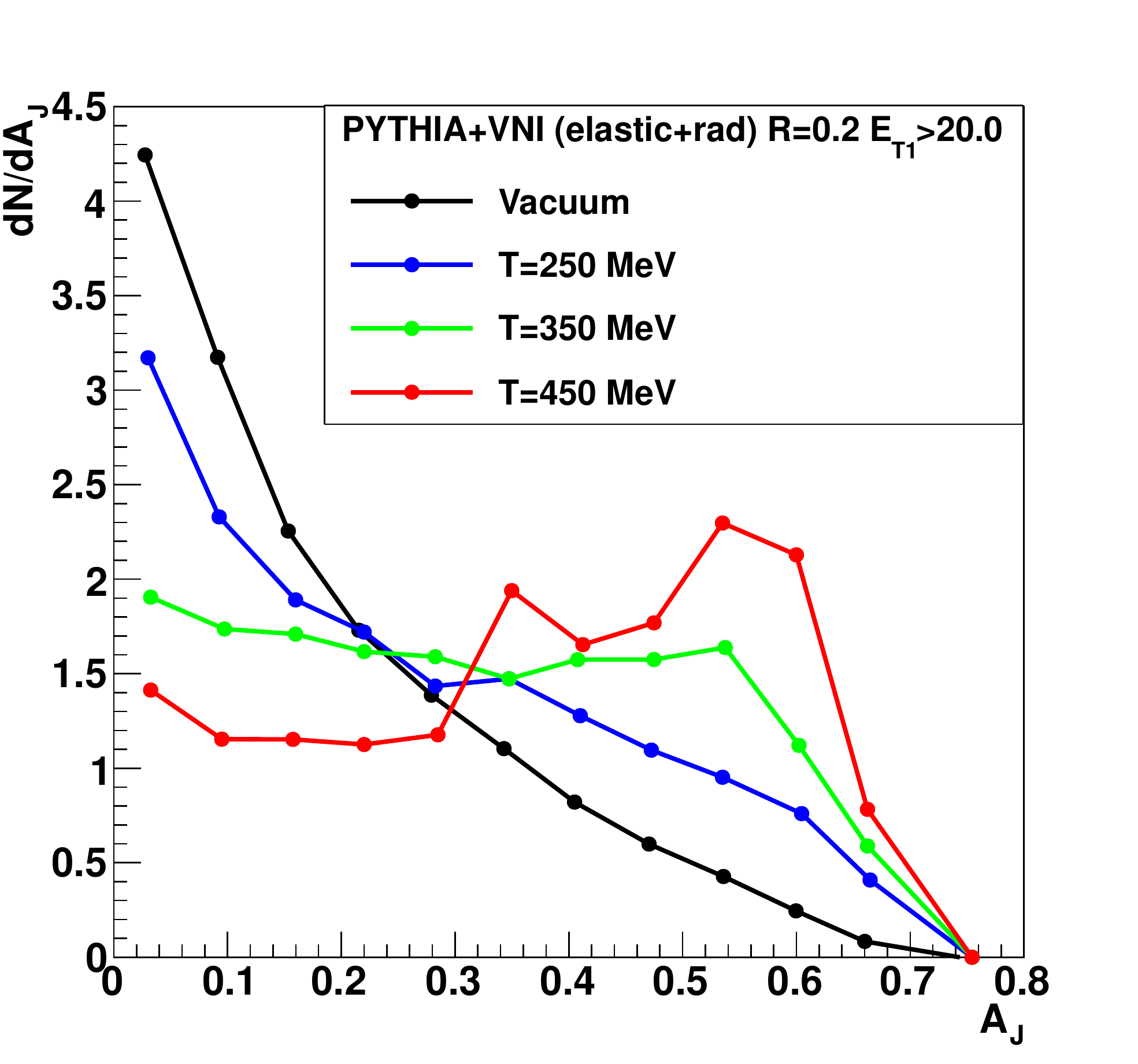}
    \hfill
    \includegraphics[trim = 2 2 2 2, clip, width=\twowidth]{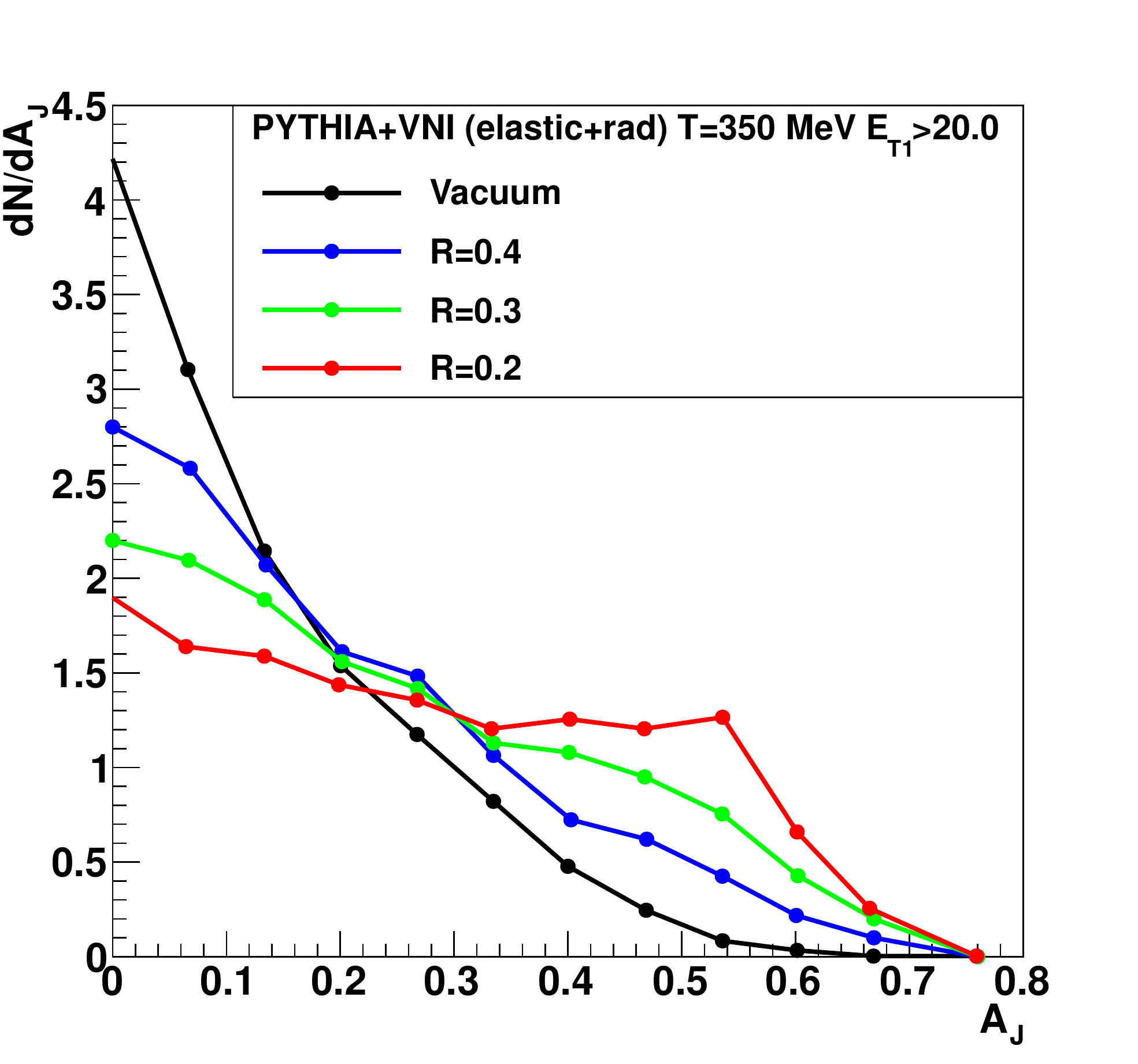}

    \caption{\label{fig:csprobescale}Calculations from
      Coleman-Smith~\protect\cite{ColemanSmith:2012vr}{} for dijets embedded into the VNI parton cascade.  
      The dijet asymmetry $A_J$ for leading jets with $E_T>20$\,GeV is shown as the medium
      temperature is varied (left panel) and as the jet cone radius is varied with fixed temperature $T=350$\,MeV (right panel).}
 \end{center}
\end{figure}

Figure~\ref{fig:csprobescale} (left panel) shows the temperature dependence of the dijet asymmetry,
now keeping the coupling $\alpha_{s}$ fixed.  One observes a similar sharp drop in the fraction
of energy balanced dijets with increasing temperature to that seen for increasing the effective coupling,
and so combining these observations with constrained hydrodynamic models and direct photon emission
measurements is important.  Given that the initial temperatures of the QGP
formed at RHIC and the LHC should be significantly different, this
plot shows that if RHIC and LHC measure the $A_J$ distribution at the
same jet energy there should still be a sensitivity to the temperature
which will lead to an observable difference.  Thus, having overlap in the measured jet energy
range at RHIC and the LHC is important, and this should be available for jet energies of 40--70\,GeV.
Figure~\ref{fig:csprobescale} (right panel) shows
the jet cone size, $R$, dependence of $A_J$ at a fixed temperature.
The narrowest jet cone $R=0.2$ has the most modified $A_J$ distribution, as partons are
being scattered away by the medium to larger angles.  


\begin{figure}[!hbt]
 \begin{center}
    \includegraphics[trim = 2 2 2 2, clip, width=0.9\linewidth]{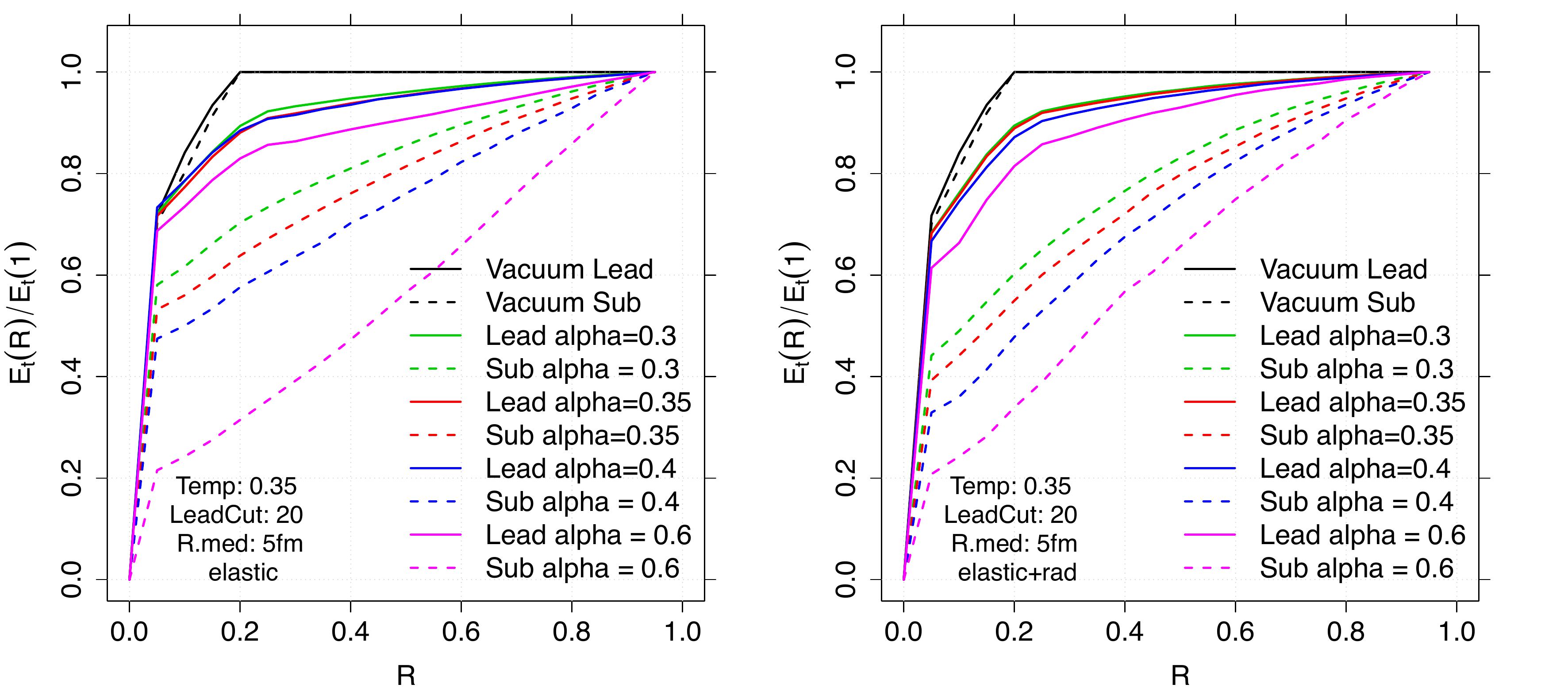}
    \caption{\label{fig:csjetprofile}Calculations from
      Coleman-Smith~\protect\cite{ColemanSmith:2012vr}{} showing the jet
      energy profile as a function of radius for leading (solid lines)
      and sub-leading (dashed lines) jets.  Leading jets have
      $E_T>20$\,GeV and sub-leading jets have $E_T>5$\,GeV.  The
      medium temperature is 350\,MeV.}
 \end{center}
\end{figure}

Complementary to measuring jets with different radius parameters is
to directly examine the profile of energy both within and outside the reconstructed jet. 
Results on the predicted distribution of energy as a function of radius are shown in 
Figure~\ref{fig:csjetprofile}.  The solid lines are for the leading jet and for different
values of the medium coupling $\alpha_s$.  The dashed lines are for the sub-leading jet.  One observes
a particularly strong dependence on the coupling in the radial energy profile of the sub-leading jet, 
as this parton is typically biased to a longer path length through the medium.  The left panel
is including only elastic collisional interactions and the right panel incorporates additional radiative
processes.  At coupling $\alpha_{s} = 0.4$ for example, the fraction of energy in the sub-leading jet
within $R < 0.2$ is 60\% with elastic collisions only and less than 50\% when including radiative
energy loss.  The experimental extraction of these two contributions is a critical step towards
extracting a microscopic picture of the QGP.


\begin{figure}[!hbt]
 \begin{center}
    \includegraphics[trim = 2 2 2 2, clip, width=0.8\linewidth]{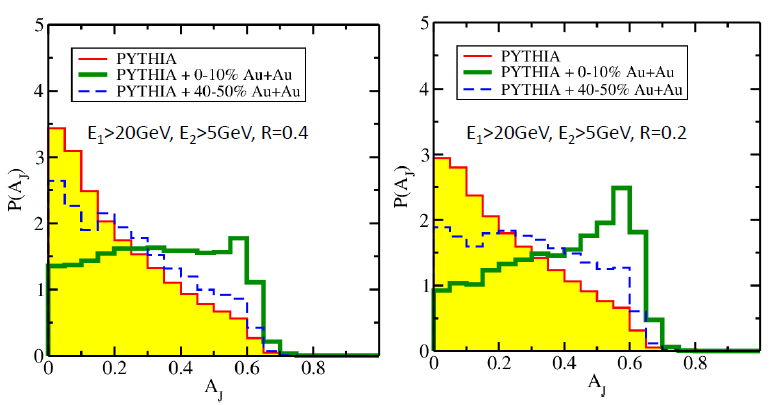}
    \caption{\label{fig:qinaj1} Calculations from Qin  et
        al.~\protect\cite{qin_privatecomm}{} of dijet $A_{J}$ for
      $E_{T,1}>20$\,GeV and $E_{T,2}>5$\,GeV for $R=0.4$ jets (left)
      and $R=0.2$ jets (right).  Central (green) and mid-central
      (blue) distributions are shown along with the initial \pythia
      distributions (red).}
 \end{center}
\end{figure}


The second results are from Qin and collaborators~\cite{Qin:2010mn,qin_privatecomm}
where they solve a differential equation that governs the evolution of the radiated gluon
distribution as the jet propagates through the medium.   Energy contained inside the jet cone is
lost by dissipation through elastic collisions and by scattering of shower partons
to larger angles.  Their calculation is able to describe the LHC measured dijet asymmetry~\cite{Qin:2010mn}.
Figure~\ref{fig:qinaj1} shows the predicted dijet asymmetry at RHIC for mid-central and central \auau
collisions for leading jets $E_{T1} > 20$\,GeV and jet radius parameter $R=0.4$ and $R=0.2$ in the
left and right panels, respectively.  Despite the calculation including a rather modest value of $\hat{q}$ and $\hat{e}$,
the modification for $R=0.2$ is as strong as the result with $\alpha_{s} = 0.6$ from Coleman-Smith and collaborators
shown above in the right panel of Figure~\ref{fig:colemansmith1}.  Calculations of $\gamma$-jet correlations indicate similar level modifications.
It is also notable that Qin and collaborators 
have calculated the reaction plane dependence of the dijet $A_J$ distribution and find negligible differences.  This observable
will be particularly interesting to measure at RHIC since these calculations have difficultly reproducing the high $p_T$
$\pi^{0}$ reaction plane dependence ($v_2$) as discussed in the previous section.

\begin{figure}[t]
 \begin{center}
    \includegraphics[trim = 2 2 2 2, clip, width=\onewidth]{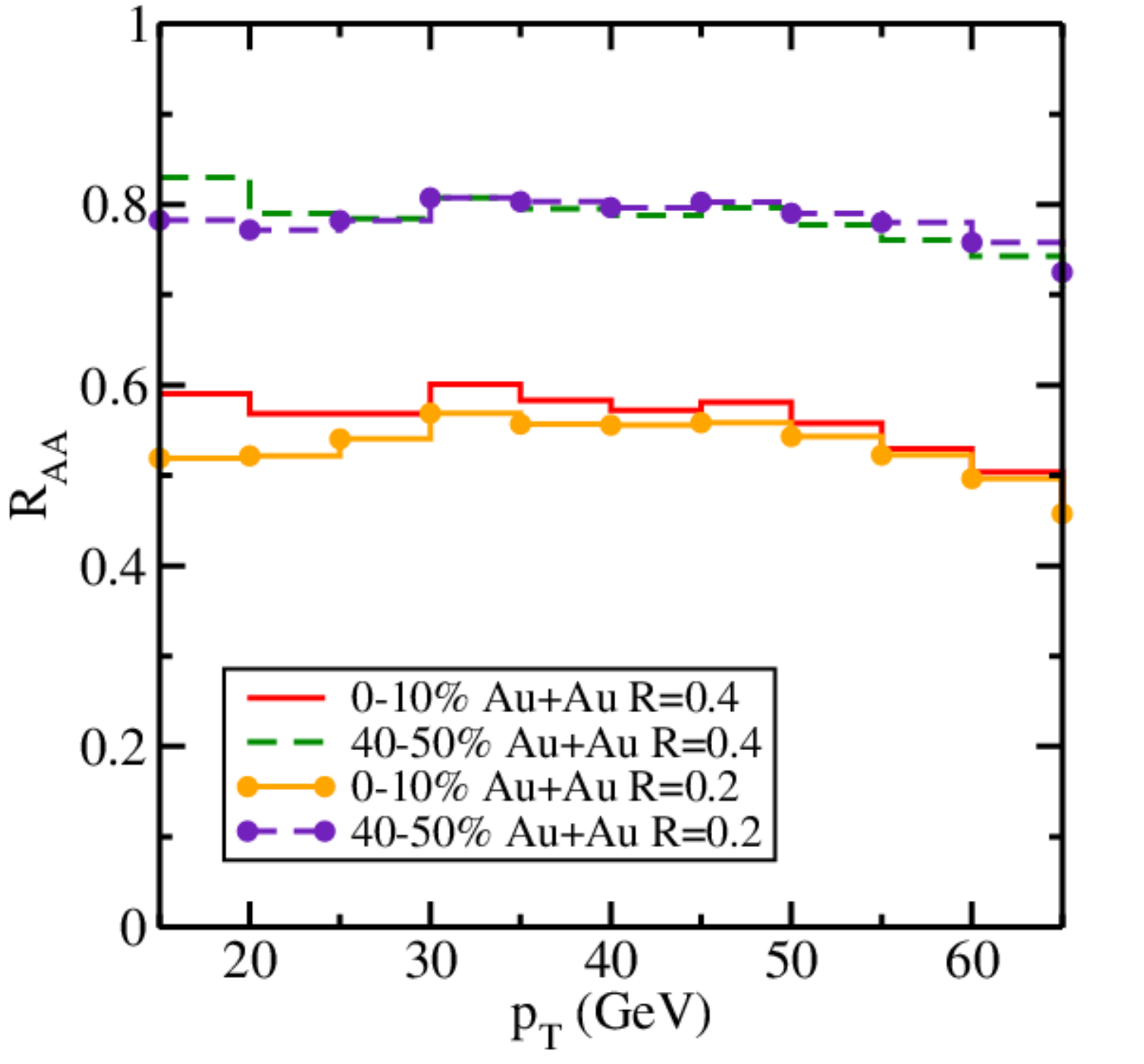}
    \caption{\label{fig:qinraa} Calculations from Qin et
      al.~\protect\cite{qin_privatecomm}{} for jet $R_{AA}$ for central 
      (solid lines) and mid-central collisions (dashed lines) for
      $R=0.2$ and 0.4 jets.}
 \end{center}
\end{figure}

Figure~\ref{fig:qinraa} shows results for the inclusive jet $R_{AA}$ as a function of $p_T$ for jet radius parameters
$R=0.2$ and $R=0.4$.  It is striking that the modification is almost independent of $p_T$ of the jet and there is very
little jet radius dependence.  The modest suppression, of order 20\%, in mid-central \auau collisions is of great interest as
previous measurements indicate modification of single hadrons and dihadron correlations for this centrality category.  
Measurements of jets with a broad range of radius parameters are easier in the lower multiplicity mid-central collisions.




\begin{figure}[t]
 \begin{center}
    \includegraphics[trim = 40 0 40 0, clip, width=\twowidth]{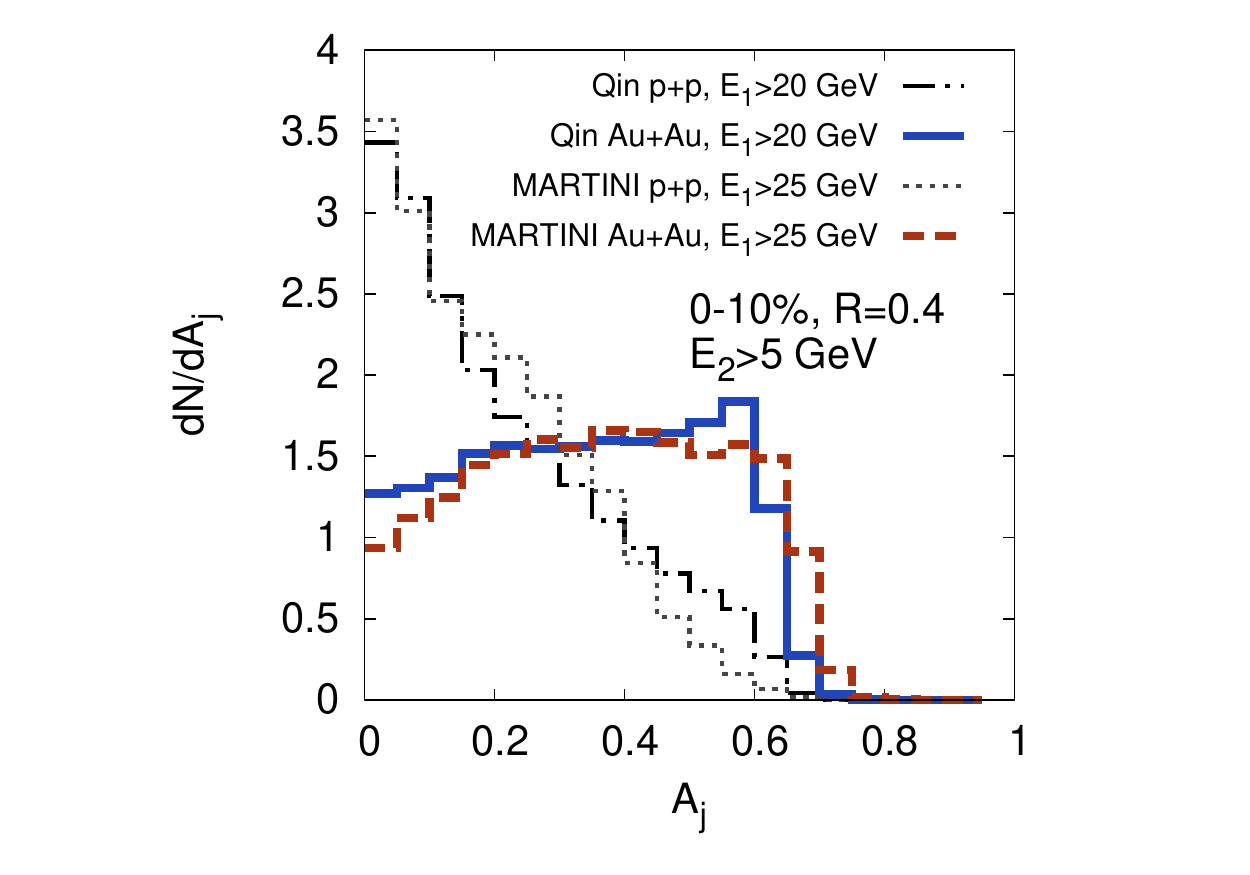}
    \hfill
    \includegraphics[trim = 40 0 40 0, clip, width=\twowidth]{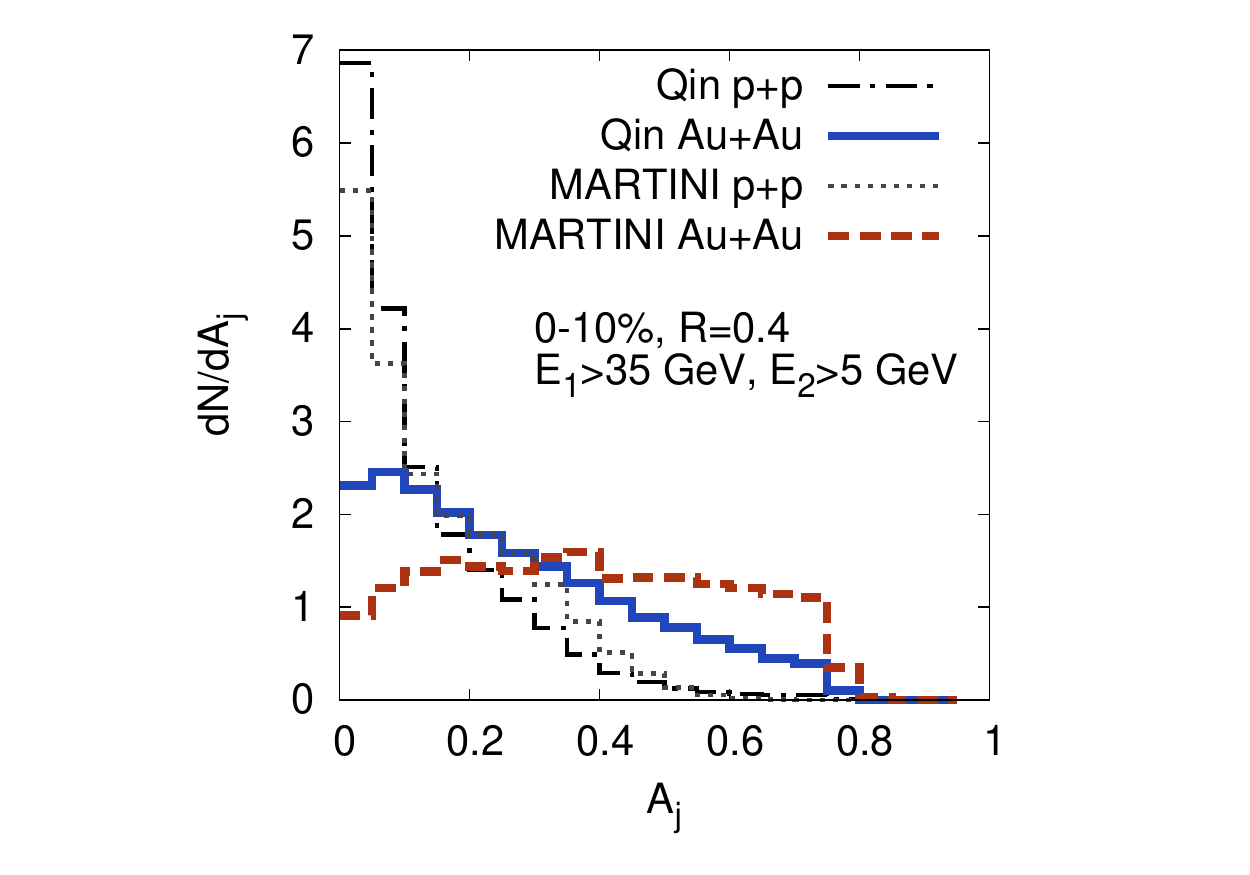}
    \caption{\label{fig:martiniaj} $A_J$ distributions in
      \martinimusic~\protect\cite{Young_privatecomm}{} and the model of Qin 
        et al.~\protect\cite{qin_privatecomm}{}.  (Left) Comparison of
      \martinimusic and Qin et al. $A_J$ calculations for
      leading jet $E_T>20$\,GeV (blue line, Qin et al.) and
      25\,GeV (red dashed line, \martinimusic).  Both calculations
      show a similar broad $A_J$ distribution.  (Right) Same as left
      panel, but with leading jet $E_T>35$\,GeV.  Here a difference
      in shape is observed between the two models with the Qin et
        al. model developing a peak at small $A_J$ while the
      \martinimusic calculation remains similar to the lower jet
      energy calculation.}
 \end{center}
\end{figure}

The third results are from Young and Schenke and collaborators~\cite{Young:2011va}.
These calculations utilize a jet shower Monte Carlo, referred to as \martini~\cite{Schenke:2009vr},
and embed the shower on top of a hydrodynamic space-time background, using the model
referred to as \music~\cite{Schenke:2010nt}.
Figure~\ref{fig:martiniaj} shows the jet
energy dependence of $A_J$ for RHIC energy dijets, $E_{T1}>25$\,GeV
and $E_{T1}>35$\,GeV in the left and right panels, respectively.  These results
are directly compared to the calculations from Qin and collaborators and indicate
a substantially different modification for the higher energy dijets.


\begin{figure}[!hbt]
 \begin{center}
    \raisebox{0.02in}{\includegraphics[trim = 2 2 2 2, clip, width=\twowidth]{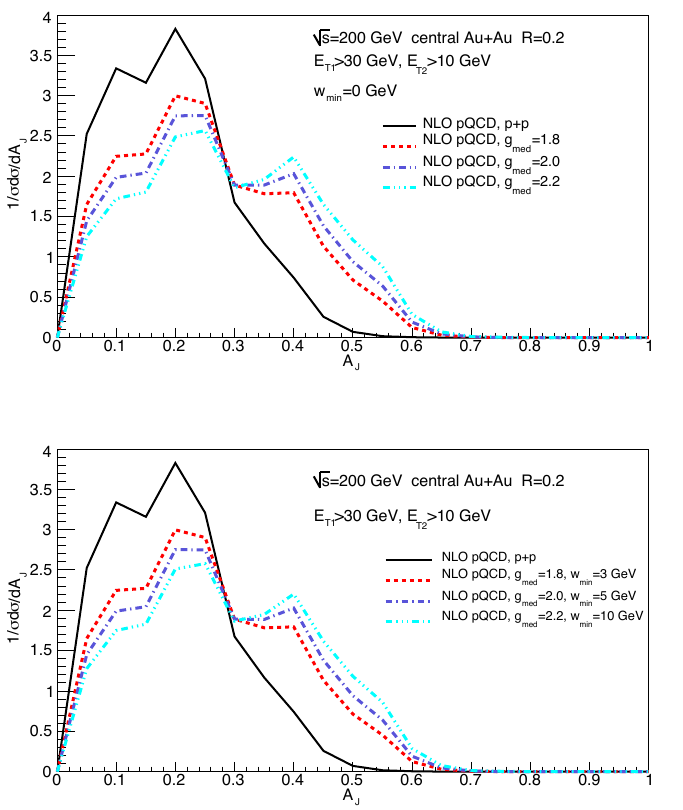}}
    \hfill
    \includegraphics[trim = 2 2 2 2, clip, width=\twowidth]{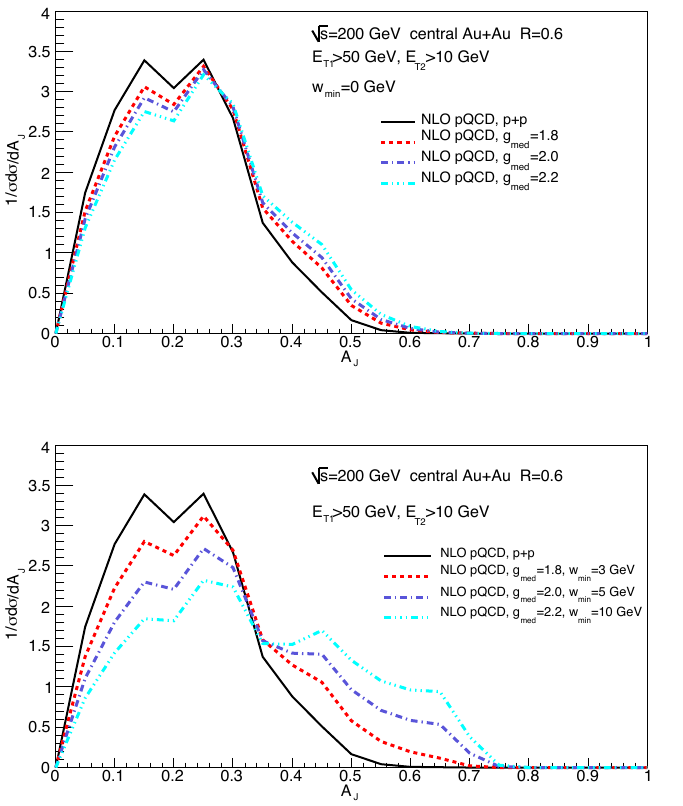}
        \caption{\label{fig:vitevaj} $A_J$ distributions calculated by
          Vitev et al.~\protect\cite{He:2011pd,Neufeld:2011yh,Vitev:2009rd}{} for two sets of kinematic cuts
          and jet cone radii.  The upper plots are for radiative energy loss only, and the lower plots include collisional energy loss as well.}
 \end{center}
\end{figure}

Our final set of illustrative theory calculations come from Vitev and
collaborators~\cite{He:2011pd,Neufeld:2011yh,Vitev:2009rd} where they
utilize a Next-to-Leading-Order (NLO) calculation and consider not only
final-state inelastic parton interactions in the QGP, but also initial-state cold
nuclear matter effects.  Figure~\ref{fig:vitevaj} shows the dijet asymmetry $A_J$ for jets 
with $E_{T1} > 30$\,GeV and $R=0.2$ (left panels) and $E_{T1} > 50$\,GeV and $R=0.6$ (right panels). 
The upper plots are for radiative energy loss only and the lower plots are including collisional energy loss as well,
and then the different colors are varying the probe-medium coupling by $\pm$10\%.   There is 
sensitivity even to these 10\% coupling modifications, and for the higher energy jets there is a
dramatic difference predicted from the inclusion of collisional energy loss.


For the inclusive jet suppression, these calculations predict a significant jet radius $R$ dependence to the modification,
in contrast to the result from Qin and collaborators.  In addition, Vitev and collaborators 
hypothesize a substantial cold nuclear matter effect of initial state parton energy loss.  Because the
high energy jets originate from hard scattering of high Bjorken $x$ partons, a modest energy loss of these
partons results in a reduction in the inclusive jet yields.
At RHIC with \dAu~running we will make cold nuclear matter measurements at the same collision energy
and determine the strength of these effects as a baseline to heavy ion
measurements. 



\section{Measuring jets, dijets, and $\gamma$-jet correlations at RHIC}

Jet and $\gamma$-jet measurements at RHIC are particularly appealing for the number of
reasons previously detailed.  In order to make these observations, one requires both sufficient rate and
acceptance for jets, dijets, and $\gamma$-jet events and a detector with large and uniform acceptance
to measure them.  The performance of the proposed sPHENIX detector is described in later chapters.  
Here we highlight the large rate of such events available at RHIC energies.

\begin{figure}[!hbt]
 \begin{center}
    \includegraphics[width=\onewidth]{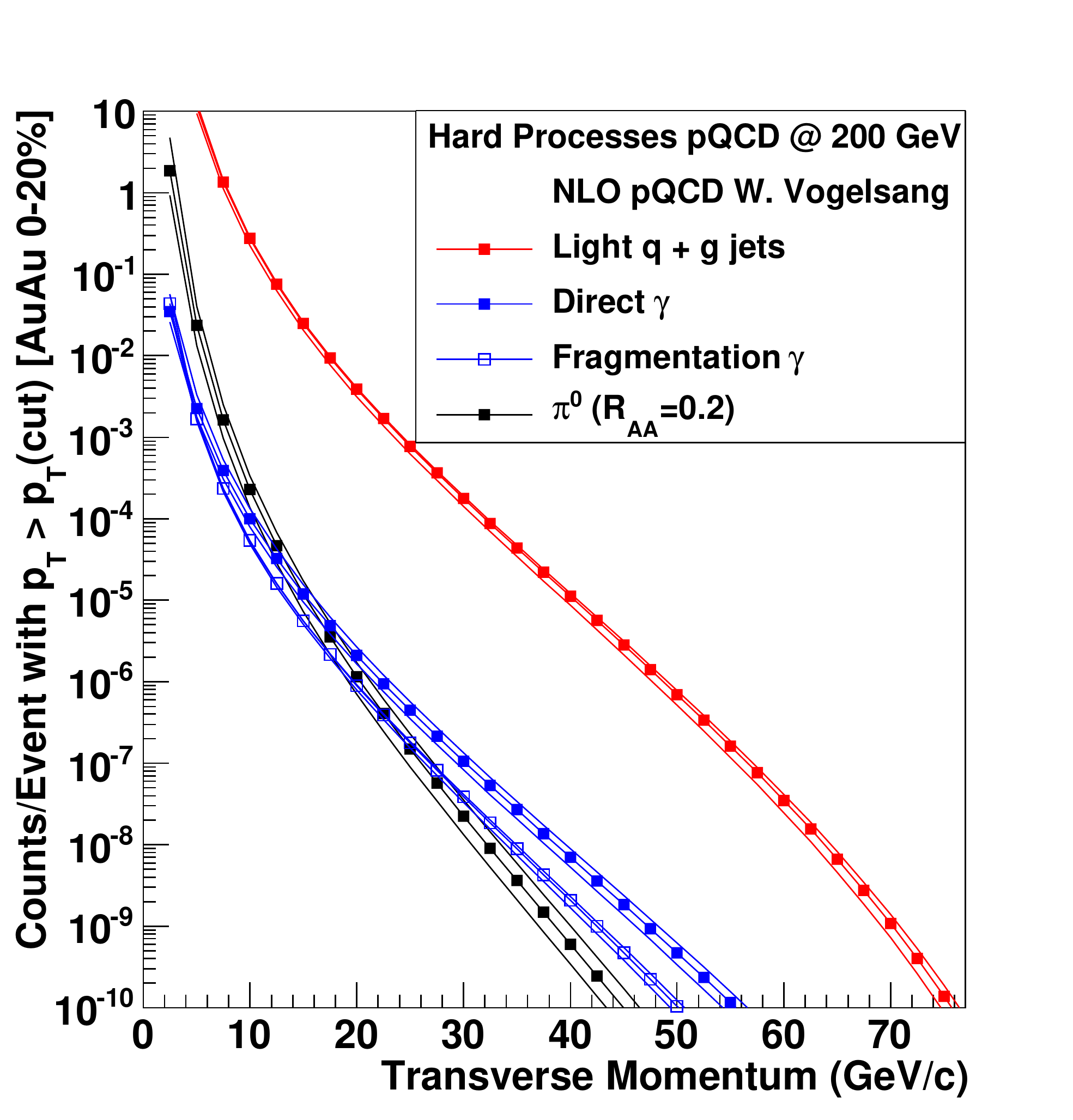}
    \caption{\label{fig:nlo_jetrates}Jet, photon and $\pi^{0}$ rates
      with $|\eta|<1.0$ from NLO pQCD~\protect\cite{Vogelsang:NLO}{}
      calculations scaled to \AuAu~ central collisions.  The scale uncertainties
      on the pQCD calculations are shown as additional lines.  
      Ten billion \AuAu~ central collisions correspond to one count at $10^{-10}$
      at the bottom of the y-axis range.}
 \end{center}
\end{figure}

The inclusive jet yield within $|\eta| < 1.0$ in 0--20\% central \auau collisions at 200\,GeV has been 
calculated for \pp collisions by Vogelsang in a Next-to-Leading-Order (NLO) 
perturbative QCD formalism~\cite{Vogelsang:NLO} and then scaled up by the
expected number of binary collisions, as shown in Figure~\ref{fig:nlo_jetrates}.  Also shown are calculation results
for $\pi^{0}$ and direct and fragmentation photons.  The bands correspond to the
renormalization scale uncertainty in the calculation (i.e., $\mu, \mu/2, 2\mu$).

The completion of the stochastic cooling upgrade to the RHIC accelerator~\cite{Fischer:2010zzd}
has been incorporated into the RHIC beam projections~\cite{RHICBeam}.  Utilizing these
numbers and accounting for accelerator and experiment uptime and the fraction of collisions
within $|z| < 10$ cm, the nominal full acceptance range for the detector, the sPHENIX
detector can sample 50 billion \auau minimum bias collisions in a one-year 20 week run.
Note that the PHENIX experiment has a nearly dead-timeless high-speed data acquisition and trigger
system that has already sampled tens of billions of \auau minimum bias collisions, 
and maintaining this high rate performance with the additional sPHENIX components is an
essential design feature.


Figure~\ref{fig:nlo_jetrates} shows the counts per event with $p_T$ larger than the value on the
x-axis for the most central 20\% \auau of events.  With 10 billion events for this centrality selection,
this translates into jet samples from 20--70\,GeV and
direct photon statistics out to 40\,GeV.  The statistical sample of jets and direct photons 
measurable in one year with sPHENIX is shown in Table~\ref{tab:nlo_jetrates}.  It is notable
that within the acceptance of the sPHENIX detector, over 80\% of the inclusive jets will also be
accepted dijet events.  

Also shown in Table~\ref{tab:nlo_jetrates} are the jet and direct photon samples in \pp~and \dAu~
collisions at the same collision energy per nucleon pair.  
The number of jets expected in the three systems are similar, meaning that
good control measurements in \pp~ and \dAu~events will be available 
on the same timescales to quantify baseline expectations and initial state effects.
Additionally, new geometries can be explored with precision utilizing asymmetric heavy ion reactions, such
as Cu$+$Au, and non-spherical geometries with U$+$U beams, now available with the RHIC EBIS upgrade~\cite{Pikin:2010zz}.
Control measurements with different geometries with high statistics are particularly interesting since
current theoretical calculations are challenged by the path length dependence of the energy lost by the
parton probe.

\renewcommand{\arraystretch}{1.9}\addtolength{\tabcolsep}{-0.5pt}
\begin{table}[!hbt]
  \centering
  \begin{tabular}[c]{r l l l}
    & \multicolumn{1}{c}{{\renewcommand{\arraystretch}{1.2}
\begin{tabular}[t]{@{}c@{}}\auau\\(central 20\%)\end{tabular}} } &
\multicolumn{1}{c}{\pp} & \multicolumn{1}{c}{\dAu} \\
    \toprule
    \multirow{2}{*}{$ > 20$\,GeV} & $10^7$ jets & $10^6$ jets & $10^7$ jets \\
    & $10^4$ photons & $10^3$ photons & $10^4$ photons \\
    \midrule
    \multirow{2}{*}{$ > 30$\,GeV} & $10^6$ jets & $10^5$ jets & $10^6$ jets \\
    & $10^3$ photons & $10^2$ photons & $10^3$ photons \\
    \midrule
    $ > 40$\,GeV & $10^5$ jets & $10^4$ jets & $10^5$ jets \\
    \midrule
    $ > 50$\,GeV & $10^4$ jets & $10^3$ jets & $10^4$ jets 
  \end{tabular}
  \caption{Table of jet rates for different systems. Each column shows
    the number of jets or direct photons that would be measured within
    $|\eta|<1$ in one 20 week running period.\label{tab:nlo_jetrates}}
\end{table}

\begin{figure}[!hbt]
 \centering
 \begin{minipage}[c]{0.67\linewidth}
   \includegraphics[trim = 20 0 45 40, clip, width=\linewidth]{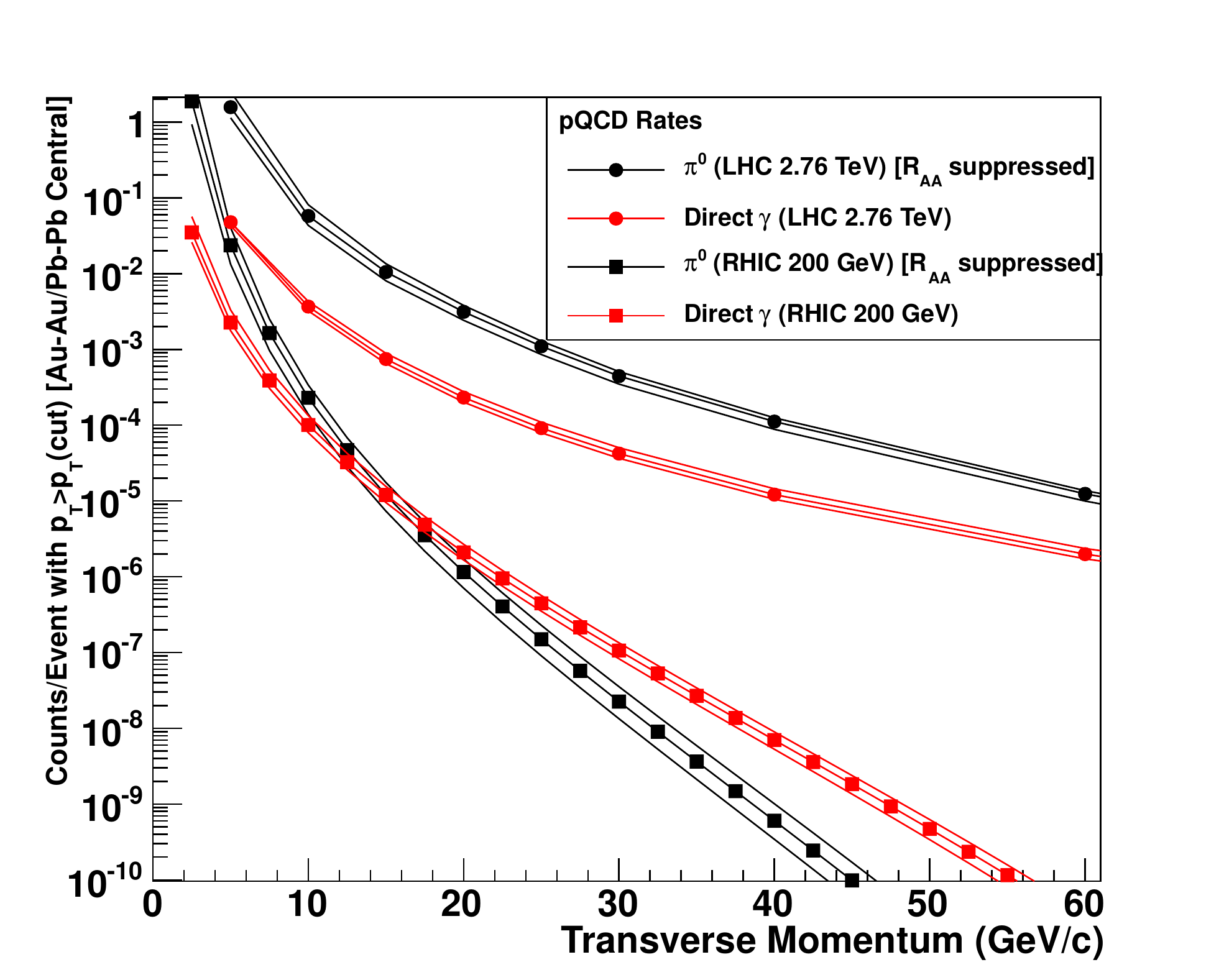}
 \end{minipage}
 \hfill
 \begin{minipage}[c]{0.32\linewidth}
   \includegraphics[trim = 8 0 45 40, clip, width=\linewidth]{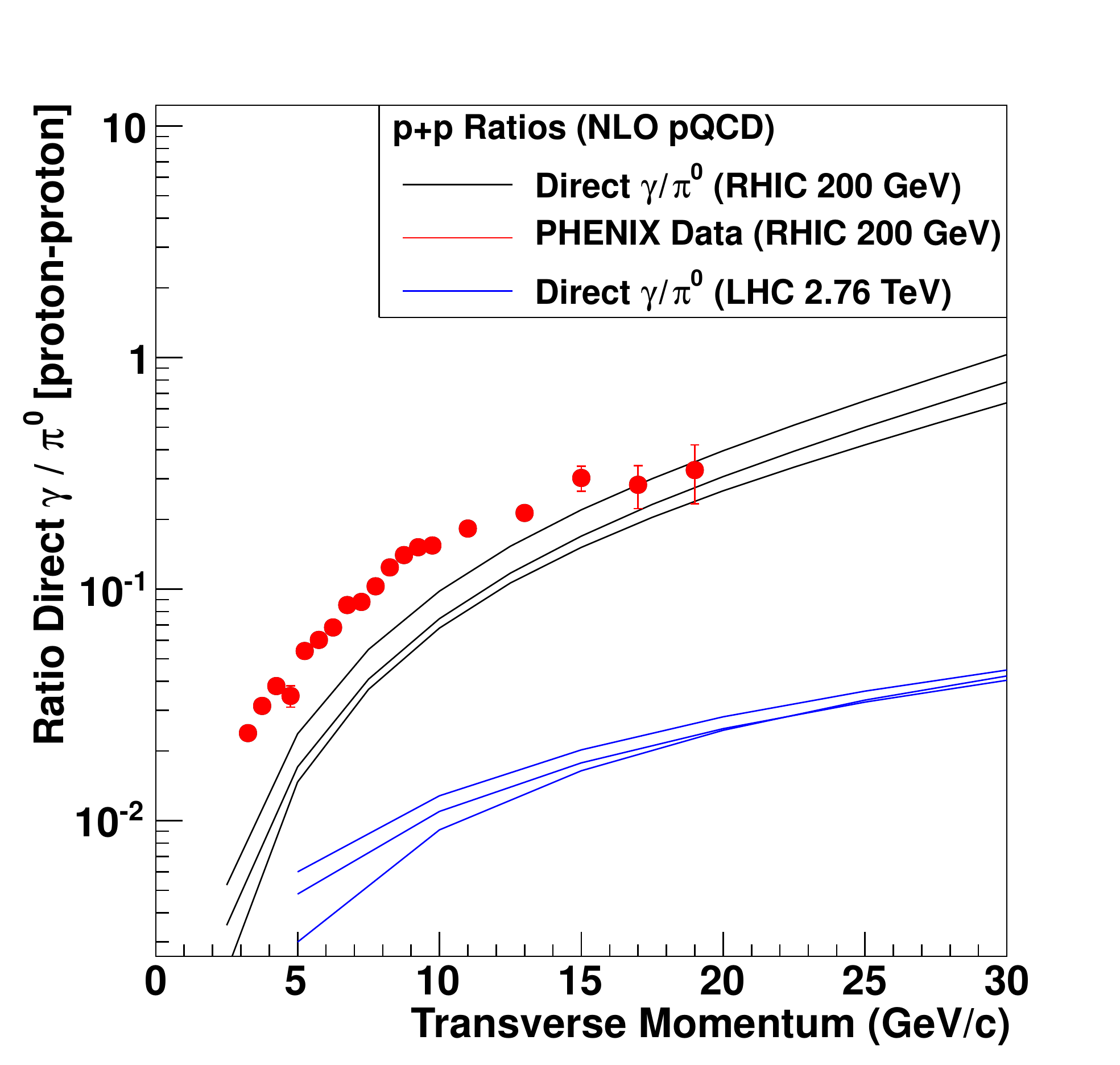}
   \\
   \includegraphics[trim = 8 0 45 40, clip, width=\linewidth]{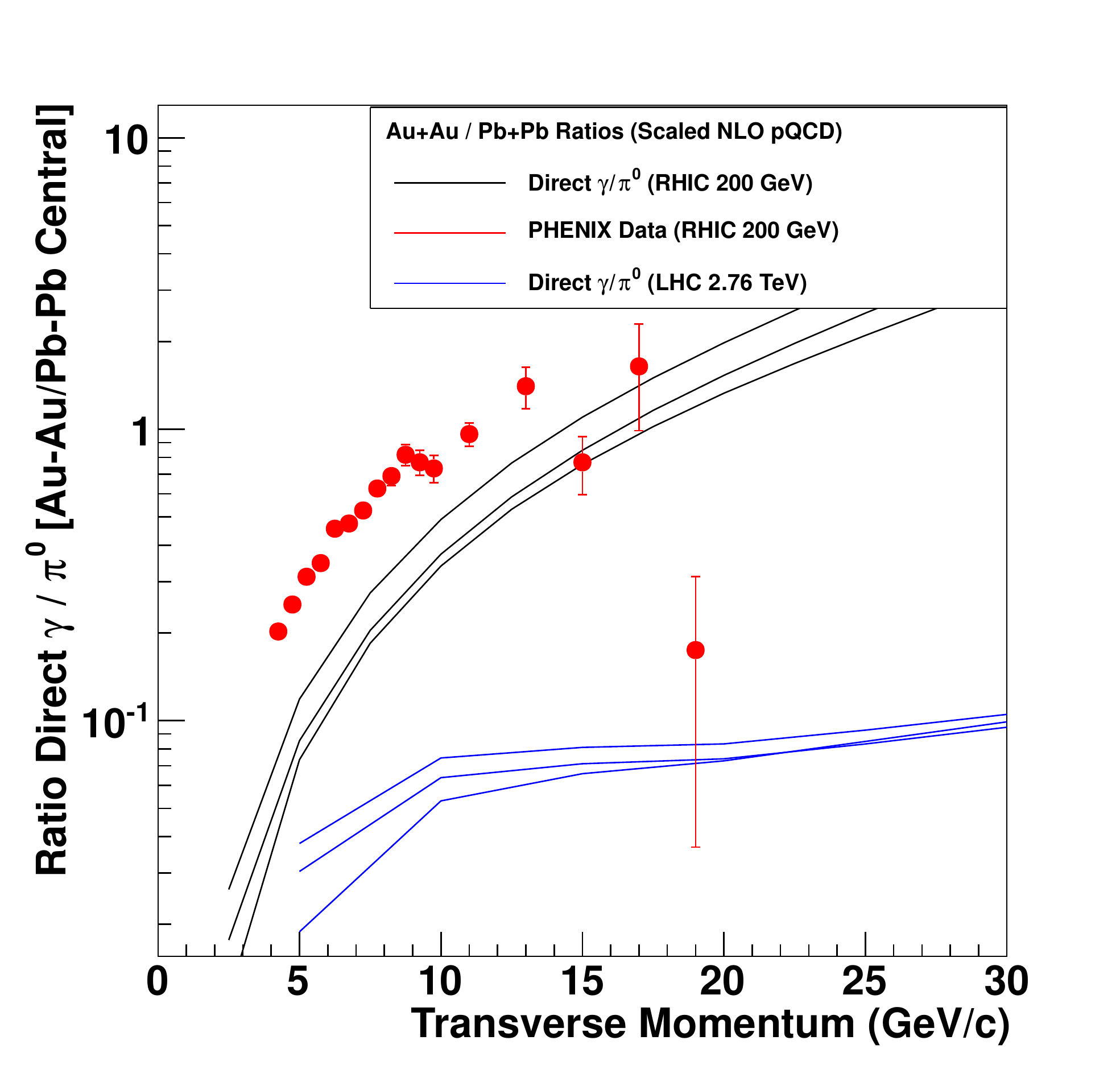}
 \end{minipage}
 \caption{\label{fig:nlo_gammarates}NLO pQCD calculations of direct
   photons and $\pi^{0}$ for RHIC and LHC.  The plot on the left shows
   the counts per event in \auau or \pbpb collisions (including the
   measured $R_{AA}$ suppression factor for $\pi^{0}$).  The upper
   (lower) panel on the right shows the direct $\gamma$ to $\pi^{0}$
   ratio in \pp (\auau or \pbpb) collisions, in comparison with
   measurements from the PHENIX experiment at
   RHIC~\cite{Afanasiev:2012dg,Adare:2012yt}.}
\end{figure}

Measurement of direct photons requires them to be separated from the 
other sources of inclusive photons, largely those from $\pi^0$ and $\eta$ meson decay.  The
left panel of Figure~\ref{fig:nlo_gammarates} shows the direct photon
and $\pi^0$ spectra as a function of transverse momentum for both $\sqrt{s}=200$\,GeV and 2.76\,TeV \pp
collisions.  The right panels show the $\gamma/\pi^0$ ratio as a
function of $p_T$ for these energies with comparison PHENIX measurements at RHIC.  
At the LHC, the ratio remains below 10\% for $p_T < 50$\,GeV while at RHIC the ratio rises sharply
and exceeds one at $p_T\approx 30$\,GeV/c.  In heavy ion collisions
the ratio is further enhanced because the $\pi^0$s are significantly
suppressed.  Taking the suppression into account, the $\gamma/\pi^0$
ratio at RHIC exceeds one for $\pT > 15$\,GeV/c.  The large signal to
background means that it will be possible to measure direct photons
with the sPHENIX calorimeter alone, even before applying isolation cuts.
Beyond measurements of inclusive direct photons, this enables measurements of $\gamma$-jet correlations and
$\gamma$-hadron correlations.


\section{Summary}
\label{sec:physicscasesummary}

Detailed information about the \qgp properties,
dynamics, time evolution, and structure at 1--2 $T_{c}$ is accessible 
at RHIC through the extensive set of reconstructed jet measurements
proposed here. The theoretical bridgework needed to connect these
measurements to the interesting and unknown medium characteristics 
of deconfined color charges is under active construction by many 
theorists. Combining this work with the flexible and high luminosity
RHIC accelerator facility can produce new discoveries in heavy ion 
collisions with an appropriate set of baseline measurements 
provided a suitable detector apparatus is constructed. Our proposed
design for a jet detector at RHIC that is best able to make use of these
opportunities is given in the following chapter.



\chapter{sPHENIX Detector Requirements}
\label{chap:detector_requirements}

In order to perform the precision jet measurements outlined in Chapter~\ref{chap:physics_case}, 
there is a set of detector requirements that must be met.  
In addition, as outlined in the Executive Summary, 
this sPHENIX upgrade serves as the foundation for future
potential upgrades including a detector for the Electron-Ion Collider (referred to as ePHENIX),
and those requirements must also be met.
In this Chapter we detail the basic sPHENIX detector design and the
requirements on the detector performance.  Details of the specific
detector and \geant simulations are given in Chapter~\ref{chap:detector_concept}.

\section{Detector Overview}

Based on the physics requirements, detector constraints, and cost considerations, a
baseline conceptual design has evolved.  Here we describe the basic features and the key design
parameters for the detector.  The basic components are:


\begin{description}

\item[Magnetic Solenoid] A thin superconducting solenoid with a 2\,T field and an inner radius
of 70\,cm.  The integrated field strength is driven by the tracking resolution requirements.  The 
radius allows sufficient space for future upgrades with high
resolution tracking and preshower detectors (as detailed in Appendix~\ref{chap:barrel_upgrade}) and particle
identification detectors for a future ePHENIX (as detailed in Appendix~\ref{chap:ePHENIX}).  
The cost of the solenoid scales approximately linearly with the inner radius~\cite{green:2008xx}, and thus 
a small radius must be maintained.  This constraint is also to fit the entire detector onto the rail
system in the PHENIX experiment hall.
Minimizing the number of radiation lengths in
the cryostat and coil allows the electromagnetic calorimeter to be
placed outside the solenoid, simplifying the deployment of the electronics.

\item[Electromagnetic Calorimeter] A compact tungsten-scintillator
sampling calorimeter outside the cryostat read out with silicon
photo-multipliers.  The small Moli\`ere radius and short radiation length
allows for a highly segmented calorimeter ($\Delta \eta \times \Delta
\phi \sim 0.024 \times 0.024$) at a radius of about 100\,cm from the beam
axis, which results in about 25,000 electronic channels.

\item[Hadronic Calorimeter] An iron-scintillator sampling calorimeter
outside the electromagnetic calorimeter.  In order to minimize the
mass and bulk, the calorimeter doubles as the flux return for the
solenoid.  A thickness of $5\lambda_\mathrm{int}$ combined with the electromagnetic
calorimeter in front is sufficient to
fully contain the energies of interest, and provide more than enough
iron for the full flux return.
The hadronic calorimeter is divided into two longitudinal compartments
consisting of plates running parallel to the beam axis with
scintillator plates interleaved, then read out via embedded wavelength
shifting fiber.  The hadronic calorimeter will use the
same silicon photomultiplier sensors as the electromagnetic
calorimeter and similar electronics.  The coarser segmentation
($\Delta \eta \times \Delta \phi \sim 0.1 \times 0.1$) results in an
electronic channel count of about 10\% that of the electromagnetic
calorimeter.

\item[Readout electronics] Bias voltage and analog signal processing
for silicon photo-multipliers in physical proximity to the
sensors, with a number of options for the digitization and buffering
using either commercial components or integrated circuits adapted from
existing experimental projects.

\end{description}

\begin{figure}[hbt!]
 \begin{center}
  \includegraphics[width=0.7\linewidth]{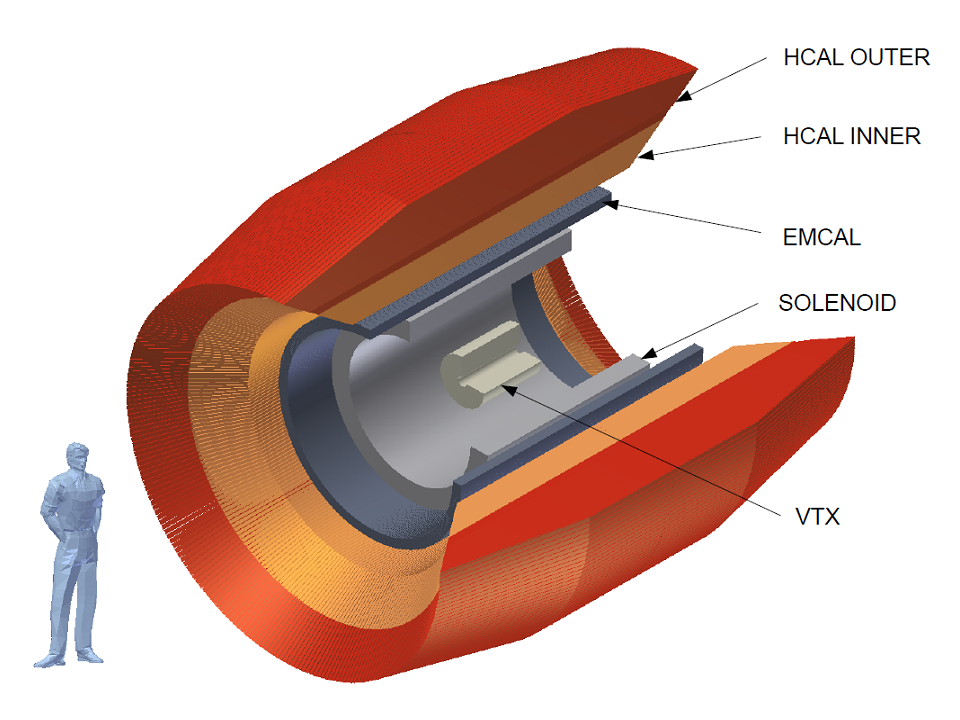}
  \caption{\label{fig:OverviewCutaway} Cutway view of the detector.}
 \end{center}
\end{figure}

\begin{figure}[hbt!]
 \begin{center}
  \includegraphics[width=0.8\linewidth]{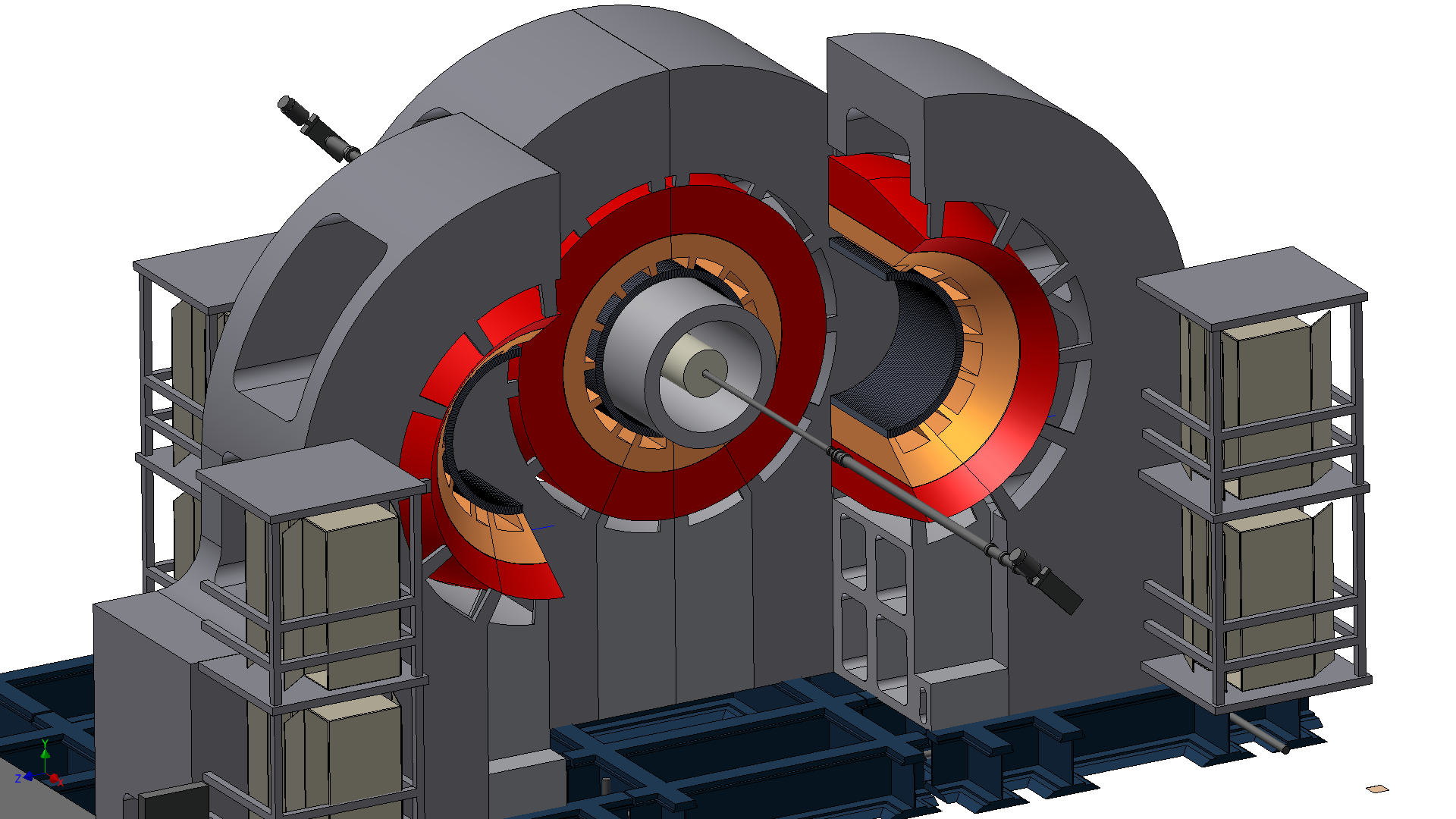}
  \caption{\label{fig:ExplodedView} View of the sPHENIX detector in the collision hall with conceptual design for structural support.}
 \end{center}
\end{figure}

\renewcommand{\arraystretch}{1.5}
\begin{table}[p]
\centering
\begin{tabular}{lllc}
Physics & Detectors & Requirements & \\
\hline
\multirow{4}{*}{Full jet reconstruction} & EMCal & $\sigma/E<20\%/\sqrt{E}$      & \multirow{4}{*}{sPHENIX} \\
                        & \multirow{2}{*}{HCal}  & $\sigma/E<100\%/\sqrt{E}$      &     \\
                        &       & $\Delta\eta \times \Delta\phi \sim 0.1\times 0.1$   &     \\
                        &       & uniform within $|\eta|<1$             &     \\
\hline
\multirow{2}{*}{Direct $\gamma$, $p_T > 10$\,GeV/c}   & \multirow{2}{*}{EMCal} & $\sigma/E\simeq15\%/\sqrt{E}$  & sPHENIX  \\
                        &       & $\Delta\eta \times \Delta\phi \sim 0.03\times 0.03$ &     \\
\hline
\multirow{2}{*}{Jet-hadron}              & VTX 4 layers            & \multirow{2}{*}{tracking $p_T< 4$\,GeV/c} & Current PHENIX \\
                        & Solenoidal field &                                        &  sPHENIX    \\
\hline
 \multirow{2}{*}{High-$z$ FFs} & Jets as above      & EMCal and HCal                   & sPHENIX    \\
                                   &  Tracking &  $\Delta p/p \simeq 2\%$ & Future Option \\
\hline
\multirow{3}{*}{Tagged HF jets}       & Jets as above      & EMCal and HCal                   & {sPHENIX } \\
                       & DCA capability     & Current PHENIX VTX               &  Current PHENIX   \\
                       &  Tracking &  $\Delta p/p \simeq 2\%$ & Future Option    \\
\hline
Heavy quarkonia                 & Electron ID   &                                       &     \\
\multirow{6}{*}{Separation of $\Upsilon$ states} & \multirow{2}{*}{EMCal} &  $\sigma/E\simeq15\%/\sqrt{E}$   & \multirow{2}{*}{sPHENIX } \\
                                &               & $\Delta\eta \times \Delta\phi \sim 0.03\times 0.03$  &     \\
                                & \multirow{2}{*}{Preshower}     & $e/\pi$ rejection  & Future Option\\ 
                                &                                & fine segmentation  &     \\
                                & Tracking & $B = 2 T$                          & sPHENIX  \\
                                &                    &  $\Delta p/p \simeq 2\%$ & Future Option    \\
\hline
\multirow{4}{*}{$\pi^{0}$ to $p_T = 40$\,GeV/c}   & \multirow{2}{*}{EMCal}             &  $\sigma/E\simeq15\%/\sqrt{E}$  & sPHENIX  \\
                                 &                   &  $\Delta\eta \times \Delta\phi \sim 0.03\times 0.03$ &     \\
                                 & \multirow{2}{*}{Preshower} & $2 \gamma$ separation   & Future Option \\
                                 &                            &   fine segmentation  & \\
\end{tabular}
\caption{\label{tbl:detector_requirements} Summary of detector requirements, showing the  capabilities needed for various physics observables, and whether those capabilities are part of the current proposal or are possible additions to the detector through other means.  Those items labeled sPHENIX are included with the
detector upgrades in this proposal.  Those items labeled ``Future Option'' are detailed in Appendix~\ref{chap:barrel_upgrade} and require modest additional
upgrades.} 
\end{table}

The detector concept that has resulted from these considerations is
shown in Figure~\ref{fig:OverviewCutaway} and Figure~\ref{fig:ExplodedView} 
and will be described in detail in Chapter~\ref{chap:detector_concept}.
Taking advantage of both technological developments in the era of
RHIC and LHC experiments, and building on equipment already in place  
in PHENIX, the detector is both compact, which plays a large role in
keeping costs under control, and much larger in solid angle than current PHENIX experiment.
The outer radius of the hadronic calorimeter is about 200 cm and there are approximately 27,000 electronic
channels for the two calorimeters combined.

The physics requirements that drive the design are summarized in 
Table~\ref{tbl:detector_requirements} and will be discussed in the 
following section.

\afterpage{\clearpage}

\section{Design Goals}

\subsection{Coverage}

The total acceptance of the detector is determined by the requirement of high statistics jet measurements
and the need to fully contain both single jets and dijets.
To fully contain hadronic showers in the detector requires both large solid angle
coverage and a calorimeter deep enough to fully absorb the
energy of hadrons up to 70\,GeV.

\begin{figure}[hbt!]
 \begin{center}
   \raisebox{0.5mm}{\includegraphics[trim = 2 2 2 2, clip, width=0.53\linewidth]{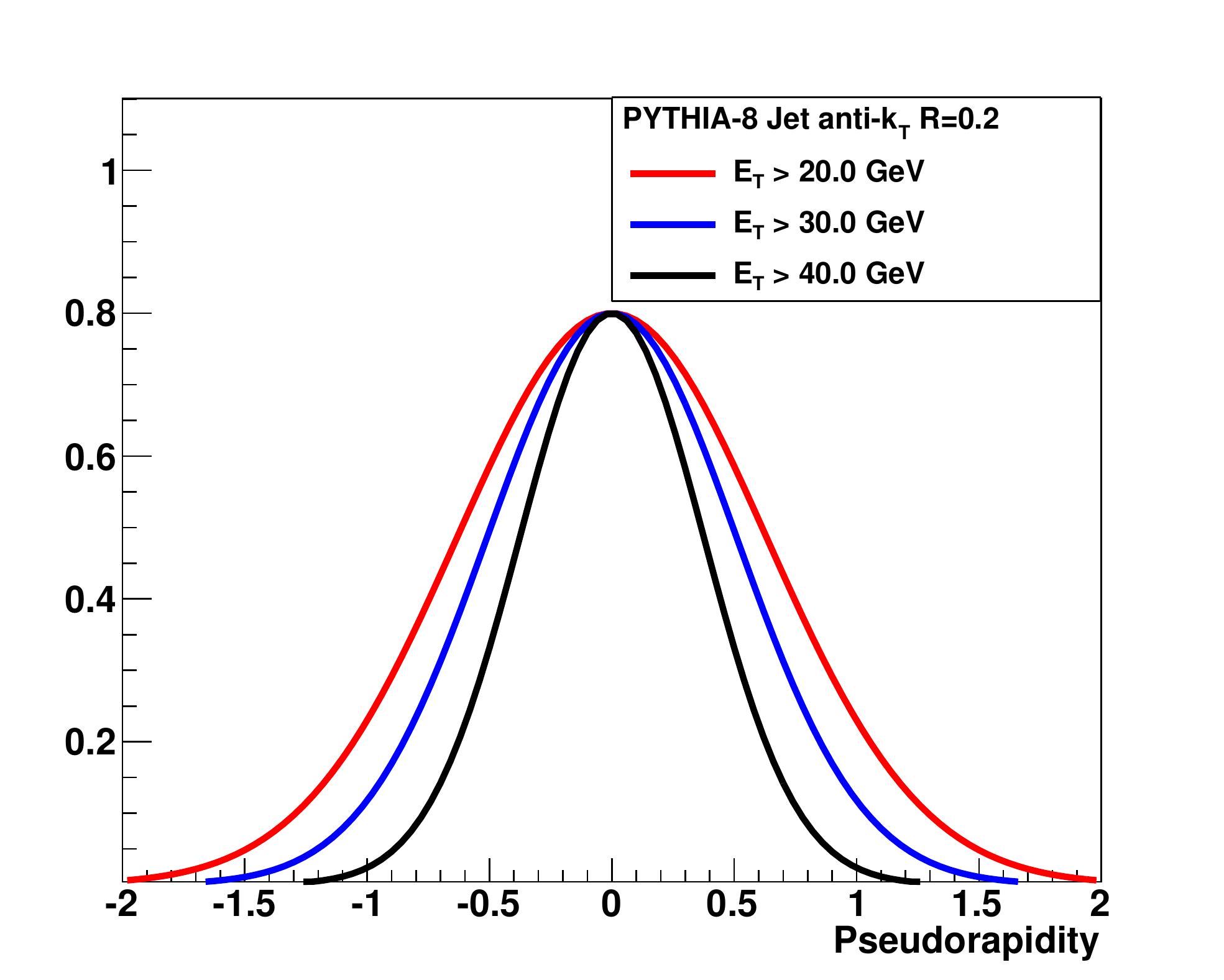}}
  \hfill
  \includegraphics[trim = 2 2 2 2, clip, width=0.45\linewidth]{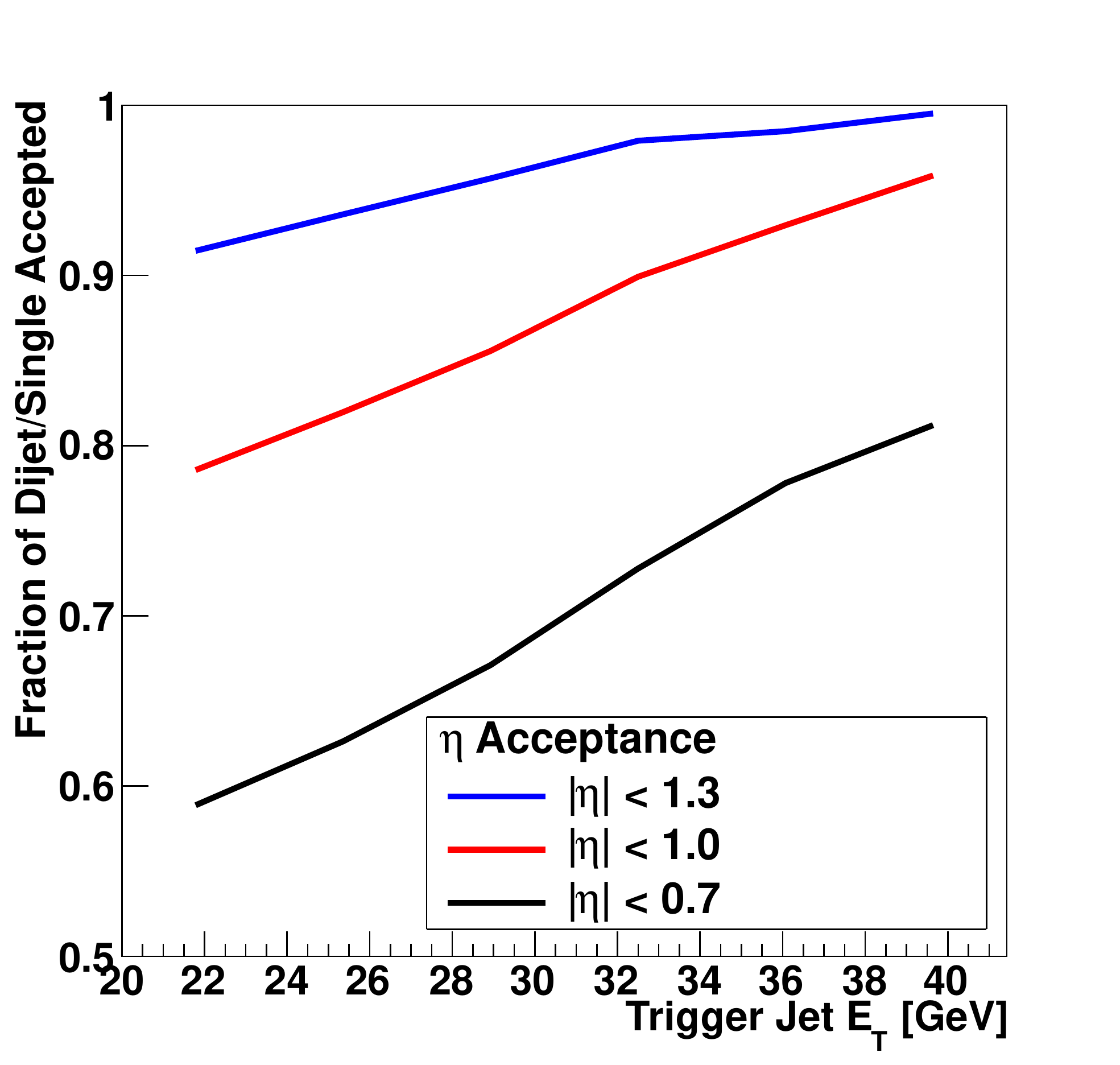}
  \caption{\label{fig:pythia_dijet_accept}(Left) Pseudorapidity
    distribution of \pythia jets reconstructed with the \fastjet
    anti-k$_{T}$ and R=0.2 for different transverse energy selections.
    (Right) The fraction of \pythia events where the leading jet is
    accepted into a given pseudorapidity range where the opposite side
    jet is also within the acceptance.  Note that the current PHENIX
    acceptance of $|\eta|<0.35$ corresponds to a fraction below 30\%.}
 \end{center}
\end{figure}

\begin{figure}[hbt!]
 \begin{center}
   \includegraphics[width=0.6\linewidth]{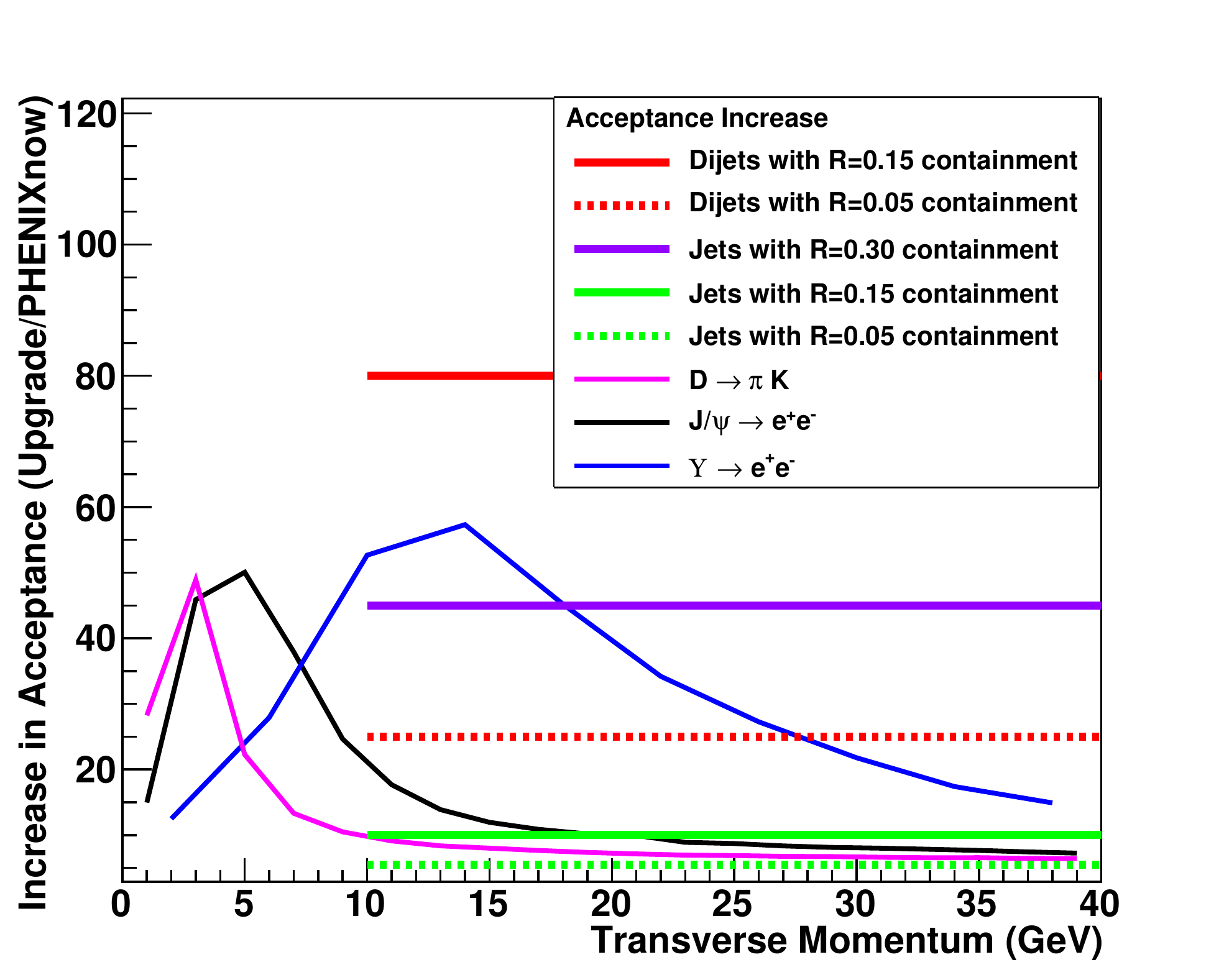}
   \caption{\label{fig:acceptincrease}Acceptance increase for various
     processes (as modeled using the \pythia event generator) for the
     proposed sPHENIX barrel detector compared with the current PHENIX
     central arm spectrometers.}
 \end{center}
\end{figure}

The \pythia event generator has been used to generate a sample of \pp at 200\,GeV
events which can be used to demonstrate the pseudorapidity distribution
of jets.
The left panel in Figure~\ref{fig:pythia_dijet_accept} shows the pseudorapidity distribution
of jets with $E_{T}$ above 20, 30, and 40\,GeV.
The right panel in Figure~\ref{fig:pythia_dijet_accept} shows the fraction of events where a
trigger jet with $E_{T}$ greater than a given value within a pseudorapidity range
has an away side jet with $E_{T} > 5$\,GeV accepted within the same
coverage.
In order to efficiently capture the away side jet, the detector should cover
$|\eta|<1$, and in order to fully contain hadronic showers within this
fiducial volume, the calorimetry should cover slightly more than that.
Given the segmentation to be discussed below, the calorimeters are required to
cover $|\eta|<1.1$.

It should be noted that reduced acceptance for the away-side jet
relative to the trigger suffers not only a reduction in statistics for
the dijet asymmetry and $\gamma$-jet measurements but also results in a
higher contribution of low energy \fake jets (upward fluctuations
in the background) in those events where the away side jet is out of
the acceptance.   For the latter effect, the key is that both jet axes
are contained within the acceptance, and then events can be rejected
where the jets are at the edge of the detector and might have partial
energy capture.  

Compared to the current PHENIX acceptance (the central
arms cover $|\eta|<0.35$ and $\Delta \phi = \pi$), 
full azimuthal coverage with $|\eta|<1.0$ results in a 
very substantial increase in the acceptance of single jets
and an even larger increase in the acceptance of dijets
as shown in Figure~\ref{fig:acceptincrease}.  Also shown in Figure~\ref{fig:acceptincrease} 
is the substantial increase in acceptance for other observables including
heavy quarkonia states.  Thus, the large acceptance and high rate detector
with incremental additional upgrades enables a much broader program as detailed
in Appendix~\ref{chap:barrel_upgrade}.

\subsection{Segmentation}

Jets are reconstructed from the four-vectors of the particles or
measured energies in the event via different algorithms (as described
in Chapter~\ref{chap:jet_performance}), and with a typical size $R =
\sqrt{\Delta\phi^{2} + \Delta\eta^{2}}$.  In order to reconstruct jets
down to radius parameters of $R = 0.2$ a segmentation in the hadronic
calorimeter of $\Delta \eta \times \Delta \phi = 0.1 \times 0.1$ is
required.  The electromagnetic calorimeter segmentation should be
finer as driven by the measurement of direct photons for
$\gamma$-jet correlation observables.  The compact electromagnetic
calorimeter design being considered for sPHENIX has a Moli\`ere
radius of $\sim 15$\,mm, and with a calorimeter at a radius of about
100\,cm, this leads to an optimal segmentation of $\Delta \eta \times
\Delta \phi = 0.024 \times 0.024$ in the electromagnetic section.

\subsection{Energy Resolution}

The requirements on the jet energy resolution are driven by considerations
of the ability to reconstruct the inclusive jet spectra and dijet asymmetries 
and the fluctuations on the \fake jet background (as detailed in Chapter~\ref{chap:jet_performance}.
The total jet energy resolution is typically driven by the hadronic calorimeter
resolution and many other effects including the bending of  charged particles bending in the magnetic field
out of the jet radius.  Expectations of jet resolutions approximately 1.2 times worse than the
hadronic calorimeter resolution alone are typical (see a more detailed discussion in Chapter~\ref{chap:jet_performance}).

In a central \AuAu event, the average energy within a jet cone of radius
$R = 0.2$ ($R=0.4$) is approximately 10\,GeV (40\,GeV) resulting in an
typical RMS fluctuation of 3\,GeV (6\,GeV).  This sets the scale for
the required reconstructed jet energy resolution, as a much better resolution
would be dominated by the underlying event fluctuations regardless.
A measurement of the jet energy for $E = 20$\,GeV with $\sigma_{E} = 100\% \times \sqrt{E}
= 4.4$\,GeV gives a comparable contribution to the underlying event fluctuation.
A full study of the jet energy resolution with a \geant simulation of the detector configuration is required and
is presented in Chapter~\ref{chap:jet_performance}.


Different considerations set the scale of the energy resolution
requirement for the EMCal. The jet physics requirement is easily met
by any EMCal design.  For the direct $\gamma$-jet physics, the photon
energies being considered are $E_{\gamma} > 10$\,GeV where even a modest 
$\sigma_{E}/E = 20\%/\sqrt{E}$ represents only a blurring of 0.6\,GeV.
In \auau central events, the typical energy in a $3 \times 3$ tower 
array is approximately 400\,MeV.  These values represent a negligible performance
degradation for these rather clean photon showers even in central \auau
events.

The energy resolution is driven by physics enabled by the future option upgrades to the sPHENIX
detector.  These future options upgrades described in Appendix~\ref{chap:barrel_upgrade}
incorporate electron identification using a combination of a
preshower detector and energy in the EMCal matching with the momentum
from charged particle tracking.  These set a more stringent
requirement on the energy resolution of the EMCal, and the trade-off
determines how low in $p_T$ electrons can be identified without
requiring additional detectors for electron identification.  As
detailed in Appendix~\ref{chap:barrel_upgrade}, for the quarkonia
measurements in the Upsilon family, an EMCal resolution of order
$15\%/\sqrt{E}$ is required, along with the preshower for
electron-pion separation.   A similar EMCal resolution requirement must
be met in a future ePHENIX as described in 
Appendix~\ref{chap:ePHENIX}

Most of these physics measurements require complete coverage over a large range of rapidity and
azimuthal angle ($\Delta\eta\le$ 1.1 and $\Delta\phi$ = 2$\pi$) with
good uniformity and minimal dead area. The calorimeter should be
projective (at least approximately) in both $\eta$ and $\phi$. 
For a compact detector design there is a trade-off in terms of thickness of the calorimeter and Moli\`ere radius
versus the sampling fraction and, therefore, the energy resolution of
the device.  Further optimization if these effects will be required as
we work towards a final design.

\subsection{Triggering}

The jet energy should be available at the Level-1 trigger as a standard
part of the PHENIX dead-timeless Data Acquisition and Trigger system.
This triggering ability is important
as one requires high statistics measurements in proton-proton,
proton-nucleus, light nucleus-light nucleus, and heavy nucleus-heavy
nucleus collisions with a wide range of luminosities.   It is important
to have combined EMCal and HCal information available so as to avoid
a specific bias on the triggered jet sample.

\subsection{Tracking}

Tracking capabilities are critical both in the sPHENIX upgrade and
for the physics enabled by future option upgrades.  The sPHENIX
detector with a reconfiguration of the existing PHENIX barrel silicon vertex
detector will be able to track charged hadrons up to $p_T \approx
5$\,GeV/c, which is important for understanding how the soft part of
parton showers is modified and potentially completely equilibrated in
the quark-gluon plasma.  A number of physics measurements are enabled
by additional tracking layers which are not part of this proposal.
These are described in more detail in Appendix~\ref{chap:barrel_upgrade}.

\chapter{sPHENIX Detector Concept}
\label{chap:detector_concept}

In this Chapter we detail the sPHENIX detector design including the
magnetic solenoid, electromagnetic and hadronic calorimeters, and
readout electronics.  Detector performance specifications are checked
using a full \geant simulation of the detector.  Full physics
performance measures are detailed in
Chapter~\ref{chap:jet_performance}.

The sPHENIX detector concept takes advantage of technological
developments to enable a compact design with excellent performance.  A
tungsten-scintillator electromagnetic calorimeter read out with
silicon photomultipliers (SiPMs) or avalanche photodiodes (APDs)
allows for a physically thin device which can operate in a magnetic
field, without the bulk of photomultiplier tubes and the need for high
voltage distribution.  The smaller electromagnetic calorimeter also
allows the hadronic calorimeter to be less massive, and the use of
solid-state sensors for the hadronic calorimeter allows for nearly
identical electronic readout for the two major systems.  A
superconducting magnet coupled with high resolution tracking detectors
provides good momentum resolution inside a compact solenoid.  The
detector has been designed from the beginning to minimize the number
of distinct parts to be simpler to manufacture and assemble.  The use
of components insensitive to magnetic fields enables the hadronic
calorimeter to double as the flux return for the solenoid, reducing
both mass and cost.  Adapting existing electronic designs for the
readout allows for reduced development cost and risk, and leverages a
decade and a half of experience at PHENIX.  We now detail each
subsystem in the following Sections.

\section{Magnet and Tracking}
\label{sec:magnet}

The magnet and tracking system should ultimately be capable of order 1\%
momentum resolution at 10\,GeV/c, cover the full $2\pi$ in azimuth and
$\left| \eta \right| < 1.1$.  Achieving this will require a central
field of up to 2\,T, and in order to minimize the material in front of
the calorimeters, a thin superconducting solenoid is the clear choice
for the magnet.  

A natural question is whether or not an existing superconducting
solenoid might be repurposed for use in sPHENIX.   
The physical constraints of the PHENIX interaction region (IR) have a strong influence on
the dimensions of the sPHENIX design.  The RHIC beamline is 444.8\,cm
above the tracks that are used to move detectors into the collision
hall and 523.2\,cm above the floor, and we propose to keep the track
system in place for maneuvering detectors in and out.  Thus, the
detector should have an outer radius of no more than 400\,cm in order
to provide room for support.  The flux return and
calorimeters require a radial extent of at least 125\,cm, and the
support structure and rollers allowing disassembly of the detector
require on the order of 50\,cm.  Instrumentation in the forward and
backward direction is not part of this proposal (see Appendix~\ref{chap:fsPHENIX}), but the space
available is approximately the same as the present muon tracker
systems.  The north muon magnet was assembled in place and is fixed to
the floor, while the south muon magnet is movable on the rail system.
With the south muon magnet fully retracted to the muon identifier
steel (also fixed to the floor), there is about 520\,cm of open space
available along the beam line.

We have surveyed several completed experiments that used solenoid magnets to see if
their magnet would be suitable.  The basic facts we uncovered in this
survey are are shown in Table~\ref{solenoids-tbl}.  The CDF and BaBar
solenoids have been considered, but they are either too large or
unavailable.  The CLEO solenoid could be available, and is in good
condition, but its size and thickness would also make for an extremely
massive hadronic calorimeter.  The D0 solenoid is slightly smaller in
radius than is desirable to allow for high resolution tracking and
other possible detectors in the future, but it would be a reasonable
candidate for use in sPHENIX.  However, removal of the solenoid from
inside the D0 liquid Argon calorimeter is a sizable task, and after
visiting Fermilab and consulting the D0 decommissioning coordinator,
it was found to be impractical. The result of this exercise drives us
to consider a new magnet, the specifications of which will be
discussed in the next section.

\renewcommand{\arraystretch}{1.9}
\addtolength{\tabcolsep}{-0.5pt}
\begin{table}[hbt!]
\centering
\begin{tabular}{lcccl}
Experiment & Inner radius (cm) & Field strength (T)& Length (cm)& $X/X_0$ \\ 
\toprule
CDF   & 150 & 1.5  & 507 & 0.84 \\
BaBar & 150 & 1.5  & 346 & large \\
CLEO  & 155 & 1.45 & 380 & 2.5 \\
D0    & 55  & 2.0  & 273 & 0.9 \\
\bottomrule
\end{tabular}
\caption{Existing solenoids considered for use in this proposal.}
\label{solenoids-tbl}
\end{table}

\subsection{Solenoid Magnet Specifications}

The basic features of the solenoid were determined by the present and
anticipated future physics needs of sPHENIX to require an
inner radius of at least 70\,cm with a length of 187\,cm to cover a
pseudorapidity of $\left|\eta\right| < 1.1$.  
The thickness of the solenoid and cryostat is to be less than one radiation length
at normal incidence, with a radial thickness of less than 20 cm, design parameters
similar to the D0 solenoid.
The design of the solenoid
will be done by the vendor, and several vendors responded to a
preliminary request for information prepared with Brookhaven National Laboratory's
Procurement Department.

\subsection{Magnetic Field Calculations}

\begin{figure}[htb!]
 \begin{center}
    \includegraphics[width=0.8\linewidth]{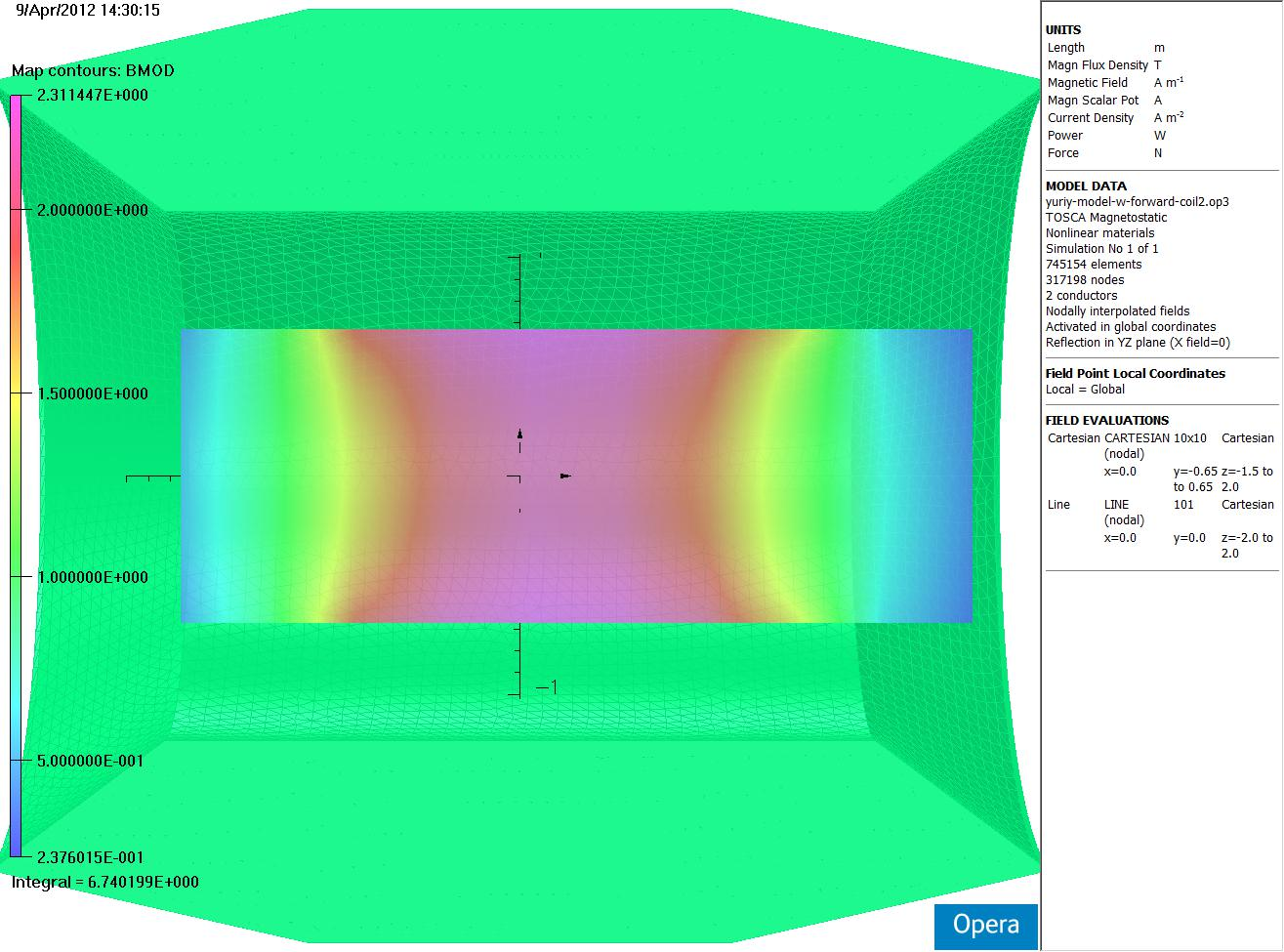}
    \caption{Calculation of the magnetic field from the solenoid with
      the iron flux return.  The coil has been removed for clarity.}
    \label{fig:field-map} 
 \end{center}
\end{figure}

Magnetic field calculations of the solenoid coil and a three
dimensional model of the return steel were carried out with OPERA.  A
field map is shown in Figure \ref{fig:field-map}.

\section{The Electromagnetic Calorimeter}
\label{sec:emcal}

The concept for the sPHENIX electromagnetic calorimeter follows from
the physics requirements outlined earlier in this proposal.  These
requirements lead to a calorimeter design that is compact (i.e. has a
small Moli\`{e}re radius and short radiation length), has a high
degree of segmentation ($0.024\times 0.024$ in $\eta$ and $\phi$), and
can be built at a reasonable cost. Since the calorimeter will be
located just outside the coil of the solenoid, it will also have to
operate in the rather strong fringe field of the magnet.  This has led
us to a so-called optical accordion design, which is a descendant of
the design of the ATLAS lead-liquid argon calorimeter
\cite{ATLASCalorimeters:2008}, but uses tungsten as the absorber
material and scintillating fibers as the active medium. This has the
advantage of being very compact, as described below, and being able to
be read out with silicon photomultipliers (SiPM's), which provide high
gain, similar to conventional phototubes, but can work inside the
magnetic field; avalanche photodiodes (APD's) offer many of the same advantages. 
It is similar to other scintillating fiber
calorimeters which have been built using lead as an
absorber~\cite{Adinolfi:2002zx}. Recently, very good resolution
($\sim$12\%/$\sqrt{E}$) has been obtained with a fiber calorimeter
using tungsten as an absorber~\cite{McNabb:2009dz}.

\begin{figure}[hbt!]
 \centering
    \includegraphics[trim = 0 0 0 550, clip, width=0.9\linewidth]{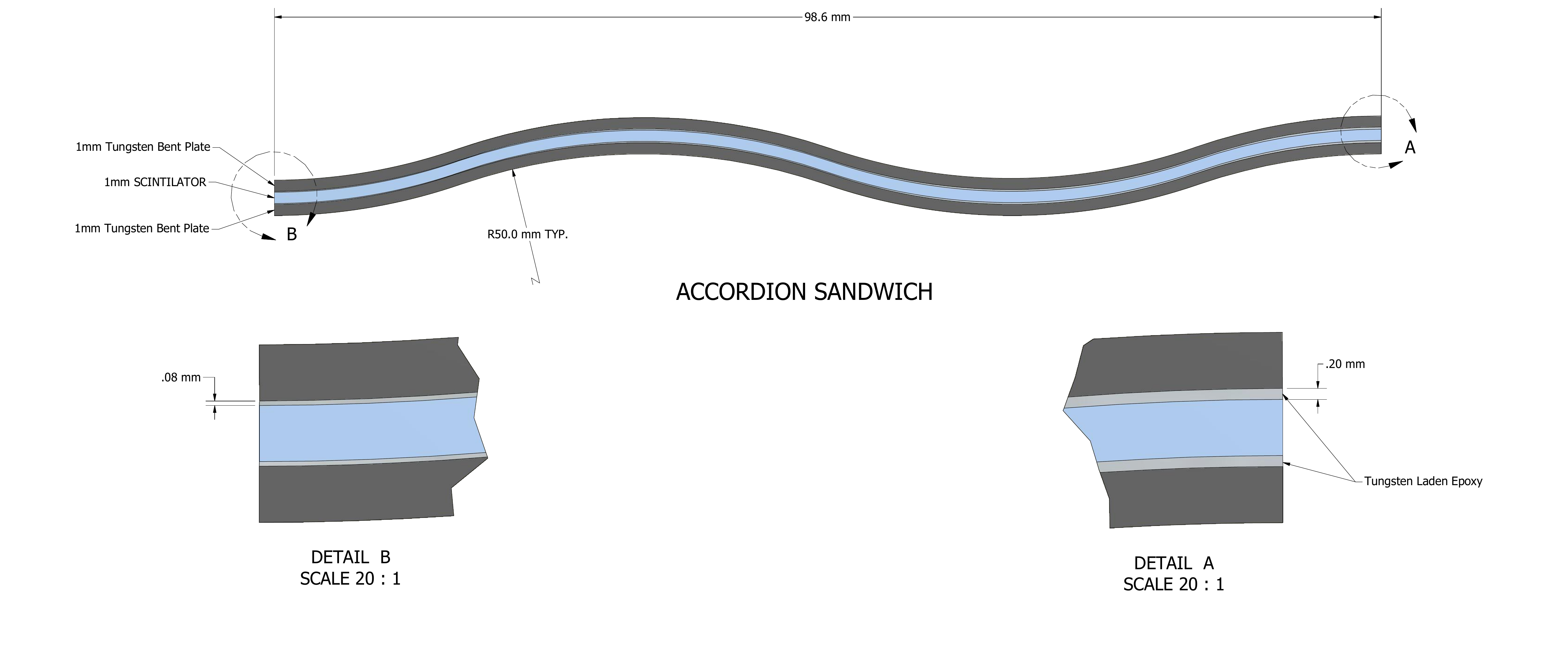}
    \caption{Optical accordion ``sandwich'' consisting of two tungsten 
             plates in an accordion shape (1 mm thickness) and a layer of 1 mm scintillating fibers with
             tungsten powder and epoxy filling the gaps.
             The characteristic accordion-like undulations prevent
             channeling of particles through the scintillator layers alone.}
    \label{fig:Accordion_stack} 
\end{figure}

The EMCal optical accordion will consist of alternating layers of thin
tungsten sheets glued onto cast composite layers consisting of
scintillating fibers embedded in a matrix of tungsten powder and
epoxy. The basic structure is shown in Figure~\ref{fig:Accordion_stack}.
The undulations, characteristic of the accordion design, provide a
more uniform response for particles incident at various positions and
angles by preventing channeling of particles through the
calorimeter---something which could occur if the plates were flat and
a particle traversed the calorimeter interacting only with
scintillator. This design can be made projective in the azimuthal
direction by having the thickness of the layers increase as a function
of radius.  It is not possible to vary the fiber thickness, so one
must increase the absorber thickness, either by increasing the
thickness of the tungsten plates, or by increasing the thickness of
the tungsten-epoxy layer.

\begin{figure}[hbt!]
  \centering
  \includegraphics[width=0.7\linewidth]{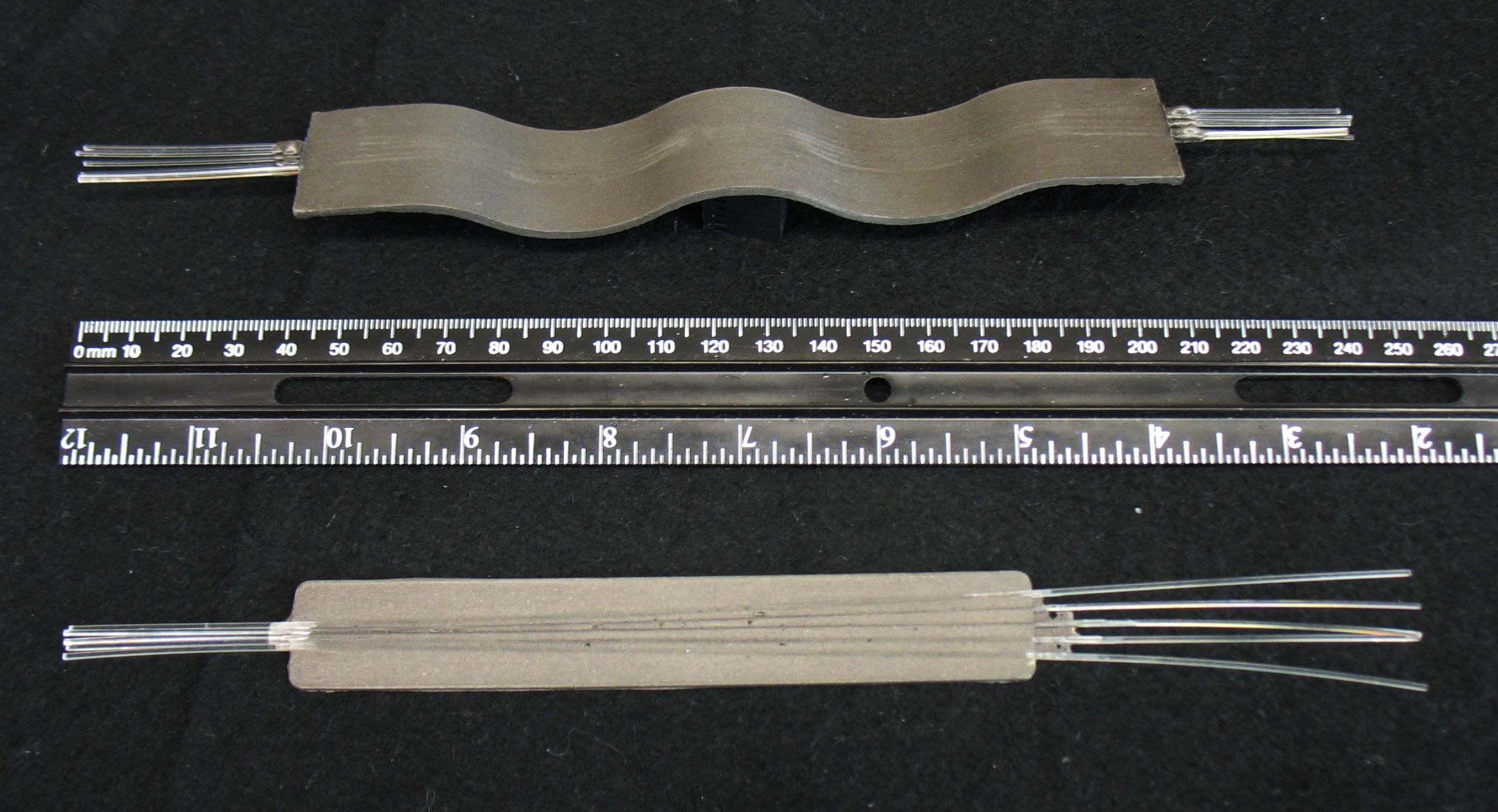}
  \caption{Samples of scintillating fiber embedded in a formed
    tungsten epoxy mixture.  Produced by Tungsten Heavy Powder.}
  \label{fig:THP_samples}
\end{figure}

\begin{figure}[!hbt]
 \begin{center}
   \includegraphics[width=0.7\linewidth]{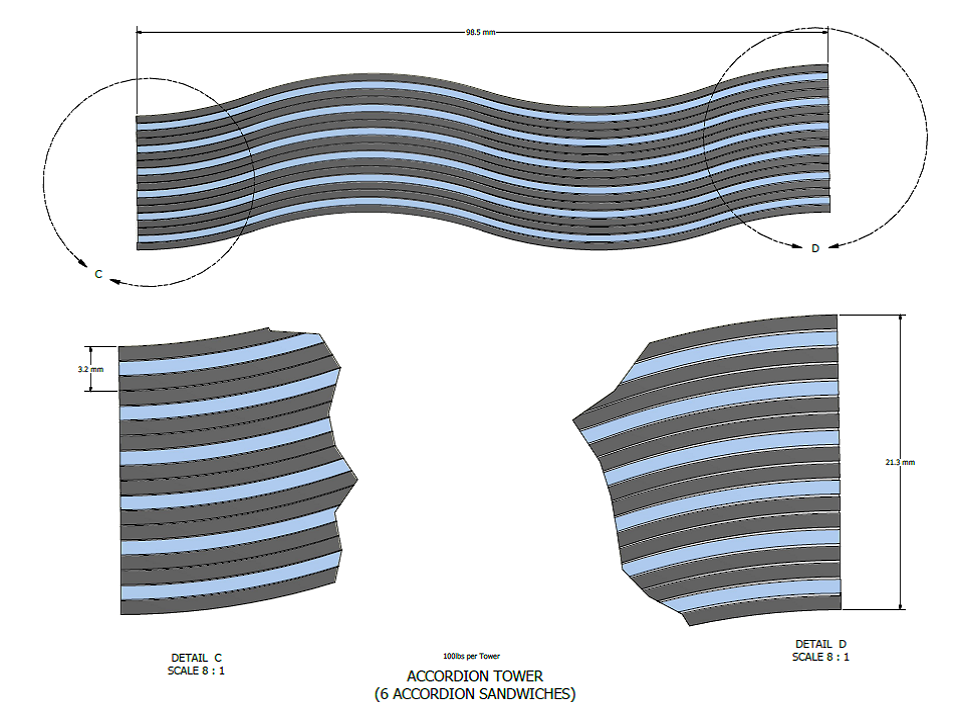}
   \caption{ Cross section of the accordion calorimeter in the plane
     normal to the beam direction, showing how the single layers seen
     in Figure~\ref{fig:Accordion_stack} are stacked. Scintillating
     fibers are embedded in tapered and undulating layers of tungsten
     and epoxy mixture and are approximately projective towards the
     interaction region, which has an extent of $\pm$ 30\,cm along
     the beam direction.}
   \label{fig:Accordion_assembly}
 \end{center}
\end{figure}

\begin{figure}[hbt!]
  \centering
  \includegraphics[width=0.7\linewidth]{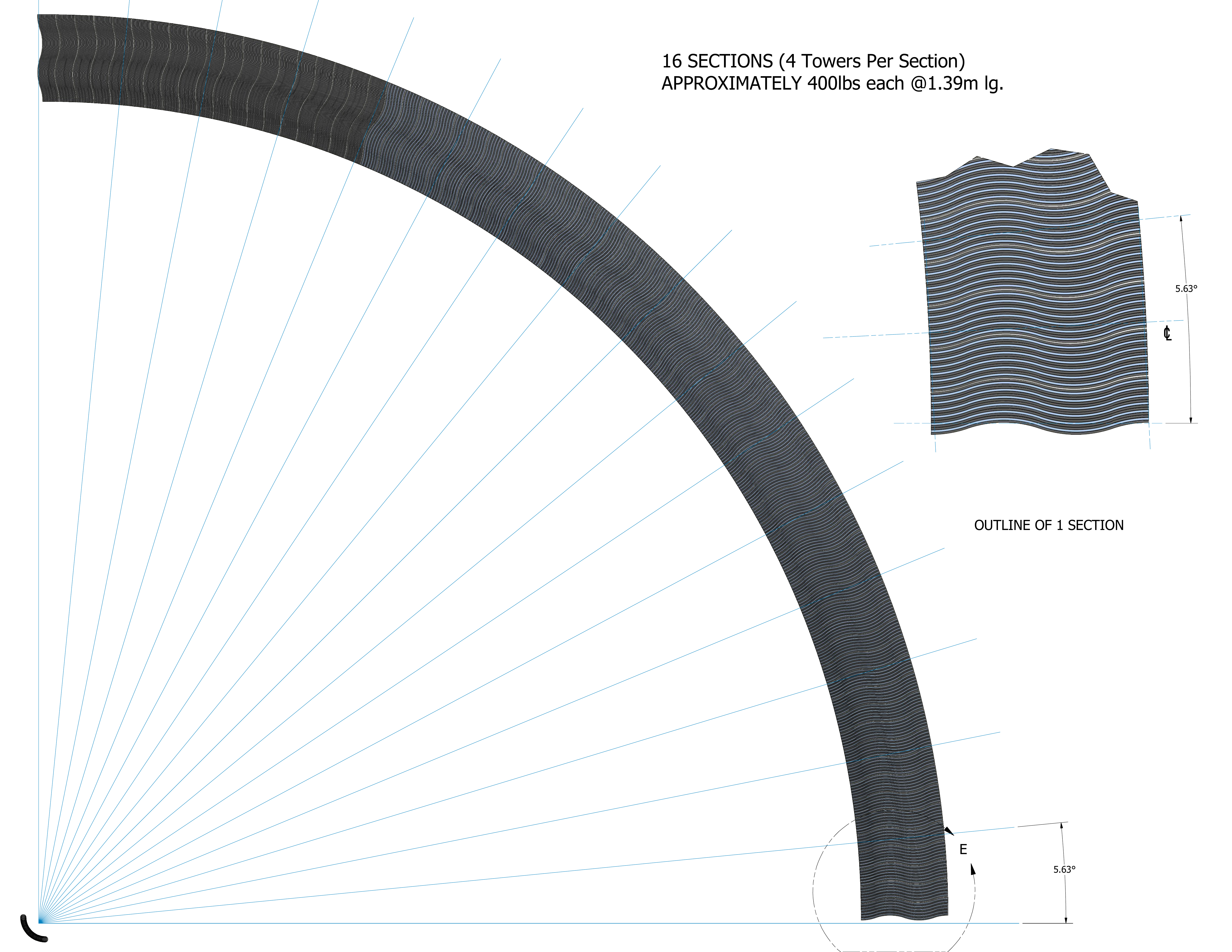}
  \caption{Tower modules combined into sections azimuthally to form a ring.}
  \label{fig:AccordionEMcalLayoutsPage3}
\end{figure}

\begin{figure}[hbt!]
  \centering
  \includegraphics[angle=90, width=0.7\linewidth]{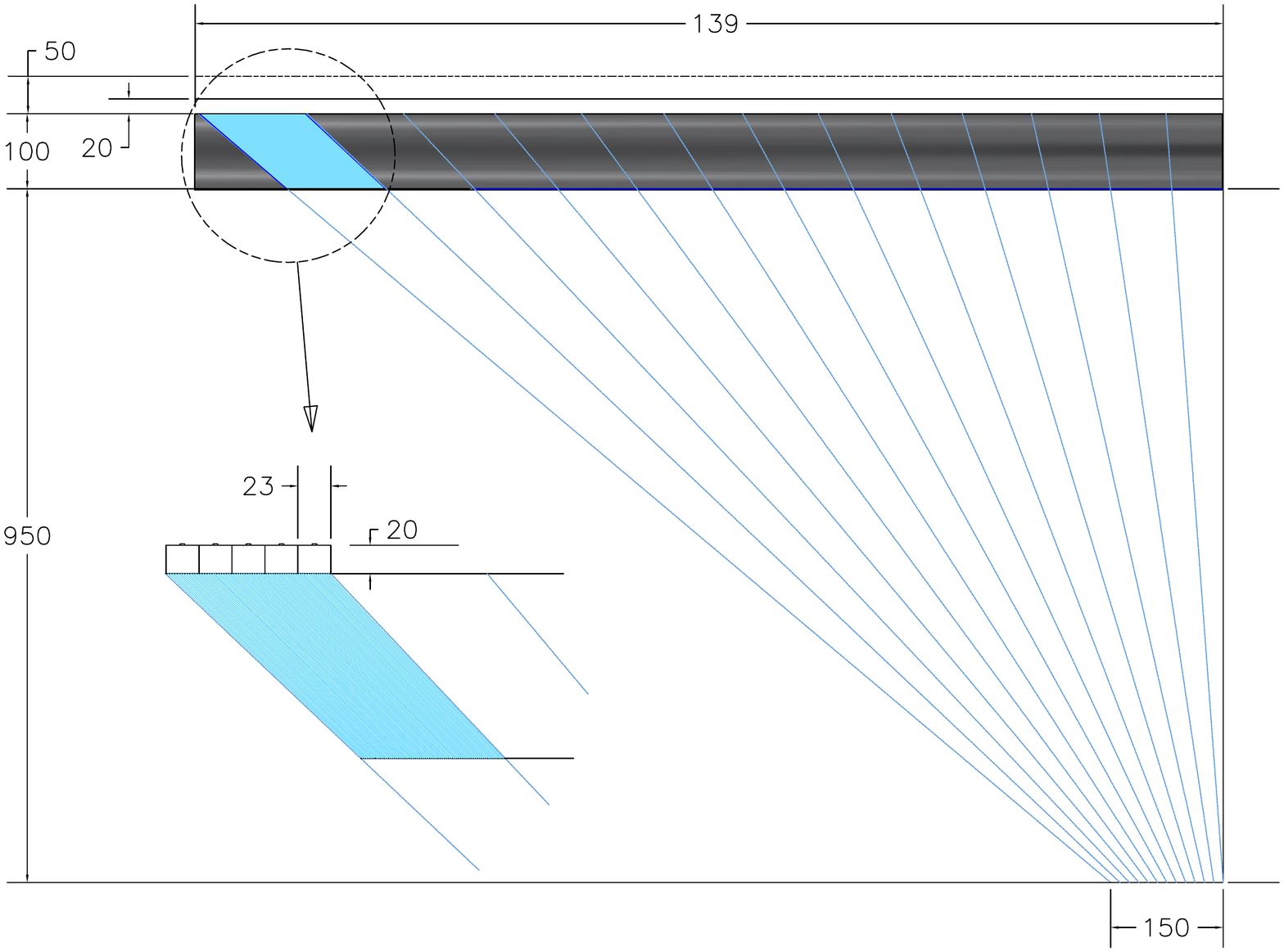}
  \caption{View of the calorimeter along the beam direction showing approximately 
    projective towers back to the interaction region, which has an extent of $\sim \pm$ 30 cm.}
  \label{fig:AccordionEMcalLayoutsPage6}
\end{figure}

Until recently, it had only been possible to achieve an accordion
shape for absorbers made of highly malleable materials such as lead.
New technology now makes it possible to achieve this with tungsten,
which results in a higher density, and hence more compact,
calorimeter. We have been working with a company, Tungsten Heavy
Powder~\cite{THP}, that fabricates a wide variety of tungsten
components, to produce a practical, cost effective design for the
calorimeter that would allow it to be manufactured in private
industry. 
Based on our discussions with them, it appears that the simplest and
most economical method to achieve this design is to use thin, uniform
thickness tungsten plates and scintillating fibers with a tungsten
powder epoxy polymer to fill the gaps. In this design, shown in Figure~\ref{fig:Accordion_stack},
two uniform thickness tungsten plates
with a thickness $\sim$ 1 mm would be pre-bent into the accordion
shape and cast together with a layer of scintillating fibers and a
mixture of tungsten powder and epoxy in a mold to form a ``sandwich''
with the desired shape. 
Figure~\ref{fig:THP_samples} shows an example of scintillating fibers embedded into an 
epoxy layer with tungsten powder and formed into an accordion shape.
Six layers of these sandwiches would be glued
together to form a tower module measuring $\sim 2.1$\,cm in the $\phi$
dimension and half the length of the calorimeter
($\sim 1.39$\,m) along the beam direction, as shown in Figure~\ref{fig:Accordion_assembly}. 
Four tower modules would then
be combined into sections weighing about 180\,kg each and arranged
azimuthally to form a ring, as shown in Figure~\ref{fig:AccordionEMcalLayoutsPage3}.
Figure ~\ref{fig:AccordionEMcalLayoutsPage6} shows a cross sectional view of
the calorimeter along the beam direction. The fibers
are arranged in a radial pattern projective to the vertex.
The fibers are closely spaced together
at the front of the calorimeter and flare out slightly towards the
back in order to make the device projective. The fibers are grouped at
the back into individual towers (corresponding to the $\eta$ and
$\phi$ segmentation as discussed below in the readout section) and
coupled to a light mixer box that randomizes and collects the light
from all of the fibers of a given tower onto a single SiPM.

\subsection{Segmentation and readout}

The segmentation of the calorimeter is determined by a number of
different requirements. One primary factor is the occupancy of the
individual readout towers in heavy ion collisions, which determines
the ability to resolve neighboring showers and to measure the energy
in the underlying event.  In addition, the degree of segmentation also
determines the ability to measure the transverse shower shape, which
is used in separating electrons from hadrons (e/$\pi$ rejection). All
of these capabilities could be improved with the addition of a finely
segmented preshower detector (as detailed in Appendix~\ref{chap:barrel_upgrade}), 
but we believe the segmentation chosen
for the baseline detector will provide the capability to perform the
physics program of this proposal.

The calorimeter will be divided into individual towers corresponding
to a segmentation in $\eta$ and $\phi$ of approximately
$0.024\times0.024$ and would result in about 25,000 readout channels 
(256 in $\phi \times$ 96 in $\eta$). 
The fibers from the back of the calorimeter will be
grouped into towers measuring $\sim$ 2$\times$2 cm where the light
from $\sim$ 125 fibers will be collected and randomized using a small
light mixer box and read out with a single SiPM.  It has not yet been
decided how this light collection and mixing will be accomplished, but
a number of options are being explored. These include a small
reflecting and diffusing cavity, or possibly a wavelength shifting
block. We will keep the thickness of the mixer as thin as possible in
order to minimize the radial space required by the mixer, SiPM and
readout electronics.

The light yield resulting from the mixer configuration is of primary
importance in determining the photostatistics for the
readout. Fortunately, with an energy resolution of 15\%/$\sqrt{E}$,
the requirement on the light yield is not very severe. We have made a
number of measurements with scintillating fibers that have been
embedded into various mixtures of tungsten powder and epoxy to
determine their light yield, and have obtained light yields $\sim$ 100
photoelectrons/MeV of energy deposit in the scintillator with a SiPM reading out
the fibers directly With the thicknesses of the tungsten plates,
scintillator and tungsten powder epoxy in the current design, the
sampling fraction for the energy deposited in the scintillator is
$\sim$ 4\%, so this corresponds to $\sim$ 4000 photoelectrons/GeV of energy
deposit in the calorimeter, which would have a negligible effect on
the calorimeter energy resolution.  This number will be reduced by the
light collection efficiency of the mixer, but with this
high initial light yield, it should be possible to maintain sufficient
photostatistics so as not to affect the overall energy resolution of
the calorimeter.

The PHENIX collaboration has been working with the company Tungsten Heavy Powder~\cite{THP}
on the design and fabrication of actual calorimeter components.  Tungsten Heavy Powder
has also recently received a Phase I Small Business Innovation Research (SBIR) grant to study
and develop materials and components for compact tungsten based calorimeters for nuclear physics
applications.  Members of the sPHENIX group, as part of a broader collaboration, have submitted
a ``Joint Proposal to Develop Calorimeters for the Electron Ion Collider'' for EIC research and development funds.
Thus, this technology is an area of very active work and for which test beam results for
the sPHENIX type design should be available soon.



\section{The Hadronic Calorimeter}
\label{sec:hcal}

The hadronic calorimeter is a key element of sPHENIX and many of the
overall performance requirements are directly tied to performance
requirements of the HCal itself.  The focus on measuring jets and
dijets in sPHENIX leads to a requirement on the energy resolution of
the calorimeter system as a whole---the particular requirement on the
HCal is that it have an energy resolution better than $\sigma_E/E =
100\%/\sqrt{E}$.
The jet measurement requirements
also lead to a transverse segmentation requirement of $\Delta \eta
\times \Delta \phi \approx 0.1 \times 0.1$ over a rapidity range of
$|\eta|<1.1$ with minimal dead area.

\begin{figure}[htb!]
 \begin{center}
    \includegraphics[width=0.7\linewidth]{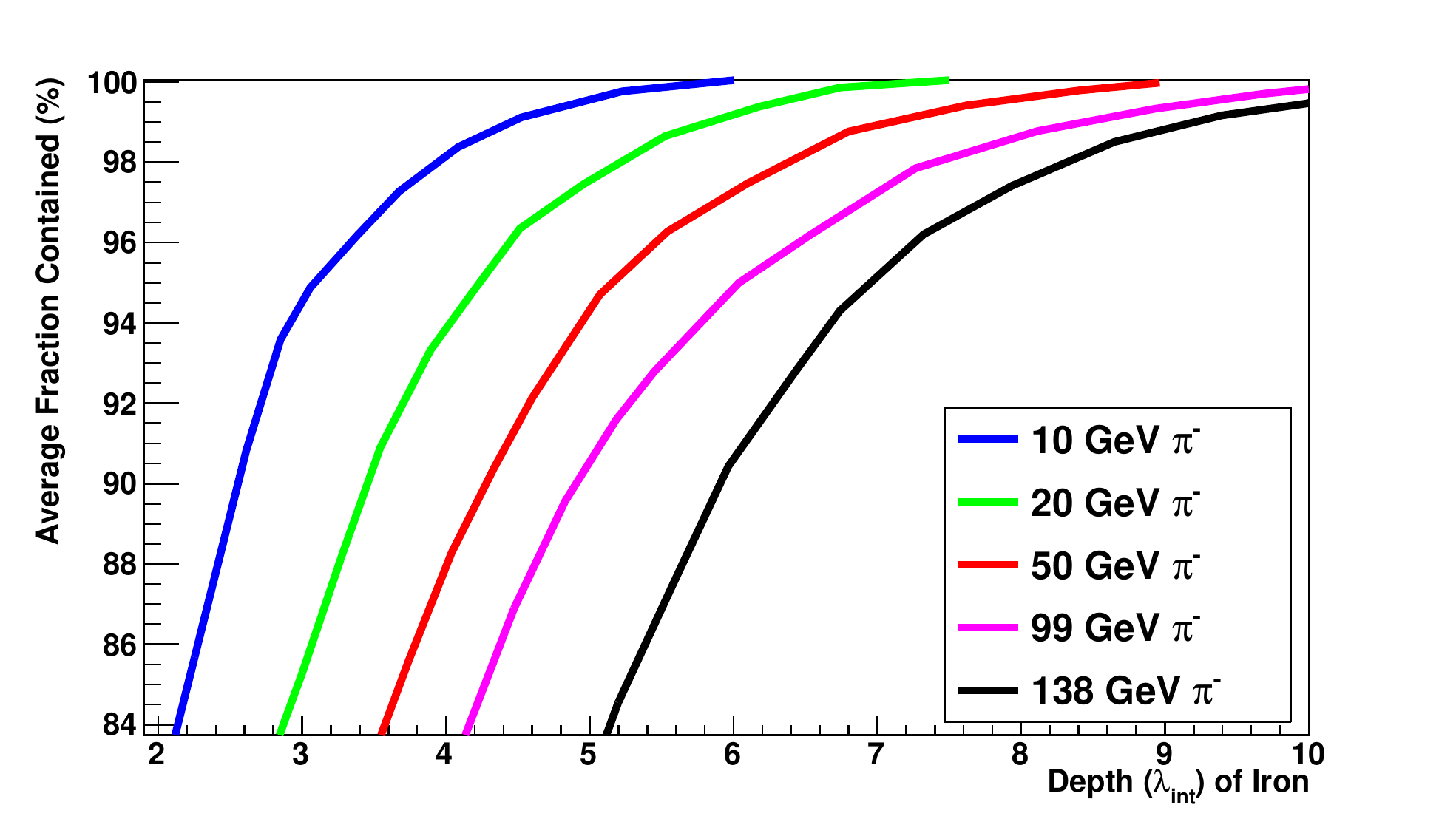}
    \caption{Average energy fraction contained in a block of iron with
      infinite transverse dimensions, as a function of the thickness
      of the block.  Figure adapted from Ref.~\protect\cite{Wigmans:2000vf}{}.}

    \label{fig:HC-containment} 

 \end{center}
\end{figure}

The combination of the EMCal and the HCal needs to be at least
$\sim6\lambda_\mathrm{int}$ deep---sufficient to absorb $\sim 97\%$ of the energy
of impinging hadrons with momenta below 50\,GeV/c, as shown in
Figure~\ref{fig:HC-containment}.  The electromagnetic calorimeter is
~$\sim 1\lambda_\mathrm{int}$ thick, so an iron-scintillator hadronic
calorimeter should be~$\sim5\lambda_\mathrm{int}$ deep. The thickness of the
HCal is driven by physics needs, but these needs dictate a device of
sufficient thickness that, with careful design, the hadronic
calorimeter can also serve as the return yoke for the solenoid.

\begin{figure}[htb!]
  \begin{center}
    \includegraphics[trim=20 20 20 20, clip, width=0.8\linewidth]{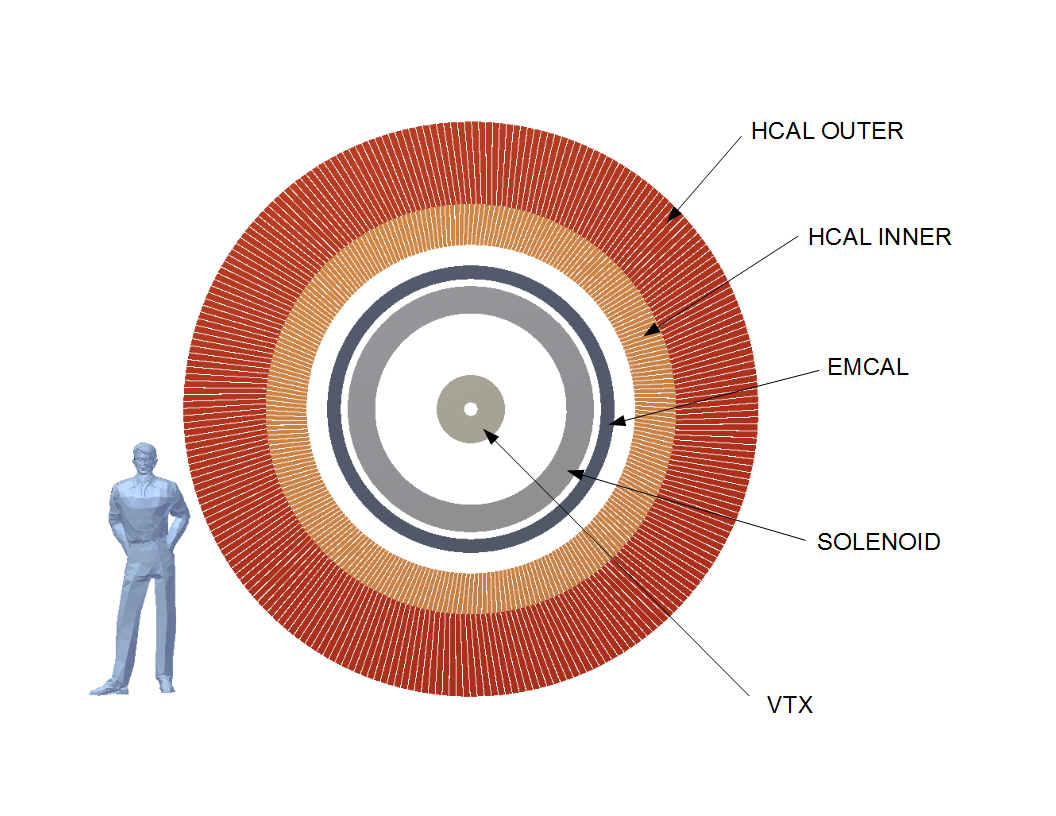}
    \caption{Cross section of sPHENIX, showing a typical collider
      detector structure.  The calorimeters sit just outside the
      solonoid. }

    \label{fig:sPHENIX-quadrant}

  \end{center}
\end{figure}

The hadronic calorimeter as shown in Figure~\ref{fig:sPHENIX-quadrant} surrounds the electromagnetic calorimeter
with an active volume which extends from a radius of ~112\,cm to
212\,cm and is segmented longitudinally (i.e., along a radius vector)
into two compartments of 1.5 and 3.5 interaction lengths deep (at
normal incidence).  The inner and outer sections share the total
energy of a shower approximately equally.

\begin{figure}[htb!]
 \begin{center}
   \includegraphics[trim = 0 0 0 100, clip, width=0.7\linewidth]{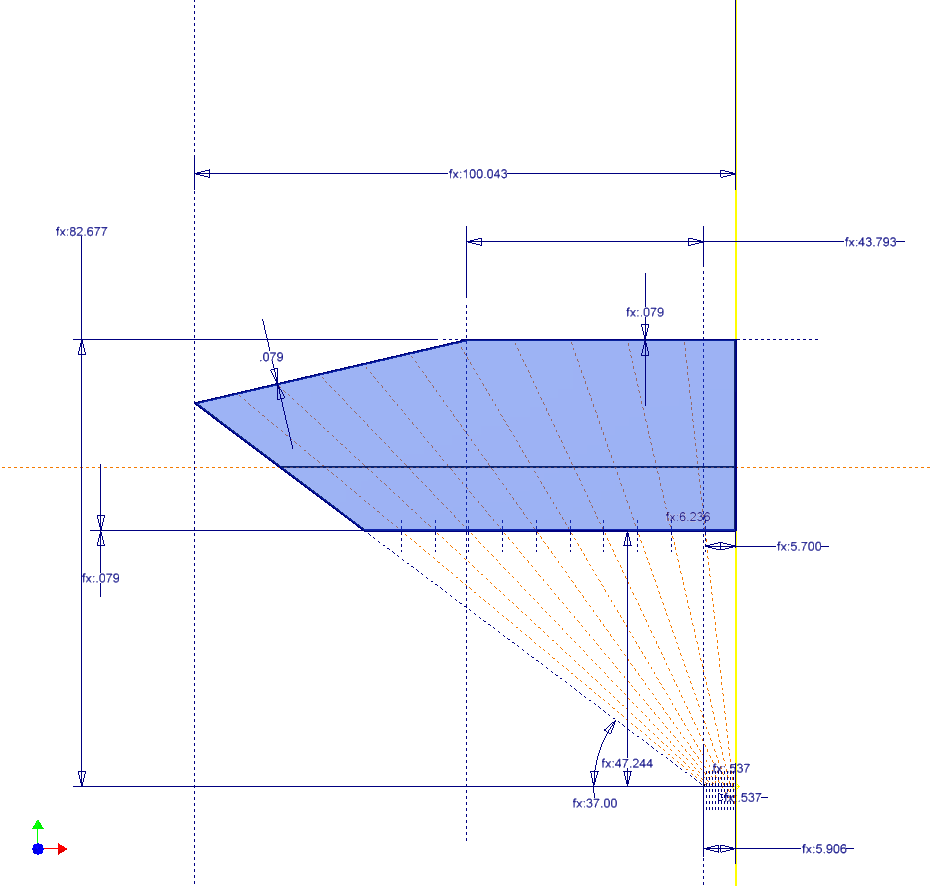}
   \caption{Scintillating tiles in the sampling gap of sPHENIX
     hadronic calorimeter, showing the transverse segmentation into
     element 0.1 units of pseudorapidity wide.}
   \label{fig:HC-TiledLayer} 
 \end{center}
\end{figure}

Both the inner and outer longitudinal segments of the calorimeter are
constructed of tapered absorber plates, creating a finned structure,
with each fin oriented at an angle of $\pm 5^\circ$ with respect to a
radius vector perpendicular to the beam axis.  There are 256 fins in
each of the inner and outer segments.  The fins in the inner and outer
segments are radially tilted in opposite directions resulting in a 10
degree angle with respect to each other.  They are also staggered by
half a fin thickness.  The gaps between the iron plates are 8\,mm wide and contain
individually wrapped 7\,mm thick scintillating tiles with a diffuse
reflective coating and embedded wavelength shifting fibers following
a serpentine path.  
The slight tilt and the azimuthal staggering of steel fins and scintillating
tiles prevents particles from traversing the depth of the calorimeter
without encountering the steel absorber.  
The benefits of two longitudinal segments include a further reduction in the 
channeling of particles in the scintillator, shorter scintillators with embedded fibers for
collecting the light, and shower depth information.

With plates oriented as described, particles striking the calorimeter at normal incidence will,
on average, cross 22.5\,cm of steel in the inner and 57.5\,cm of steel
in the outer sections resulting in total probability for the punch
through of particles with momenta above $\sim 2$\,GeV/c of only 1\% as 
confirmed with \geant simulations.  This punch through probability
varies from 0.93--1.07 depending on the incident angle across the sampling cell.
This design has a very small number of distinct components which is
designed to make it simple to fabricate, assemble, and to model.

Within each gap, there are 22 separate scintillator tiles of 11
different shapes, corresponding to a detector segmentation in
pseudorapidity of $\Delta\eta\sim0.1$ (see
Figure~\ref{fig:HC-TiledLayer}).  Azimuthally, the hadronic
calorimeter is divided into 64 wedges ($\Delta\phi\simeq 0.1$).  Each
wedge is composed of four sampling cells (steel plate and
scintillating tile) with the scintillating tile edges pointing towards
the origin.  The 22 pseudorapidity slices result in towers about
10\,cm$\times 10$\,cm in size at the inner surface of the
calorimeter. The total channel count in the calorimeter is
$1408 \times 2$.

\begin{figure}[htb!]
  \begin{center}
    \includegraphics[width=0.7\linewidth]{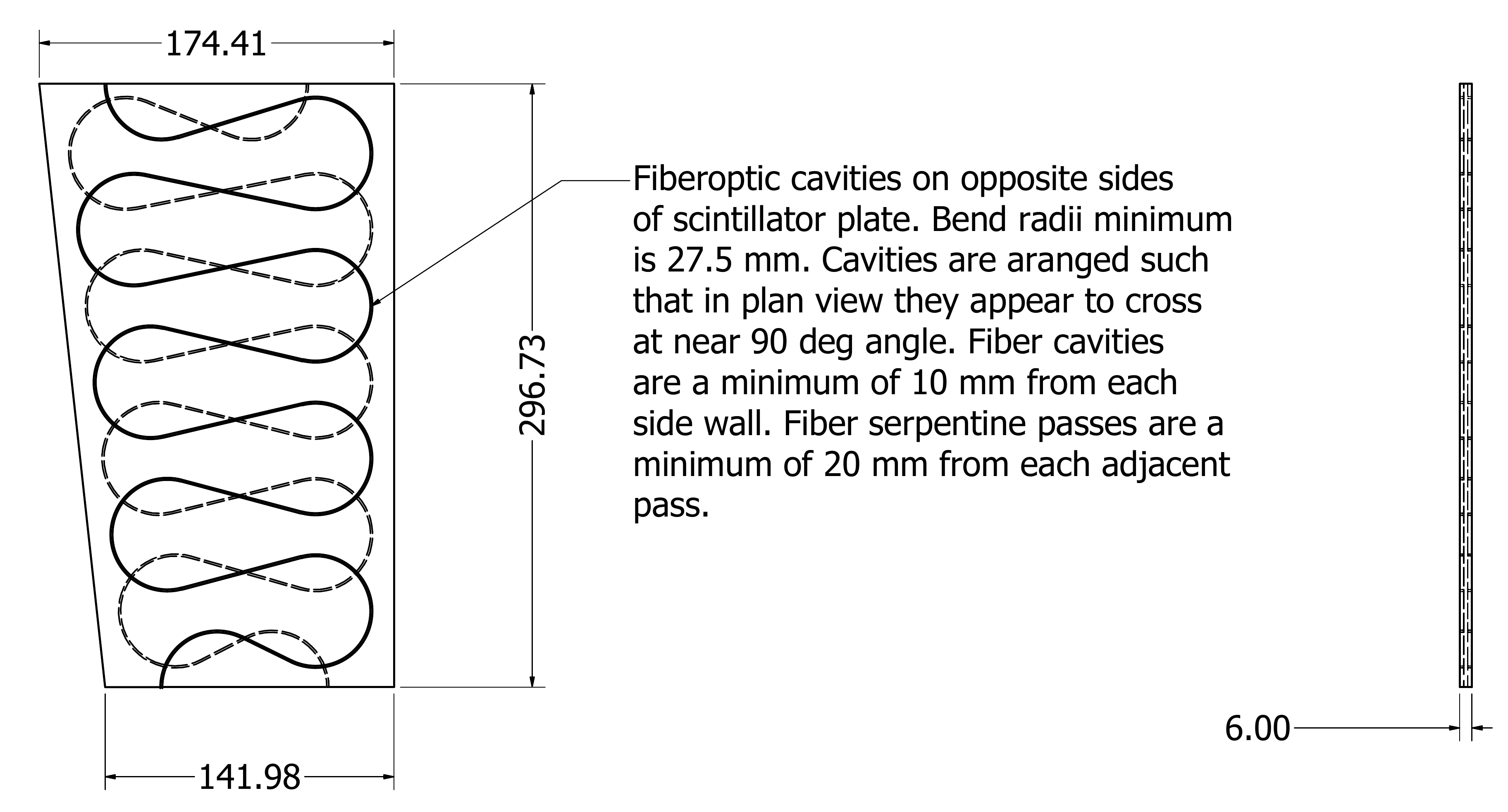}
    \caption{Grooved scintillating tiles for inner HCal section,
      showing the serpentine pathway the fiber will follow and the
      uniform thickness of the tiles.}
    \label{fig:HC-Tiles} 
  \end{center}
\end{figure}

The light from the scintillating tiles between the steel fins is
collected using wavelength shifting fibers laid along a serpentine
path as shown in Figure~\ref{fig:HC-Tiles}.  This arrangement provides
relatively uniform light collection efficiency over the whole tile.
We have considered two fiber manufacturers: (1) Saint-Gobain (formerly
BICRON), product brand name BCF91A ~\cite{Saint-Gobain} and (2)
Kuraray, product name Y11~\cite{Kuraray}.  Both vendors offer single
and double clad fibers.

The calorimeter performance is determined by the sampling fraction and
the light collection and readout efficiency.  The readout contributes
mostly to the stochastic term in calorimeter resolution through
Poisson fluctuations in the number of photoelectrons on the input to
the analog signal processing.  Factors contributing to those
fluctuations are luminous properties of the scintillator, efficiency of
the light collection and transmission, and the photon detection
efficiency of the photon detector.

The scintillating tiles are based on the design of scintillators for
the T2K experiment by the INR group (Troitzk, Russia) who designed and
built 875\,mm long scintillation tiles with a serpentine wavelength
shifting fiber readout~\cite{Izmaylov:2009jq}.  The T2K tiles are
injection molded polystyrene tiles of a geometry similar to those
envisioned for sPHENIX, read out with a single serpentine fiber, with
each fiber viewed by an SiPM on each end.  The measured light yield
value was 12 to 20 photoelectrons/MIP at
20$^\circ$C~\cite{Mineev:2006ek}.  With 12\,p.e./MIP measured by T2K
for 7\,mm thick tiles (deposited energy $\sim1.4$\,MeV) and an average
sampling fraction of 4\% estimated for the sPHENIX HCal we expect the
light yield from the HCal to be about 400\,p.e./GeV.  A 40\,GeV hadron
will share its energy nearly equally between the inner and outer HCal
segments so the upper limit of the dynamic range of the HCal can be
safely set to $\sim 30$\,GeV which corresponds to a yield of
12000\,p.e. on the SiPM.  To avoid signal saturation and ensure
uniform light collection, care will be required to both calibrate the
light yield (possibly requiring some attenuation) and randomize it.

\begin{figure}[tb]
  \begin{center}
    \includegraphics[width=0.95\linewidth]{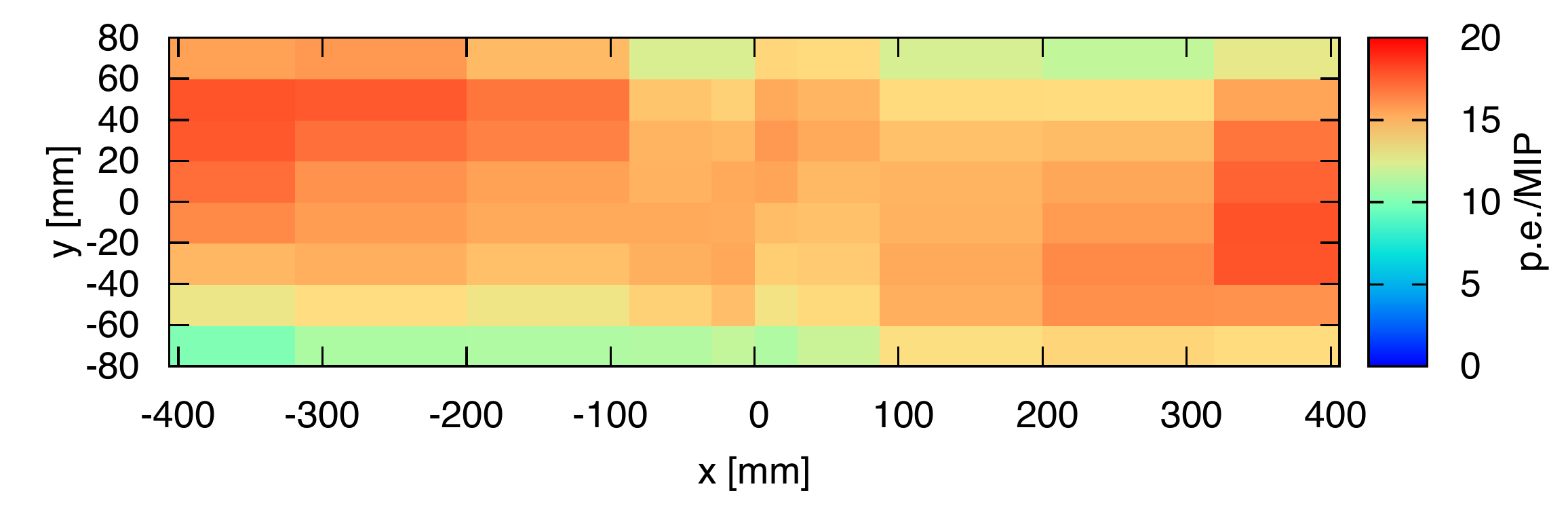}
    \caption{ Light yield (photoelectrons) profile for the T2K
      scintillation tile, showing the degree of uniformity achieved
      using a serpentine fiber layout~\protect\cite{Mineev:2006ek}{}.}
    
    \label{fig:HC-LYProfile}
  \end{center}
\end{figure}

The uniformity of light collection as measured by T2K using the serpentine fiber
arrangement can be judged from Figure~\ref{fig:HC-LYProfile} (data
from Ref.~\cite{Mineev:2006ek}).  The largest drop in the light yield is
along the tile edges and in the corners farthest from the fibers.

We note that this design is optimized for simplicity of manufacturing,
good light yield, and also acting as the flux return for the magnetic.
As such, it has a manifestly non-uniform sampling fraction as a
function of depth and is not highly compensated.  However, the
performance specifications are quite different from particle physics
hadronic calorimeters, particularly with a limited energy measurement
range (0--70\,GeV).  \geant simulations described in the next Section
indicate a performance better than the physics requirements, and full
test beam results are necessary as a validation.

\section{Simulations\label{sec:g4sim}}

We have employed the \geant simulation toolkit
\cite{Agostinelli:2002hh} for our full detector simulations. It
provides collections of physics processes suitable for different
applications. We selected the QGSP\_BERT list which is recommended for
high energy detector simulations like the LHC experiments.  We have
integrated the sPHENIX simulations with the PHENIX software framework,
enabling us to use other analysis tools we have developed for PHENIX.

The detectors and readout electronics and support structures are
currently implemented as cylinders.  The setup is highly configurable,
making it easy to test various geometries and detector concepts.  The
simulation is run using a uniform solenoidal field of 2\,T. We keep
track of each particle and its descendants so every energy deposition
can be traced back to the original particle from the event generator.

The existing PHENIX silicon vertex detector (VTX) consists of four inner silicon layers
at a radius of 2.5\,cm (200\,$\mu$m), 5\,cm (200\,$\mu$m), 10\,cm
(620\,$\mu$m), 14\,cm (620\,$\mu$m).
These are followed by a one radiation thick layer of aluminum which
represents the effects of the superconducting magnet.  The
electromagnetic and hadronic calorimeters are implemented as tungsten
and iron cylinders respectively in which 1792 (EMCal) and 320 (HCal)
scintillator plates of 1\,mm (EMCal) and 6\,mm (HCal) thickness with a
configurable tilt angle (currently 5$^\circ$) are embedded. The
readout electronics for the EMCal is approximated by 5\,mm of Teflon
following the EMCal.

All tracks which reach a layer 10\,cm behind the HCal are aborted to
prevent particles which are curled up by the 2\,T field from
re-entering the detector.  Adding up the energy of those aborted
tracks yields an estimate of the energy which is leaked from the back
of the HCal.

We have two algorithms to account for the granularity of the detectors
and their readout. For the silicon layers the deposited energy is
summed using a given strip or pixel size. The dimensions for the inner
2 layers are 0.05\,mm$\times$0.425\,mm, layer 3 and 4 are
0.08\,mm$\times$1\,mm and layers 5 and 6 are 0.08\,cm$\times$3\,cm.
The energy deposited in the scintillators of the calorimeters is
summed in equal sized bins of pseudorapidity and azimuthal angle.  The
size for the EMCal is $0.024\times0.024$, the size for the HCal is
$0.1\times0.1$.

\subsection{Electromagnetic Calorimeter Simulation}

\begin{figure}[htb!]
 \begin{center}
    \includegraphics[width=0.7\linewidth]{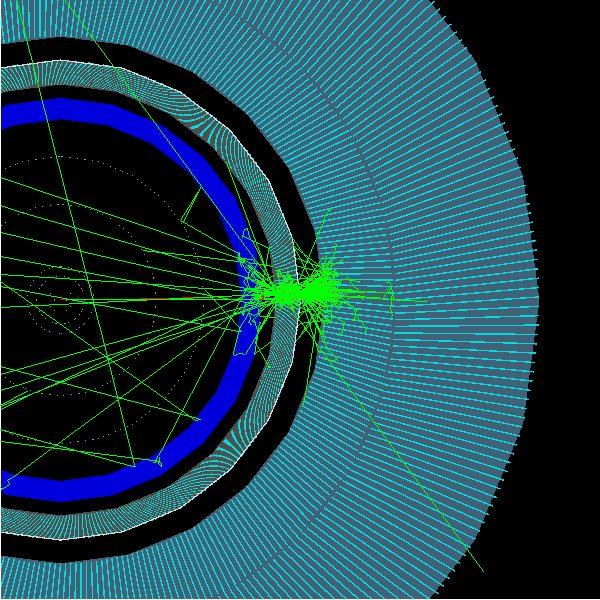}
    \caption{\label{fig:elec} Transverse view of a 10\,GeV/c electron in
      sPHENIX. It penetrates the magnet (blue) and showers mainly in
      the EMCal.}
 \end{center}
\end{figure}

\begin{figure}[htb!]
 \begin{center}
    \includegraphics[trim = 2 2 2 2, clip, width=\twowidth]{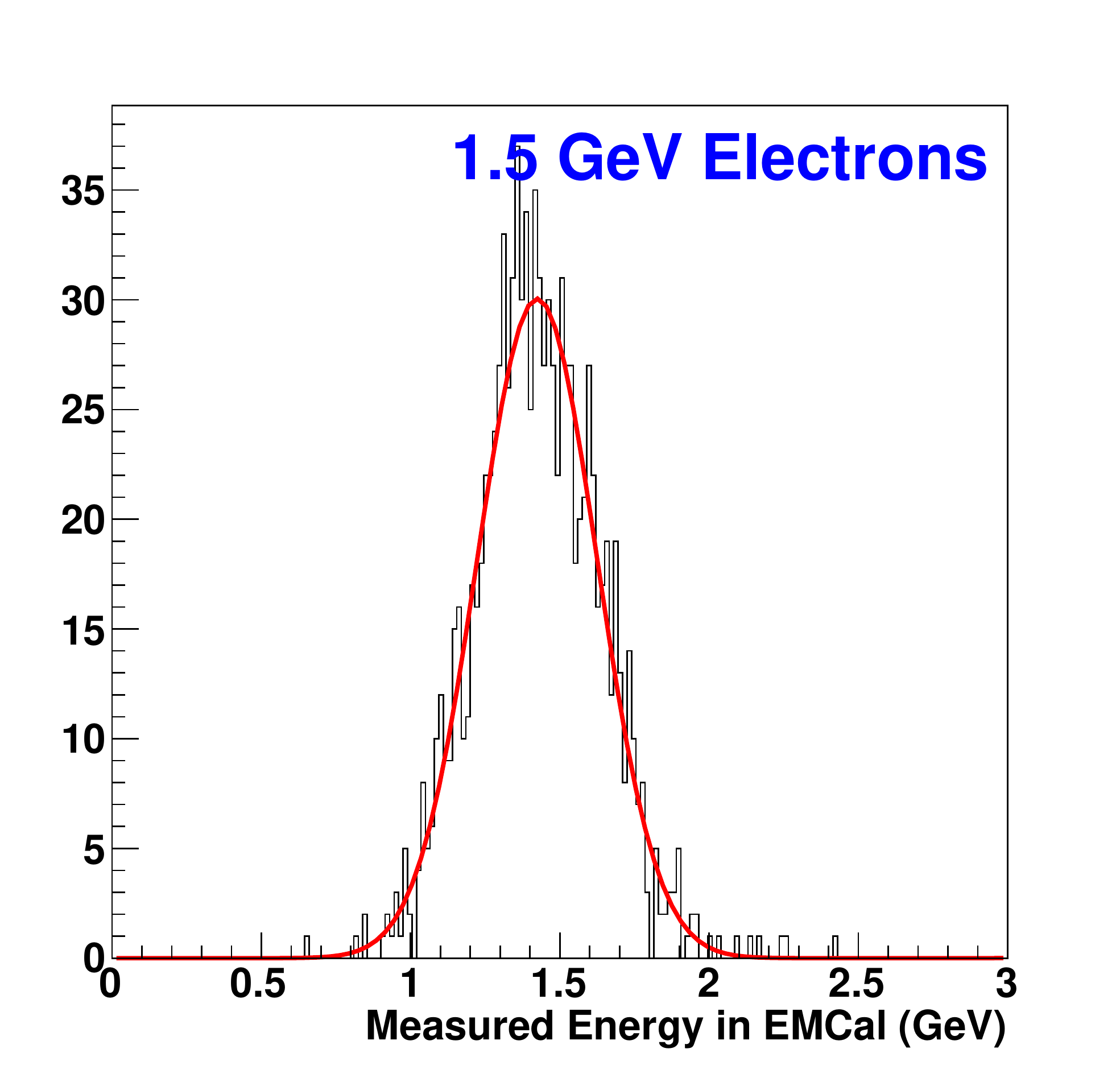}
    \hfill
    \includegraphics[trim = 2 2 2 2, clip, width=\twowidth]{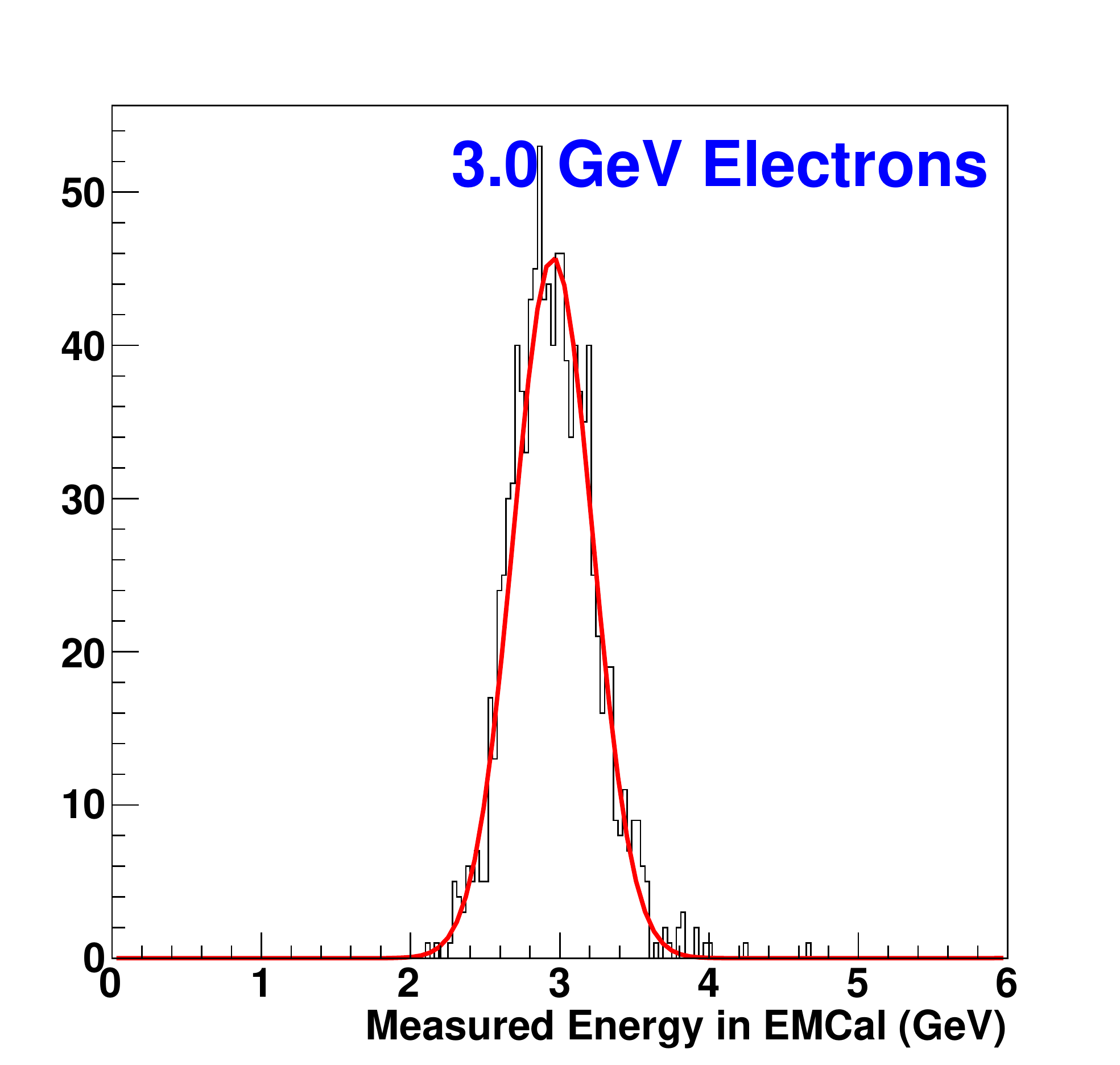}
    \\
    \includegraphics[trim = 2 2 2 2, clip, width=\twowidth]{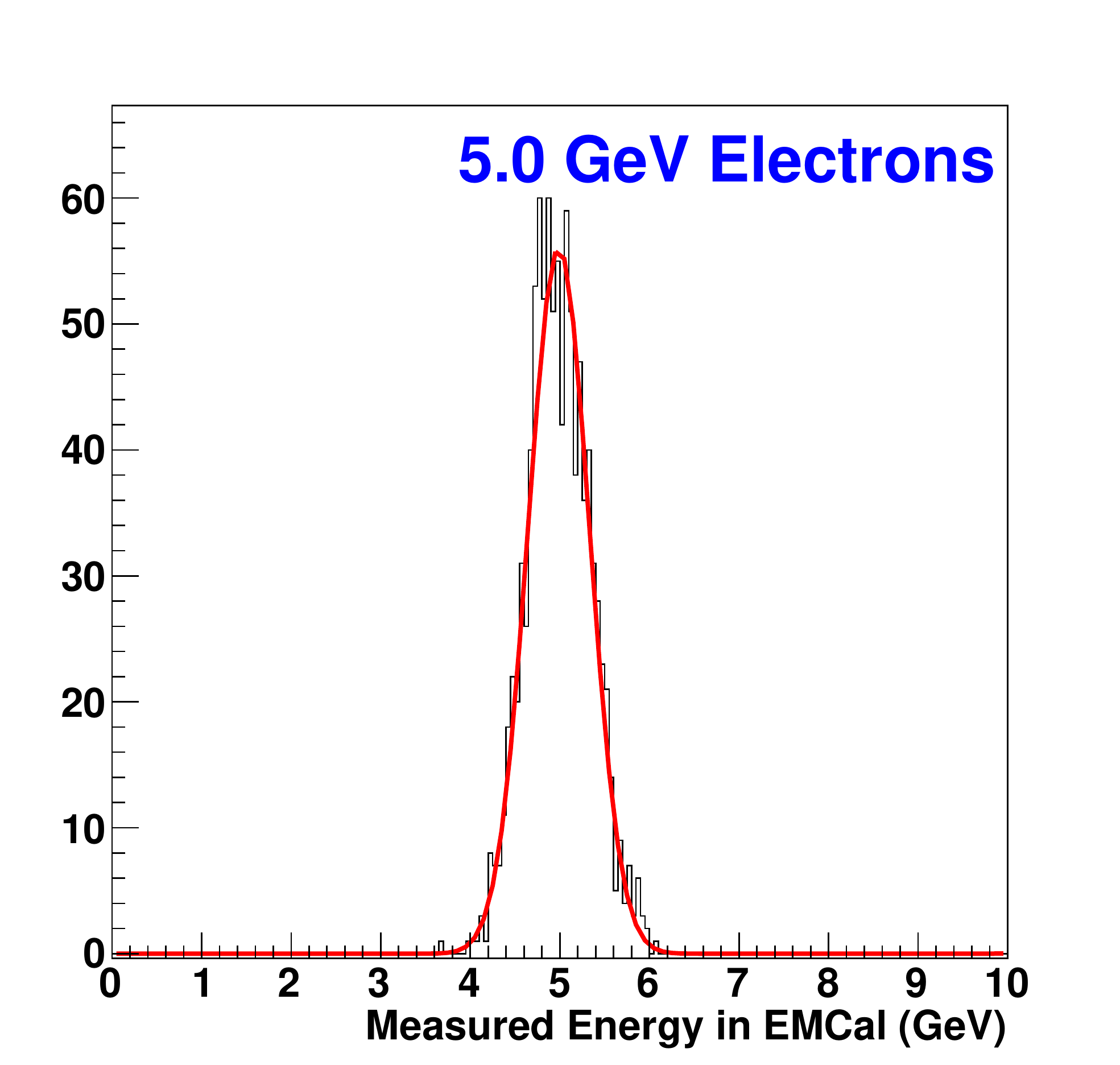}
    \hfill
    \includegraphics[trim = 2 2 2 2, clip, width=\twowidth]{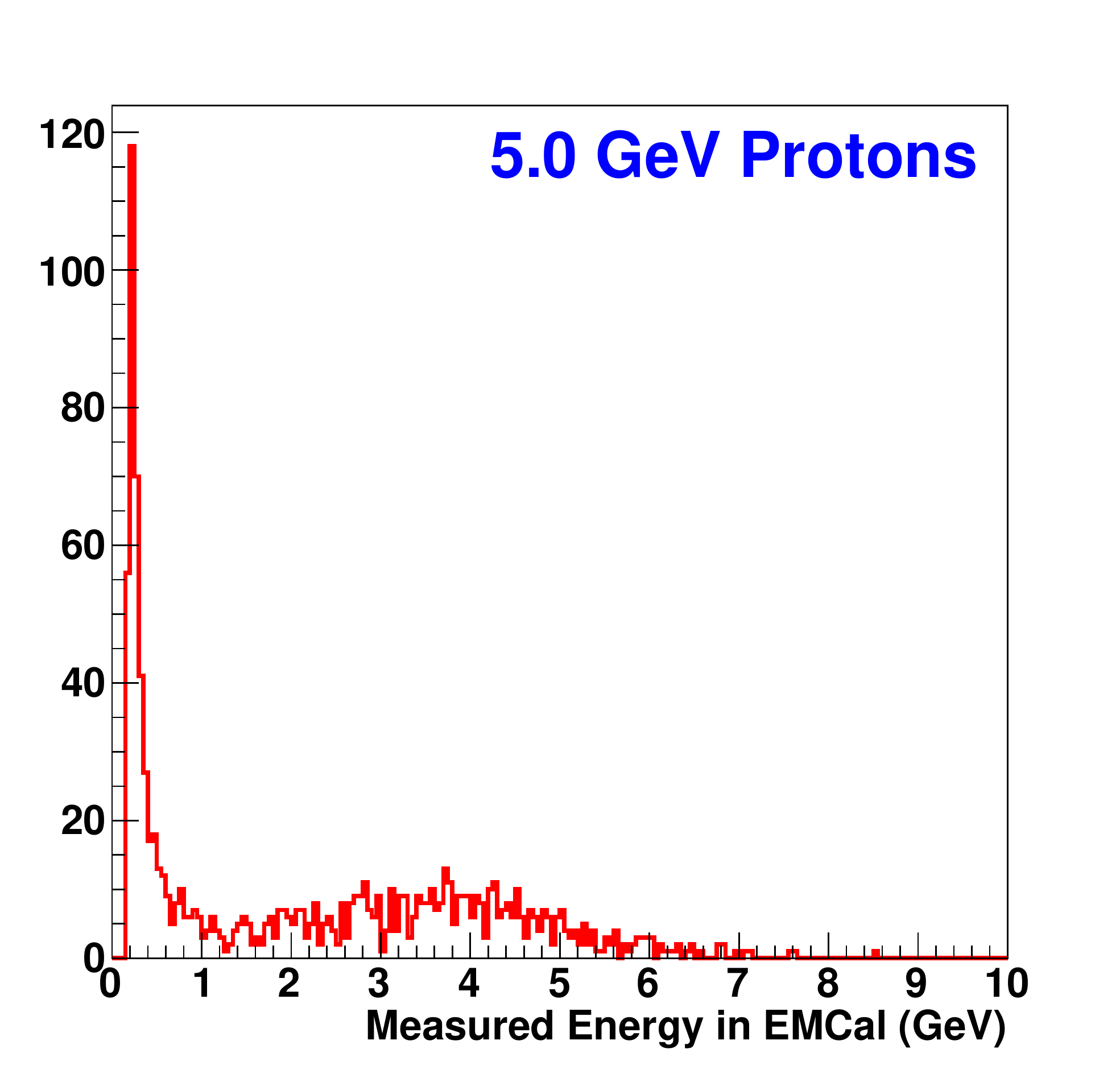}
    \caption{\label{fig:emcal_fullG4} Response of the electromagnetic
      calorimeter to 1.5,~3, and~5\,GeV electrons and 5\,GeV
      protons.  For the protons there is a large minimum ionizing particle (MIP) peak and a broad
distribution corresponding to cases where the proton induces an hadronic shower at some depth
into the EMCal.}
 \end{center}
\end{figure}

\begin{figure}[htb!]
 \begin{center}
    \includegraphics[width=\onewidth]{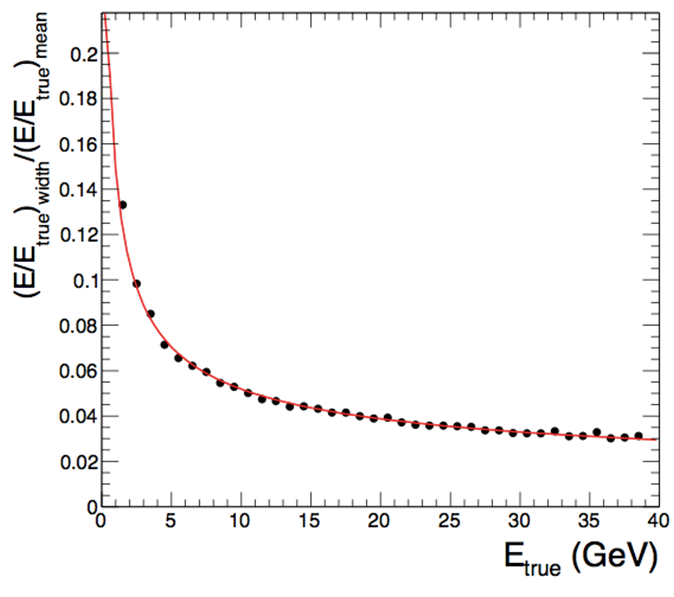}
    \caption{\label{fig:EMCAL_Eres_normal} Energy resolution of a
      tungsten-scintillator sampling calorimeter with the same
      sampling fraction as the sPHENIX tungsten-scintillator accordion
      calorimeter. The data are obtained for electrons at normal
      incidence with energies indicated. The fit indicates an energy
      resolution of $\sigma_{E}/E=$ 14.2\%/$\sqrt{E}$+0.7\%.}
 \end{center}
\end{figure}

The electromagnetic calorimeter has been simulated using the \geant tools
described above.  Figure~\ref{fig:elec} shows a typical \geant event
in which a 10\,GeV/c electron hits the calorimeter.  Most of the
shower develops in the EMCal.  The response of the electromagnetic
calorimeter to electrons and protons at normal incidence is shown in
Figure~\ref{fig:emcal_fullG4}.  The resolution of the electromagnetic
calorimeter for electrons at normal incidence is summarized in
Figure~\ref{fig:EMCAL_Eres_normal}.  The single particle energy
resolution at normal incidence is determined to be 14.2\%/$\sqrt{E} + 0.7\%$.

\begin{figure}[htb!]
 \begin{center}
    \includegraphics[trim = 15 0 30 0, clip, width=0.8\linewidth]{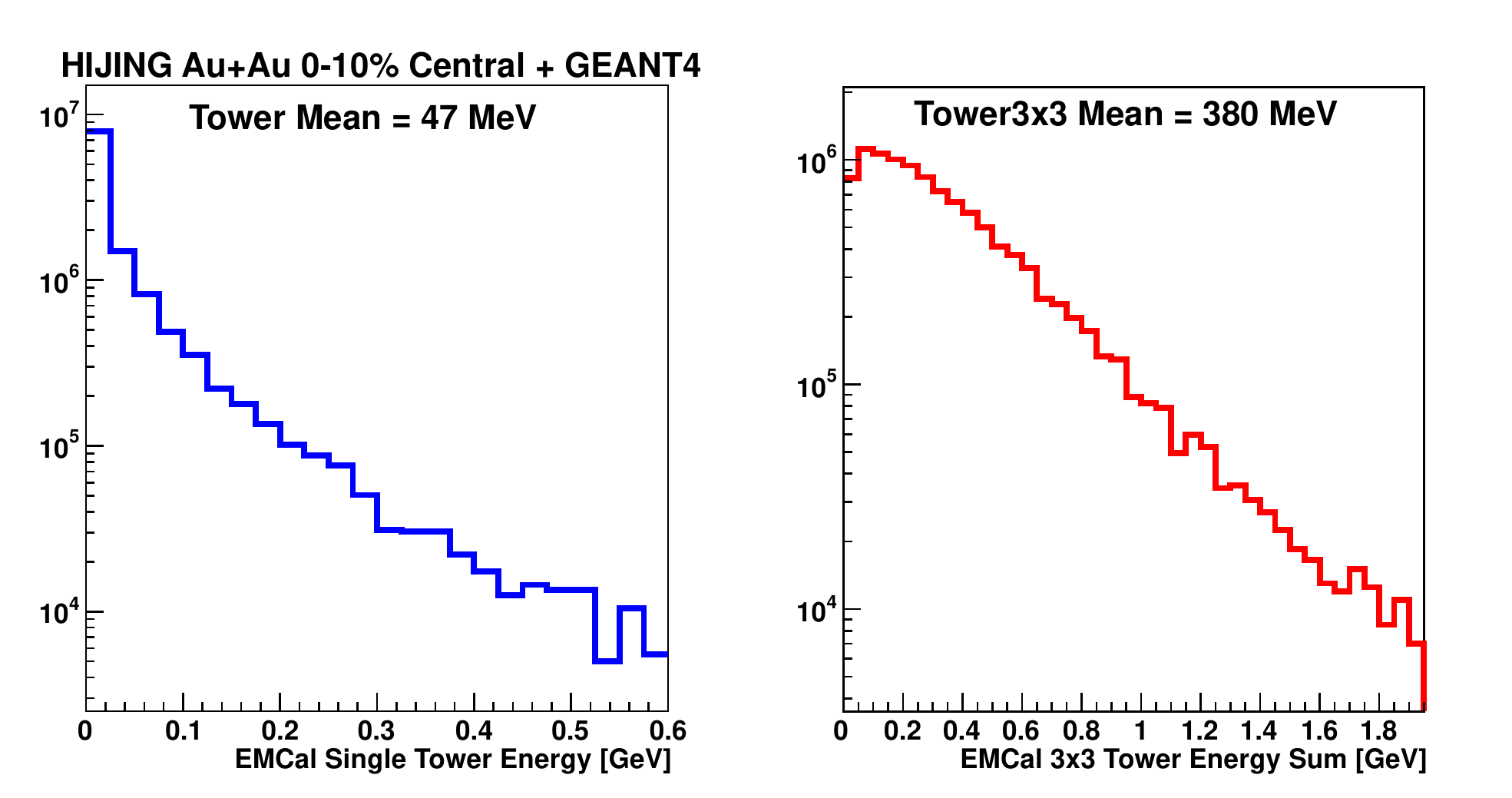}
    \caption{\label{fig:g4emcal_hijingoccup} Distribution of energy
      deposited in the electromagnetic calorimeter for single towers
      (left panel) and in $3\times3$ arrays of towers (right panel) in
      central \hijing events.}
 \end{center}
\end{figure}

The energy deposited in the electromagnetic calorimeter in central
\hijing events is shown in Figure~\ref{fig:g4emcal_hijingoccup}.
The mean energy deposited in any single tower is estimated to be 
47\,MeV.  The existing PHENIX electromagnetic calorimeter cluster finding
algorithm has been adapted for the sPHENIX EMCal specifications.  Initial
results indicate that for a 10\,GeV photon there is an extra 4\% of 
underlying event energy in the cluster and a degredation of approximately 10\% in the energy resolution 
when embedded in a central \auau event (simulated with the \hijing event generator).  

\subsection{Hadronic Calorimeter Simulation}

\begin{figure}[tb!]
 \begin{center}
  \includegraphics[width=0.7\linewidth]{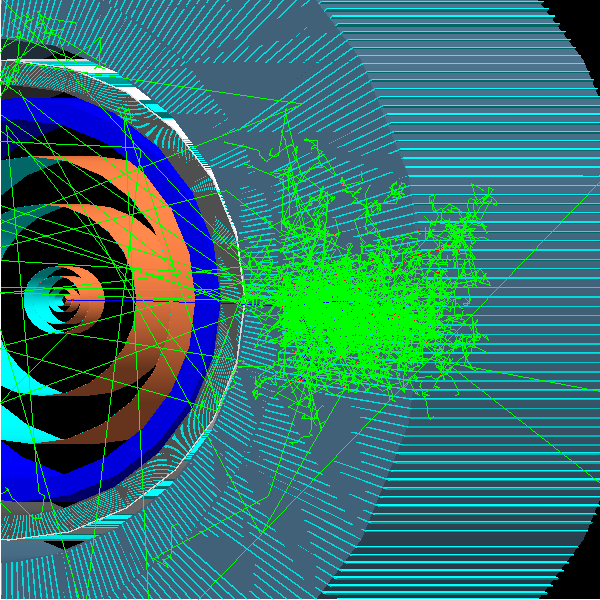}
  \caption{\label{fig:piplus} Transverse view of a 10\,GeV/c $\pi^+$ in 
sPHENIX. It penetrates the magnet (blue) and the EMCal and showers 
in the first segment of the HCal.}
 \end{center}
\end{figure}

The hadronic calorimeter has been simulated using the \geant tools
described above.  Figure~\ref{fig:piplus} shows a typical \geant event
in which a 10\,GeV/c $\pi^+$ incident on the calorimeter showers in the Hcal.

\begin{figure}[tb!]
  \begin{center}
    \includegraphics[trim = 2 2 450 2, clip, width=\twowidth]{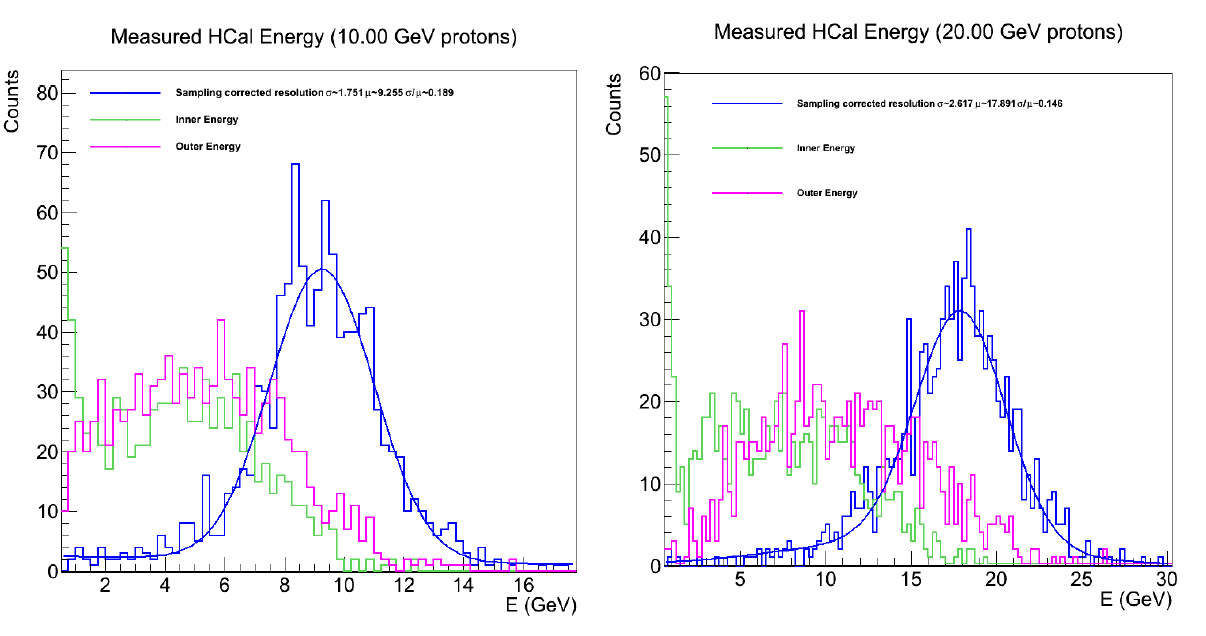}
    \hfill
    \includegraphics[trim = 450 2 2 2, clip, width=\twowidth]{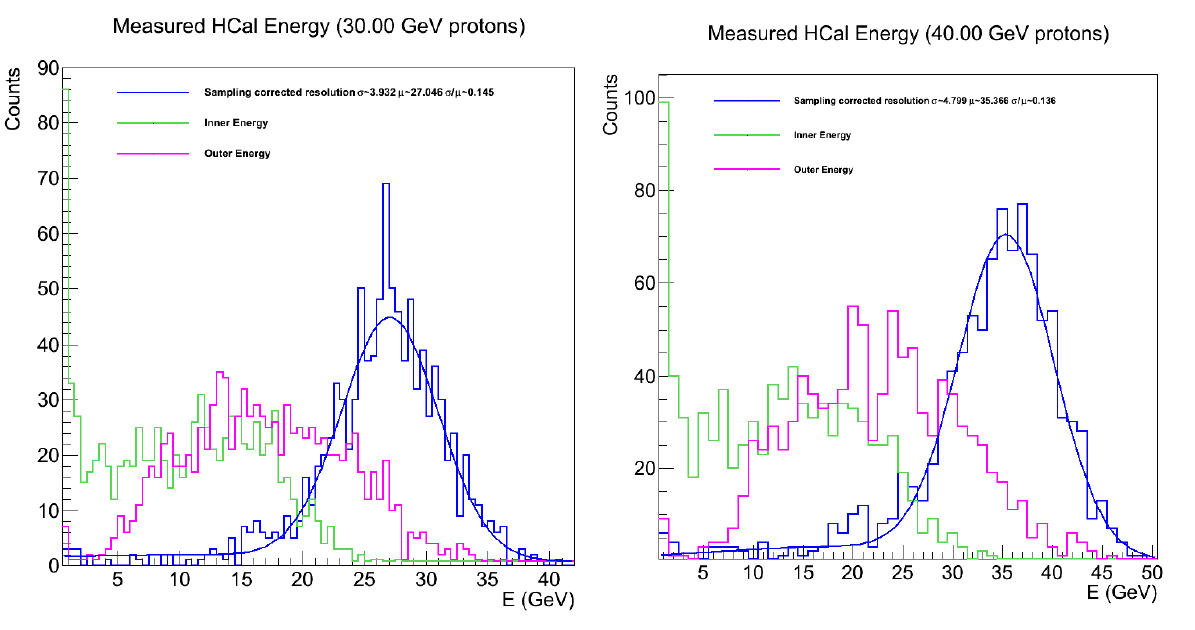}
    \caption{Energy deposited in the hadronic calorimeter by 10 (left
      panel) and 40 (right panel) GeV/c protons, showing the good
      containment and Gaussian response of the calorimeter.}
    \label{fig:hcal_proton} 
  \end{center}
\end{figure}

\begin{figure}[tb!]
  \begin{center}
    \includegraphics[width=1.0\linewidth]{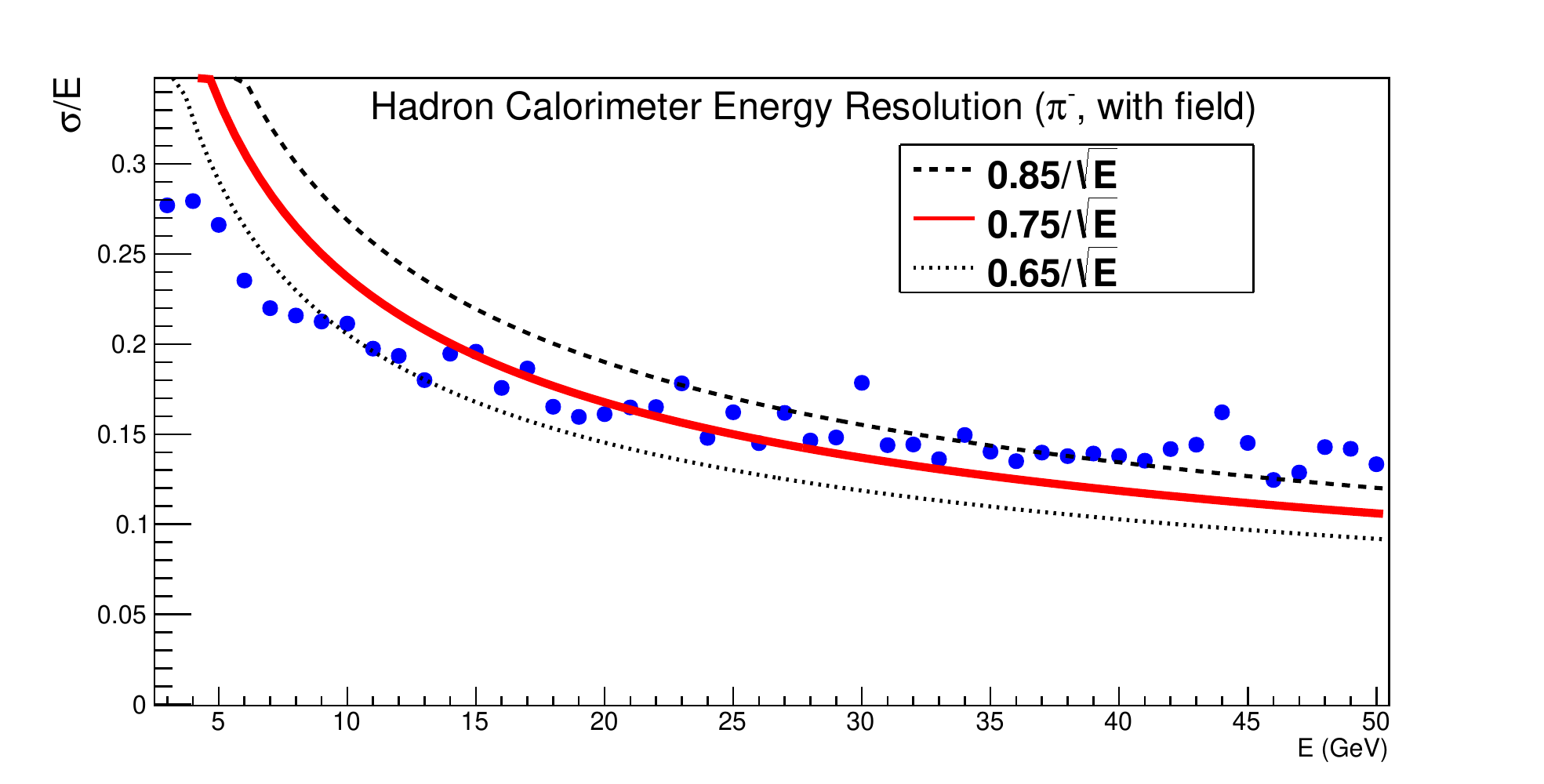}
    \caption{Energy resolution of the hadronic calorimeter as one might
      measure in a test beam.  The HCal is isolated, with nothing in
      front of it, and is illuminated by pions.}
    \label{fig:hcal_testbeamG4_resolution} 
  \end{center}
\end{figure}

\begin{figure}[tb!]
 \begin{center}
  \includegraphics[width=1.0\linewidth]{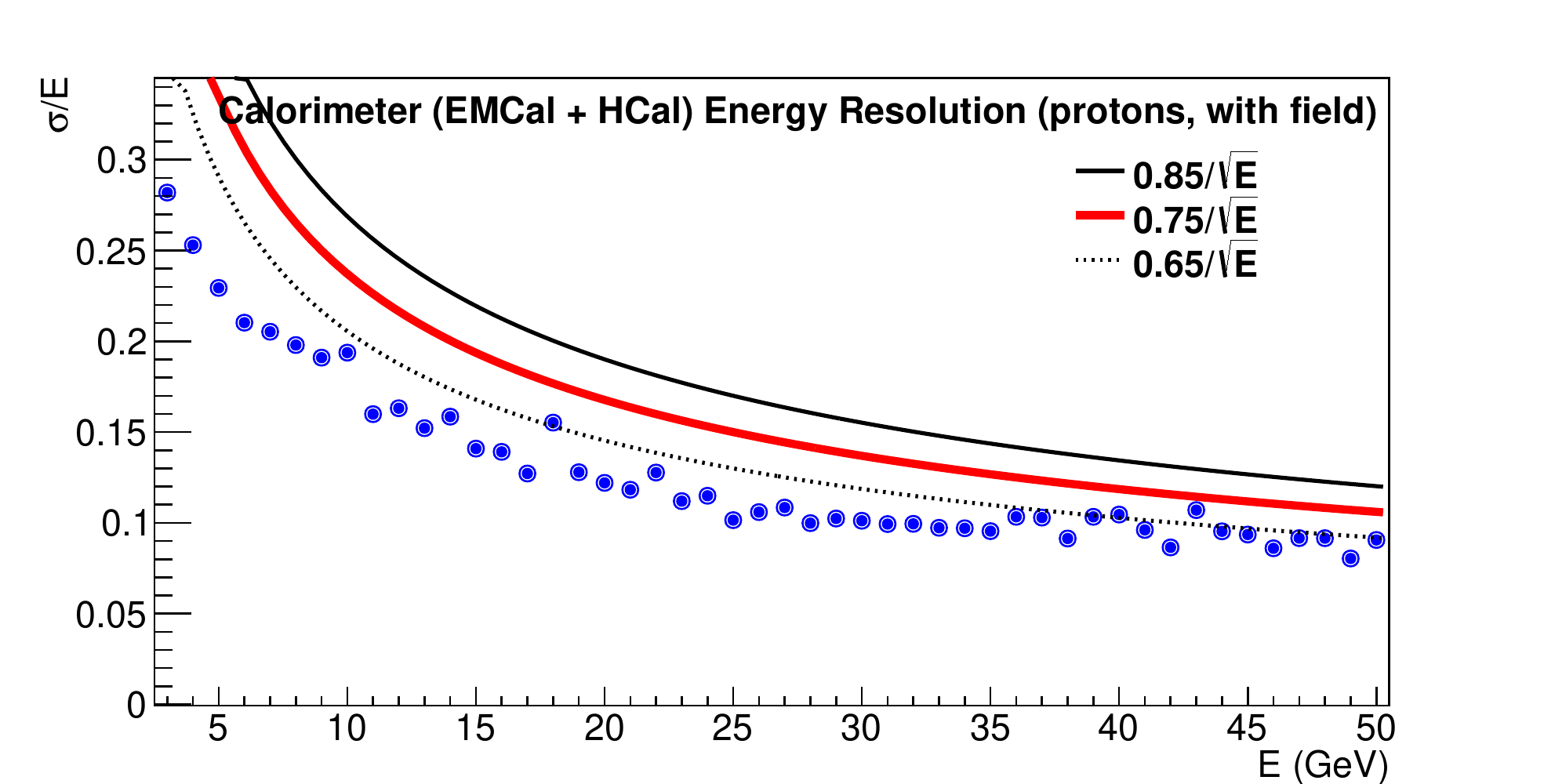}
  \caption{\label{fig:emcal_hcal_G4_resolution} Energy resolution of
    the combined system of EMCal and HCal.  In this case, the
    calorimeters sit behind the VTX and the solenoid magnet.}
 \end{center}
\end{figure}

The single particle energy resolution in the HCal has been determined
using a full \geant description of the calorimeters.  The energy
deposition in the scintillator is corrected for the average sampling
fraction of the inner and outer sections separately, calculated to be
18.2\% for the inner and 27\% for the outer.
The calorimeter response to single protons is shown in Figure~\ref{fig:hcal_proton}.
Figure~\ref{fig:hcal_testbeamG4_resolution} shows the resolution of
just the HCal itself when illuminated by a beam of $\pi^-$. In this
case, there is nothing in front of the HCal, it is just an isolated
device.  Figure~\ref{fig:emcal_hcal_G4_resolution} shows the energy
resolution of the combined system of EMCal and HCal when illuminated
by a beam of protons.  In this case, the full \geant description of
sPHENIX is in place. 

The mean and standard deviation from a Gaussian fit to the measured
energy distribution are used to calculate the nominal detector
resolution.  In both Figure~\ref{fig:hcal_testbeamG4_resolution}
and Figure~\ref{fig:emcal_hcal_G4_resolution}, the resolution
determined from simulation is compared to curves of $0.85/\sqrt{E}$,
$0.75/\sqrt{E}$, and $0.65/\sqrt{E}$ as reference for the
simulated resolution.  These indicate a \geant performance level better
than the physics requirements.

As mentioned above, the proposed sPHENIX calorimeter system is about
$6\lambda_{\mathrm int}$ deep, and one expects some leakage of energy
out of calorimeter. The amount of this leakage and its energy
dependence can be estimated from literature
Figure~\ref{fig:HC-containment} above or from simulation which is tuned
to available experimental data.  
The probability for a                                 
proton to go through the whole depth of calorimeter without an hadronic interaction is about 0.6\% (verified
with full \geant simulations).
Energy leakage out the back is thus not expected to be a serious problem for this calorimeter.

\section{Electronics}
\label{sec:electronics}

For the readout of both the EMCal and HCal a common electronics design will 
be used to reduce the overall cost and minimize the design time. 
Two viable alternatives for reading out the sPHENIX calorimeters have been 
identified. The first approach is based on electronics developed
for the PHENIX Hadron Blind Detector (HBD) and Resistive Plate Chambers (RPC), 
and uses the current PHENIX DAQ as the backend readout.  
The second approach is based on the \beetle chip developed for the
LHCb experiment and the SRS DAQ developed at
CERN. The following sections describe both approaches and how they
could be implemented in sPHENIX. Both approaches will be evaluated in
terms of performance and cost to enable an eventual selection of a
readout system for the sPHENIX calorimeters.

\subsection{Sensors}

For both the electromagnetic and hadronic calorimeters, we are
currently considering as sensors 3\,mm$\times$3\,mm silicon photomultipliers
(SiPMs), such as the Hamamatsu S10362-33-25C MultiPixel Photon
Counters (MPPC).  These devices have 14,400 pixels, each
25\,$\mu$m$\times$ 25\,$\mu$m. Any SiPM device will have an intrinsic
limitation on its dynamic range due to the finite number of pixels,
and with over 14K pixels, this device has a useful dynamic range of
over 10$^4$.  The saturation at the upper end of the range is
correctable up to the point where all pixels have fired.  The photon
detection efficiency is $\sim36\%$ and it should therefore be possible
to adjust the light level to the SiPM using a mixer to place the full
energy range for each tower ($\sim$ 5\,MeV--50\,GeV) in its useful
operating range. For example, if the light levels were adjusted to
give 10,000 photoelectrons for 50\,GeV, this would require only 200 photoelectrons/GeV,
which should be easily achieved given the light level from
the fibers entering the mixer.

While we believe that the SiPMs are  likely the most suitable
sensor for the calorimeters, we are also considering avalanche
photodiodes (APDs) as an alternative. They have much lower gain
($\sim$50--100 compared to $\sim 10^5$ for SiPMs), and therefore would
require lower noise and more demanding readout electronics, but they
do provide better linearity over a larger dynamic range. In addition,
while the gain of both SiPMs and APDs depend on temperature, SiPMs
have a stronger gain variation than APDs (typically 10\%/$^\circ$C for
SiPMs vs 2\%/$^\circ$C for APDs). Thus, we are considering APDs as an
alternative solution as readout devices pending further tests with
SiPMs and our light mixing scheme.

\subsection{All Digital Readout [Option 1]}

\subsubsection {SiPM Preamplifier Circuitry}

The requirements of the sPHENIX calorimeter preamplifier circuit board
are to provide localized bias/gain control, temperature compensation,
signal wave shaping and differential drive of the SiPM signal to an
ADC for acquisition. Gain adjustment and temperature compensation are
performed as part of the same control circuit. Signal wave shaping is
performed by the differential driver to satisfy the sampling
requirements of the ADC.

\subsubsection{Temperature Compensation}

\begin{figure}[htb]
  \centering
  \includegraphics[width=0.7\linewidth]{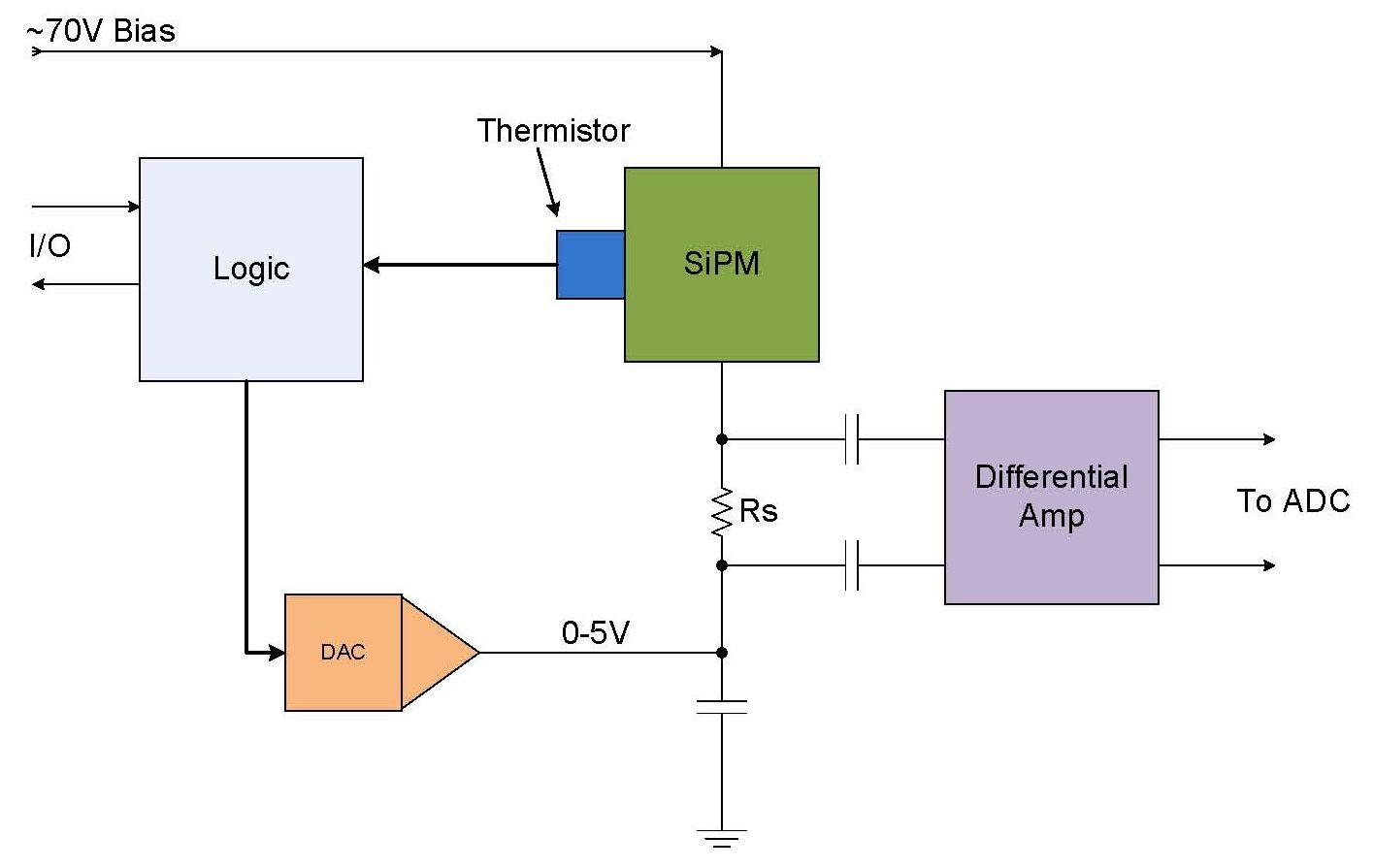}
  \caption{Block diagram of a temperature compensating circuit for SiPMs}
  \label{Fig:TempComp}
\end{figure}

The reverse breakdown voltage V$_{\mathrm br}$ for the Hamamatsu
S10362-33-25C device is nominally 70 Volts. As the bias is increased
over the value of V$_{\mathrm br}$ the SiPM begins to operate in
Geiger mode with a gain of up to $2.75 \times 10^{5}$. The range of this
over-voltage (V$_{\mathrm ov}$) is typically 1--2 Volts and represents
the useful gain range of the device.  The V$_{\mathrm br}$ increases
by 56\,mV/$^\circ$C linearly with temperature and must be compensated to
achieve stable gain. This compensation is achieved using a closed
feedback loop circuit consisting of a thermistor, ADC, logic and DAC
voltage control as shown in Figure~\ref{Fig:TempComp}.

The thermistor is fixed to the back of the SiPM and provides a
significant voltage variation over temperature when used as part of a
voltage divider, thereby easing temperature measurement over a length
of cable. The bias supply for an array of SiPMs is fixed nominally at
V$_{\mathrm br} + 2.5$V. The DAC in each SiPM circuit then outputs a
subtraction voltage of~\,0\,V to 5\,V to provide a full range of gain
control over the device temperature range. The SiPM gain may then be
adjusted externally through an interface to the logic.

\subsubsection{Preamplifier-Shaper-Driver}

\begin{figure}[htb]
  \centering
  \includegraphics[width=0.7\linewidth]{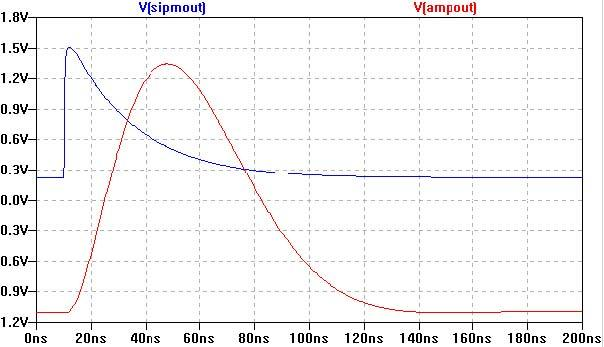}
  \caption{SPICE simulation of a prototype temperature compensating circuit 
for SiPM readout of the sPHENIX EMCal and HCal.}
  \label{Fig:Shaper}
\end{figure}

The SiPM current develops a voltage across the load resistor $R_s$
proportional to the number of pixels fired. To avoid the region of
greatest non-linearity due to saturation of the SiPM, the maximum
signal level is optically adjusted to 10K out of 14.4K pixels fired.
Simulations of the SiPM indicate that the current could be as much as
several tenths of an ampere at this maximum level. Results of a SPICE 
simulation are shown in Figure~\ref{Fig:Shaper}. Such a large current
affords the use of a small value for $R_s$ which virtually eliminates the
contribution of $R_s$ to non-linearity. This signal voltage is sensed
differentially, amplified and filtered by a low power, fully
differential amplifier. For sampling by a 65MSPS ADC, a peaking time
of approximately 35\,ns is achieved through the use of a second order
Butterworth filter implemented in the differential driver circuit.

\subsubsection{Signal Digitization}
One solution for the readout of the EMCal and HCal detectors for sPHENIX is the
direct digitization of the SiPM signal. The signals from the SiPM are
shaped to match the sampling frequency, and digitized using a flash
ADC. The data are stored in local memory pending a Level-1 (L1) trigger
decision. After receiving an L1 trigger decision, the data are read out
to PHENIX Data Collection Modules (DCM II).  These second generation Data Collection
Modules would be the identical design as those developed and implemented for reading
out the current PHENIX silicon detectors.
One advantage of direct
digitization is the ability to do data processing prior to sending
trigger primitives to the L1 trigger system. The data processing can
include channel by channel gain and offset corrections, tower sums,
etc. This provides trigger primitives that will have near offline
quality, improved trigger efficiency, and provide better trigger
selection.

\begin{figure}[htb]
  \centering
  \includegraphics[width=0.7\linewidth]{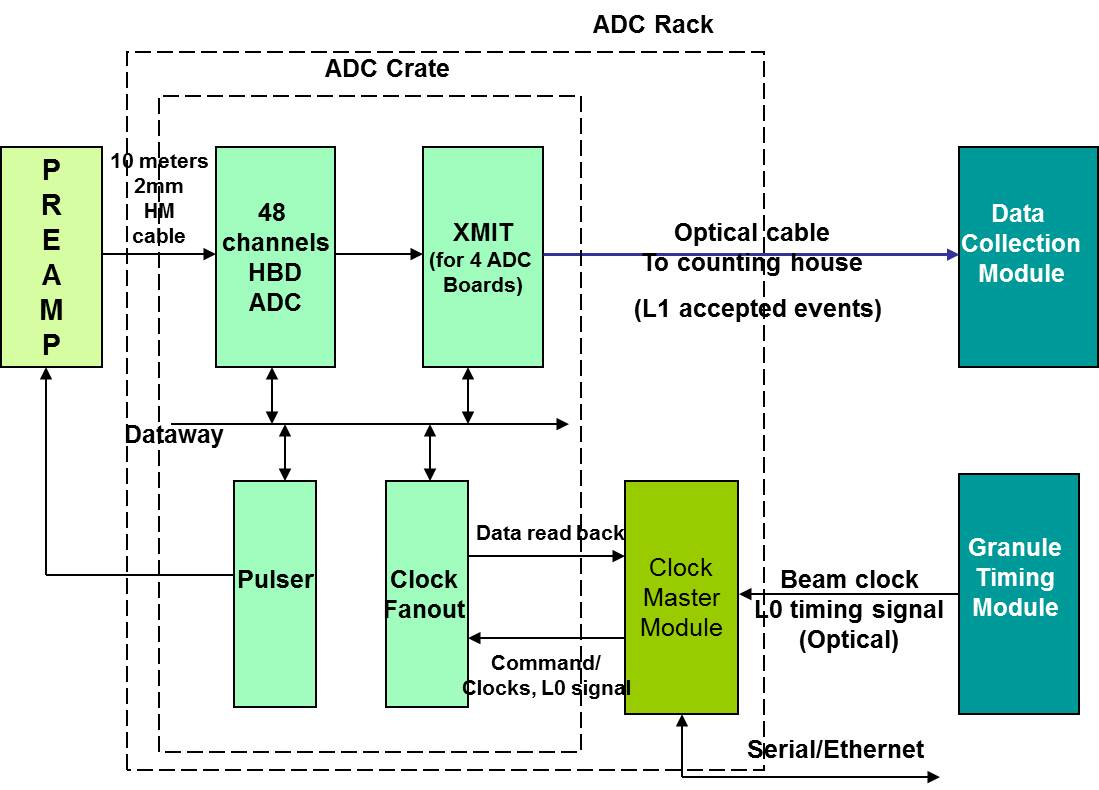}
  \caption{Block diagram of the HBD read out electronics}
  \label{Fig:HBD}
\end{figure}

\begin{figure}[htb]
  \centering
  \includegraphics[width=0.7\linewidth]{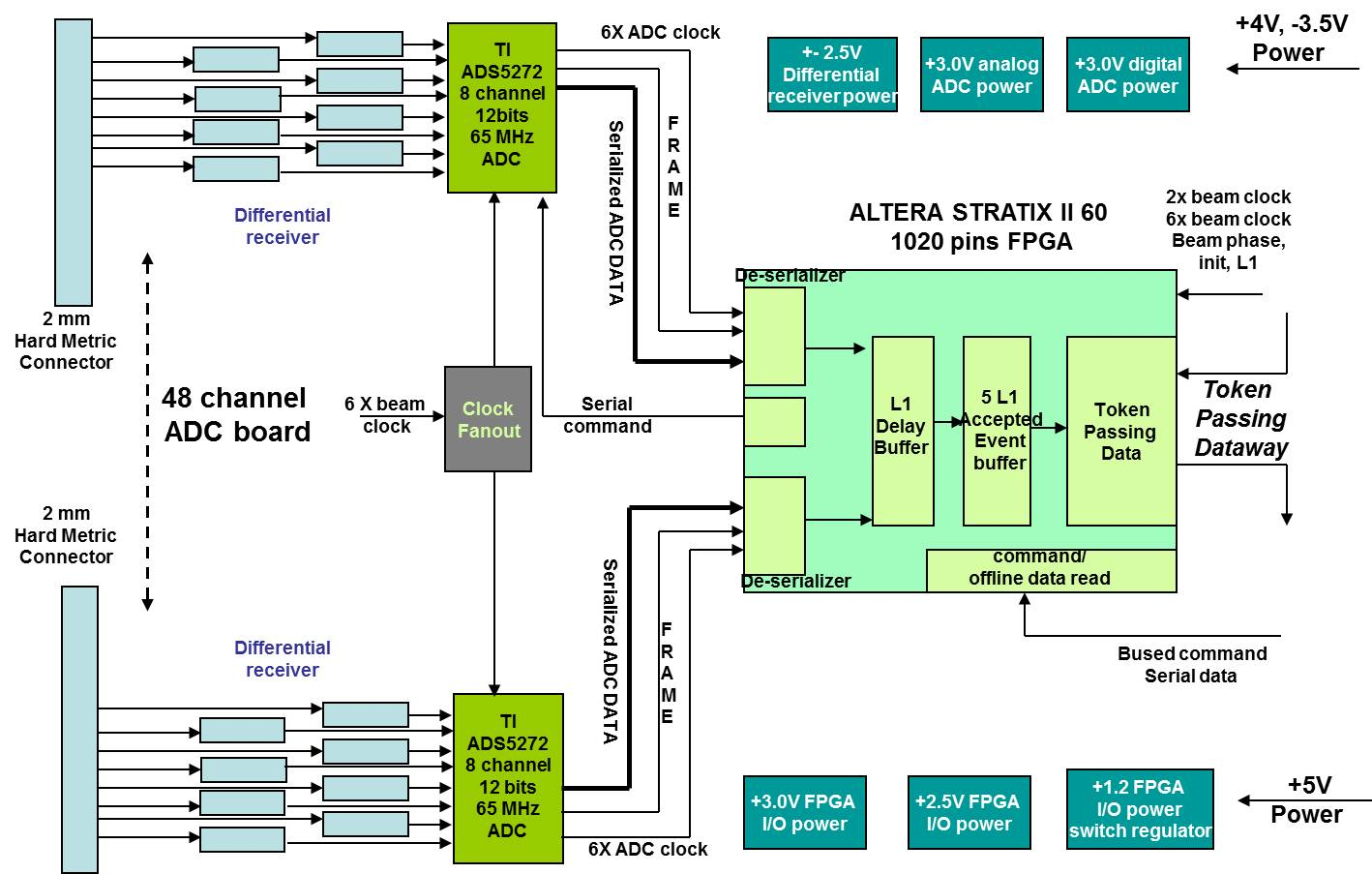}
  \caption{Block diagram of the HBD FEM electronics}
  \label{Fig:HBD_ADC}
\end{figure}

A readout system based on this concept was implemented for the Hadron
Blind Detector (HBD) for the PHENIX experiment as shown in
Figure~\ref{Fig:HBD} and subsequently modified for the PHENIX
Resistive Plate Chamber (RPC) system.  The block diagram of the
Front-End Module (FEM) is showed in Figure~\ref{Fig:HBD_ADC}.  In the
HBD system, the discrete preamplifier-shaper is mounted on the detector and
the signals are driven out differentially on a 10 meter Hard Metric
cable. The signals are received by Analog Device AD8031 differential
receivers which also serves as the ADC drivers. Texas Instruments
ADS5272 8 channel 12 bit ADCs receive the differential signals from 8
channels and digitize them at 6x the beam crossing clock . The 8
channels of digitized data are received differentially by an Altera
Stratix II 60 FPGA which provides a 40 beam crossing L1 delay and a 5
event L1 triggered event buffer.

The L1 triggered data from 4 FEMs is received by an XMIT board using 
token passing to control the readout of the FEMs. The data is then sent 
by 1.6 GBit optical links to the PHENIX DAQ.  A ClockMaster module interfaces 
to the PHENIX Granulate Timing Manager (GTM) system and fans out the clocks, L1 triggers and test enable 
signals to the FEMs and XMIT modules. The ClockMaster module also receives 
slow control signals for configuring the readout.

Although not shown in the block diagram, the FEM has 4 LVDS outputs
that can be used to bring out L1 trigger primitives at 800 Mbits/sec. 
This feature was not used for the  HBD readout, however it has been 
implemented for the RPC detector. A trigger module for the RPC system based on 
the Altera Arria FPGA receives the trigger primitives from the FEMs, 
combines them and sends them to the PHENIX L1 trigger system through two 
3.125 GBit optical links.

For implementation in sPHENIX, two possible implementations are under
consideration.  The first design would place the analog and digital
electronics directly on the detector. All control and clock signals
would be brought in and L1 trigger primitives and triggered digital
data transmitted out via high-speed optical fibers.

The second approach has the temperature compensating preamplifier mounted 
on the detectors and the shaped
and amplified signals driven differentially to the digital modules
located in racks near the detector using shielded differential cables.
High speed fiber optic cables bring in all control and clock signals and 
transmit L1 trigger primitives and triggered data to the PHENIX DAQ.

\subsection{Mixed-Mode Readout [Option 2]}

\subsubsection{Preamp ASIC}

A preamp ASIC appropriate for readout of sPHENIX calorimeters has been
identified.  This custom ASIC is being developed at ORNL for front end
readout of a new forward calorimeter (FoCal) under consideration as an
upgrade for ALICE at CERN.  This ASIC, or a very close variant, is
appropriate for front end readout of the sPHENIX EMCal, HCal, (and
possible future strip-pixel preshower as discussed in Appendix~\ref{chap:barrel_upgrade}) detectors.  The ORNL ASIC
development is funded as part of a multi-disciplinary DOE SC LAB
11-450 project which is in its first year.  The ORNL team is in close
communication with colleagues at BNL and are working to coordinate
simulation and actual testing of the ORNL ASIC with appropriate
Hamamatsu silicon photomultipliers (SiPM) for the sPHENIX EMCal and
HCal.  The already-funded first year of LAB 11-450 work at ORNL will
generate first round ASIC chips this summer for testing.

Traditional charge-sensitive preamplifiers (CSP) are commonly used for
readout of capacitive detectors (silicon pads, strips, etc.) for two
reasons. First, all the charge generated in a detector due to a
radiation event is ultimately collected by the preamplifier
irrespective of the detector capacitance. Higher detector capacitance
may slow the preamplifier bandwidth such that it takes many
microseconds to collect the charge but it will ultimately be
collected. Second, the ratio of the output voltage to the input charge
(charge gain) is determined by the feedback capacitor used in the CSP
and not the detector. Since $Q/C=V$, this will allow a small charge
signal to be processed by a small feedback capacitor on the CSP
instead of that same small charge on a much larger detector
capacitance. This results in a proportionally larger voltage signal
for subsequent processing.

\begin{figure}[htb]
  \centering
  \includegraphics[width=0.7\linewidth]{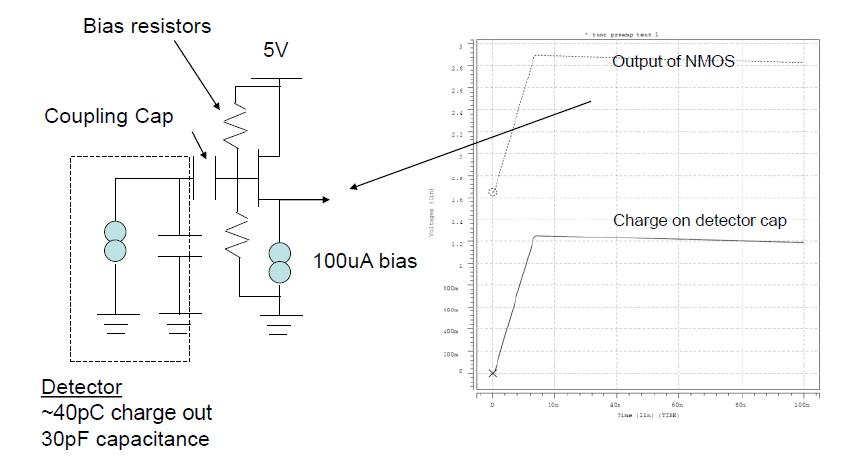}
  \caption{ASIC follower schematic and output signal}
  \label{Fig:Preamp}
\end{figure}

\begin{figure}[htb]
  \centering
  \includegraphics[width=0.7\linewidth]{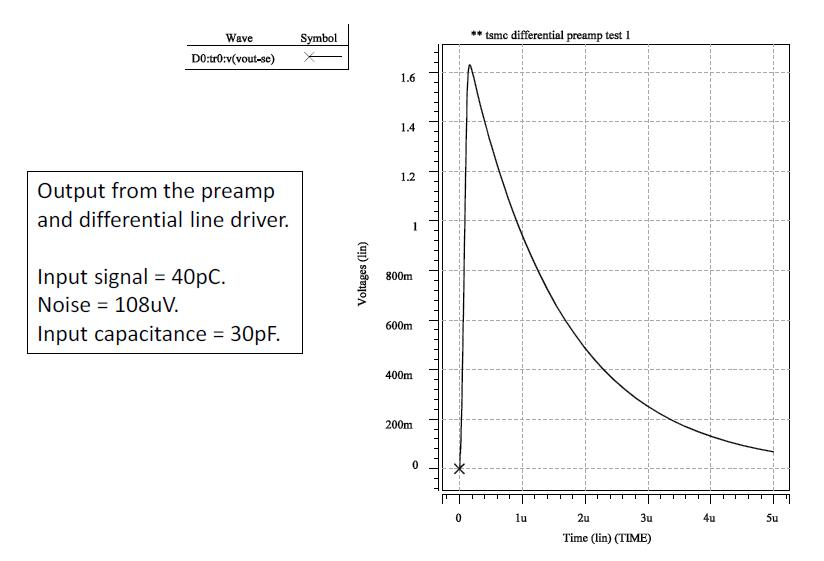}
  \caption{Simulation ASIC preamp output voltage versus time.}
  \label{Fig:ASIC}
\end{figure}

Because of the large amount of charge per event available from an SiPM
and the need for a fast trigger signal (fast preamplifier response), a
traditional CSP is likely not ideal or needed. Therefore, a truly
application-specific approach to on-chip readout is proposed. For
simplicity, we can utilize a very fast high-speed follower topology
similar to that used on a photomultiplier tube. This will allow us to
maintain high speed, low noise, and simplicity at the front end
detector. With a follower, we will have sufficient bandwidth to
provide a fast trigger without having to maintain a high bandwidth
closed loop CSP. Processing electronics can be placed away from the
detector thus somewhat mitigating heat and power-distribution
problems. The follower, shown in Figure~\ref{Fig:Preamp}, is very
straightforward. Simulations in Figure~\ref{Fig:ASIC} show
that if we design the detector/follower such that our input maximum
charge results in approximately 1.6~V output, we can develop a circuit
which will exhibit noise of approximately 108 $\mu$V RMS, a peak/RMS
ratio of 14,800. This shows that we will likely not be limited by
noise, but by inter-channel crosstalk. The follower requires a
buffered output, preferably differential to minimize crosstalk. The
output of the differential buffer will drive the signal to an area
with more available space, where it will be connected to processing
electronics (shaper, trigger processor, ADC), simplifying their
requirements.

\begin{figure}[htb]
  \centering
  \includegraphics[width=0.7\linewidth]{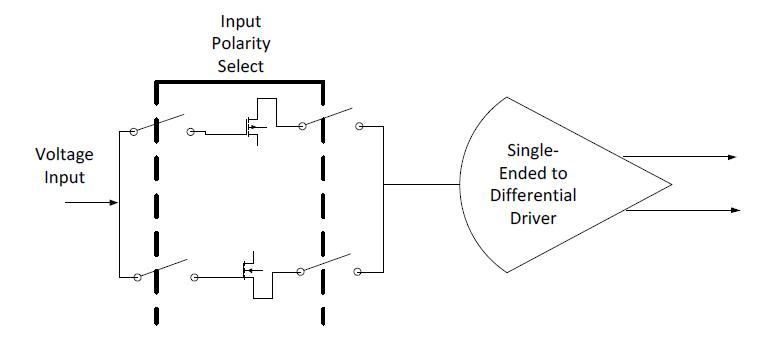}
  \caption{A block diagram for the proposed preamplifier/driver.}
  \label{Fig:ASIC_Buffer}
\end{figure}

A block diagram for the proposed preamplifier/driver is shown in
Figure~\ref{Fig:ASIC_Buffer}.  The preamplifier connects to the detector as
shown in Figure~\ref{Fig:Preamp} (through a coupling capacitor if
needed) and can utilize either polarity of charge input.  There are
bias setting resistors on the chip that set the quiescent input
voltage.  When an event occurs, the charge is collected on the
detector capacitance and the voltage output is buffered and sent to
the single-ended-to-differential driver.  This driver is designed to
drive a 100-ohm differential line.  The power dissipation is currently
under 10 mW for the entire circuit which operates on 2.5 V.  The
preamplifier is presently under design in the TSMC $0.25\mu$m CMOS
process.  A layout estimate results in an expected chip area of under
2\,mm$\times$2\,mm for four channels.

This electrical engineering design and development of the ASIC is
undertaken as part of a separate ongoing DOE LAB 11-450
project. Fabrication and testing of 120 ASIC chips is scheduled for
summer 2012.  The chip bench testing will be performed at ORNL and
include tests of basic functionality to ensure essential operation of
the device such as amplification, rise time, power dissipation,
channel-to-channel gain variation, noise, and chip-to-chip variations.
 
We have obtained a MOSIS quotation for fabrication and packaging of a
sufficient number of 4-channel preamp ASICs plus spares for the
sPHENIX EMCal and HCal for a cost of \$2.80/channel.  This price does
not include the testing which can be accomplished very cost effectively by
EE and physics graduate students with direct supervision by electrical
engineers.

\subsubsection{Front-End Readout Design using the CERN SRS}

In this Section we present a design outline for sPHENIX calorimeter
readout based on the already-existing CERN Scalable Readout System
(SRS) which has been developed as part of the CERN RD51 project~\cite{RD51:SRS,RD51:2012}.

\begin{figure}[htb]
  \centering \includegraphics[width=0.7\linewidth]{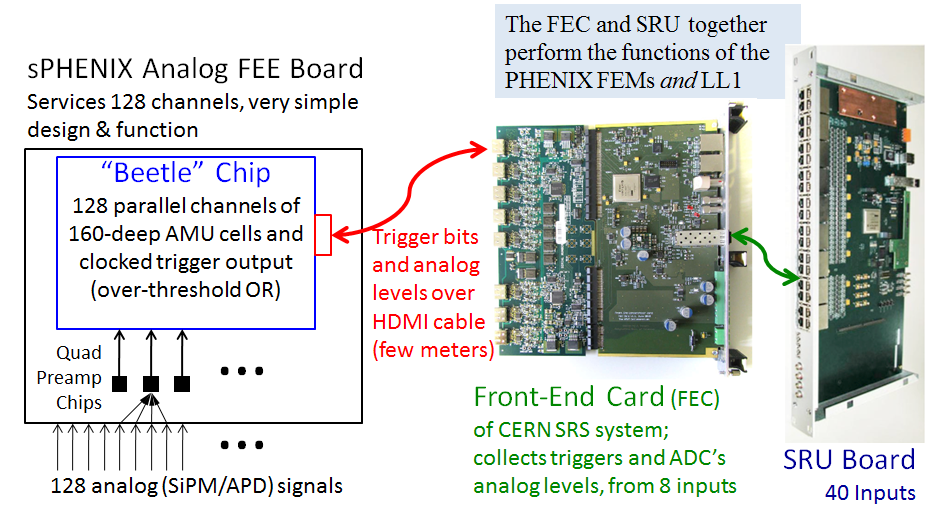}
  \caption{The SRS topology: The analog FEE board sits on the detector
    and buffers analog levels, which are then transferred to, and
    digitized on, the Front-End Card (FEC); the array of FECs are
    controlled by a Scalable Readout Unit (SRU) board.  Only the
    analog FEE board is specific to the detector; the FEC and SRU are
    already-existing components of the SRS.}
  \label{Fig:SRS}
\end{figure}

The SRS architecture consists of three stages, as shown in
Figure~\ref{Fig:SRS}.  Signals from the detector elements are
conditioned and analog buffered on an analog FEE board (see below for
more detail), which also generates trigger primitives. When an event
is read out, the FEE board transmits analog levels to the front-end
card (FEC), where they are digitized and assembled as sub-events.  The
transfer from the FEE board to the FEC is carried across commercial
standard HDMI-format cables, which can accommodate a separation of
several meters from a detector-mounted board to crate-mounted FECs.
The FECs receive trigger primitives from the FEE boards along the same
HDMI cables.
 
With existing implementations, each FEC can service eight FEE boards.
Continuing hierarchically, up to 40 FECs can be gathered through
standard network connections, to one Scalable Readout Unit (SRU)
component of the SRS system.  The SRU gathers the real-time trigger
information from the whole system and fulfills the same function as
the existing PHENIX Local Level-1 (LL1) system.  The SRU also serves
as the overall controller/director for the FECs and fulfills the same
function as the existing the PHENIX Granule Timing Module (GTM) to
pass down readout and control instructions.  When an event is
processed, the FECs can put out sub-event data on standard network
connections directly to an Event Builder; and thus the FEC fulfills
the function of both the Data Collection Module (DCM II) and Sub-Event
Builder (SEB) of the PHENIX architecture.

\subsubsection{The sPHENIX Analog FEE Board}

The advantage of adopting the SRS, for any large-scale system,
is that only the analog FEE board needs to be designed specifically
for the detector in question, and its functionality is relatively simple.
It only needs to buffer and transmit analog levels; all the ADC and
digital processing functions are carried out on existing FECs.

In the SRS-based readout design, we would use an existing circuit for
the analog buffering function: the \beetle chip, designed for use
in the LHCb experiment\cite{Lochner:2006sq}.
An SRS FEE card based on the \beetle chip is being developed by a group
from the Weizmann Institute for use in an ATLAS upgrade.  The \beetle
has 128 analog input channels, each of which can be buffered at up to
40 MHz in a 160-sample analog ring buffer.  On readout, the \beetle
copies the analog level from the appropriate ring cells to an on-chip
buffer, so the ring operation is not interrupted; the \beetle then
multiplexes these analog levels over to the FEC for digitization.  The
entire complement of 128 channels for one event can be transferred and
converted in slightly under one microsecond.

\begin{figure}[htb]
  \centering \includegraphics[width=0.7\linewidth]{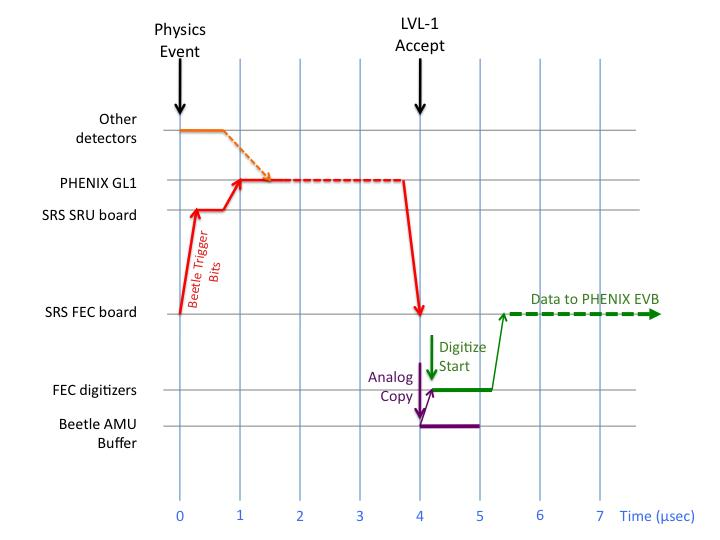}
  \caption{Timing diagram for processing one physics event, showing
    the operation of the \beetle-based analog FEE board and the FEC and
    SRU components of the SRS, staying well within the PHENIX
    specifications for digitization and readout.}
  \label{Fig:SRS_Timing}
\end{figure}

Figure~\ref{Fig:SRS_Timing} shows a timing diagram for the processing
of one physics event, with the trigger primitive bits coming up
through the FECs and the SRU to the PHENIX GL1 system, which returns
the LVL-1 accept down to the FECs.  Sampling at up to 40 MHz, the
\beetle analog ring has enough depth to accommodate the PHENIX-standard
4 microsecond latency between the crossing of interest and the arrival
of the LVL-1 accept instruction.

Trigger primitive bits are generated within the \beetle chip, and are
continually passed up to the FECs, where they are gathered in the SRU
for calorimeter-wide processing.  The trigger information provided by
each \beetle chip is essentially a channel-by-channel
voltage-over-threshold condition, of which groups of four channels are
then OR'ed together.  The simplest global condition would be a logical
OR of the over-threshold for all the towers in a fiducial portion of
either the EMCal or HCal layers of the calorimeter.

One advantage of adapting the SRS system for sPHENIX is the large
potential savings in development time and effort and procurement
costs.  The only component which needs to be specifically designed for
the detector is the analog FEE board; and in the scheme outlined here
that board is relatively simple, interfacing the ORNL preamp ASIC to
the detector and carrying the \beetle analog buffer chips.  The
digitization and digital processing are all carried out on the FECs,
which use multiplexing of analog levels for higher economy; and the
FECs are crate-mounted up to several meters away from the detector,
which would simplify the effort of deployment.  All together, the
FEC/SRU portions of the readout chain are estimated to be available
for approximately \$2/channel for large channel count systems, based
on the production costs of the first prototype SRS systems including
FEC and SRU modules with power supplies and SRS crates, and including
also FEE boards based on the AVX chip.

\subsection{sPHENIX DAQ}

The sPHENIX DAQ will be largely based on the current PHENIX DAQ. In the PHENIX DAQ, trigger 
primitives from the FEMs are transferred via optical 
fibers to the Local Level-1 (LL1) trigger system that process the signals and generates an LL1
accept if the event meets the trigger requirements. The trigger
operates in a pipeline mode with a 40 beam crossing latency, generating a 
trigger decisions fro each crossing. The Local Level-1 trigger can be configured 
to accept events with different signatures and can operate at up to 10 kHz.

The LL1 accept is transmitted to all FEMs, and the corresponding event is 
transferred to the DCM II modules via optical fibers. The DCM II modules
zero suppress the data and transmit the zero suppressed data to the 
event builder which collects the data and formats it for archiving.  The 
formatted data is buffered locally at the PHENIX experimental hall before 
being transferred to HPSS for archiving. The PHENIX Online Computing System 
(ONCS) configures and initializes the DAQ, monitors and controls the data 
flow, and provides monitoring and control of axillary systems.

For the all digital approach 48 SiPMs are readout by a single FEM and data 
from 4 FEMs is collected and readout to a single DCM II channel.  Each DCM II module
has 8 channels, so based on channel count a total of 16 DCM II modules are required for 
the EMCal and another 2 DCM II modules are required for the HCal. 

For the mix-mode approach using the SRS, the SRS replaces the DCM II modules and the 
data from the SRS would be transmitted directly to the event builder over 
high speed ethernet.

In either case, raw data manipulation, databases, logging and archiving, controls and
monitoring can be adapted from the existing PHENIX architecture with minimal upgrades,
taking advantage of a developed system which has been functioning for more than a decade.

\section{Mechanical Design and Infrastructure Concept}
\label{sec:integration}

\begin{figure}[hbt!]
  \centering
  \includegraphics[width=0.8\linewidth]{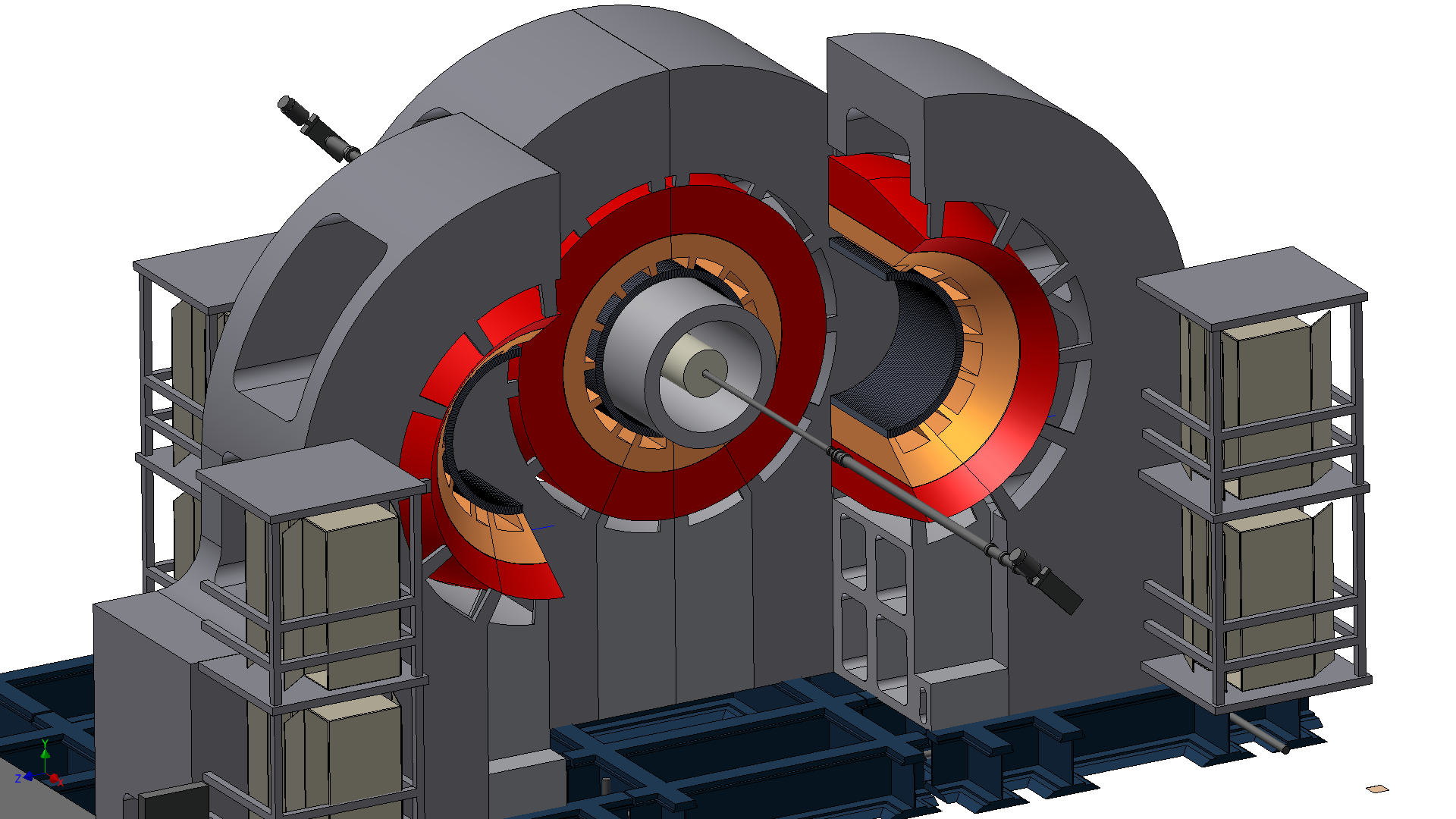}
  \caption{Illustration of sPHENIX underlying structural support,
    support equipment, overall assembly and maintenance concepts}

  \label{fig:sphenix_support_concept}
\end{figure}

sPHENIX has been designed to be straightforward to manufacture and
assemble, but it still requires significant and well thought out
infrastructure to support and service it. The overall concept for how
sPHENIX will sit in the existing PHENIX IR is shown in
Figure~\ref{fig:sphenix_support_concept}. A set of envelope dimensions
and design constraint parameters for each of the major components of
sPHENIX has been established and is discussed below.

\subsection{Beampipe} The existing PHENIX beampipe will be used with
minimal modification. The current beampipe has a 40\,mm outside
diameter in the central area, and connected on either end with
transition pipe sections from 40\,mm to 75\,mm OD and 75\,mm OD to
125\,mm OD.  A new support structure to support the beampipe inside
the superconducting solenoid will need to be designed.

\subsection{Silicon Vertex Tracker (VTX)} The support structure for the VTX, utilities supply
and readout design will need to be modified to allow the VTX to fit
within the superconducting solenoid cryostat. Existing VTX and
upgrades to detector subassembly will be integrated into a new
structural support design and mechanisms which will mount the VTX onto
rails supported by the cryostat inner surface, allowing the VTX to be
separated laterally then extracted from inside the cryostat
longitudinally parallel to the beampipe for maintenance.
The VTX electronics and services inside the cryostat will not be
serviceable during runs.  The VTX support structure will have a
clamshell design to allow the east and west halves to be opened then
extracted longitudinally on a rail system during long maintenance
shutdowns. 

\subsection{Superconducting solenoid magnet} The magnet has a 2 Tesla
solenoid field, 700\,mm inner cryostat radius, 900\,mm outer cryostat
radius, 1870\,mm cryostat length. The cryostat is not designed to be
disassembled.  The cryostat will incorporate two or more support
mounts that will fit in the clearance between the EMCal and the outer
skin of the cryostat. The cryostat will have an angled services stack
at the south end to exit beyond the end of the EMCal and HCal
detectors to cryogenic supply lines, power supplies and monitoring
equipment. Provision will also be incorporated for transport, lifting
and installation. Preliminary specifications and requirements for the solenoid
and cryostat have been developed based on a recently developed solenoid in the
BNL Magnet Division.

\subsection{Electromagnetic calorimeter} The EMCal will have a 100\,mm
radial thickness with an inside 50\,mm radial clearance from the
cryostat outer radius and a 75\,mm outer radial allowance for
electronics and services with full $2\pi$ azimuthal coverage. The
EMCal will have integral support for the Cryostat and/or clearance for
support from a lower structure. The EMCal will also incorporate
provision for support of itself in the fully assembled configuration, any
maintenance configuration, and for assembly/disassembly and integration
of component segments.  The EMCal will be constructed of tungsten (2\,mm
thick) and light fibers (1\,mm thick).
These will be grouped into 314 azimuthal segments. Details of the mechanical
design of the EMCal segments is covered in Section~\ref{sec:emcal}.

\subsection{Hadronic calorimeter} The HCal will be 900\,mm in radial
thickness, with full $2\pi$ azimuthal coverage, and with the calorimeter divided
into an inner radial section and an outer radial section.  The inner
radial section will be 300\,mm in radial thickness with a 75\,mm inner
radial allowance for readout electronics and services.  The outer
calorimeter will be 600\,mm in radial thickness with a 75\,mm outer
radial allowance for readout electronics and services.  The HCal will
have integral support for the EMCal and Cryostat and/or clearance for
support from lower structure. The HCal will also incorporate provision
for support of itself in the fullys assembled configuration, any maintenance
configuration and for assembly/disassembly and integration of
component segments.HCal will be constructed of 256 segments of 6\,mm
thick scintillator sections with embedded optical fibers to accumulate
the light energy.  The scintillator sections will be sandwiched between
tapered steel plates angled at 5 degrees from the radial direction,
with the inner steel dividers angled in the opposite direction from
the outer steel and offset by a half a segment thickness. Details of the 
mechanical design of the HCal segments is covered in Section~\ref{sec:hcal}.

\subsection{Structural support apparatus} Structural support for the
sPHENIX major components will provide appropriate structural support
for all of the equipment with the following criteria:

\begin{itemize}

\item Appropriate structural support will be provided to all
  components, with integral connections and support interfaces and/or
  clearances for support structure designed into the comprising
  detector subassemblies and the superconducting solenoid.

\item Components will be able to be completely assembled in the PHENIX
  Assembly Hall (AH) utilizing existing cranes (40 ton max.). The
  assembly will be mounted on the existing PHENIX rail system or a
  modification of the existing rail system.

\item Functional tests including pressure, and magnetic tests will be
  able to be performed in the AH.

\item The sPHENIX will have designed-in capabilities to separate into
  subdivisions to allow maintenance of any electronics, support
  services and replaceable components. This capability will be
  available with the full assembly in the AH or the Interaction Region
  (IR), with full maintenance capabilities during shutdowns between
  runs and with as much maintenance capabilities during a run as
  possible.

\item The sPHENIX assembly will be relocatable from the AH to the IR
  using the existing rail system or a modification to the existing
  rail system. This relocation may be accomplished fully assembled or
  disassembled into subdivisions which are reassembled in the IR.
  Disassembly and re-assembly will use existing AH   and IR cranes.
  
\item Support equipment for the above components and the utilities
  supplied to the above structure including provision for electronics racks, cooling
  services, cryogenics, power and signal cables, and monitoring and
  control equipment will be provided.

\item The assembled sPHENIX will allow partial disassembly during
  maintenance periods to provide access to all serviceable components,
  electronics and services. The assembled sPHENIX will provide for
  electronics racks and all other support components for operation and
  monitoring of the sPHENIX active components. Safe and efficient
  access to all service/monitoring components will be integrated into
  the design of the underlying structural support.

\item Infrastructure used successfully for the past twelve years of of PHENIX
  operation will be adapted and expanded to support sPHENIX.

\end{itemize}

\chapter{Jet, Dijet, and $\gamma$-Jet Performance}
\label{chap:jet_performance}

In this Chapter we detail the sPHENIX jet, dijet, and $\gamma$-jet reconstruction performance and demonstrate 
the ability to measure key observables that can test and discriminate different quenching
mechanisms and coupling strengths to the medium.  
The important aspects of jet performance
are the ability to find jets with high efficiency and purity, and to
measure the kinematic properties of jet observables with good
resolution.  In addition, it is necessary to discriminate between jets from
parton fragmentation and \fake jets caused by fluctuations in the underlying event background.
For the sPHENIX physics program, there are three crucial
observables that we have simulated in detail to demonstrate the jet performance: 
single inclusive jet yields, \dijet correlations, and \gj correlations.  There are other
significant observables such as the participant plane dependence (e.g. $v_{2}$, $v_{3}$, etc.) of jets and jet-hadron correlations that are also
enabled by this upgrade.  

\section{Simulations}

sPHENIX will sample jet observables from 50 billion \auau minimum bias interactions per year.
It is not possible to simulate with full \geant~\cite{Agostinelli:2002hh} the equivalent data sample.
Thus, we perform three different levels of simulations described in detail below.  
The most sophisticated and computationally intensive are full \geant simulations with \pythia~\cite{Sjostrand:2000wi} 
or \hijing~\cite{Gyulassy:1994ew}
events where all particles are traced through the magnetic field, energy deposits in the calorimeters recorded, clustering
applied, and jets are reconstructed via the \fastjet package~\cite{Cacciari:2005hq}.  We utilize this method to determine the jet
resolution in \pp collisions from the combined electromagnetic and hadronic calorimeter information.  For studies
of \fake jets in \auau central collisions, one needs to simulate hundreds of millions of events and for this we
utilize a \fast simulation where the particles from the event generator are parsed by their particle type, smeared
by the appropriate detector resolution parametrization from \geant simulations, and segmented into detector cells.
As described in detail below, a full underlying event
subtraction procedure is applied, and then jets reconstructed via \fastjet.  This method is also utilized for
embedding \pythia or \pyquen~\cite{Lokhtin:2005px} (a jet quenching parton shower model with parameters tuned to RHIC data)  
events into \auau \hijing events to study \dijet and \gj observables.
Finally, in order to gain a more intuitive understanding of the various effects, we run a \veryfast simulation
where \pythia particles are run directly through \fastjet and then the reconstructed jet energies smeared by the
parametrized resolutions and underlying event fluctuations.

The Chapter is organized as follows.  First we describe the jet
reconstruction and evaluate its performance in \pp collisions for both
an idealized detector as well as a fully simulated version.  Then we
describe our study of \fake jet contamination, which has already
been submitted for publication in Physical Review
C~\cite{Hanks:2012wv}.  Finally we show the expected performance for
sPHENIX measurements of inclusive single jet, \dijet and \gj
processes.


\section{Jet finding algorithm}

For all of the studies presented here we use the anti-$k_T$ jet
algorithm~\cite{Cacciari:2008gp} implemented as part of the \fastjet
package~\cite{Cacciari:2005hq}.  The \antikt algorithm is well suited
to heavy ion collisions and produces cone-like jets in an infrared and
collinear safe procedure.  The parameter that controls the size of the
jet in this algorithm is the jet radius, $R$.  While this is not
strictly a cone size it does specify the typical extent of the jet in
$\eta$-$\phi$ space.  High energy experiments typically use large $R$
values of 0.4--0.7 in order to come as close as possible to capturing
the initial parton energy.  In heavy ion collisions, the desire to
measure the quenching effects on the jet profile and to minimize the
effects of background fluctuations on jets has led to the use of a
range of $R$ values.  Values from 0.2 to 0.5 have been used to date in
\pbpb collisions at 2.76\,TeV at the
LHC~\cite{Aad:2010bu,Chatrchyan:2012ni}.  We note that looking at the
jet properties as a function of the radius parameter is very
interesting and potentially sensitive to modifications to the jet
energy distribution in the medium.  For the studies presented here we
use $R$ values of 0.2, 0.3, and 0.4.

\section{Jet performance in \pp collisions}
\label{sec:pp_jet_performance}

We begin by exploring the performance of the detector in \pp
collisions.  This allows us to investigate the effects of detector
resolution and to investigate how well the process of unfolding these
effects in simpler collisions works before considering the additional
effects of the underlying event and jet quenching in heavy-ion collisions.

The most realistic understanding of the sPHENIX jet reconstruction
performance comes from a full \geant simulation of the detector
response.  In this case, \pythia particles are run through a \geant
description of sPHENIX, the resulting energy deposition is corrected
for by the sampling fraction of the relevant calorimeter, binned in
cells of $\eta$-$\phi$ ($0.024\times 0.024$ for the EMCal and
$0.1\times 0.1$ for the HCal) and the resulting cells are used as
input to \fastjet, which is used to cluster the full events into jets.
Particles from \pythia events are put through \fastjet to determine
the truth jets.  The distribution of the difference between the truth
and reconstructed jet energy is used to determine biases in the jet
energy measurement as well as the jet energy resolution.

For the purposes of this proposal, it is important to extend this
characterization to include the response of the system of calorimeters
to the passage of a jet consisting of statistical ensemble of
particles of a variety of particle species.  This is a rather different study
than for the single particle response.  For example, even if one looks
only at jets of a particular energy, the fraction of electromagnetic
and hadronic energy will vary from jet to jet, which is found to make
a strong contribution to the jet energy resolution.

\begin{figure}[hbt!]
  \centering
  \includegraphics[trim = 0 0 0 0,clip,width=0.6\linewidth]{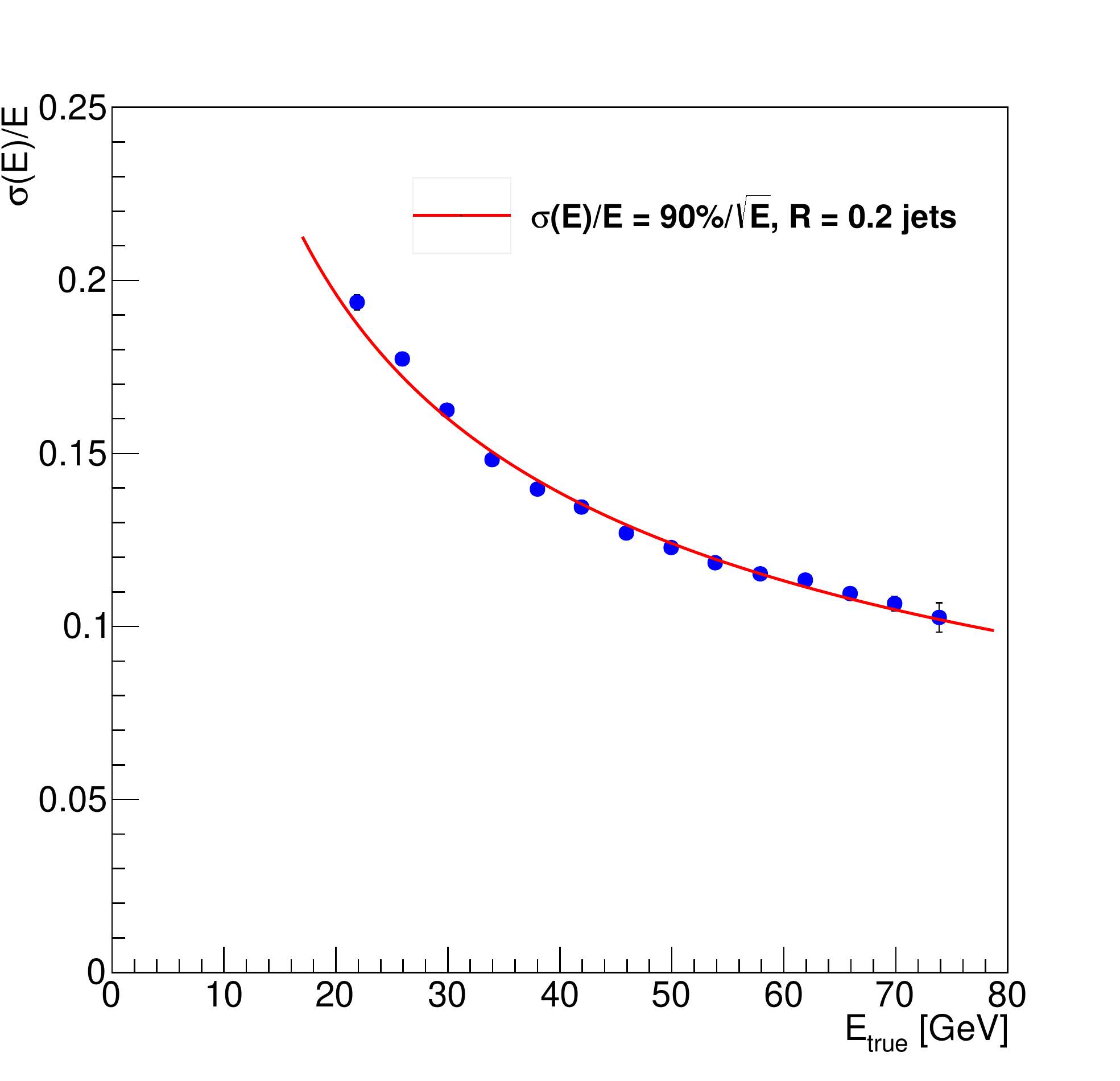}
  \caption{The energy resolution of single jets in \pp collisions
    reconstructed with the \fastjet \antikt algorithm with $R = 0.2$.
    The particles have been put through a full \geant description of
    the apparatus.}
  \label{fig:jet_energy_resolution}
\end{figure}

From the set of matched jet pairs (i.e. truth and reconstructed jets),
one next determines the difference between the energy of the
reconstructed calorimeter jets, $E_{\rm reco}$, and the particle-level
truth jets, $E_{\rm true}$.  The width of this distribution,
$\sigma(E)$, as a function of $E_{\rm true}$ is determined.  Finally
$\sigma(E)/E$ versus $E$ is fit with a functional form
\begin{equation}
  \label{eq:jet_energy_resolution}
  \frac{\sigma(E)}{E} = \frac{a}{\sqrt{E}} + b
\end{equation}
The energy resolution for jets reconstructed with \antikt and $R=0.2$
is shown in Figure~\ref{fig:jet_energy_resolution} and results in a
resolution of 90\%/$\sqrt{E}$ and a constant term of order 1\%.

It is notable that often the jet energy resolution in collider
experiments is found to be a factor of 1.2--1.3 worse than the
quoted single particle resolution of the hadronic calorimeter.  This
factor is a balance of many effects including the better resolution
for the electromagnetic part of the shower, soft particles that
deflect out of the jet cone in the magnetic field, some lost energy,
etc.   The CMS quoted jet resolution in \pp
collisions at 7.0\,TeV is approximately
120\%/$\sqrt{E}$ which is roughly 1.2 times worse
than the quoted single particle hadronic calorimeter resolution~\cite{CMSJets:2010}.
There are various methods to improve upon these resolutions, and the
value for sPHENIX of 90\%/$\sqrt{E}$ is consistent with this
expectation given the hadronic calorimeter single particle resolutions
described previously.

\subsection{\pp Inclusive Jet Spectra}
\label{sec:pp_very_fast_simulation}

In order to model the jet resolution effects described above on the
inclusive jet spectra in \pp collisions at $\sqrt{s_{NN}} = 200$\,GeV,
we have used the \veryfast simulation.  This method entails running
\pythia, sending the resulting final state particles through \fastjet
to find jets, and then blurring the energy of the reconstructed jets.
We do not impose any detector response on the particles themselves.
Instead, the jet energies have been smeared by an energy dependent
resolution consistent with values obtained from the full \geant
simulation.

\begin{figure}[ht!]
  \centering
  \includegraphics[width=\onewidth]{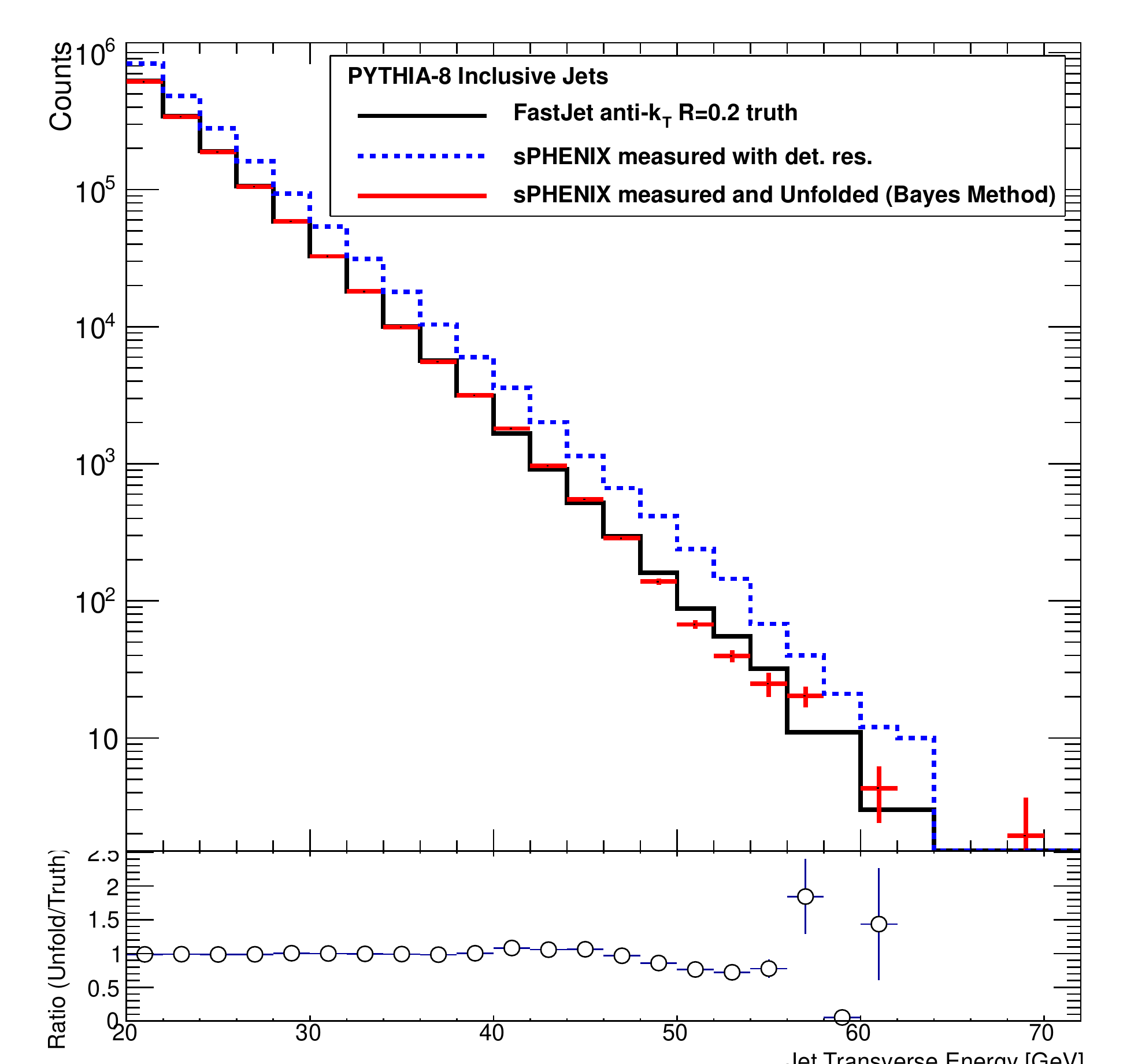}
  \caption{Unfolding the effect of finite detector resolution on jet
    reconstruction in \pp events.  The black histogram is the truth
    spectrum of jets from \pythia, the blue dotted histogram is the
    spectrum after smearing by the jet energy resolution
    and the red histogram shows the result of using
    \roounfold Iterative Bayes method to unfold the detector effects.
     The lower panel shows the ratio of the unfolded to the true $p_T$ spectrum.}
  \label{fig:pp_very_fast_inclusive}
\end{figure}

The truth spectrum of jets is obtained by using \fastjet to cluster
the \pythia~\cite{Sjostrand:2000wi} event with the \antikt algorithm.
Figure~\ref{fig:pp_very_fast_inclusive} shows the true jet \pt
spectrum as the solid histogram.  The convolution of the hard
parton-parton scattering cross section and the high-$x$ parton
distribution function results in a jet cross section that falls nearly
exponentially over the range 20--60\,GeV, before turning steeply
downward as it approaches the kinematic limit, $x = 1$.

Figure~\ref{fig:pp_very_fast_inclusive} also shows the \veryfast
simulation result for the measured jet \pt spectrum.  The main effects
of the jet resolution on the jet energy spectrum are to shift it to
higher energy and stiffen the slope slightly.  Both of these effects
can be undone reliably by a process of unfolding.  We have employed
the \roounfold~\cite{Adye:2011gm} package and for this demonstration
utilize the Iterative Bayes method with 4 iterations.  The results of
the unfolding are also shown in
Figure~\ref{fig:pp_very_fast_inclusive}, along with the ratio of the
unfolded to the true \pt spectrum, in the lower panel.  The ratio of
the two distributions demonstrates that the measurement provides an
accurate reproduction of the true jet energy spectrum.

\subsection{\pp Dijet Asymmetry}
\label{sec:dijet_pp}

\begin{figure}[hbt!]
  \centering
  \includegraphics[width=0.6\linewidth]{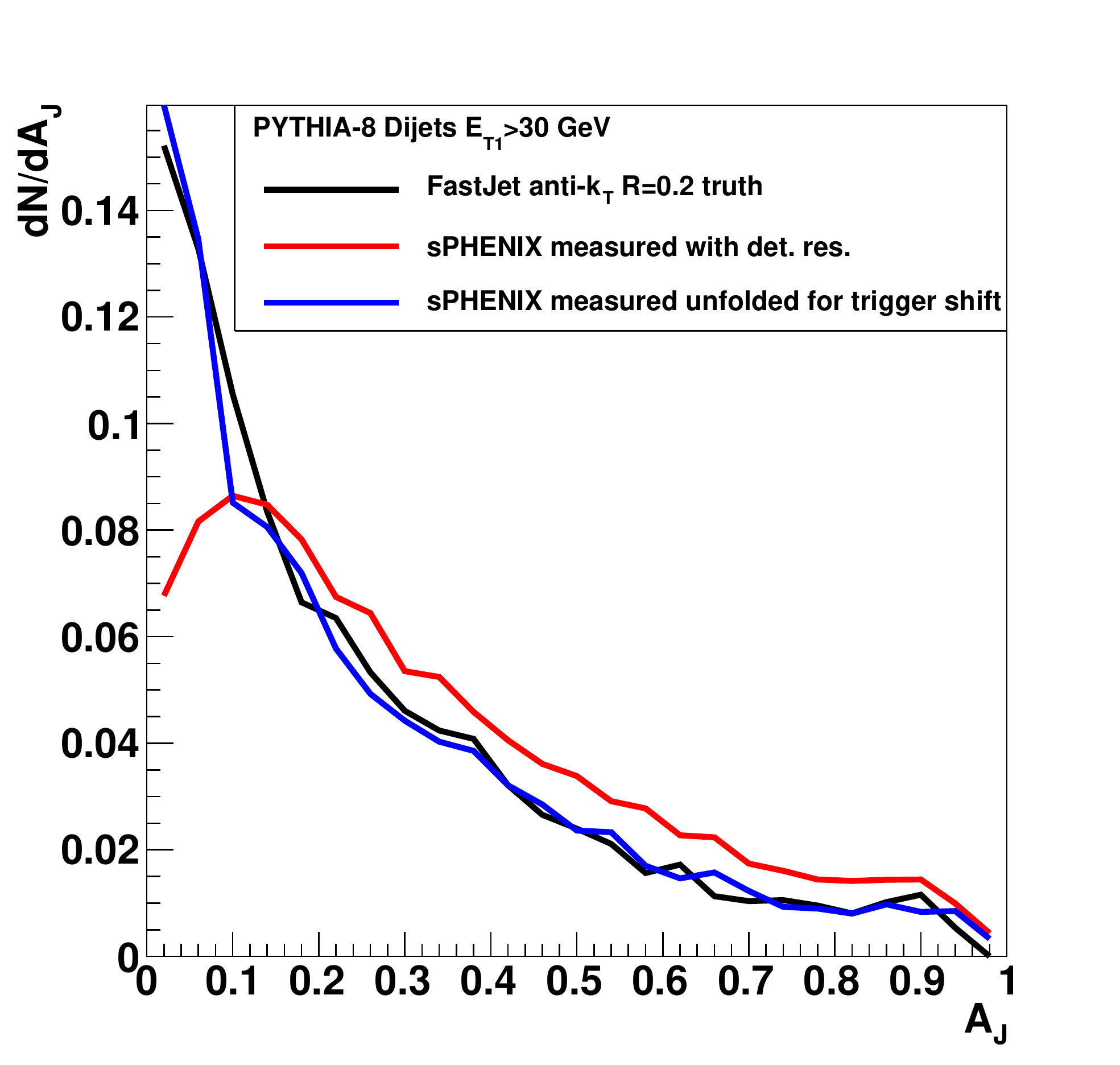}
  \caption{Dijet asymmetry, $A_J$, in \pp collisions.  The truth spectrum
    is shown in black; the spectrum measured in \pythia and smeared by
    the jet energy resolution is shown in red.  The effect of the
    unfolding of the trigger jet bias is also shown in blue.} 
  \label{fig:pp_very_fast_dijet}
\end{figure}

The \veryfast simulation is also used to establish expectations
for \dijet correlations.  Figure~\ref{fig:pp_very_fast_dijet}
shows the \dijet correlation for \pythia events reconstructed using the \antikt
algorithm with $R = 0.2$.  The highest energy jet in the event is taken
as the trigger jet and its transverse energy is compared to the transverse energy of the
highest energy jet in the opposite hemisphere.  

The jet asymmetry $A_{J} = (E_{T1} - E_{T2}) / (E_{T1} + E_{T2})$ for
the true jets, reconstructed at the particle level, is shown for
leading jets with $E_{T1} > 30$\,GeV in
Figure~\ref{fig:pp_very_fast_dijet}.  Also shown is the simulated
measurement with the smearing due to the jet resolution included.  It
is notable that this results in a significant reduction in the
fraction of events observed with balanced jet energies (i.e. near
$A_{J} \approx 0$).  To date, the ATLAS and CMS \dijet asymmetries in
Pb+Pb collisions have been published without unfolding for these
detector or underlying event
effects~\cite{Aad:2010bu,Chatrchyan:2011sx}.  A simultaneous
two-dimensional unfolding of both the jet energies (i.e., $E_{T1}({\rm
  meas}), E_{T2}({\rm meas}) \rightarrow E_{T1}({\rm true}),
E_{T2}({\rm true})$) is required in this case.  Both ATLAS and CMS
collaborations are actively working on this two-dimensional unfold,
and the sPHENIX group is as well.  At RHIC energies, the largest
effect is that the trigger jet is being selected from a steeply
falling spectrum and is biased by the resolution to be reconstructed
higher than the true energy.  If one simply shifts the trigger jet
down by this average bias (and inverts the identity of trigger and
associated jet if the trigger jet energy is then below that of the
associated jet), the original \dijet asymmetry distribution is
recovered, as shown in Figure~\ref{fig:pp_very_fast_dijet}.  This
procedure is not a replacement for the eventual two-dimensional
unfolding, but demonstrates the predominant effect.


\section{Jet Performance in \AuAu collisions}
\label{sec:aa_jet_performance}

Here we simulate the performance of inclusive jet and \dijet observables in heavy ion collisions,
where we have focused our simulations on 0--10\% central \auau collisions.
The sPHENIX trigger and data acquisition will sample jets from the full \auau minimum
bias centrality range, resulting in key measurements of the full centrality dependence of jet quenching
effects.    Finding jets and dealing with the rate of \fake jets
become much easier as the multiplicity drops, and so we have concentrated
on showing that we have excellent performance in central \auau collisions
(i.e., in the most challenging case).

The effective jet resolution also has a significant contribution from
fluctuations in the underlying event in the same angular space as the
reconstructed jet.  In addition, fluctuations in the underlying event
can create local maxima in energy that mimic jets, and are often
referred to as \fake jets.  While resolution effects can be accounted
for in a response matrix and unfolded, significant contributions of
\fake jets cannot be since they appear only in the measured
distribution and not in the distribution of jets from real hard
processes.  Thus, we first need to establish the range of jet
transverse energies and jet radius parameters for which \fake jet
contributions are minimal. Then within that range one can benchmark
measurements of the jet and \dijet physics observables.

\subsection{Jet and Fake Jet Contributions}

In this section we discuss both the performance for finding true jets
and estimations based on \hijing simulations for determining the contribution
from \fake jets.  
It is important to simulate very large event samples in order to evaluate the 
relative probabilities 
for reconstructing \fake jets compared to the rate of true high $E_T$ jets.  Thus, we employ 
the \fast simulation method and the \hijing simulation model for \auau collisions.
The ATLAS collaboration has found that the energy fluctuations in the heavy ion data are well matched by \hijing at 
$\sqrt{s_{NN}}=2.76$\,TeV~\cite{Cole:2011zz}.
We have also added elliptic flow to the \hijing events used here.
The \fast simulation takes the particles from the event generator and parses them by their particle type.
 The calorimeter energies are summed into cells based on the detector segmentation 
and each tower is considered as a four-vector for input into \fastjet.

\begin{figure}[hbt!]
  \centering
  \includegraphics[width=\onewidth]{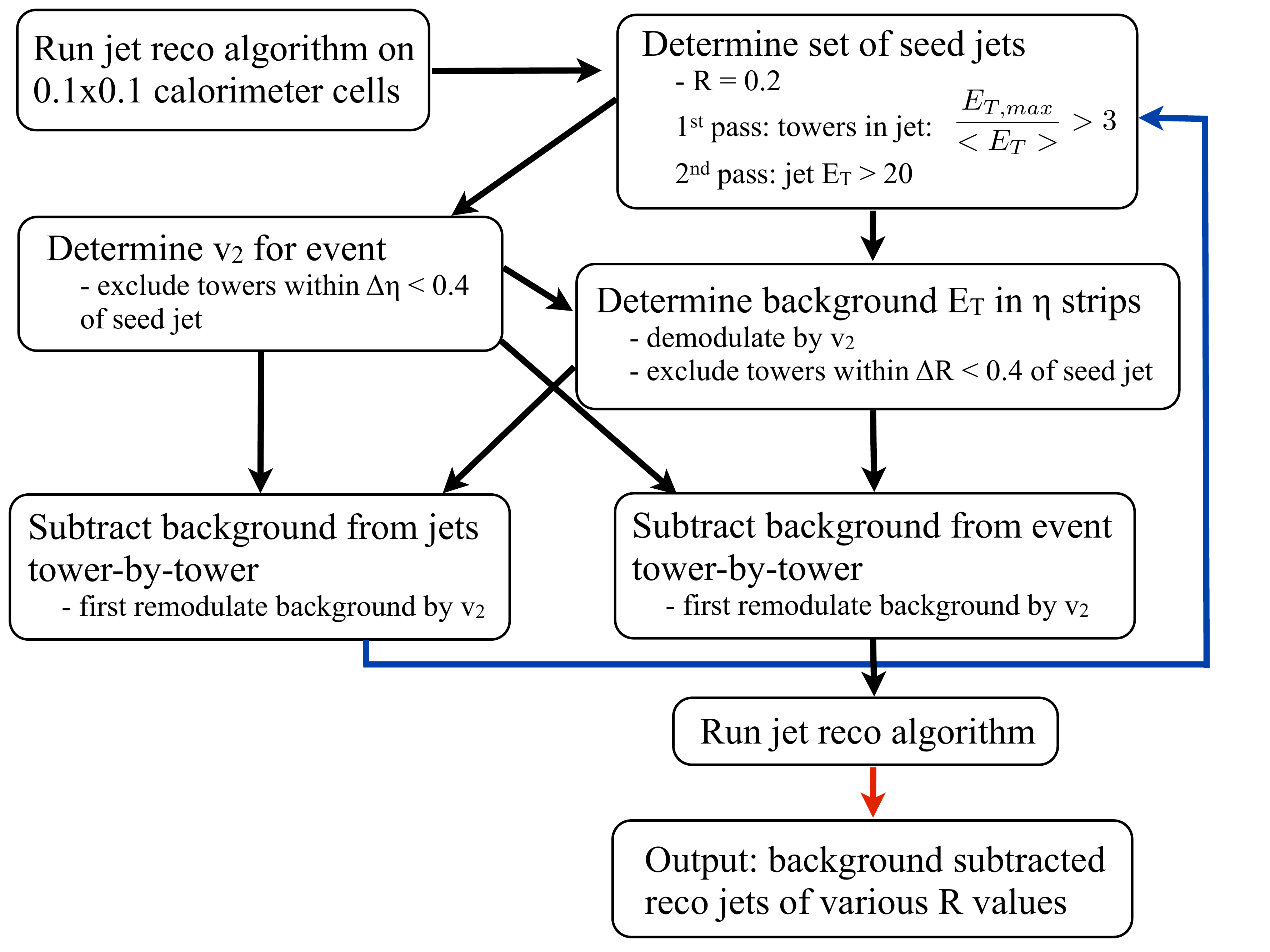}
  \caption{
Candidate jets are found and temporarily masked out of the event.  The
remaining event background is then characterized by the strength of
its $v_2$ (the effect of higher Fourier coefficients was not included in this study)
and overall background level.  A second set of candidate jets are determined
and the background and $v_2$ are recalculated.  The jet finding algorithm
is then re-run on the background subtracted event to determine
the collection of final reconstructed jets. For a detailed description see Ref.~\protect\cite{Hanks:2012wv}{}.}
  \label{fig:flowchart}
\end{figure}

Any jet measurements in heavy ion collisions must remove the
uncorrelated energy inside the jet cone from the underlying event.  Various
methodologies have been applied to this problem.
The approach developed in our studies is described in detail in Ref.~\cite{Hanks:2012wv}.
A schematic diagram of the algorithm (based on the ATLAS heavy ion method) is shown in Figure~\ref{fig:flowchart}.
Candidate jets are found and temporarily masked out of the event.  The
remaining event background is then characterized by the strength of
its $v_2$ (the effect of higher Fourier coefficients was not included in this study)
and overall background level in individual slices in pseudorapidity.  
New candidate jets are determined
and the background and $v_2$ are recalculated.  The jet finding algorithm
is then re-run on the background subtracted event to determine
the collection of final reconstructed jets.  
This process is then run iteratively to a convergent result.

In order to distinguish true jets from \fake jets we have augmented the \hijing code to run
the \fastjet \antikt algorithm with the output of each call to the fragmentation routine (HIJFRG).
In this way the true jets are identified from a single parton
fragmentation without contamination from the rest of the simulated event.
The reconstructed jets can then be compared to these true jets.  Reconstructed jets which are within
$R = \sqrt{\Delta\eta^2 + \Delta\phi^2} < 0.25$ of a true jet with $E_T>5$\,GeV are considered to be matched
and those which are not are classified as \fake jets.  Other estimates of \fake jet rates in heavy ion collisions
have failed to take into account how the structure of the background fluctuations and the detector
granularity affects the probability of any particular fluctuation being reconstructed as a jet.    Note that
simply blurring individual particles by a Gaussian with an underlying event fluctuation energy results in a substantial
overestimate of the \fake jet rate, and is not a replacement for a complete event simulation incorporating \fastjet reconstruction
with a full jet and underlying event algorithm implementation.
Thus, we believe these studies provide an accurate assessment of the effect of \fake jets.

\begin{figure}[hbt!]
  \centering
  \begin{minipage}[c]{0.47\linewidth}
    \includegraphics[width=\textwidth]{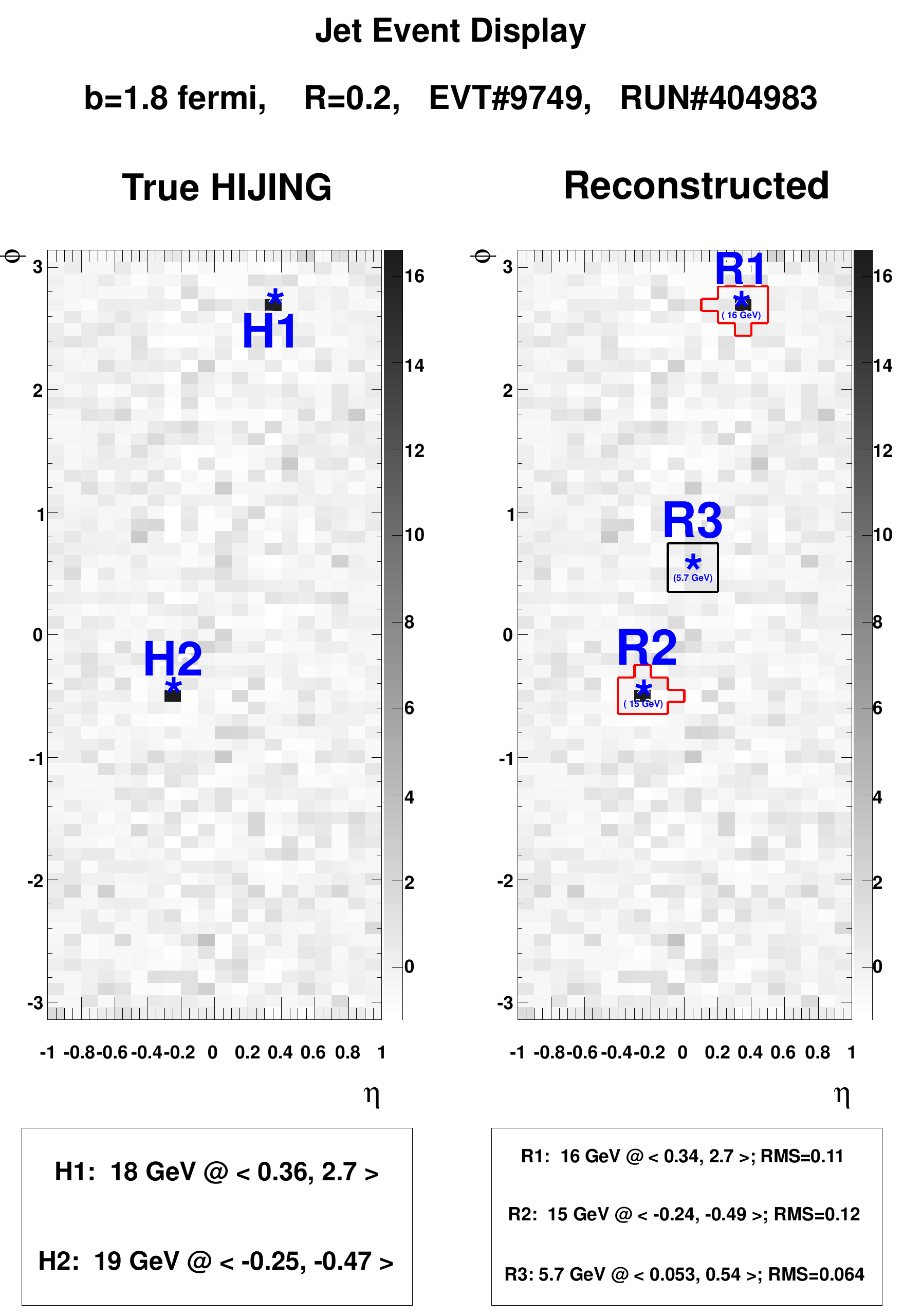} 
  \end{minipage}
  \begin{minipage}[c]{0.47\linewidth}
     \includegraphics[width=\textwidth]{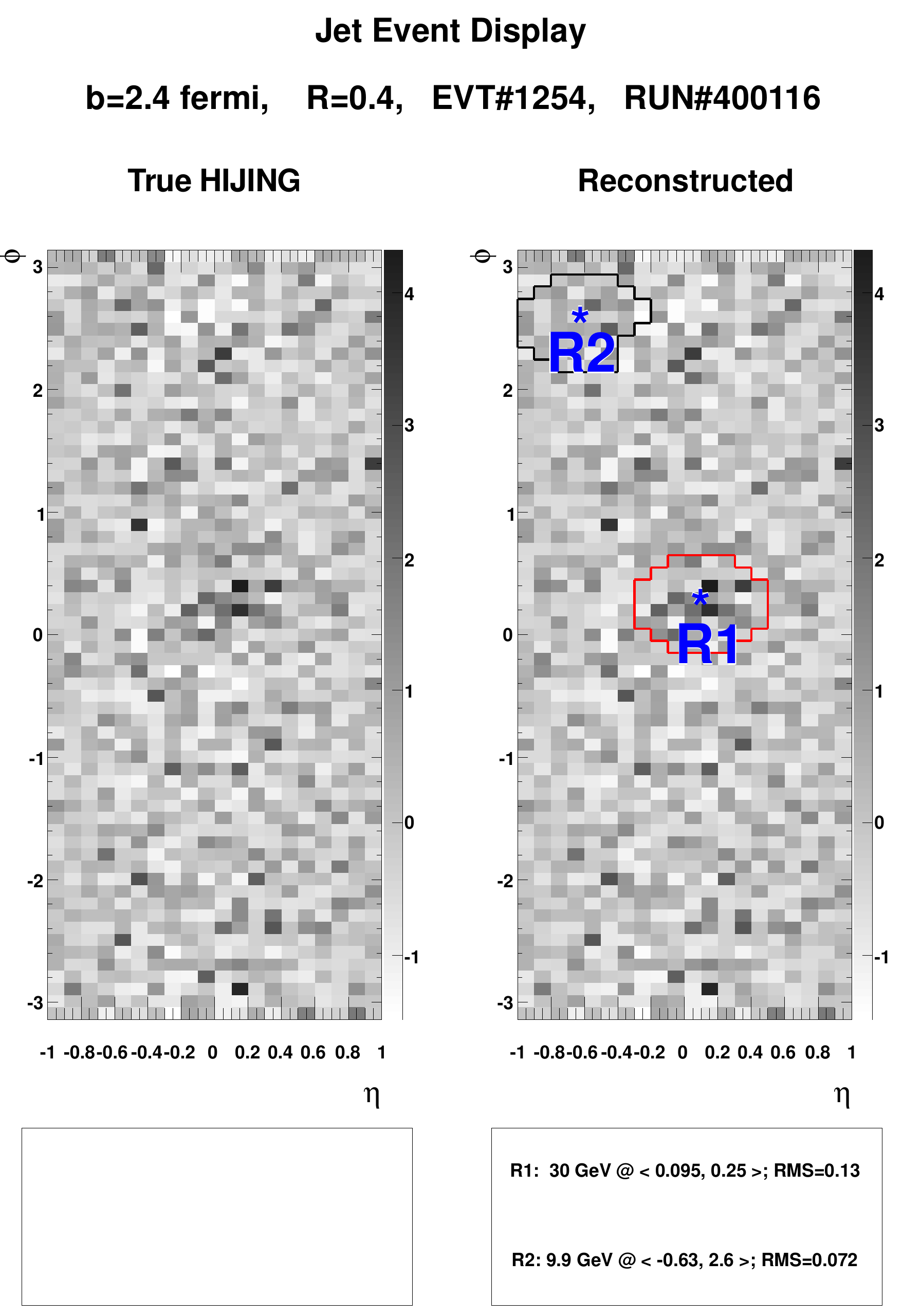}
\end{minipage}
  \caption{Event displays of true and reconstructed jets shown overlaid on
background subtracted calorimeter towers from fast simulations.  The left
event shows a \hijing \dijet event where both dijets (labeled H1 and H2) are 
reconstructed and matched (R1 and R2).  A third jet, not matched to a true jet,
is also reconstructed (R3).  The right event shows a \hijing event with no
true jets with $E_T>5$\,GeV.  Two \fake jets are reconstructed, one with
$E_T=30$\,GeV.  From Ref.~\protect\cite{Hanks:2012wv}{}.}
  \label{fig:eventdisplays}
\end{figure}

As an illustration of true and \fake jets we show two calorimeter event displays in 
Figure~\ref{fig:eventdisplays}.
True jets at high $E_T$ are a rare occurrence.  A large energy background
fluctuation at high $E_T$ that mimics a jet is also a rare occurrence.
Thus the only way to quantify the impact of \fake jets on the jet performance
is to run a large sample of untriggered simulated events and assess the relative
probability of true and \fake jets as a function of $E_T$ and R.

\begin{figure}[hbt!]
  \centering
  \begin{minipage}[c]{0.67\linewidth}
    \includegraphics[width=\textwidth]{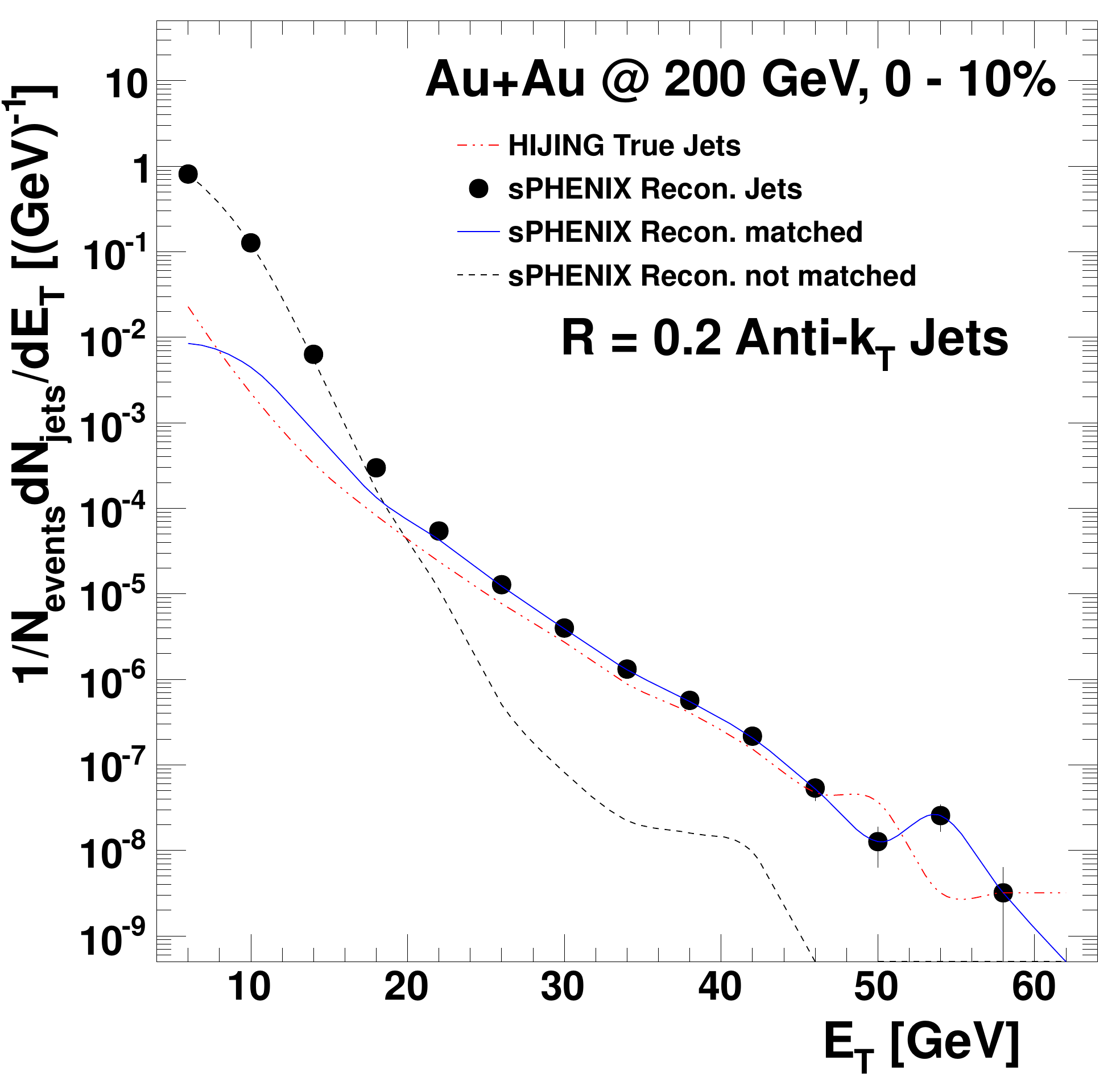} 
  \end{minipage}
  \begin{minipage}[c]{0.32\linewidth}
    \includegraphics[trim=0 0 40 0,clip,width=\textwidth]{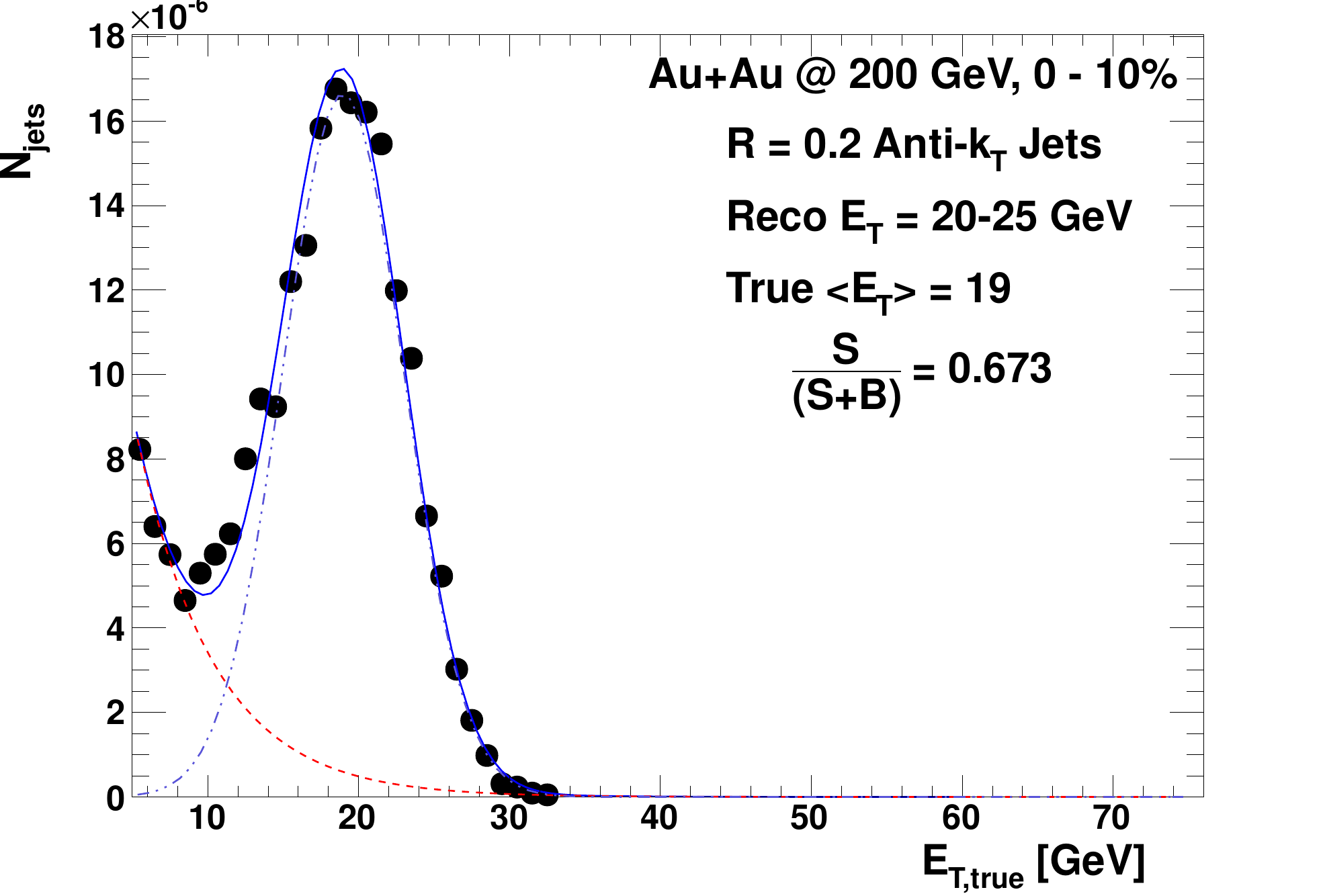}
    \\
    \includegraphics[trim=0 0 40 0,clip,width=\textwidth]{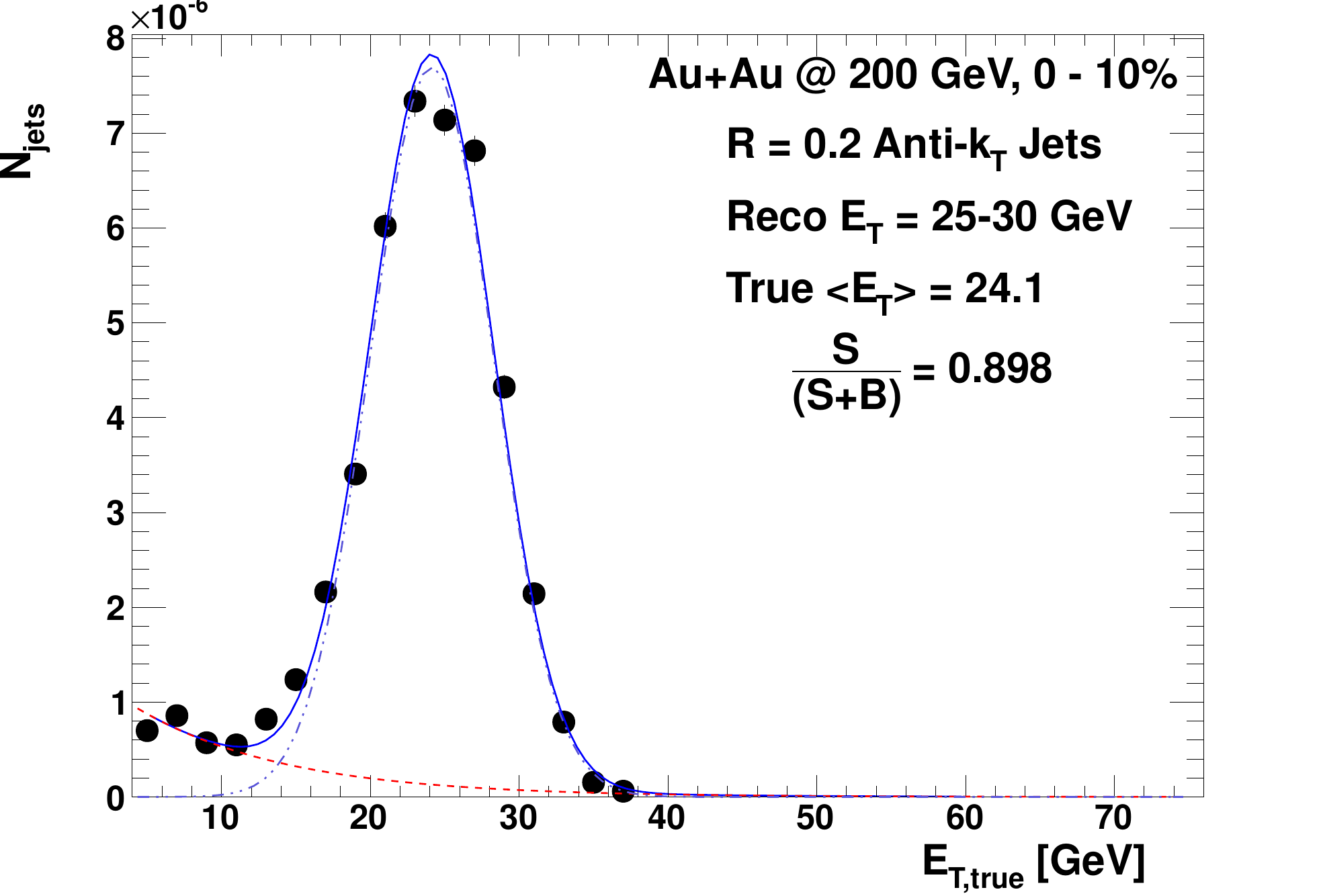}
    \\
    \includegraphics[trim=0 0 40 0,clip,width=\textwidth]{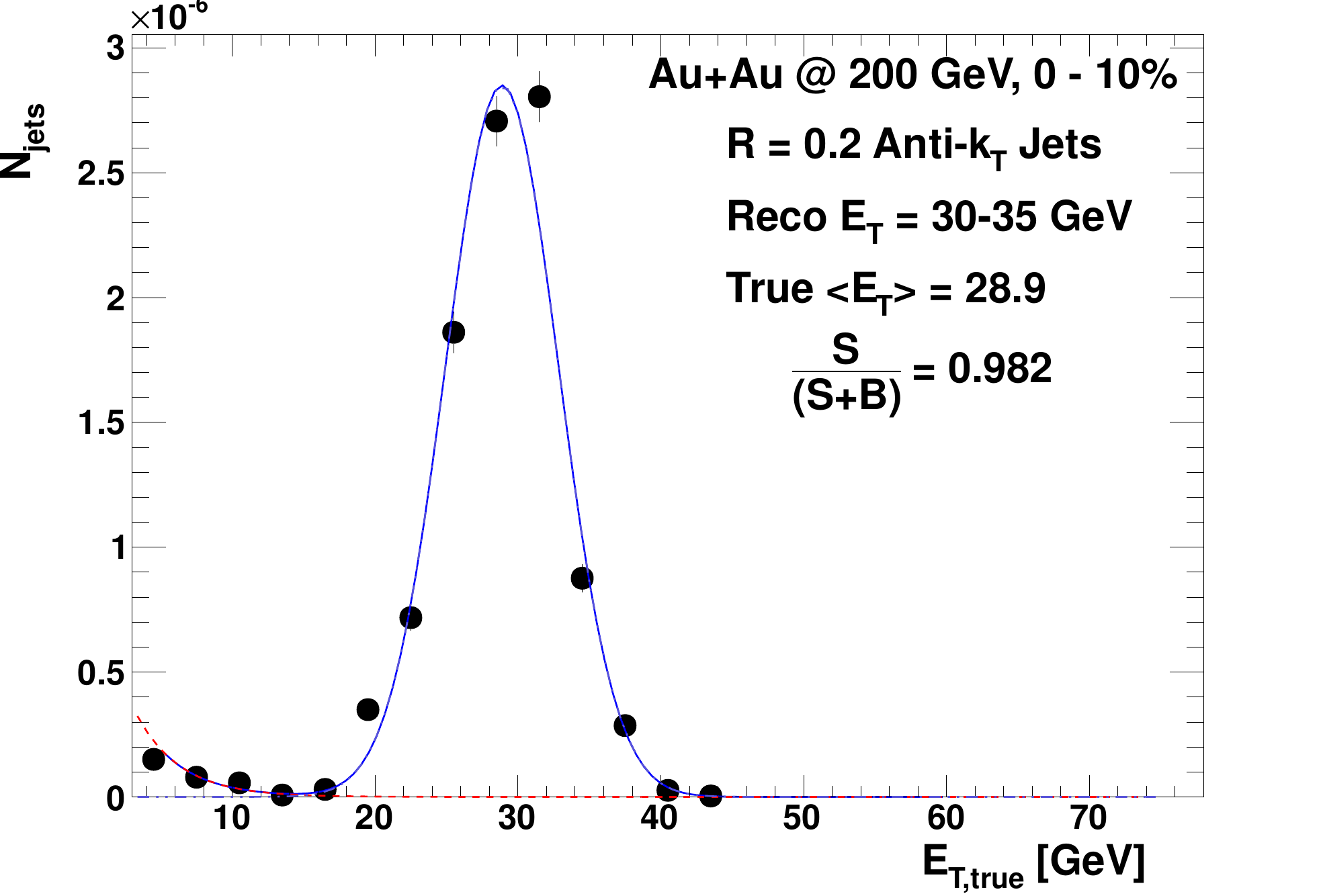}
  \end{minipage}
  \caption{The composition of the jet spectrum in central 0-10\% \auau based on
750M \hijing events.  The
    full spectrum is shown in the left plot as solid points.  The
    spectrum of those jets that are successfully matched to known real
    jets is shown as a blue curve.  That curve compares very well with
    the spectrum of true jets taken directly from \hijing.  The jets
    which are not matched with known jets are the \fake jets, and the
    spectrum of those jets is shown as the dashed curve. For $R = 0.2$,
    real jets begin to dominate over \fake jets above 20\,GeV.  The panels on
    the right are slices in true jet energy showing the distribution
    and make up of the reconstructed jet energy.  At low $E_{\rm
      true}$, \fake jets can be seen encroaching on the low energy side of
    the distribution.  For higher $E_{\rm true}$ the \fake jets are
    negligible.}
  \label{fig:auau_fake_rate}
\end{figure}

\begin{figure}
\centering
\begin{minipage}[c]{0.49\linewidth}
\includegraphics[width=\textwidth]{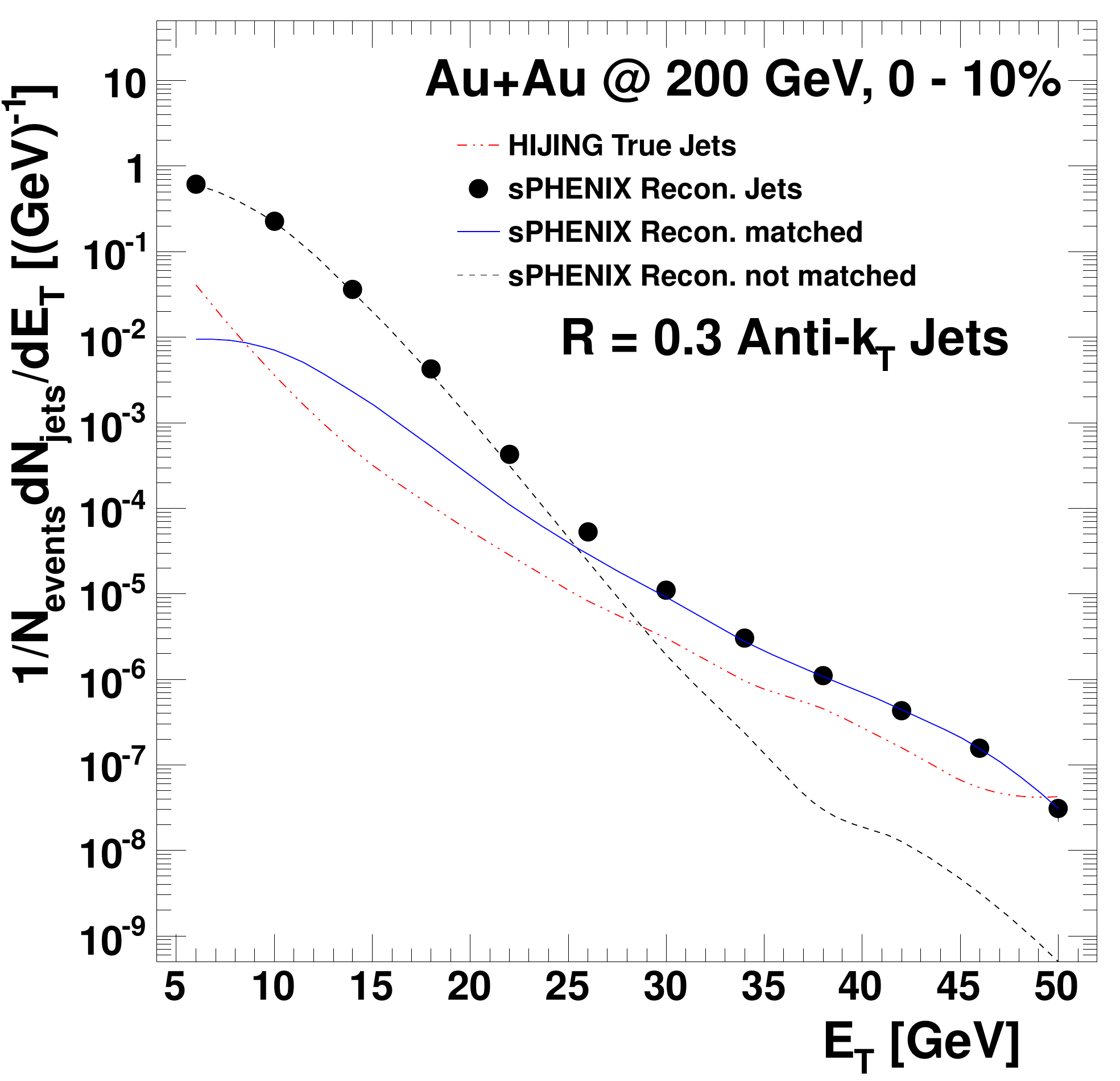}
\end{minipage}
\begin{minipage}[c]{0.49\linewidth}
\includegraphics[width=\textwidth]{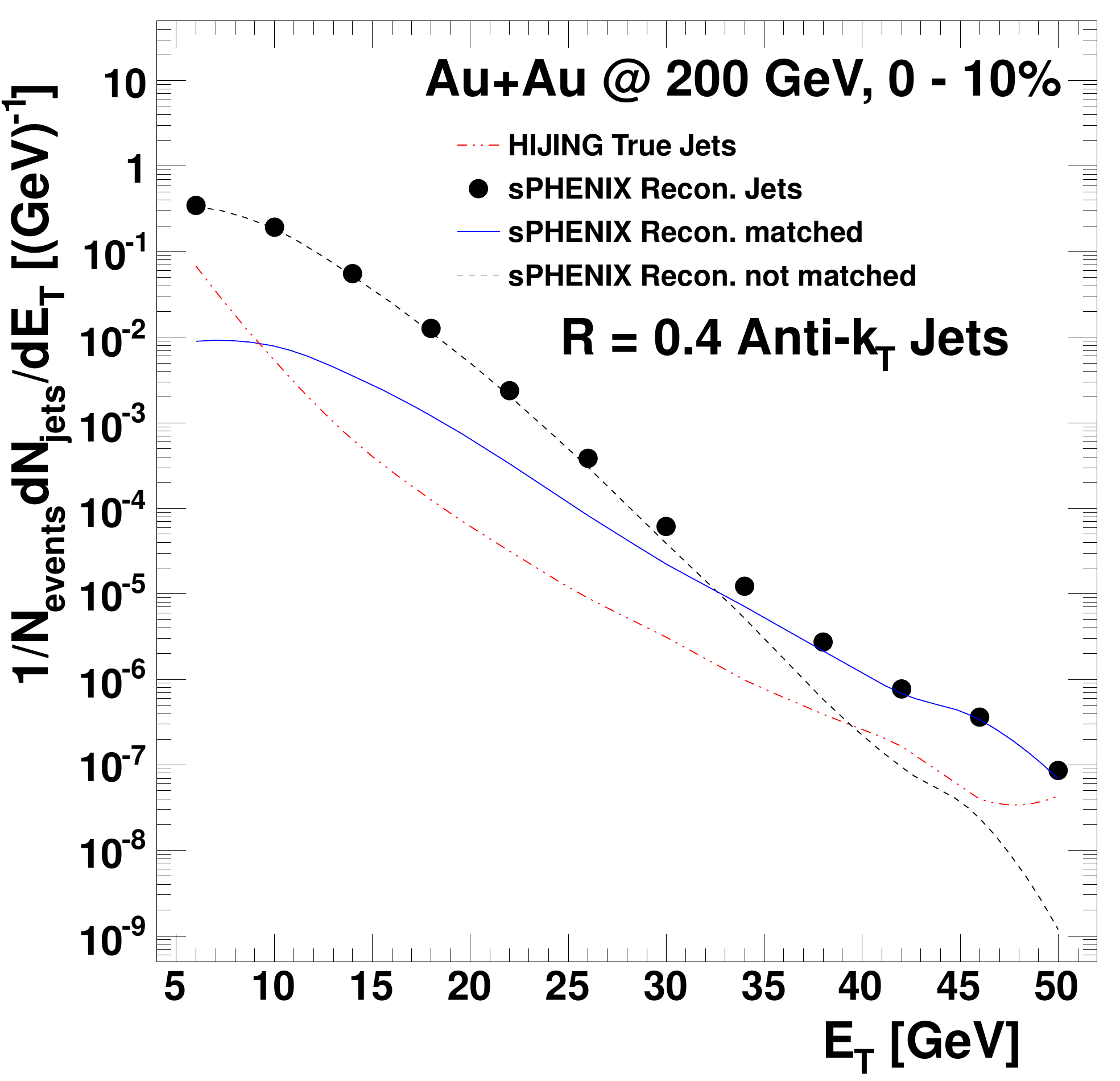}
\end{minipage}
\caption{Composition of the jet spectra in central 0-10\% \auau based on 750 million
\hijing events for $R=0.3$ (left) and $R=0.4$ (right) jets.}
\label{fig:auau_fake_rate_wide}
\end{figure}

A sample of over 750 million minimum bias \hijing events with
quenching turned off was used in these studies~\cite{Hanks:2012wv}.
The observable particles are binned in $\eta$-$\phi$ cells of size
$\Delta \eta \times \Delta \phi = 0.1 \times 0.1$.  In these
studies~\cite{Hanks:2012wv}, we have not included smearing due to
detector resolution as it is expected to be a sub-dominant effect and
we want to isolate the effects of the underlying event.  At the end of
this Section we present results including detector resolution that do
not change the key conclusions of these studies.  The \fast simulation
result for $R = 0.2$ jets without including detector-level smearing of
the jet energies is shown in Figure~\ref{fig:auau_fake_rate}.  The
full spectrum is shown on the left as solid points.  The spectrum of
those jets that are successfully matched to true jets is shown as a
blue curve.  That curve compares very well with the spectrum of true
jets taken directly from \hijing.  The jets which are not matched with
true jets are the \fake jets, and the spectrum of those jets is shown
as the dashed curve. For $R = 0.2$, real jets begin to dominate over
\fake jets above 20\,GeV.  The panels on the right of
Figure~\ref{fig:auau_fake_rate} are slices in reconstructed jet energy
showing the distribution and make up of the true jet energy.  For
reconstructed jets with $E_T = $25--30\,GeV, a contribution of \fake
jets can be seen encroaching on the low energy side of the
distribution.  For $E_{\rm reco}>25$\,GeV \fake jets are at the 10\%
level and for $E_{\rm reco}>30$\,GeV \fake jets are negligible.
Contributions from \fake jets for larger jet cones are shown in
Fig~\ref{fig:auau_fake_rate_wide}.  The true jet rate becomes large
compared to the \fake jet rate at 30\,GeV for $R=0.3$ and 40\,GeV for
$R=0.4$.  We note that in one year of RHIC running, sPHENIX would
measure $10^{5}$ jets with $E_T > 30$\,GeV and $10^{4}$ jets with
$E_T > 40$\,GeV.

There are various algorithms for rejecting \fake jets based on the jet
profile or the particles within the jet.  These methods applied by the
ATLAS experiment significantly reduce the \fake rate by an order of
magnitude or more, increasing the energy and $R$ values over which it
is possible to measure jets.  We are currently studying the utility of
\fake jet rejection, including a simple requirement that there be a
charged particle track within the jet radius with $p_T > 2$\,GeV/c.
These methods may extend the sPHENIX jet measurements to significantly
lower $E_T$, and also provide a tool for evaluating the extent of
\fake jet contributions.

\begin{figure}[hbt!]
  \centering
  \includegraphics[angle=0,trim = 0 0 0 0,clip,width=\onewidth]{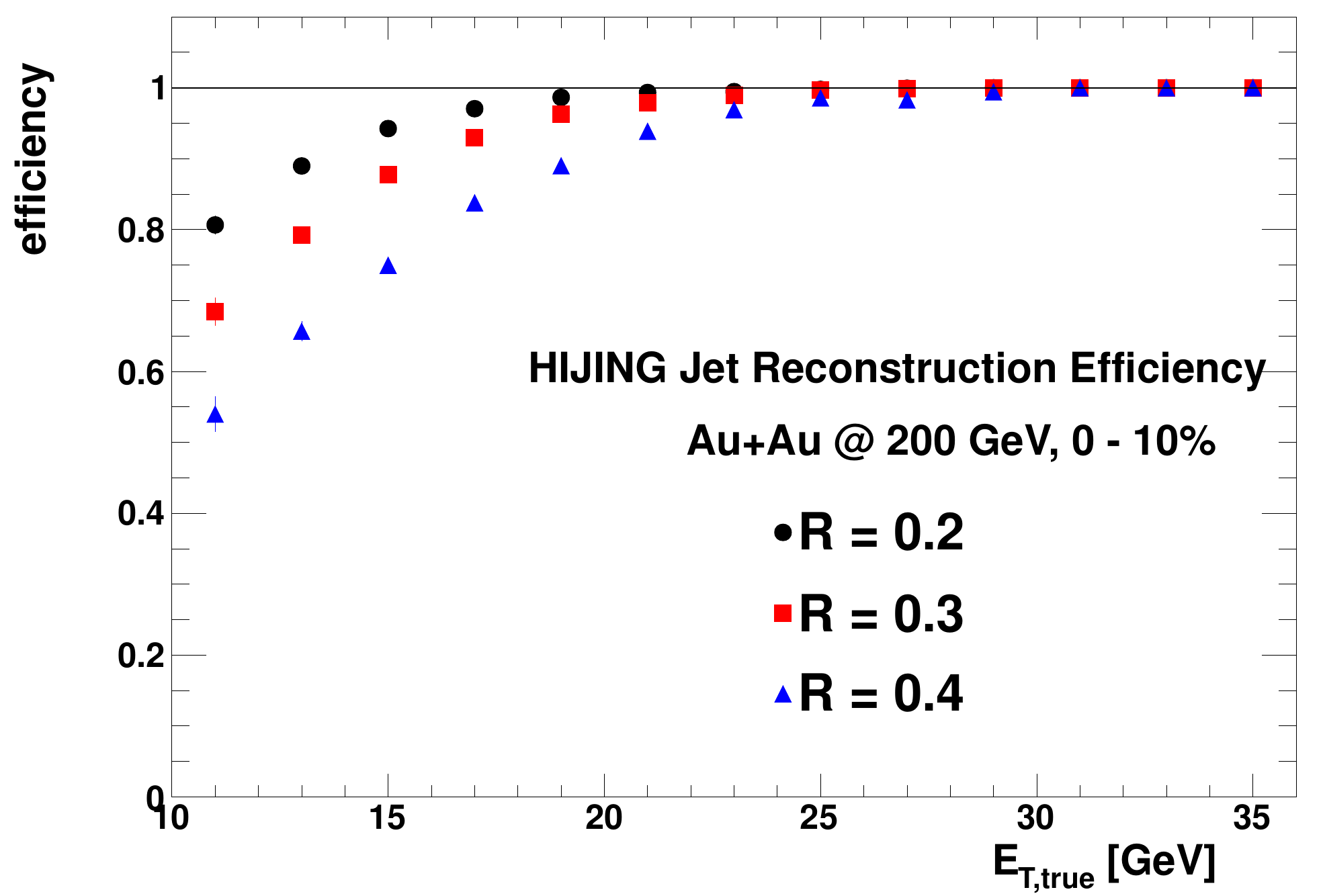}
  \caption{The efficiency for finding jets in central \auau collisions
    as a function of true jet energy and for $R = 0.2$, 0.3 and 0.4.}
  \label{fig:reconstruction_efficiency}
\end{figure}

The efficiency of finding true jets is shown in 
Figure~\ref{fig:reconstruction_efficiency}. We find $>95$\%
efficiency for finding jets above 20\,GeV reconstructed with $R =
0.2$ or 0.3 and above 25\,GeV for jets reconstructed using $R = 0.4$.

\begin{figure}[hbt!]
  \centering
  \includegraphics[width=0.8\textwidth]{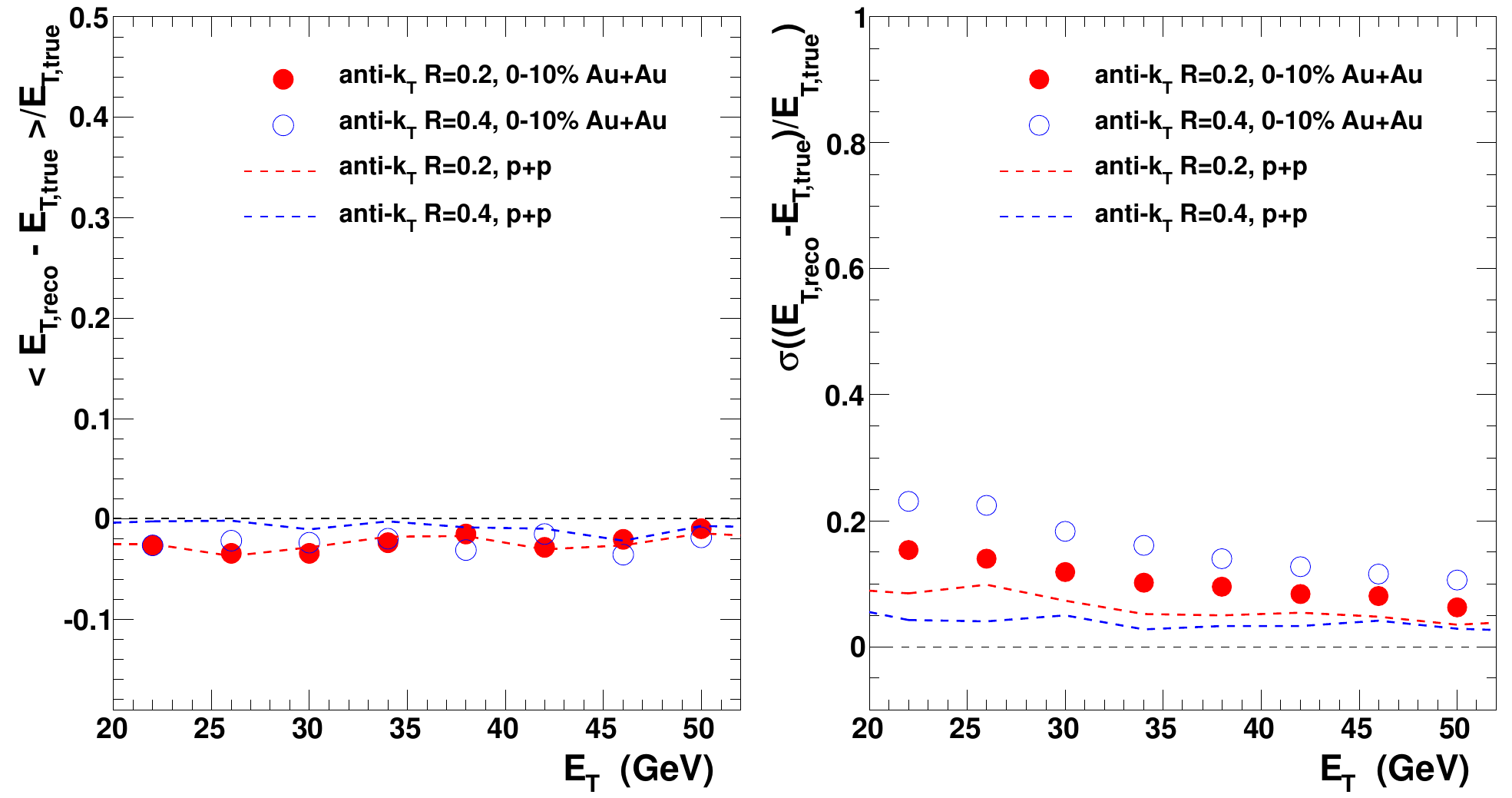}
  \caption{The energy scale of reconstructed jets in \auau
    collisions. The left plot shows the shift in the mean energy of
    the reconstructed jets compared to the true value.  There is only a
    few percent shift in the energy and no apparent dependence on jet
    cone size.  The right plot shows the jet energy
    resolution.  From Ref.~\protect\cite{Hanks:2012wv}{}.}
  \label{fig:auau_jet_energy_scale}
\end{figure}

Having found the jets in \auau with good efficiency and having
established that the rate of \fake jets coming as a result of
background fluctuations are understood and under control, we also need
to show that we can reconstruct the kinematics of jets accurately and
precisely. This is quantified by the jet energy scale, the average shift
of the jet energy between reconstructed and true jets and the
jet energy resolution which shows the relative width of the 
difference between the true and reconstructed jet energies. 
Results from $R = 0.2$ and 0.4 are shown in Figure~\ref{fig:auau_jet_energy_scale}.  
For both jet radii the
jets are reconstructed within 4\% of the true energy over the 
measured range.  The jet energy resolution shown in the right panel
only includes effects due to the detector segmentation applied
and the underlying event resolution.  In \pp collisions
the resolution for $R = 0.4$ jets is better than for $R = 0.2$ jets
because the segmentation can cause jet splitting with the smaller
jet cones.  In \auau collisions the order is swapped because the 
dominant effect is the additional smearing due to the underlying
event.

\begin{figure}[hbt!]
  \centering
  \includegraphics[trim=0 0 0 0,clip,width=\onewidth]{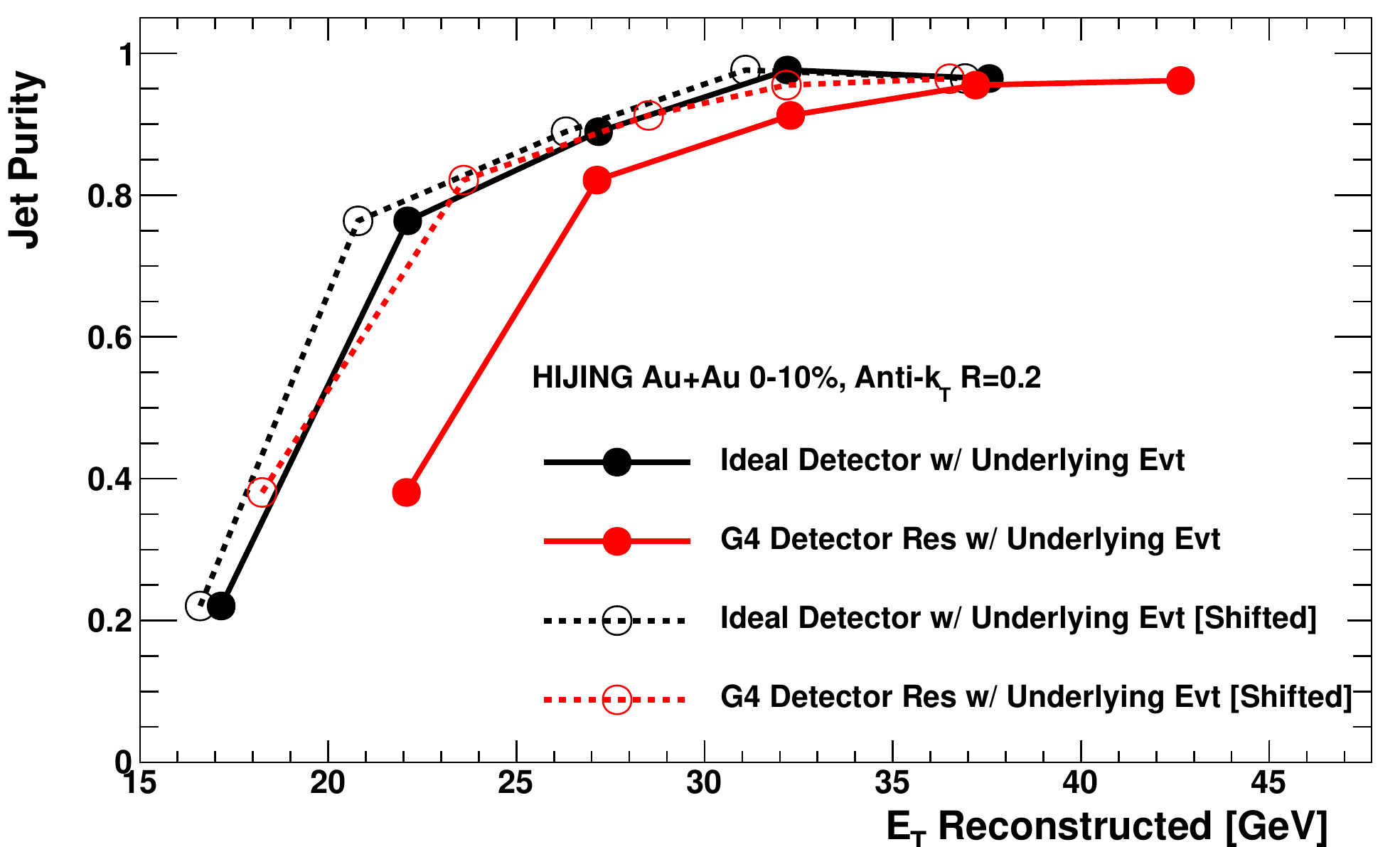}
  \caption{Results for the jet purity ($S/(S+B)$) in terms of matched true and \fake jets
in 0-10\% \auau collisions from \hijing.  The purity values are for a ideal detector (i.e, 
sPHENIX segmentation with perfect resolution) and then including the \geant parametrized
EMCal and HCal resolutions.  Both results are then shifted down in $E_T$ by the 
reconstructed energy bias.
  \label{fig:fakestudy_detres}}
\end{figure}
%
%
The \fast simulation results described above have been re-run with the
inclusion of the detector resolutions as parametrized from the single
particle \geant results -- detailed in Section~\ref{sec:g4sim}.  The
results shown in Figures~\ref{fig:auau_fake_rate} and
~\ref{fig:auau_fake_rate_wide} remain quite similar with the detector
resolution included, though with an overall shift of all the
distributions to higher $E_T$ due to the additional blurring on
falling spectra.  For $R=0.2$ jets, the smearing due to detector
resolution is comparable to the effect of the underlying event and for
larger jet cones the effect of the underlying event is found to be
much larger than detector resolution effects.
Figure~\ref{fig:fakestudy_detres} shows the jet purity (i.e., $S/(S+B)$)
for $R=0.2$ jets as a function of reconstructed $E_T$.  The solid
black (red) points correspond to the cases without (with) detector
resolution effects.  Also shown as open points are both results
shifted down in energy by the average reconstructed energy bias as
determined from the reconstructed matched jet sample.  One observes
that the relative true and \fake jet contributions are the same for
the equivalent true jet energy ranges.

\subsection{Inclusive Jet Yield in \AuAu Collisions}
\label{sec:auau_inclusive_jets}

\begin{figure}[hbt!]
  \centering
  \includegraphics[trim=0 0 0 20,clip,width=\onewidth]{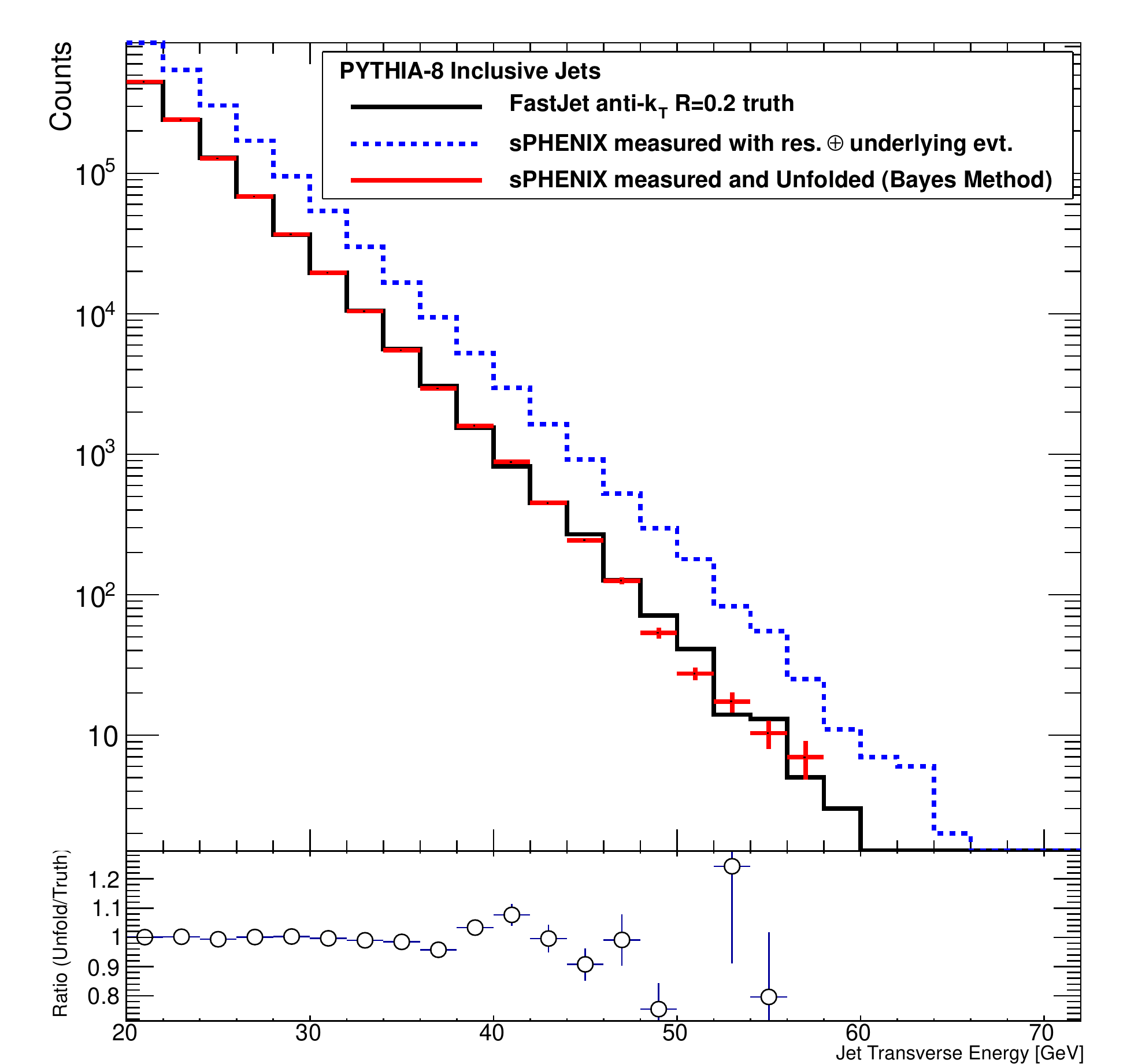}
  \caption{Effect of smearing the inclusive jet spectrum in \auau
    collisions.  The jets found by \fastjet are smeared by the jet
    resolution contributions from the detector and the underlying
    event fluctuations.  The unfolded spectrum from the Iterative
    Bayes method is shown and the ratio of the unfolded to the true
    \pt spectrum (lower panel).}
  \label{fig:auau_smearing_and_unfolding}
\end{figure}

\begin{figure}[hbt!]
  \centering
  \includegraphics[width=0.8\linewidth]{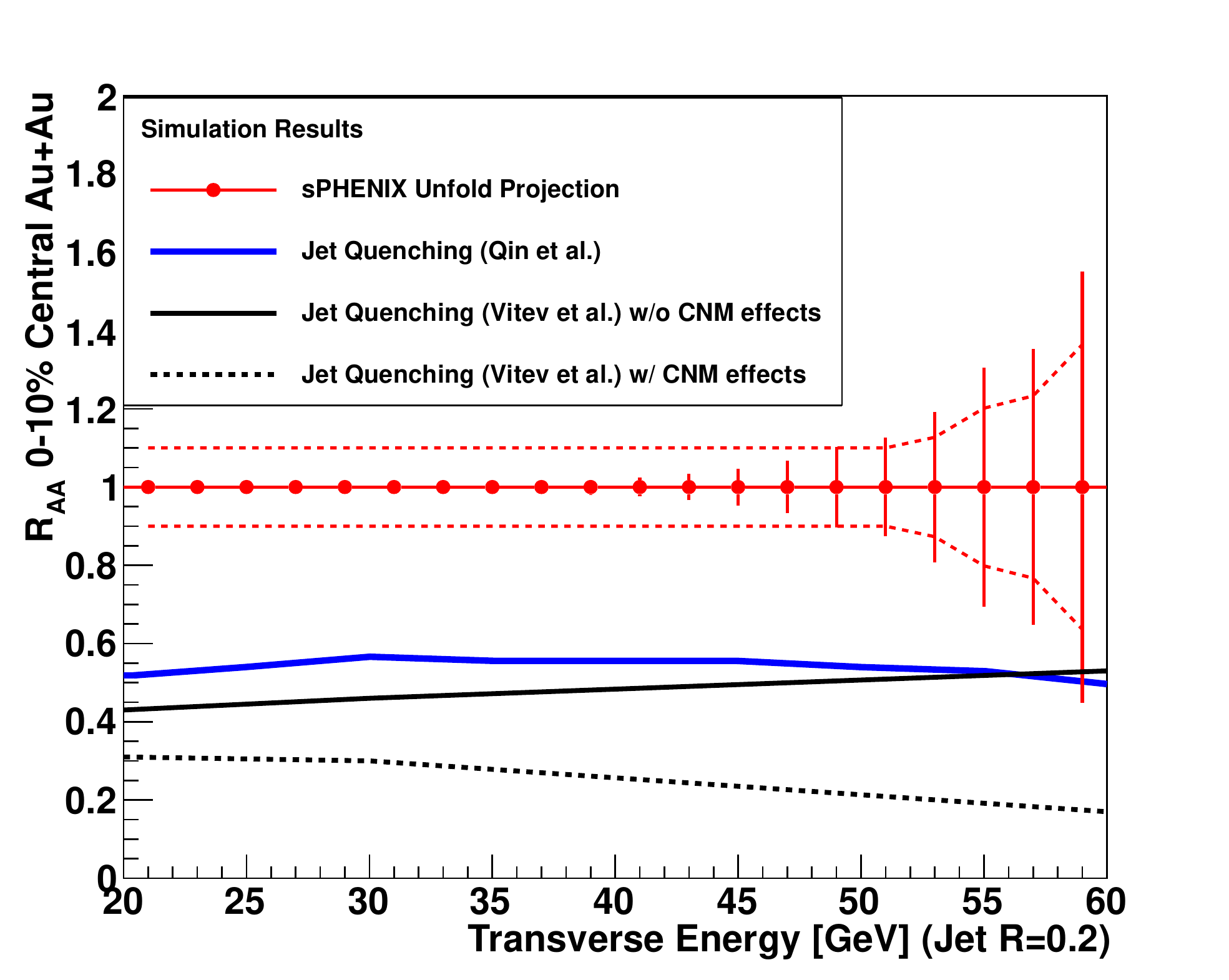}
  \caption{Single inclusive jet $R_{AA}$ with $R=0.2$ for \auau
    central events from the unfolding of the \pp and \auau spectra
    with an estimated systematic uncertainty as a multiplicative
    factor of approximately $\pm$ 10\%.  Also shown are the
    predictions from a calculation including radiation and collisional
    energy loss and broadening~\cite{qin_privatecomm} and another with
    and without cold nuclear matter
    effects~\cite{He:2011pd,Neufeld:2011yh,Vitev:2009rd} (as discussed
    in Section~\ref{sec:jetcalculations}).}
  \label{fig:jet_aa}
\end{figure}

The inclusive jet spectrum is the most important first measurement to
assess the overall level of jet quenching in RHIC collisions.  The
results shown in Figure~\ref{fig:auau_smearing_and_unfolding} were
obtained by the \veryfast simulation approach described above.
\pythia was used to generate events and the final state particles were
sent to \fastjet in order to reconstruct jets.  The resulting jet
energy spectrum was smeared by the jet resolution determined for \pp
collisions from \geant, and an additional smearing by the underlying
event fluctuations (determined from the full 0--10\% central \hijing
\fast simulation).  Finally, an unfolding procedure was used to
recover the truth spectrum.  The ratio shown at the bottom of the plot
shows that the unfolding is very effective.

As an estimate of the uncertainties on a jet $R_{AA}$ measurement from one year of RHIC running, the uncertainties
from Figures~\ref{fig:pp_very_fast_dijet} and \ref{fig:auau_smearing_and_unfolding}
are propagated and shown in Figure~\ref{fig:jet_aa}.  For $E_T<50$\,GeV the point to point
uncertainties are very small.   Also shown is an estimated systematic uncertainty including the effects
from unfolding.  All points are shown projected at $R_{AA} = 1$, and we show for comparison the predicted jet $R_{AA}$ including 
radiative and collisional energy loss and broadening from Ref.~\cite{qin_privatecomm}.

\subsection{Dijets in \AuAu collisions}
\label{sec:auau_dijets}



Fake jets contaminate \dijet observables much less than they do the
inclusive jet measurement.  In the case of inclusive jets, one is
working with a sample of $10^{10}$ central \auau events in a typical
RHIC year, so even if it is only a rare fluctuation in the background
that will be reconstructed as a real jet, there is a huge sample of
events in which to look for such fluctuations.

The case of \dijet correlations is very different.  There are $10^6$
clean trigger jets above $E_T = 30$\,GeV in central \auau collisions
in a RHIC year -- detailed in Figure~\ref{tab:nlo_jetrates}.  This
means there is a factor of $10^4$ fewer chances to find the rare
background fluctuation that appears to be a true jet in the opposite
hemisphere.  Also, the presence of a high energy jet, for which the
fake rate is known to be low, tags the presence of a hard process
occurring in the event, and thus dramatically reduces the probability
of a jet in the opposite hemisphere being a fake.  Because of these
considerations, one can go to much lower \pt for the away side partner
of a \dijet pair.  Studies presented here include away side jets down
to 5\,GeV, and we have found that the fake jet rate remains small for
the associated jets, even at these low jet energies.

\begin{figure}[hbt!]
  \centering
  \includegraphics[width=0.5\linewidth]{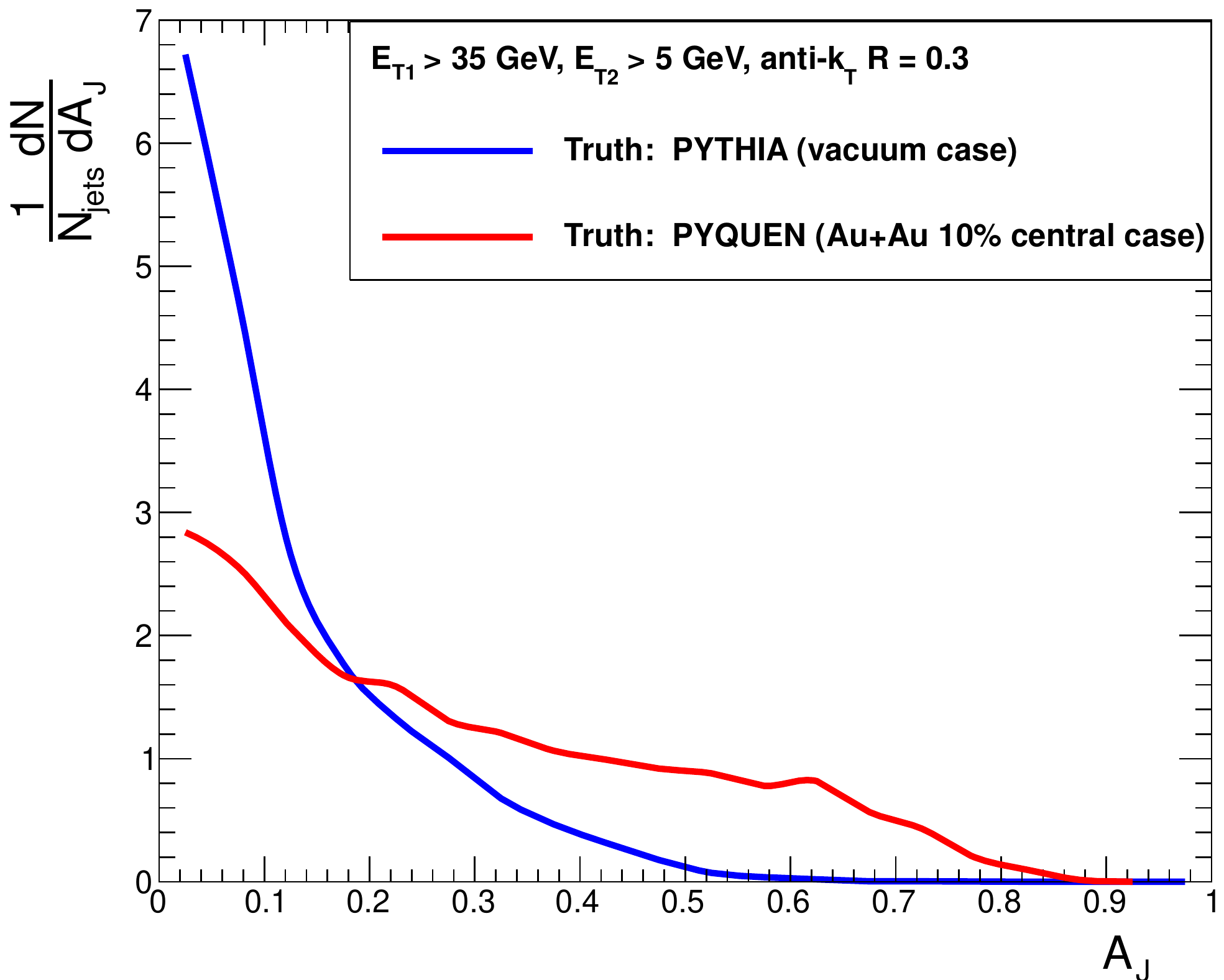}
  \hfill
  \includegraphics[width=0.5\linewidth]{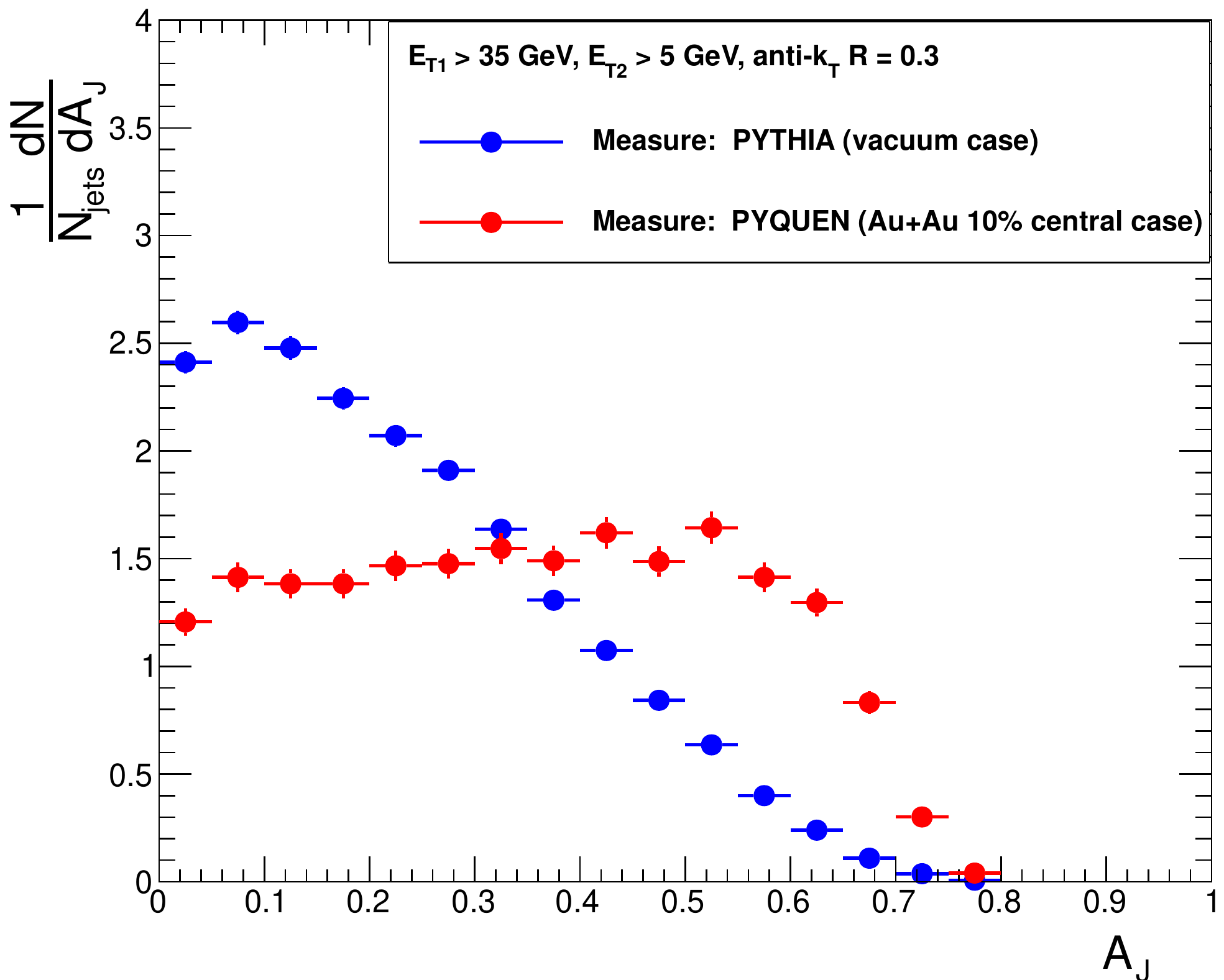}
  \hfill
  \includegraphics[width=0.5\linewidth]{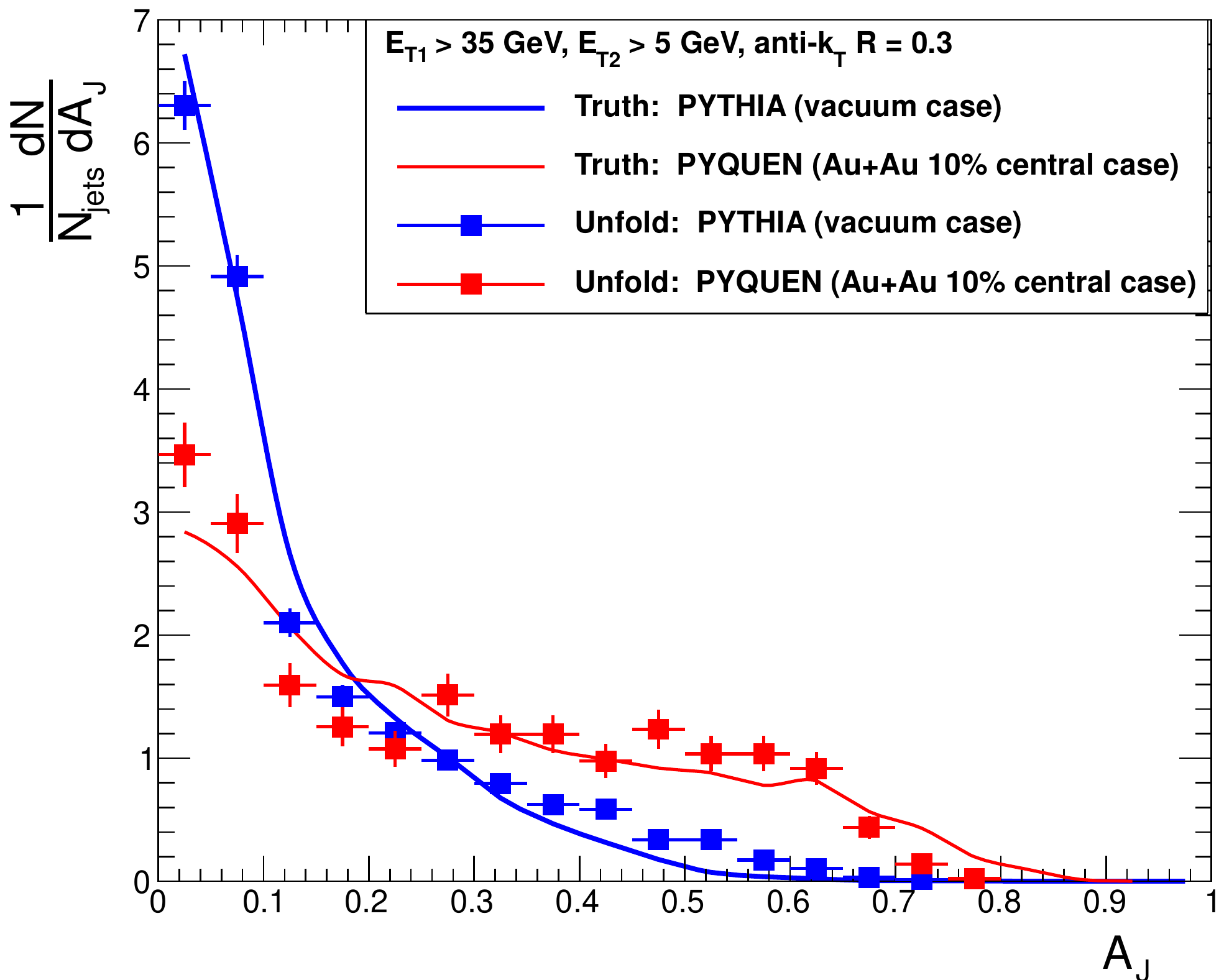}
  \caption{The effect of smearing on $A_J$ for  $R = 0.3$ jets.  
The upper panel shows the ratio expected in
    \pythia and \pyquen, showing the effect of quenching.  
    The middle panel shows
    the effect of smearing on the ratio determined from jets
    reconstructed after embedding in \auau events.  Although smeared,
    the reconstructed data still show a distinct difference between
    the quenched and unquenched results.  The bottom panel shows
the results of the ``unfolding'' procedure discussed in Section~\ref{sec:dijet_pp}.}
  \label{fig:auau_dijet_smearing}
\end{figure}

In order to address the sensitivity to modifications of the $A_J$
distributions that might be expected at RHIC here we compare \pythia
simulations with those from \pyquen~\cite{Lokhtin:2005px} (a jet
quenching parton shower model with parameters tuned to RHIC data).
All the \pyquen events generated are for central \auau events with
$b=2$\,fm. Figure~\ref{fig:auau_dijet_smearing} shows the particle
level (i.e truth) $A_J$ distributions and how they are reconstructed
after being embedded in a central \auau event with a parametrized
detector smearing and segmentation applied.  As described above, the
full iterative underlying event subtraction method is applied.  The
simulated measured distributions (middle panel of
Figure~\ref{fig:auau_dijet_smearing}) show the effects of the
smearing; and the distinction between the \pythia and \pyquen
distributions remain large.  An unfolding procedure can be applied to
these embedded distributions to regain the true distributions.
However, as in the \pp case discussed in Section~\ref{sec:dijet_pp}
this should involve a full two-dimensional unfolding.  Applying the
same ``unfolding'' applied to the \pp case where the smearing of the
trigger jet is taken as the dominant effect recovers most of the
original distribution, as shown in the lower panel of
Figure~\ref{fig:auau_dijet_smearing}.  Again, this does not replace a
full unfolding procedure, but it does show that the reconstruction is
well under control and unfolding will be possible despite the presence
of a large fluctuations in the underlying event, after baseline and
flow subtraction.

\subsection{$\gamma+$jets in \AuAu collisions}
\label{sec:auau_gamma_jet}

The rate for \gj events is lower than the rate for \dijet events by
about $\alpha_{EM}/\alpha_{s}$.  In a canonical RHIC year of running
one would expect more than 20k direct photons above 20\,GeV/c.  As
shown earlier in Figure~\ref{fig:nlo_gammarates}, at \pt = 20\,GeV the
fraction of direct photons in the inclusive photon sample is large and
$\gamma$-jet measurements will be possible without employing isolation
cuts.  The $\gamma$ measurement is very clean as \fake jets are not an
issue for trigger photons.  We show \fast simulation results for 20k
$\gamma$ triggers embedded in central \auau events.

\begin{figure}[hbt!]
  \centering
  \includegraphics[width=0.5\linewidth]{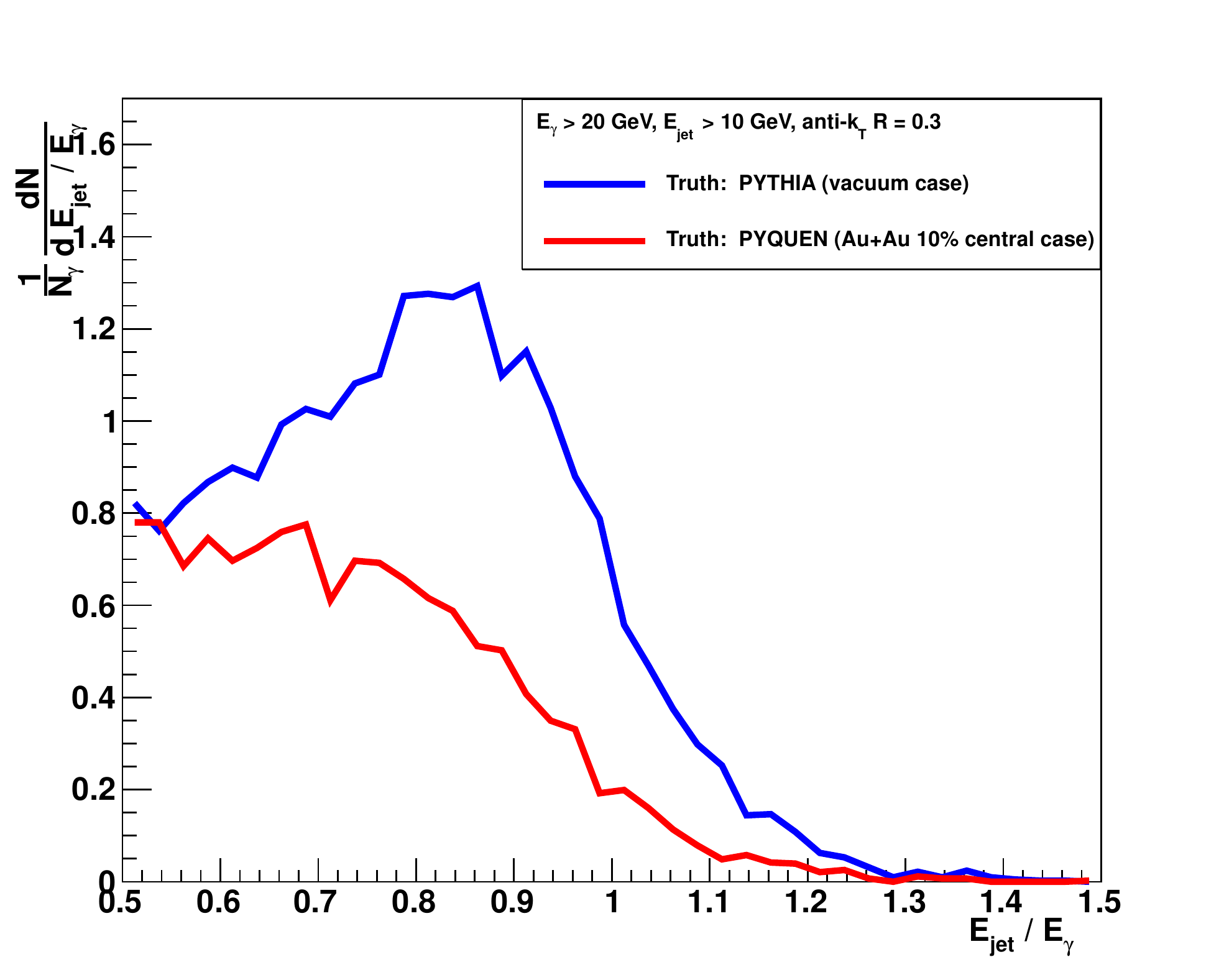}
  \hfill
  \includegraphics[width=0.5\linewidth]{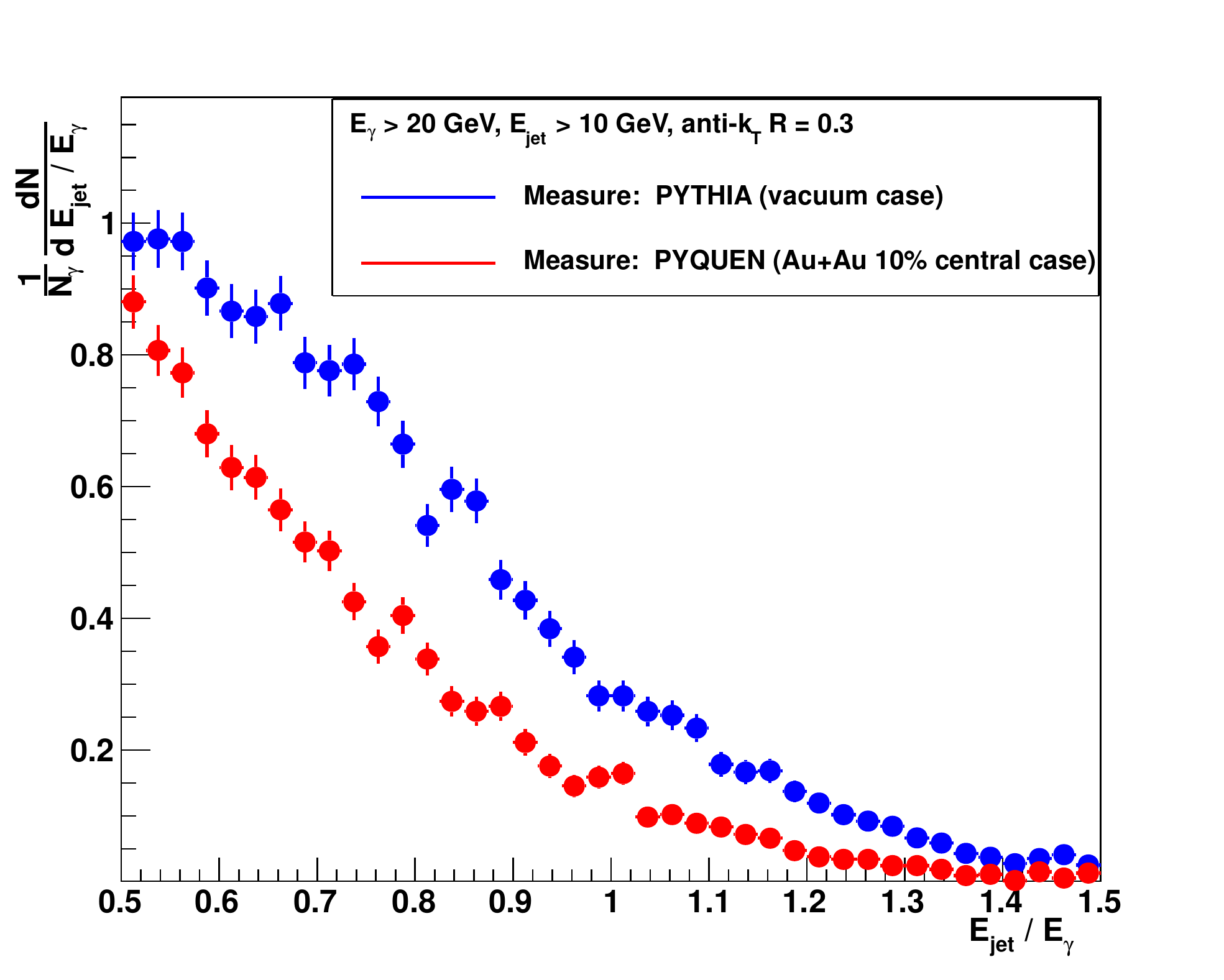}
  \hfill
  \includegraphics[width=0.5\linewidth]{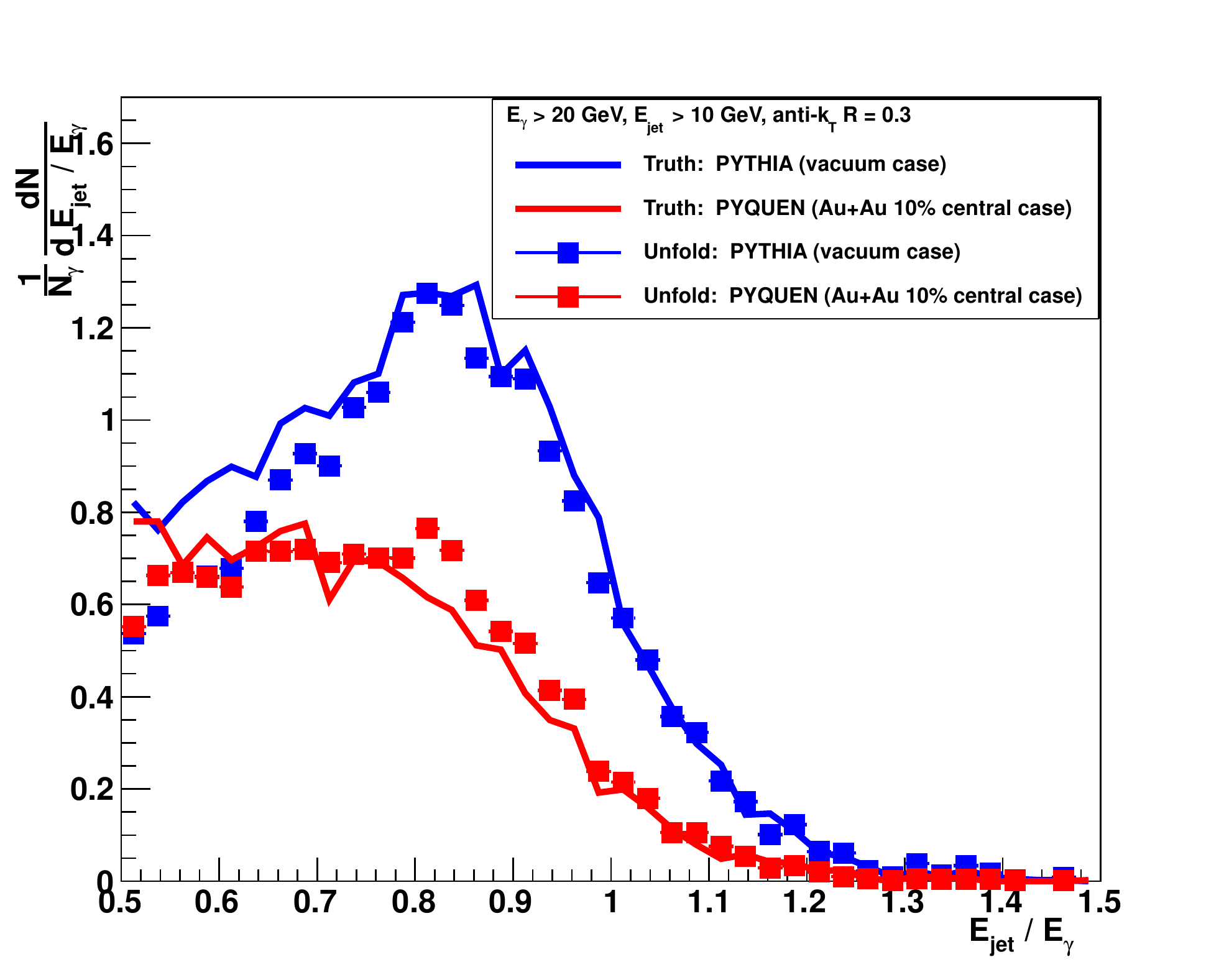}
  \caption{The effect of smearing on energy ratio $E_{\rm
      jet}/E_\gamma$ for $R = 0.3$ jets.  The upper panel shows the ratio expected in
    \pythia and \pyquen, showing the effect of quenching. 
     The middle panel shows
    the effect of smearing on the ratio determined from jets
    reconstructed after embedding in \auau events.  Although smeared,
    the reconstructed data still show a distinct difference between
    the quenched and unquenched results.  Results of a one dimensional
unfolding are compared with the truth particle level distributions in the
bottom panel.}
  \label{fig:auau_gammajet_smearing}
\end{figure}

In contrast to the \dijet case studied above, the $\gamma$-jet
measurements do not compare two similar objects with the same effects
from the underlying event.  The $\gamma$ is always the trigger.  In
this case it makes sense to measure $x\equiv E_{\mathrm
  {jet}}/E_{\gamma}$ rather than $A_J$.  While in a leading order QCD
picture the $\gamma$ and the jet should exactly balance in energy, in
reality this is not the case, especially when higher-order diagrams
are taken into account.  For small jet sizes there is a significant
probability that the away side parton shower is split into more than
one jet by the reconstruction procedure, with each carrying a fraction
of the energy needed to balance that of the $\gamma$.  This can be
seen in the \pythia truth curves in the top panels of
Figure~\ref{fig:auau_gammajet_smearing}.  The smeared and embedded
results are shown in the middle panel.  Again the smearing has a
significant effect, but the distinction between the \pythia and
\pyquen results is retained.

In the $\gamma$-jet case, the unfolding is to a very good approximation one-dimensional.  This is
because the dominant smearing effect is on the 
jet energy since the $\gamma$ is measured in the EMCal
which has a very good energy resolution compared to the jet.
We have applied a one dimensional Iterative Bayes 
unfolding procedure to the $\gamma$-jet
$x$ distributions for the $R=0.3$ jets in the bottom panel of
 Figure~\ref{fig:auau_gammajet_smearing}.
The unfolded results compare well with the particle level distributions
for both \pythia and \pyquen.

\subsection{\gh correlations in \auau collisions}
\label{sec:auau_jet_hadron}

sPHENIX will be able to track charged particles in addition to its
calorimetric jet finding capabilities, and this can be used to
construct \gh and jet+hadron correlations.  This is particularly
appealing as a complement to the \dijet measurements.  At sufficiently
low \pt, the background of \fake jets for the away-side jet in a
\dijet analysis becomes problematic.  At that same low \pt, one can
turn to the capabilities of the existing PHENIX vertex detector
tracking system (with moderate momentum resolution) to extend the
measurement.  Also, one can use \gh correlations to study the
redistribution of energy lost by the opposite going parton.  Results
from CMS~\cite{Chatrchyan:2011sx} on jet+hadron correlations indicate
that, at the LHC, this energy is spread over a wide angular range.
Measurements at RHIC of \gh correlations have not had the statistical
precision or the acceptance necessary to make comparable statements
about the modification to jet fragmentation.  In order to recover the
energy using the standard jet reconstruction, one would have to use an
extremely wide jet cone, and in a heavy-ion collision this presents a
problem, as it exposes the away side jet finder to a very large
contamination of energy coming from the underlying event.  Precisely
because of this difficulty, one could instead use correlations
between a trigger $\gamma$ and an away side hadron.

\begin{figure}[hbt!]
  \centering
  \includegraphics[width=1.0\textwidth]{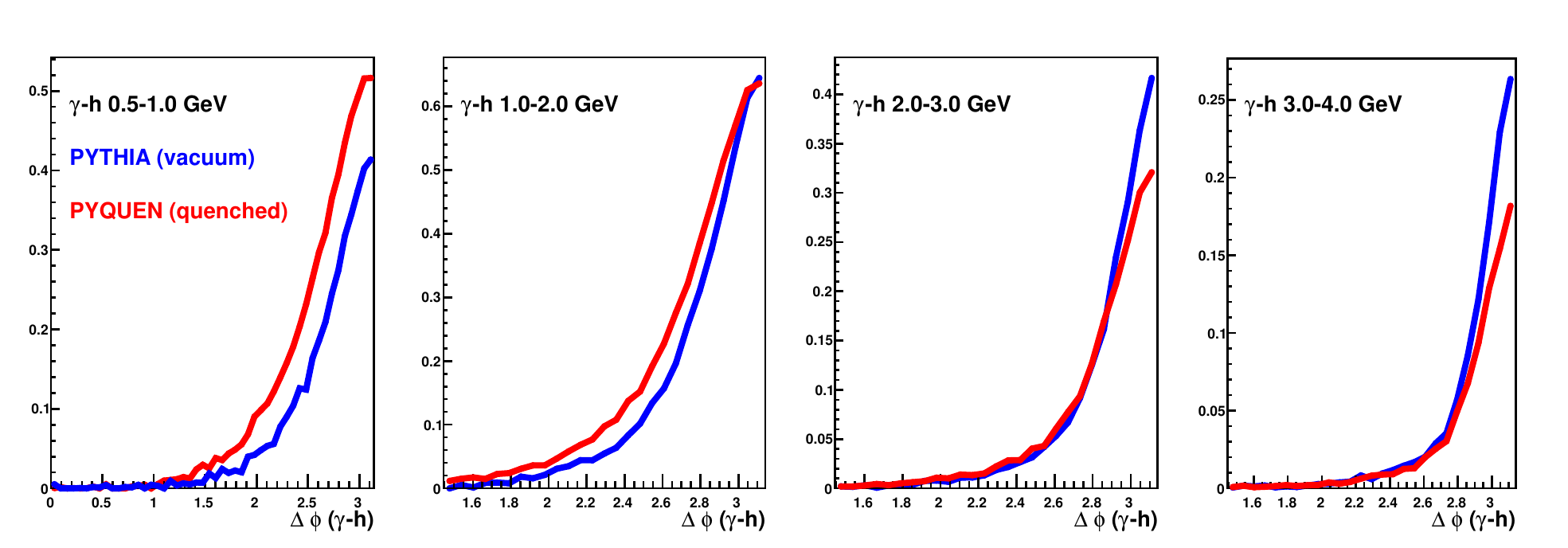}
  \hfill
  \includegraphics[width=0.6\textwidth]{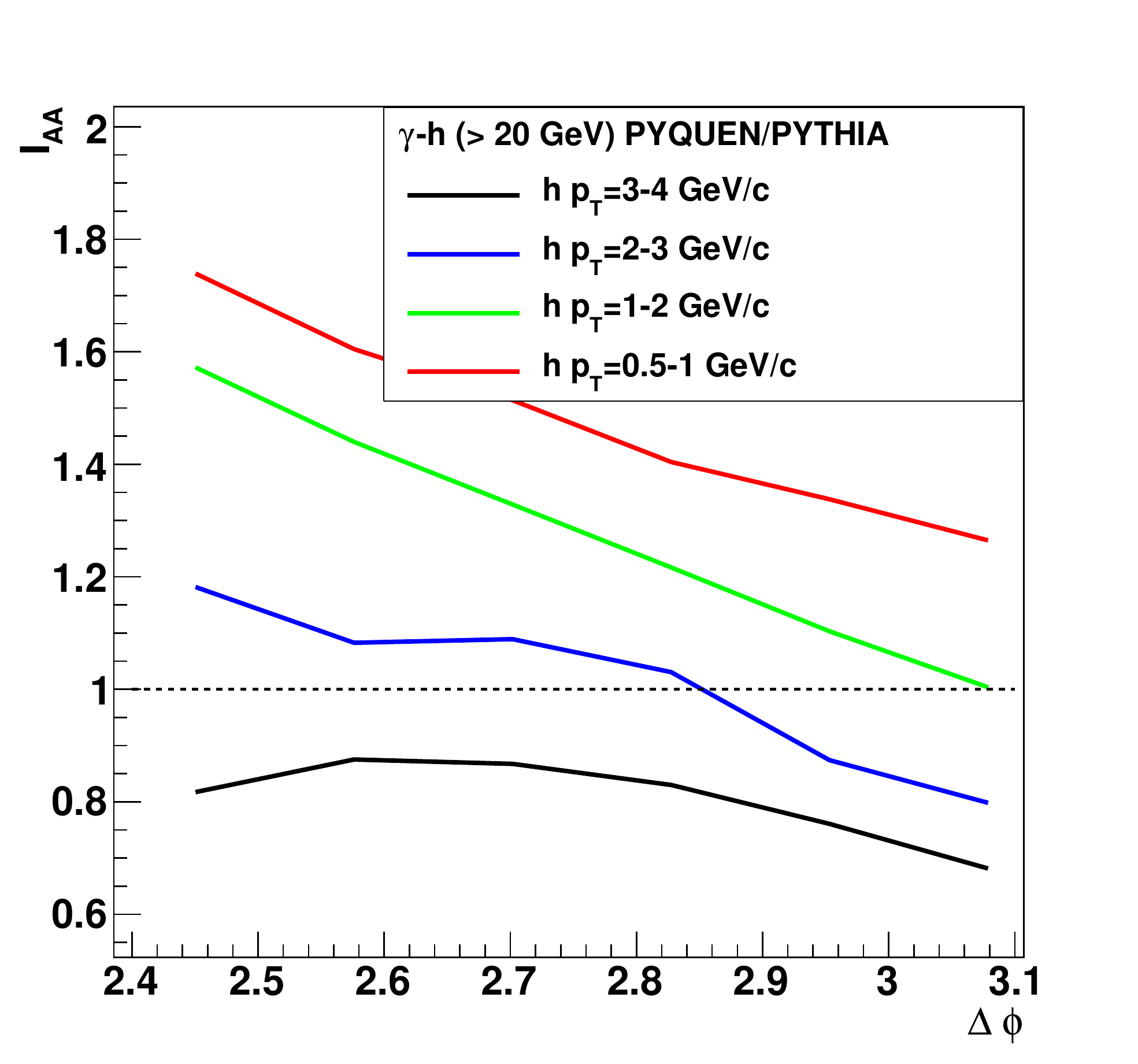}
  \caption{A simulation of the \gh angular correlation for \pythia and
  for \pyquen events for hadron $p_T$ ranges shown in the Figure.  These
  $p_T$s would be accessible with the current PHENIX silicon tracker.  
  The lower panel shows the nuclear modification $I_{AA}$ between \auau central
with PYQUEN and \pp with PYTHIA as a function of hadron \pt and $\Delta\phi$.}
  \label{fig:auau_jet_hadron}
\end{figure}

Figure~\ref{fig:auau_jet_hadron} shows \gh correlations for photons with $E_T > 20$ GeV
from \pythia and \pyquen in the hadron \pt range of 0.5--4.0\,GeV/c.  This
hadron \pt range will be accessible with no additional tracking beyond the existing VTX. 
The \pyquen distributions are broader and have a larger yield at lower $p_T$, and would be easily measured by sPHENIX.

\section{Summary}
\label{sec:jet_performance_summary}

Overall we conclude that a robust jet, dijet, and $\gamma$-jet program
with high statistics is achievable with the sPHENIX detector upgrade.
These observables indicate excellent discriminating ability between
scenarios with different medium coupling strengths and jet quenching
mechanisms.



\cleardoublepage

\appendix

\chapter[Midrapidity Upgrades and Physics]{sPHENIX Midrapidity Future Option Upgrades and  Physics}
\label{chap:barrel_upgrade}

\begin{figure}[!hbt]
 \begin{center}
    \includegraphics[width=0.7\linewidth]{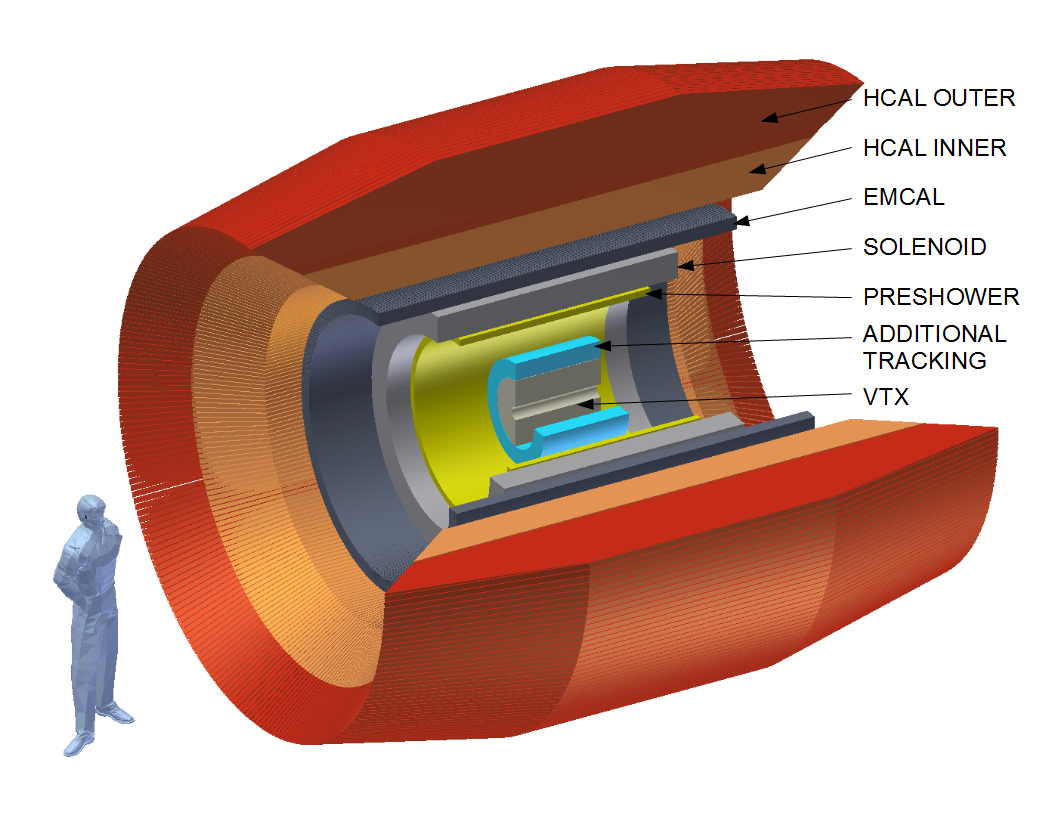}
        \caption{\label{fig:eng_drawing_options}Engineering drawing of the sPHENIX upgrade including two future option upgrades (additional
tracking and a preshower detector inside the magnetic solenoid and electromagnetic calorimeter).}
 \end{center}
\end{figure}

The sPHENIX midrapidity magnetic solenoid with electromagnetic and
hadronic calorimeters can be substantially augmented in physics
capabilities through modest incremental upgrades that have been
considered from the beginning of the sPHENIX design.  In
this Appendix we discuss two specific future option upgrades: (1)
additional charged particle tracking layers outside the existing
PHENIX silicon vertex detector (VTX) and (2) a preshower with fine
segmentation just inside the magnetic solenoid.  An engineering
drawing of the location of these future option upgrades is shown in
Figure~\ref{fig:eng_drawing_options}.  We then detail how these
additions expand the sPHENIX physics program to include the following:
(a) heavy quarkonia suppression via the three $\Upsilon$ states, (b)
tagging of charm and beauty jets, (c) jet fragmentation function
modifications, (d) nuclear suppression of $\pi^{0}$ yields up to
$p_{T} = 40$\,GeV/c, and (e) a possible low to intermediate mass
dilepton program.

\section[Tracking Upgrade]{Charged Particle Tracking Extension Upgrade}

The current PHENIX silicon vertex tracker (VTX) consists of two inner
layers (pixels) at radii 2.5 and 5\,cm from the beamline and two
outer layers (strip-pixels) at radii of 10 and 14\,cm.  Currently
in PHENIX standalone tracks (in the VTX only) are determined with a
momentum resolution of $\Delta p/p \approx 0.1 + 0.02 \times p$~[GeV/c].  The
sPHENIX magnetic field will have an appreciably larger strength (2.0
Tesla) than the current PHENIX axial field magnet (0.8 Tesla).
However, simulations from the current PHENIX VTX indicate that even
with the larger sPHENIX magnetic field, there will be significant
\fake track contributions (i.e. picking up incorrect VTX hits and thus
reconstructing to the incorrect momentum vector) for $p_{T} >
5$\,GeV/c.  In addition, with only four hits, reconstructed tracks at
large Distance of Closest Approach (DCA) have substantial \fake track
contributions.  In the current PHENIX detector, these \fake
contributions are removed by the required matching to the outer
tracking Drift Chamber and Pad Chamber hits.  In the
sPHENIX proposed in this document with only the VTX for tracking, one will be limited to charged
particle tracks with $p_{T} < 5$\,GeV/c and without heavy flavor
tagging via displaced vertices (the VTX by itself will not be able to
discriminate sufficiently against fake tracks).

\begin{figure}[!hbt]
 \begin{center}
    \includegraphics[width=0.7\linewidth]{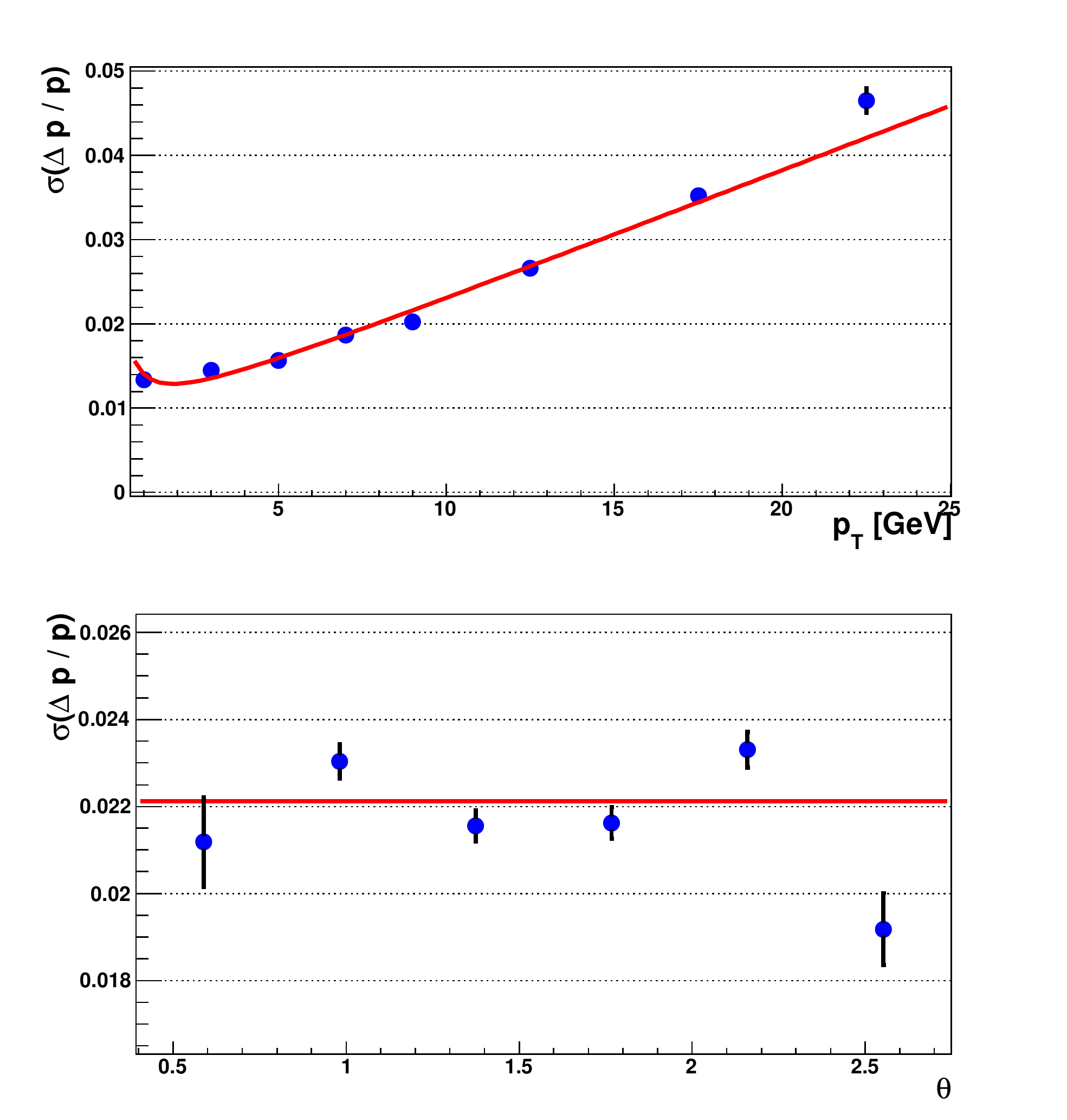}
        \caption{\label{fig:mom_res}\geant and track model evaluation of single particle momentum resolution.  From a fit to the data in the upper panel, shown
as a red line, we determine the momentum resolution to be $\Delta p / p = 0.007 + 0.0015 \times p$.  The lower panel shows the momentum resolution 
as a function of the polar angle of the track.}
 \end{center}
\end{figure}

Thus, the sPHENIX future option upgrade incorporates additional precision
tracking in the radial space from 15--65\,cm (inside the new magnetic
solenoid).  The technology and exact number of layers or space-points
has not been determined at this time.  In order to design in this
upgrade option, we have done a full \geant simulation with two
additional silicon tracking layers at radii of 40 and 60\,cm.  We have
assumed a strip design with $80 \mu$m $\times 3$ cm, which results in
1.0 (2.2) million channels in the inner (outer) layer.  The material
thickness of the intermediate layer at 40\,cm must be thin (of order
0.03 radiation lengths) to reduce multiple scattering and deliver good
momentum resolution.  We have implemented a full pattern recognition
algorithm and track reconstruction model based on software development
for the existing VTX.  The momentum resolution shown in
Figure~\ref{fig:mom_res} has an RMS $\Delta p / p = 0.007 + 0.0015
\times p$ for momentum with $p_{T} > 1$\,GeV/c.  Also shown is the
momentum averaged resolution as a function of polar angle $\theta$.
In order to have good separation of the three $\Upsilon$ states
($\Upsilon(1s), \Upsilon(2s), \Upsilon(3s)$) --- crucial to the
physics of the color screening length --- we need the term linear in
the momentum to be less than 0.002.

The inner four VTX layers are currently arranged without full $2\pi$
coverage, and would need to be re-configured and augmented to do so.
The outer layers in principle could be a similar silicon design to the
outer two VTX layers.  The exact number of layers and technology
choice required in terms of \auau central pattern recognition
efficiency, \fake track rates, and charm/beauty tagging via displaced
vertices is currently under study.

\section{Preshower Detector}

The sPHENIX proposed electromagnetic calorimeter has a segmentation of
$\Delta \eta \times \Delta \phi = 0.024 \times 0.024$ and thus has
relatively good separation of single photons from $\pi^{0} \rightarrow
\gamma\gamma$ decays up to approximately 10\,GeV.  A preshower layer in
front of the magnetic solenoid and the electromagnetic calorimeter can
extend this separation up to $p_{T} > 50$\,GeV/c (essentially the
entire kinematic range of measurements possible within the luminosity
limits).  In addition to separating single from double overlapping
showers, the preshower can provide significant additional electron
identification capability.  As we discuss later, the combined pion
rejection (i.e. electron identification) from the sPHENIX
electromagnetic calorimeter and the preshower are sufficient for
excellent $\Upsilon$ measurements.

\begin{figure}[!hbt]
 \begin{center}
    \includegraphics[width=0.45\linewidth]{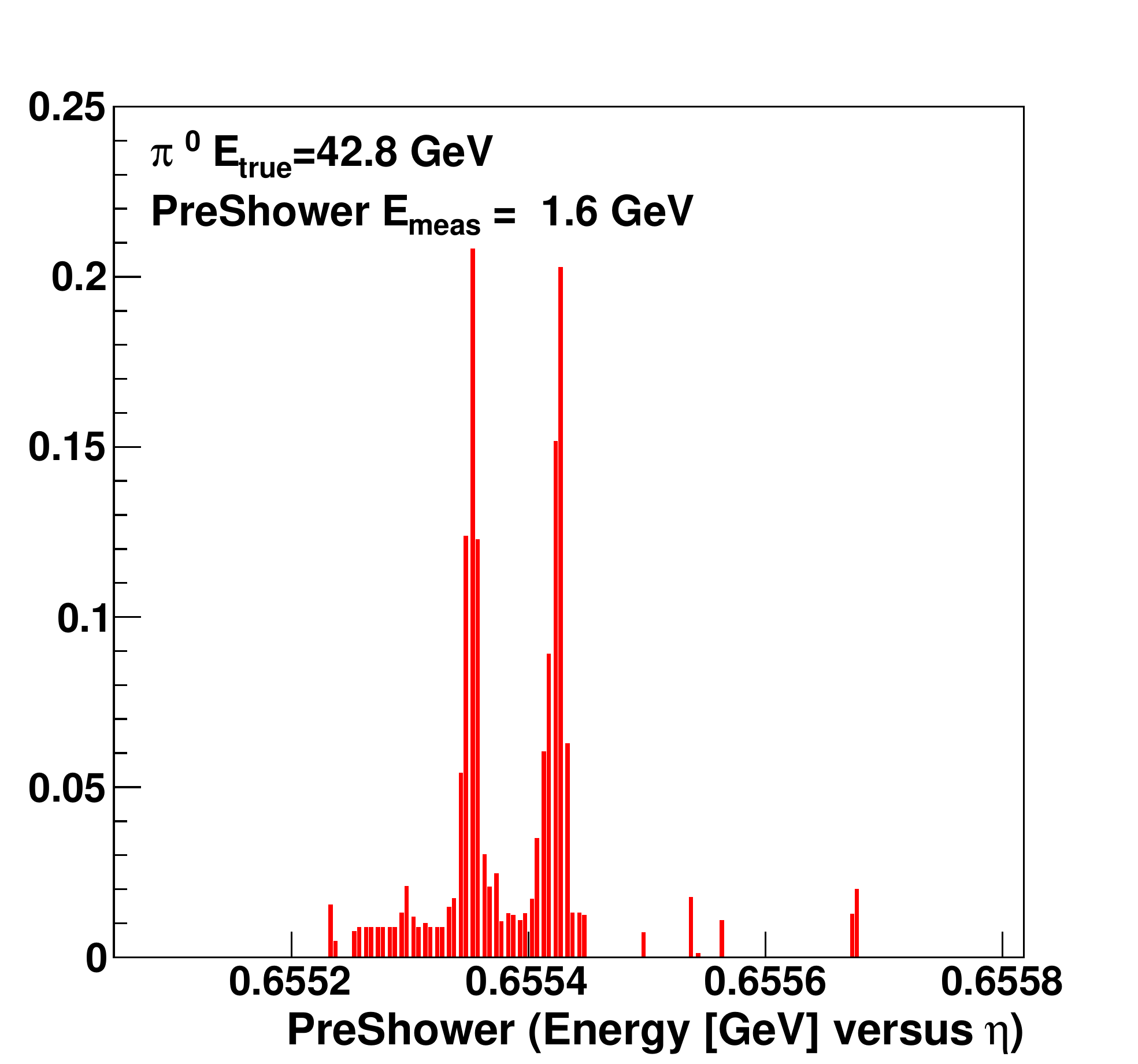}
    \hfill
    \includegraphics[width=0.45\linewidth]{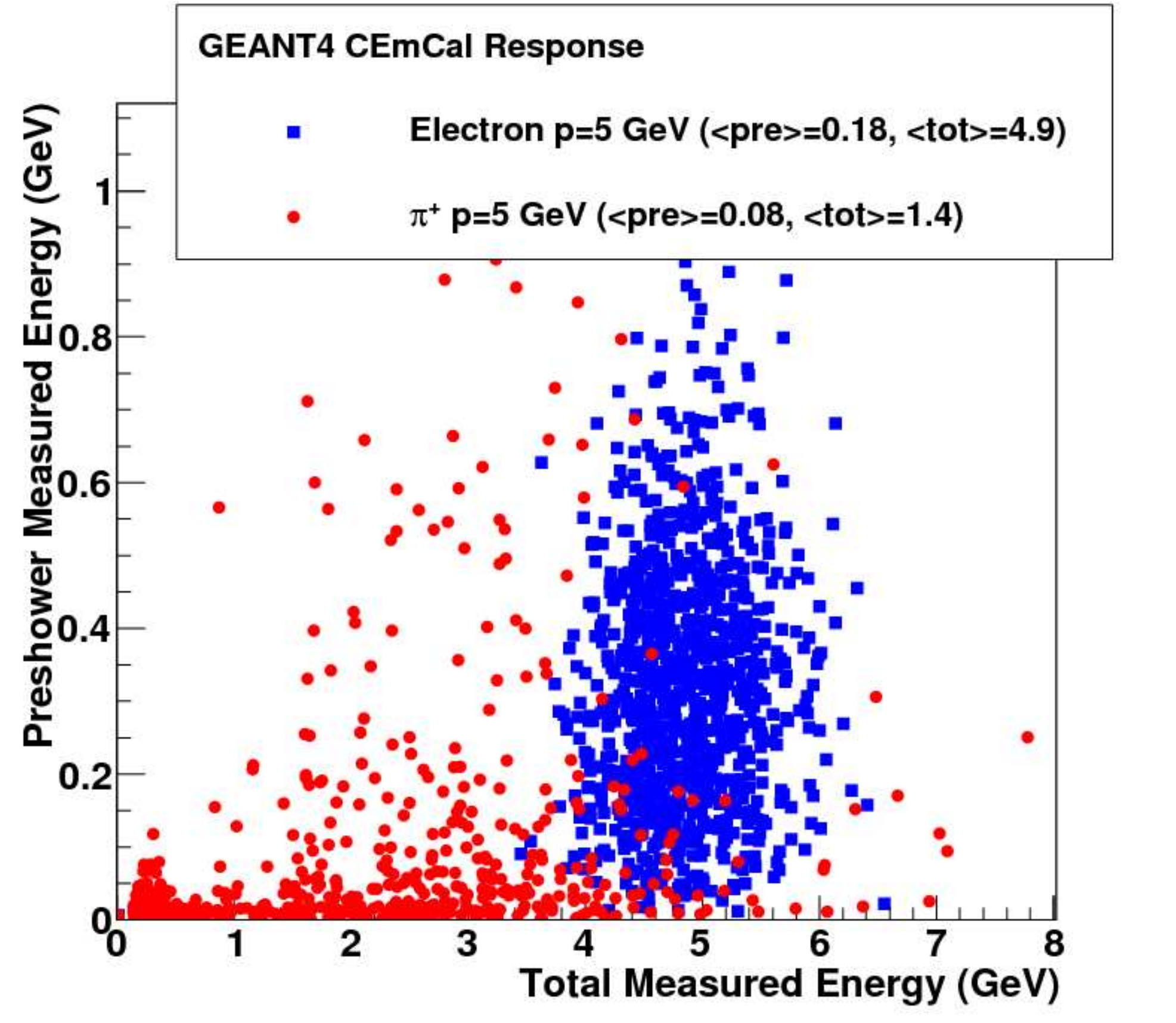}
        \caption{\label{fig:preshower}(Left Panel) \geant example preshower energy distribution for a single 42.8\,GeV $\pi^{0}$.
(Right Panel) \geant simulation examining the electron to $\pi^{-}$ separation for $p_{T} =  5$\,GeV/c.
}
 \end{center}
\end{figure}

Again, the exact design and technology for this preshower detector is
under active investigation and simulation.  For the purposes of
understanding the basic performance and design constraints on the
sPHENIX upgrade (e.g. the magnetic solenoid radius), we have
implemented a \geant configuration with a 2.3 radiation length
thickness of tungsten backed by a silicon layer with strips 300~$\mu$m
$\times 6$ cm as a pre-sampler.  The detector sits just after the
outermost tracking layer and before the magnetic solenoid.  The
segmentation corresponds to $\Delta \eta \times \Delta \phi = 0.0005
\times 0.1$.  We are still investigating whether two layers of perpendicular strips are
necessary for the physics performance in all channels (particularly
the efficiency for tagging two photons from a very high $p_{T}$
$\pi^{0}$ decay).  Shown in Figure~\ref{fig:preshower} (left panel) is
an event display of the energy deposition from a 42.8\,GeV $\pi^{0}$ in
the preshower, with clear separation of the two initiated photon
showers.  Shown in Figure~\ref{fig:preshower} (right panel) is the
response of the electromagnetic calorimeter total energy versus the
preshower energy for electrons and charged pions.  The combination of
information provides a powerful discriminator for electron
identification.  Even if the charged pion induces a hadronic shower in
the electromagnetic calorimeter, it has a much lower probability for
that interaction occurring in the first layer of tungsten of the
preshower.  Initial studies indicate a charged pion rejection of order
100--200 with good electron efficiency for $p_{T} > 2$--3\,GeV/c.

\section[Upsilon Spectroscopy]{Quarkonia Spectroscopy of the Upsilon States}

We have investigated the feasibility of using the sPHENIX detector,
with the addition of outer tracking layers and additional electron
identification capability, to make high quality $\Upsilon$
measurements at $\sqrt{s_{NN}} = 200$\,GeV.  We conclude that a world
class $\Upsilon$ measurement is possible, with separation of the three
states and statistical precision comparable with that of the LHC
experiments.  In this section we discuss the physics motivation for
these measurements, and summarize the expected performance.


\label{sec:quarkonia}

\subsection{Physics Motivation}
\label{sec:quarkonia_introduction}

An extensive program of $J/\psi$ measurements in A$+$A collisions has
been carried out at the SPS ($\sqrt{s_{NN}} = 17.3$\,GeV) and RHIC
($\sqrt{s_{NN}} = 200$\,GeV) and has now begun at the LHC
($\sqrt{s_{NN}} = 2.76$\,TeV). These measurements were motivated by a
desire to observe the suppression of $J/\psi$ production by color
screening in the QGP. In fact, strong suppression is observed at all
three energies, but it has become clear that the contribution of color
screening to the observed modification can not be uniquely determined
without a good understanding of two strong competing effects.

The first of these, the modification of the $J/\psi$ production cross
section in a nuclear target, has been addressed at RHIC and the SPS
using $p(d)$$+$A collisions, and will soon be addressed at the LHC
using $p$$+$Pb collisions. The second complicating effect arises from
the possibility that previously unbound heavy quark pairs could
coalesce into bound states due to interactions with the medium.  This
opens up the possibility that if a high enough density of heavy quark
pairs is produced in a single collision, coalescence of heavy quarks
formed in different hard interactions might actually increase the
production cross section beyond the initial population of bound
pairs~\cite{Zhao:2011cv}.

\begin{figure}
  \begin{center}
    \includegraphics[width=0.6\textwidth]{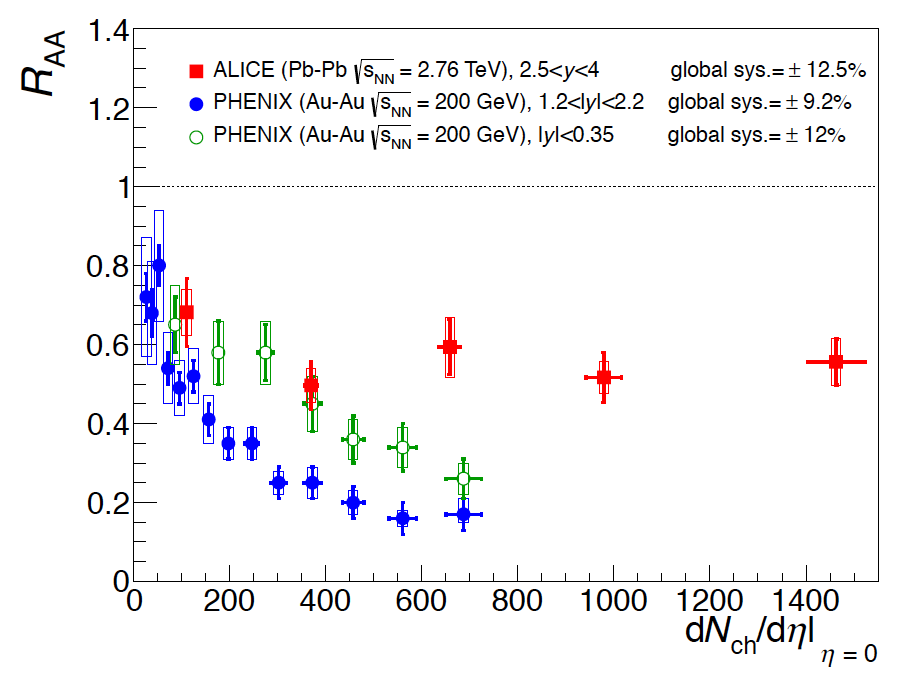}
  \end{center}
  \caption{\label{fig:quarkonia_phenix_alice_comparison}
   Comparison of nuclear modification measured by PHENIX and ALICE, showing that suppression is much stronger at the lower energy~\cite{Abelev:2012rv}. The
   modification measured by NA50 at low energy is similar to the PHENIX midrapity result.
    }
\end{figure}

Using $p$$+$Pb and $d$$+$Au data as a baseline, and under the
assumption that cold nuclear matter (CNM) effects can be factorized
from hot matter effects, the suppression in central collisions due to
the presence of hot matter in the final state has been estimated to be
about 25\% for \pbpb at the SPS~\cite{Arnaldi:2009ph}, and about 50\%
for \auau at RHIC~\cite{Brambilla:2010cs}, both measured at
midrapidity. The first $J/\psi$ data in \pbpb collisions at
$\sqrt{s_{NN}} = 2.76$\,TeV from ALICE~\cite{Abelev:2012rv}, measured
at forward rapidity, are shown alongside PHENIX data in
Figure~\ref{fig:quarkonia_phenix_alice_comparison}. Interestingly, the
suppression in central collisions is far greater at RHIC than at the
LHC. This is qualitatively consistent with a
predicted~\cite{Zhao:2011cv} strong coalescence component due to the
very high $c \overline{c}$ production rate in a central collision at LHC.
There is great promise that, once CNM effects are estimated from
$p$$+$Pb data, comparison of these data at widely spaced collision
energies will lead to an understanding of the role of coalescence.

Upsilon measurements have a distinct advantage over charmonium
measurements as a probe of deconfinement in the QGP. The
$\Upsilon$(1S), $\Upsilon$(2S) and $\Upsilon$(3S) states can all be
observed with comparable yields via their dilepton decays. Therefore
it is possible to compare the effect of the medium simultaneously on
three bottomonium states---all of which have quite different radii and
binding energies.

CMS has already shown first upsilon data from \pbpb at 2.76\,GeV
that strongly suggest differential suppression of the 2S and 3S states
relative to the 1S state~\cite{Sanchez:2011th}. With longer \pbpb
runs, and a $p$$+$Pb run to establish a CNM baseline, the LHC measurements
will provide an excellent data set within which the suppression of the
three upsilon states relative to $p$$+$Pb can be measured
simultaneously at LHC energies.

At RHIC, upsilon measurements have been hampered by a combination
of low cross sections and acceptance, and insufficient momentum
resolution to resolve the three states. So far, there are preliminary
measurements of the three states combined by
PHENIX~\cite{daSilva:2011zz} and STAR~\cite{Reed:2011fr}, including in
the STAR case a measurement for \auau.  However a mass-resolved
measurement of the modifications of the three upsilon states at
$\sqrt{s_{NN}} = 200$\,GeV would be extremely valuable for several
reasons.

 First, the core QGP temperature is approximately $2 T_c$ at RHIC at
 1\,fm/$c$ and is at least 30\% higher at the LHC (not including the
 fact that the system may thermalize faster)~\cite{muller:2012zq}.
 This temperature difference results in a different color screening
 environment. Second, the bottomonium production rate at RHIC is lower
 than that at the LHC by $\sim 100$~\cite{Brambilla:2010cs}.  As a
 result, the average number of $b \overline{b}$ pairs in a central \auau
 collision at RHIC is $\sim 0.05$ versus $\sim 5$ in central \pbpb at
 the LHC. Qualitatively, one would expect this to effectively remove
 at RHIC any contributions from coalescence of bottom quarks from
 different hard processes, making the upsilon suppression at RHIC
 dependent primarily on color screening and CNM effects. This seems to
 be supported by recent theoretical calculations~\cite{Emerick:2011xu}
 where, in the favored scenario, coalescence for the upsilon is
 predicted to be significant at the LHC and small at RHIC.

The luminosity at RHIC for \auau collisions is $\sim 30$ times that at
the LHC for \pbpb collisions, and running cycles for heavy ions are
longer at RHIC. Therefore, with large acceptance and good momentum
resolution, it is possible in one year of running to make upsilon
measurements in the sPHENIX acceptance with yields comparable to those
at the LHC.

STAR is constructing a Muon Telescope Detector (MTD) to measure muons
at midrapidity~\cite{Ruan:2009ug}. Scheduled for completion in 2014,
it will have a coverage of $|\eta| < 0.5$, with about 45\% effective
azimuthal coverage. The MTD will have a muon to pion enhancement
factor of 50--100, and the mass resolution will provide a clean
separation of the $\Upsilon$(1S) from the $\Upsilon$(2S+3S), and
likely the ability to separate the $\Upsilon$(2S) and $\Upsilon$(3S)
by fitting. While STAR will already have made upsilon measurements
with the MTD at RHIC before the upgrade to sPHENIX proposed here would
be available, the upgrade to sPHENIX would provide better mass
resolution and approximately 10 times higher yields per run for
upsilon measurements. This would substantially enhance the ability
of RHIC to provide upsilon data of comparable quality to the LHC
data.

\subsection{Detector Performance}
\label{sec:quarkonia_signal}

We report first the expected yield and line shape of the
$\Upsilon$(1S), $\Upsilon$(2S) and $\Upsilon$(3S) signal from decays
to dielectrons. The results were obtained with single simulated
$\Upsilon$ events in a GEANT 4 simulation containing the VTX detector
and two additional tracking layers at 40 and 60 cm radius. The
magnetic field was 2 tesla. The sPHENIX acceptance times tracking
efficiency for $\Upsilon(1S+2S+3S) \rightarrow e^+e^-$ decays was
found to be 0.34, in the mass window 7--11\,GeV/$c^2$.

The baseline p+p cross section for $\Upsilon(1S+2S+3S)$ of $B_{ee}
d\sigma/dy_{|y=0} = 114 \pm 40$~pb is taken from a PHENIX central arm
measurement~\cite{daSilva:2011zz}. The rapidity dependence was taken
from PYTHIA. The relative yields of the three $\Upsilon$ states were
taken from CDF measurements at 1.8\,TeV~\cite{Acosta:2001gv}.
Estimates of the \pp yields in sPHENIX are shown in Table~\ref{tab:upsilon_yields}, along
with projected yields of the three $\Upsilon$ states for a \auau run.
These assume binary scaling, and no suppression of any of the $\Upsilon$
states.

\renewcommand{\arraystretch}{1.9}
\addtolength{\tabcolsep}{-0.5pt}

\begin{table}[hbt!]
  \centering
\begin{tabular}{cccccccc} 
Species & $\mathbf{\displaystyle\int L\,dt}$ & Events &
$\left<N_{\mathrm coll}\right>$ & $\Upsilon$(1S) & $\Upsilon$(2S) & $\Upsilon$(3S)  &$\Upsilon$(1S+2S+3S)\\
\hline 
\pp 	&   		18 $pb^{-1}$		&  	756 B	&		1			&	805  & 202  & 106 & 1113
\\
\cmidrule(r){2-8} 
\auau (MB) & 				& 50 B				& 	   240.4      &   	12794	&   3217 & 1687 & 17698
\\
\cmidrule(r){3-8}
\auau  (0--10\%) & 				& 5 B			& 	   962      &    5121 	&  1288     &  675 & 7084
\\
\bottomrule
\end{tabular}
\caption{The yield of different $\Upsilon$ states obtained in one year of \pp
  or \auau RHIC running. The numbers for \auau in this table are
  calculated assuming no suppression of any of the $\Upsilon$ state yields.}  
  \label{tab:upsilon_yields}
\end{table}

A critical question is whether the proposed tracking system, with a
magnetic field of 2 tesla, is capable of adequately resolving the
$\Upsilon$(1S) from the $ \Upsilon$(2S) and $\Upsilon$(3S) states.

\begin{figure}
  \begin{center}
    \includegraphics[width=0.49\textwidth]{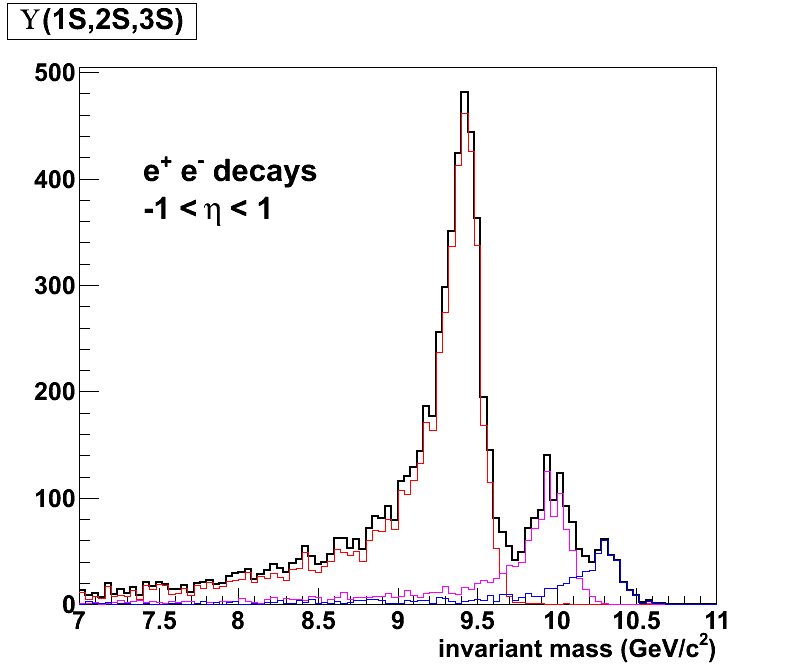}
    \includegraphics[width=0.49\linewidth]{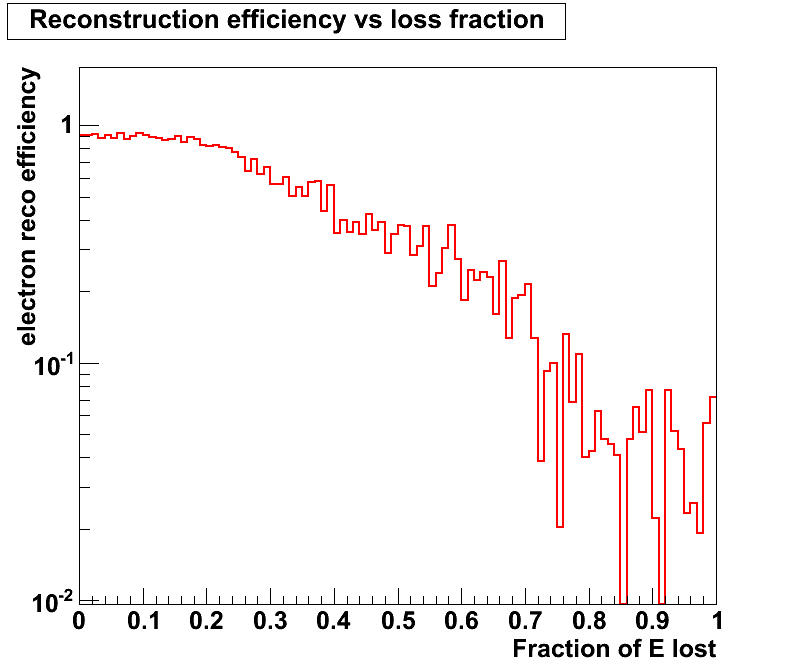} 
  \end{center}
  \caption{\label{fig:quarkonia_upsilonlineshape} Left panel: The mass
    spectrum from reconstructed electron decay tracks for the three
    Upsilon states combined.  The yield corresponds roughly to that
    for the 0--10\% centrality bin from 50B minimum bias events,
    assuming no suppression in \auau collisions.  Right panel: The
    electron track reconstruction efficiency for reconstructed
    electrons from $\Upsilon$ decays versus the radiative energy loss of the
    electron as it exits the last tracking layer.  }
\end{figure}

The reconstructed mass spectrum for dielectron decays is shown in the
left panel of Figure~\ref{fig:quarkonia_upsilonlineshape}. That spectrum
contains the number of Upsilons expected in the 0--10\% centrality bin
if there is no suppression. It can be seen that there are significant
low mass tails on the Upsilon mass peaks due to radiative energy loss
in the material of the VTX and outer tracking layers of sPHENIX.  The
radiative tails are found to be significantly (and helpfully)
suppressed by the drop in tracking efficiency with increasing energy
loss, due to the use of a circular track algorithm, as shown in the
right panel of Figure~\ref{fig:quarkonia_upsilonlineshape}.

The background under the Upsilon peaks consists of an irreducible
(physics) background due to dileptons from correlated charm,
correlated bottom and Drell Yan. There is also combinatorial
background from misidentified charged pions. The latter can be
estimated and removed by like sign subtraction, or by the mixed event
method.

To study the physics background, correlated charm and bottom
di-electron invariant mass distributions predicted by PYTHIA were
normalized to the measured charm and bottom cross-sections in \auau
collisions.  The PYTHIA Drell-Yan di-electron invariant mass
distribution was normalized to the theoretical prediction by
Vogelsang.

The combinatorial background was studied by generating events with
fake electrons due to misidentified pions, using input pion
distributions taken from measured $\pi^0$ spectra in \auau collisions.
A $p_T$-independent rejection factor was applied to the $\pi^0$
spectra to imitate fake electron spectra. In the results presented
here a rejection factor of 200 was used.

\begin{figure}
  \begin{center}
    \includegraphics[width=0.48\textwidth]{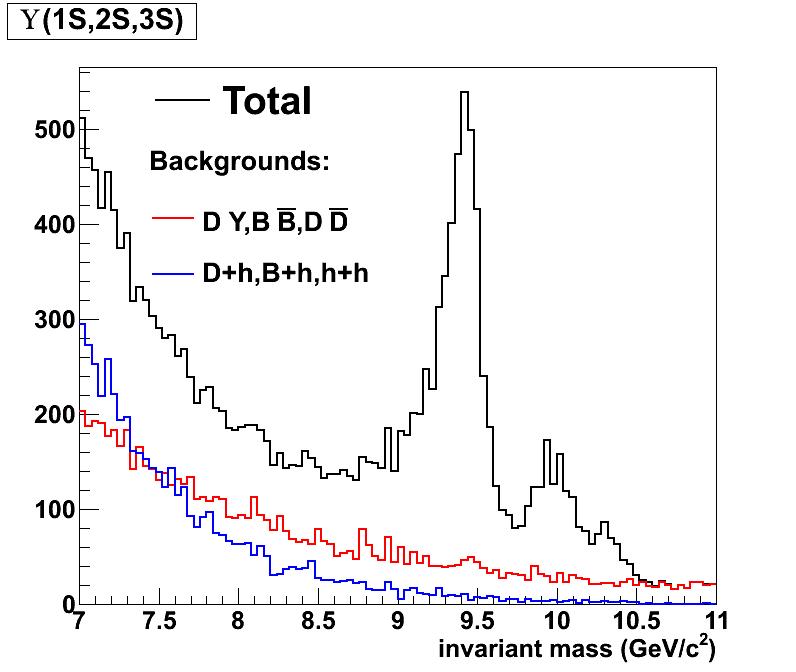}
    \includegraphics[width=0.48\textwidth]{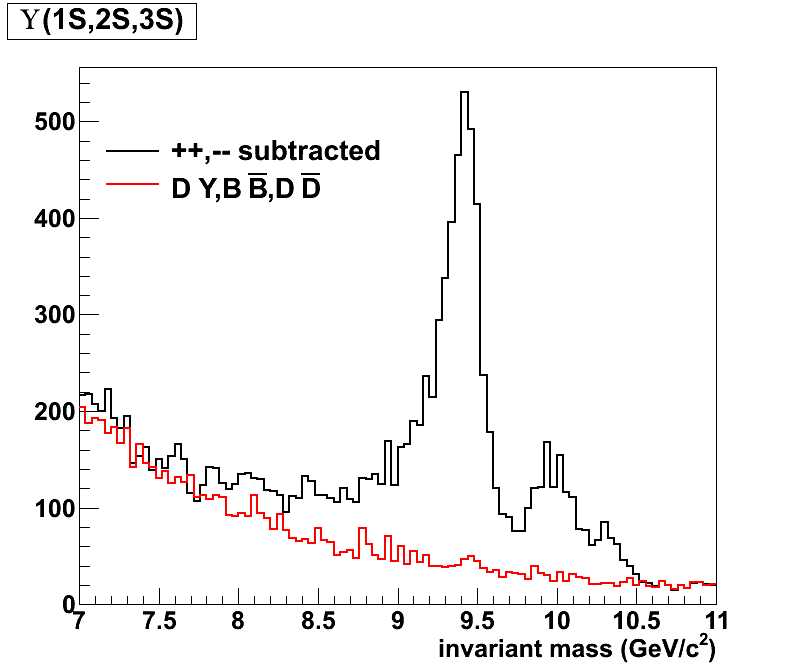}
  \end{center}
  \caption{\label{fig:quarkonia_bg} (Left) The signal plus background
    in the Upsilon mass region for 5B 0--10\% central \auau events,
    assuming a pion rejection factor of 200.  The combined backgrounds
    due to correlated bottom, correlated charm, and Drell Yan are
    shown as the red curve. The combined backgrounds due to fake
    electrons combining with themselves, bottom, and charm are shown
    as the blue line.  (Right) The expected invariant mass
    distribution for 5B 0--10\% central \auau events, after
    subtraction of combinatorial background using the like-sign
    method.  The remaining background from correlated bottom, charm
    and Drell Yan is not removed by like sign subtraction. It must be
    estimated and subtracted.  }
\end{figure}

All combinations of fake electrons from misidentified pions were made
with each other, and with high $p_T$ electrons from physics sources.
The latter turned out to be the least important source of background.
The results are summarized in Figure~\ref{fig:quarkonia_bg}(left), which
shows the signal + background in the $\Upsilon$ mass region for the 5B
0--10\% most central events, along with our estimates of the total
correlated (physics) background and the total uncorrelated
(combinatoric) backgrounds.  In Figure~\ref{fig:quarkonia_bg} (right) we
show the di-electron invariant mass distribution for 5B 0--10\%
central \auau events after the combinatorial background has been
removed by subtracting all like-sign pairs.

\begin{figure}
  \begin{center}
    \includegraphics[width=0.6\textwidth]{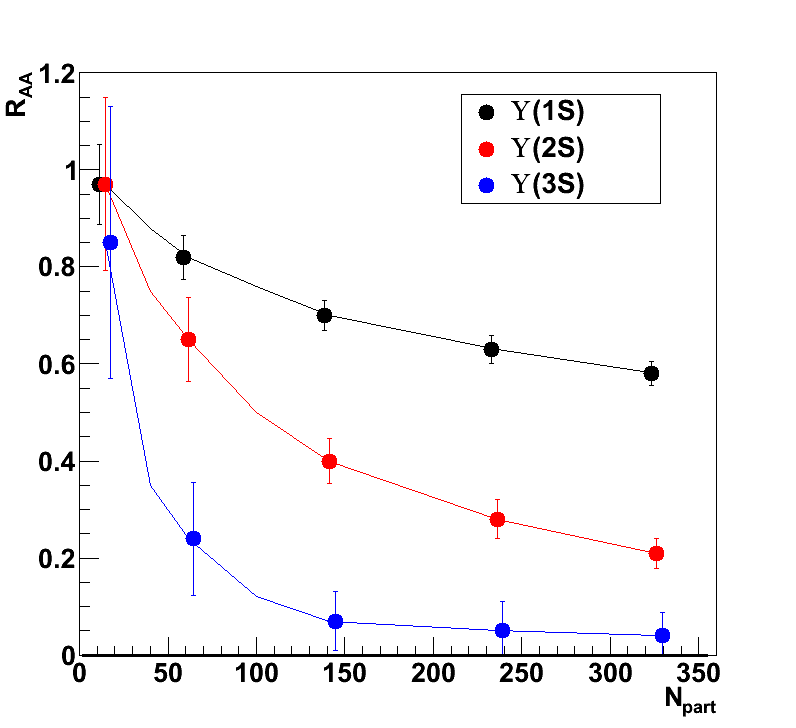}
  \end{center}
  \caption{\label{fig:quarkonia_upsilon_raa_statistics} Estimate of
    the statistical precision of a measurement of the $\Upsilon$
    states using sPHENIX, assuming that the measured $R_{AA}$ is equal
    to the results of a recent theory
    calculation~\cite{Strickland:2011aa}. The yields assume 50B
    recorded \auau events.  }
\end{figure}

From Figure~\ref{fig:quarkonia_bg} (left) we estimate that without
$\Upsilon$ suppression the S/B ratios are $\Upsilon$(1S): 2.4,
$\Upsilon$(2S): 1.4, and $\Upsilon$(3S): 0.67.  Using these estimates
as the unsuppressed baseline, we show in
Figure~\ref{fig:quarkonia_upsilon_raa_statistics} the expected
statistical precision of the measured $R_{AA}$ for 50B recorded \auau
events . For illustrative purposes, we take the measured suppression
for each state to be equal to that from a recent theory
calculation~\cite{Strickland:2011aa}.  For each state, at each value
of $N_\mathrm{part}$, both the $\Upsilon$ yield and the S/B ratio were
reduced together by the predicted suppression level.

We conclude from these results that the proposed upgrade to the
sPHENIX detector would provide a good $\Upsilon$ measurement in one
future RHIC \auau run, and would have the required mass resolution and
S/B to separate the $\Upsilon$(1S) state from the $\Upsilon$(2S) and
$\Upsilon$(3S) states. Further, we expect that by fitting a line
shape---which could be determined very well from the $\Upsilon$(1S)
peak---we could extract the $\Upsilon$(2S) and $\Upsilon$(3S) yields
separately with reasonable precision.

\section{Tagging Charm / Beauty Jets}

\begin{figure}[!hbt]
 \begin{center}
    \includegraphics[width=0.7\linewidth]{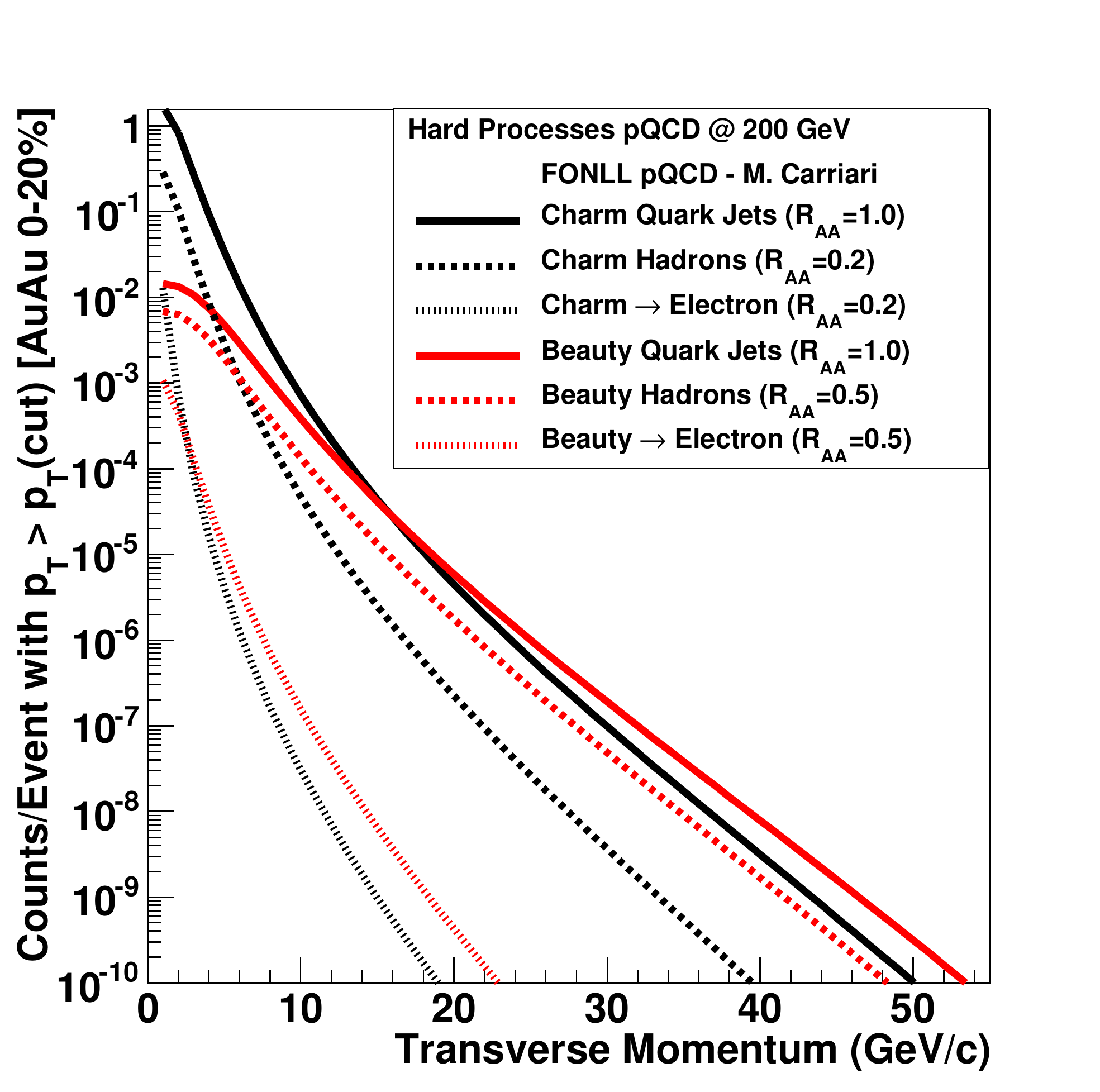}
        \caption{\label{fig:heavyrates}FONLL calculations~\cite{Cacciariprivate} for heavy flavor (charm and beauty) jets, fragmentation hadrons ($D, B$ mesons primarily), and decay electrons as a function of transverse momentum.  The rates have been scaled to correspond to counts with $p_{T} > p_{T}(cut)$
 for \auau 0--20\% central collisions.}
 \end{center}
\end{figure}

A main motivation for studying heavy flavor jets in heavy ion
collisions is to understand the mechanism for parton-medium
interactions and to further explore the issue of {\it strong versus weak} coupling~\cite{horowitz}.  There are crucial measurements
of single electrons from semileptonic $D$ and $B$ decays and direct $D$ meson reconstruction with the current PHENIX VTX and the soon to come
STAR Heavy Flavor Tracker (HFT) upgrade.  The sPHENIX program can significantly expand the experimental
acceptance and physics reach by having the ability to reconstruct full jets with a heavy flavor tag.
The rates for heavy flavor production from perturbative QCD calculations~\cite{Cacciariprivate} are shown in Figure~\ref{fig:heavyrates}.

\begin{figure}[!hbt]
 \begin{center}
    \includegraphics[width=0.7\linewidth]{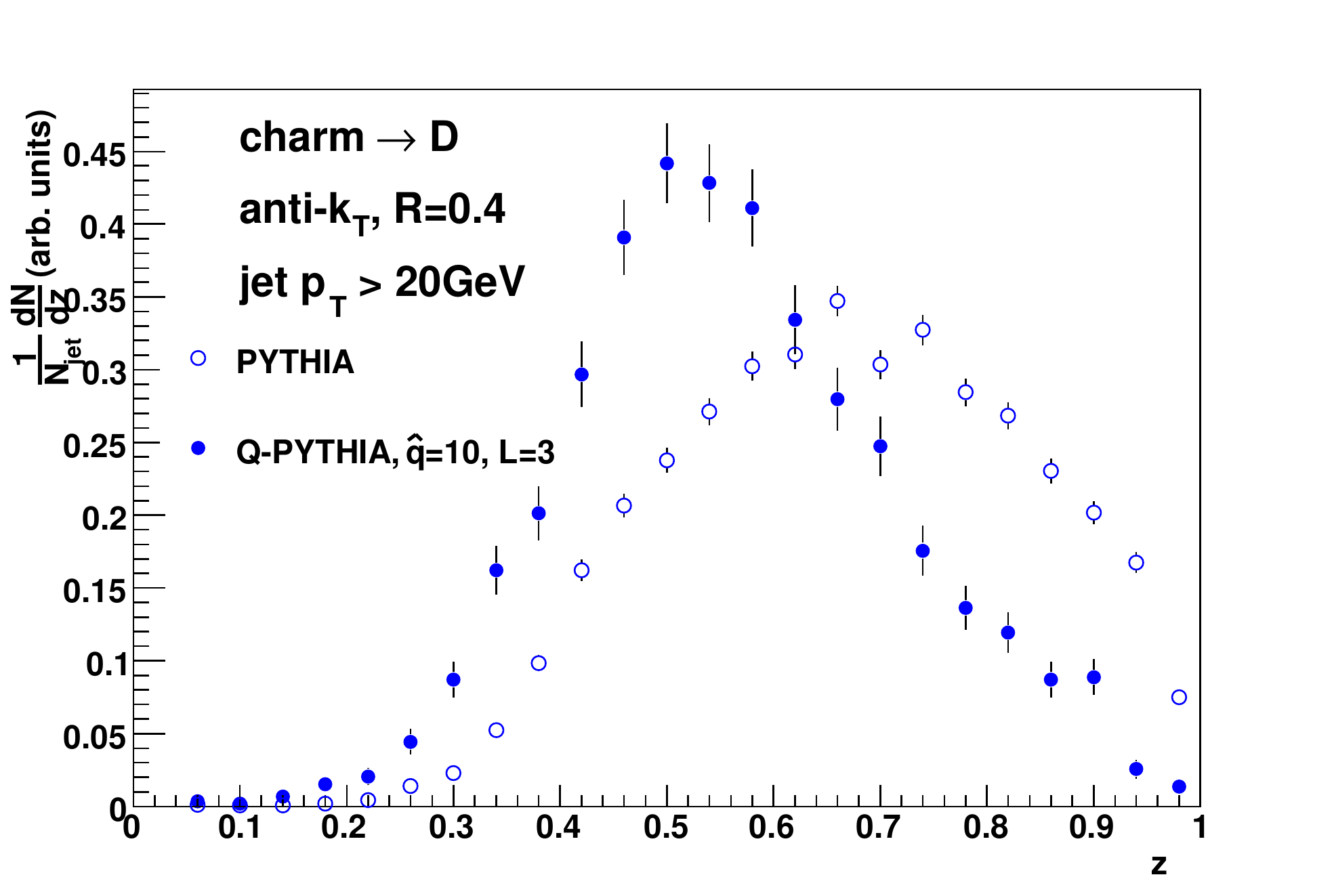}
    \caption{\label{fig:charmfrag} $D$ meson fragmentation function in
      \pythia (open points) and Q-\pythia (solid points) for
      anti-$k_{T}$ jets with $R = 0.4$ and $E_{T}(jet) > 20$\,GeV as a function
      of $z$, the fractional momentum of the $D$ meson relative to the charm quark.  }
 \end{center}
\end{figure}

One promising tool is the study of heavy flavor jet-shape modification
in \auau relative to \pp collisions.  Different mechanisms of energy
loss (radiative versus collisional) predict different re-distributions
of the jet fragments both inside and outside the jet cone.  There are
also scenarios where the heavy meson forms inside the medium and is
dissociated in the matter~\cite{Adil:2006ra,Sharma:2009hn}.  This
would lead to a nearly unmodified jet shape relative to \pp collisions
and a much softer fragmentation function for the leading heavy meson.
Figure~\ref{fig:charmfrag} shows the D meson fragmentation function in
\pythia and Q-\pythia for 20\,GeV charm jets.  The peak of the
fragmentation function is shifted in Q-\pythia from $z \approx 0.7$ to
$z \approx 0.5$.  Thus, for a given $p_{T}$, $D$ mesons are more
suppressed than charm jets.

The tagging of charm and beauty jets has an extensive history in
particle physics experiments.  Detailed studies for this tagging
within the sPHENIX upgrade with the additional tracking and electron
identification described above are underway.  There are three ways to
tag heavy flavor jets.  First is the method of tagging via the
selection of a high $p_{T}$ electron with a displaced vertex inside
the jet.  In minimum bias \auau collisions at $\sqrt{s_{NN}} = 200$\,GeV,
the fraction of inclusive electrons from $D$ and $B$ meson decays is
already greater than 50\% for $p_{T} > 2$\,GeV/c.  The VTX in
combination with the additional tracking layers can confirm the
displaced vertex of the electron from the collision point, further
enhancing the signal.  Since the semileptonic branching fraction of
$D$ and $B$ mesons is approximately 10\%, this method provides a
reasonable tagging efficiency.  Also, the relative angle of the lepton
with respect to the jet axis provides a useful discriminator for
beauty jets as well, due to the decay kinematics.  Second, the direct
reconstruction of $D$ and $B$ mesons is possible within sPHENIX, with
the additional tracking.  The current PHENIX VTX is limited in its
acceptance for $D$ decays by the need to also reconstruct the track in
the existing PHENIX central arm outer spectrometer, which has
$|\eta|<0.35$ and $\Delta \phi = 2 \times \pi/2$.  The sPHENIX
acceptance will yield a much higher (order of magnitude) yield of $D$
mesons.  The third method utilizes jets with many tracks that do not
point back to the primary vertex.  This technique is used by the $D0$
collaboration to identify beauty jets at the Tevatron~\cite{d0_nim}.  This
method exploits the fact that most hadrons with a beauty quark decay
into multiple charged particles all with a displaced vertex.  The
detailed performance metrics for tagged heavy flavor jets are being
developed in conjunction with converging on a design for the
additional tracking layers.

\section{Extending $\pi^{0}$ $R_{AA}$ to 40\,GeV/c}

\begin{figure}[!hbt]
 \begin{center}
    \includegraphics[width=0.49\linewidth]{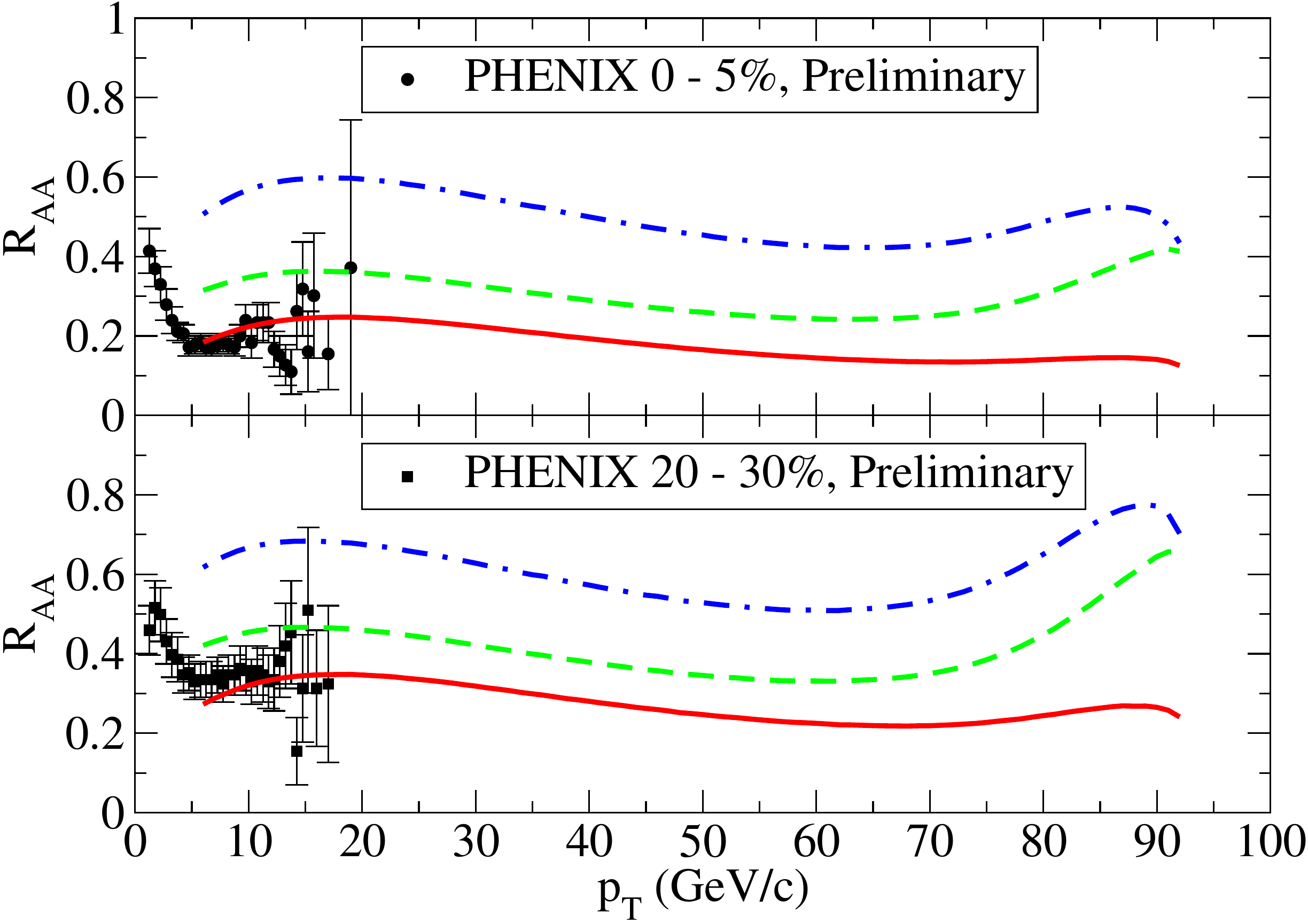}
    \hfill
    \raisebox{1mm}{\includegraphics[width=0.45\linewidth]{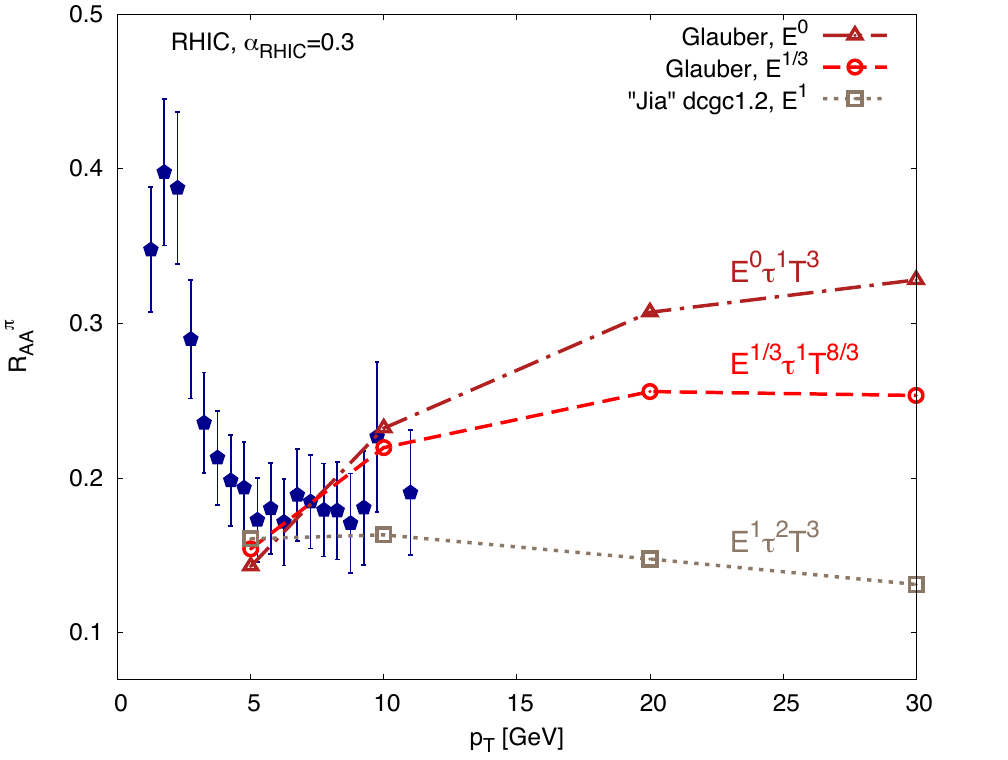}}
    \caption{\label{fig:raa_highpt} (Left) Calculations for $\pi^{0}$
      show a clear modification of $R_{AA}$ in \auau collisions at
      $\sqrt{s_{NN}} = 200$\,GeV that include collisional (blue),
      radiative (green), and both (red) energy loss mechanisms.  Also
      shown are PHENIX measured $\pi^{0}$ results. (Right) Three
      different parameterized energy loss calculation results using
      the simplified analytic ``polytrop'' jet energy-loss
      model~\cite{Betz:2012qq}.  }
 \end{center}
\end{figure}

The preshower detector will allow separation of single photon and two
photon (from $\pi^{0}$ decay) showers and thus substantially extend
the high $p_{T}$ measurement of the $\pi^{0}$ $R_{AA}$.  As shown in
Figure~\ref{fig:nlo_jetrates}, with 50 billion \auau minimum bias
collisions and the very large acceptance increase for sPHENIX, that
would permit $R_{AA}$ measurements out to $p_{T} \approx 40$\,GeV/c.
With this extended range it would be particularly interesting to see
if one observes the predicted rise in $R_{AA}$ that is a common
feature of all perturbative radiative energy loss models.  Shown in
Figure~\ref{fig:raa_highpt} (left panel) is the calculation from
Ref.~\cite{Qin:2007rn} for collisional energy loss only (blue),
radiative energy loss only (green), and both (red).  One sees good
agreement with the measured PHENIX $\pi^{0}$ data, but then no rise at
higher $p_T$ and instead a modest decrease.  In fact, the initial rise
at lower $p_T$ may be from switching from the predominance of gluon to
quark jets and then the almost exponentially falling spectra leads to
a slow decrease in the predicted $R_{AA}$.

In Ref.~\cite{Betz:2012qq} the authors utilize a simplified analytic ``polytrop''
jet energy-loss model that is used to test different jet-energy, path length,
and temperature-power dependencies.  They conclude that the experimental data indicate
an approximate 60\% reduction of the coupling $\kappa$ from RHIC to LHC.  The results from three calculations 
are shown 
in Figure~\ref{fig:raa_highpt} (right panel) and they note that ``future higher statistics measurements
at RHIC in the range $5 < p_{T} < 30$ GeV/c are obviously needed to differentiate between the energy-loss models.''
sPHENIX will make just such a set of precision measurements.

\section{High $z$ Jet Fragmentation Functions}

\begin{figure}[!hbt]
 \begin{center}
    \includegraphics[width=0.85\linewidth]{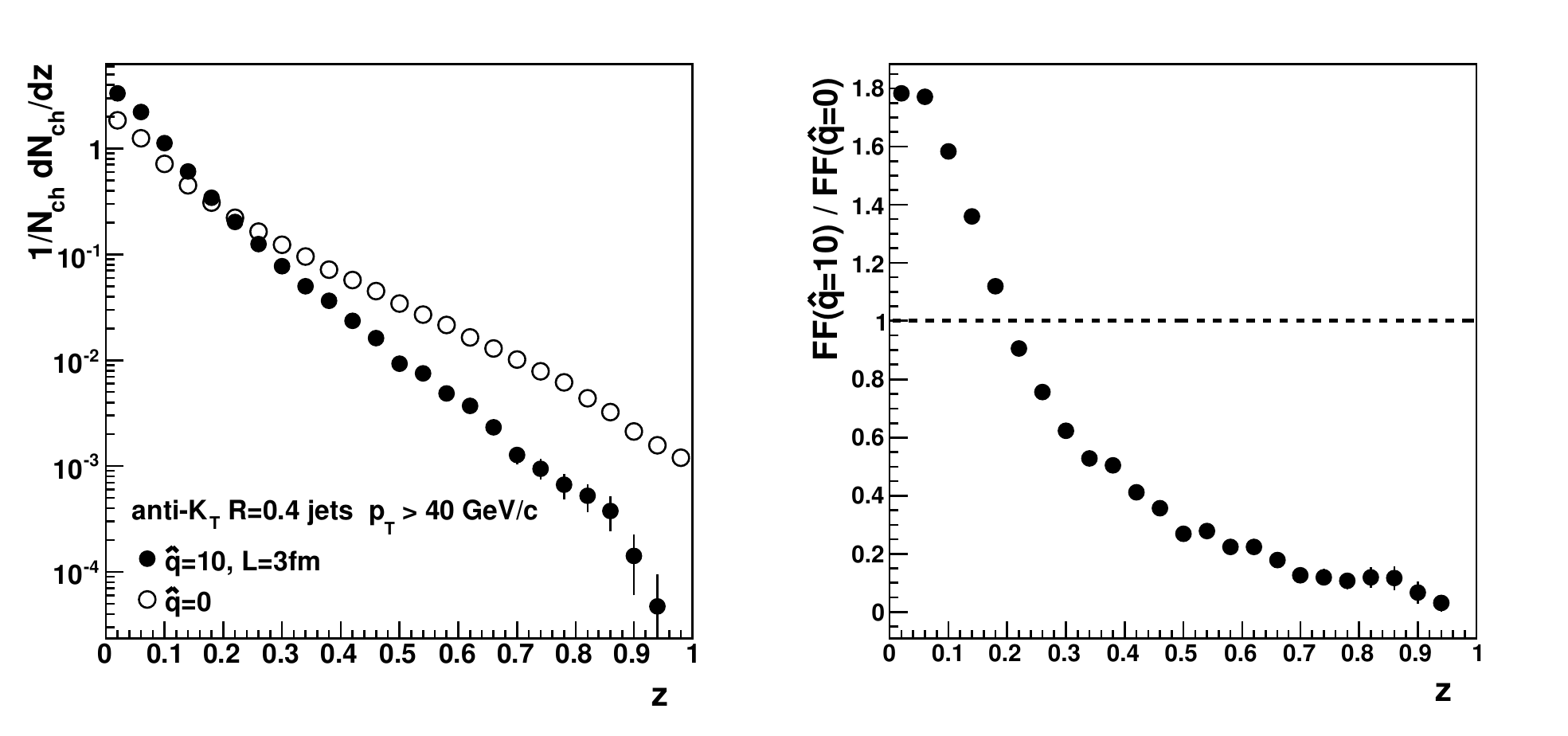}
        \caption{\label{fig:fragz}
Q-\pythia simulation with quenching parameter $\hat{q} = 0$ (i.e., in vacuum) and $\hat{q} = 10$\,GeV/c$^{2}$ for the fragmentation function
of light quark and gluon jets as a function of $z$. 
}
 \end{center}
\end{figure}

The original predictions of jet quenching in terms of induced forward
radiation had the strongest modification in the longitudinal
distribution of hadrons from the shower (i.e. a substantial softening
of the fragmentation function).  One may infer from the nuclear
suppression of $\pi^{0}$ in central \auau collisions $R_{AA} \approx
0.2$ that the high $z$ (large momentum fraction carried by the hadron)
showers are suppressed.  However, a direct measurement with
reconstructed jets and $\gamma$-jet events provides significantly more
information.  Shown in Figure~\ref{fig:fragz} is the fragmentation
function for 40\,GeV jets in vacuum (\pythia) compared with the case of
substantial jet quenching (Q-\pythia with $\hat{q} = 10$\,GeV$^{2}$/fm.
In the sPHENIX upgrade, fragmentation functions via
charged hadron measurements will be limited to the soft region ($p_{T}
\lesssim 5$\,GeV/c).  The additional tracking extends these
measurements over the full range for jets of 20--30\,GeV (with the
highest $p_T$ reach currently being evaluated).  Also, the independent
measurement of jet energy (via calorimetry) and the hadron $p_T$ via
tracking is crucial.  This independent determination also dramatically
reduces the \fake track contribution by the required coincidence with
a high energy jet.

Preliminary measurements of fragmentation functions from the CMS and
ATLAS experiments in \pbpb collisions show no modification within
uncertainties . Although one explanation is that the jets that are
reconstructed are from near the surface and thus not modified, with a
nuclear modification factor for inclusive jets $R_{AA} \approx 0.5$
that explanation is challenged. Similar measurements at RHIC energies
significantly augment the sPHENIX detector deliverables.

\section{Low and Intermediate Mass Dileptons}



Ever since the formation of \qgp in heavy ion collisions
was postulated, photons and dileptons have been considered among the
most important probes to study the QGP~\cite{Shuryak:1978ij}.  
This is because electromagnetic radiation is generated at every stage of the collision and, once
created, escapes the collision volume without any strong interactions.


The energy spectra of photons and dileptons reflect the energy density
and collective velocity of the strongly interacting matter from which
they are emitted. For matter in local equilibrium, electromagnetic
radiation provides information about the space-time evolution of
temperature and collective motion.  In this sense electromagnetic
radiation can be considered ``thermal'' radiation, though strictly
speaking it is not in equilibrium with the matter.  Even if the matter
is not in local equilibrium, the electromagnetic radiation emitted
will still carry information about the energy density.

For both real and virtual photons the temperature and collective
motion affect the momentum spectrum and it is difficult to unravel the
two contributions. However, the mass of the virtual photons is Lorentz
invariant and thus the mass distribution must be frame independent.
Consequently, the mass distribution of the virtual photons will only
be sensitive to the temperature. Measuring the virtual photon yield as
a function of mass and momentum is the next step beyond existing data
to separate temperature and flow velocity and map out the space-time
evolution of the system.

Isolating thermal radiation requires subtracting
lepton pairs from decays of pseudoscalar, vector, and
heavy flavor carrying mesons.  The CERN-SPS experiment NA60
pioneered this technique with high precision~\cite{Arnaldi:2008fw}.
NA60 results give detailed insights into the space-time evolution of
matter produced at low energies, where the emission is dominated by
$\pi$-$\pi$ annihilation from the hot hadronic phase.
At RHIC a larger fraction of
the radiation is emitted earlier in the collision; recent results
bring into question some of the current
ideas about the space-time evolution.


PHENIX developed a method using virtual
photons at low mass but at $p_T \gg m$ to extrapolate back to the real
photon point ~\cite{Adare:2008ab}. 
The $p_T$ spectrum is nearly exponential, with an inverse slope of
approximately 220\,MeV. Since the yield of thermal radiation is
expected to scale with the temperature to the fourth power, the 
yield and inverse slope suggest emission from early times. 
Hydrodynamic models that reproduce the data
indicate that the initial temperature must be between 300 and
600\,MeV,
depending on choice of time after the collision begins at 
which the system is equilibrated. 

Another important measurement established the elliptic flow ($v_2$) of
direct photons~\cite{Adare:2011zr}, which was found to be very similar in
the thermal region to that of hadrons.  The observation is now
confirmed 
with real photons which are converted in the detector
material and detected as $e^+e^-$ pairs.  
In hydrodynamical models, large anisotropies of momentum
distributions are the consequence of collective motion driven by
pressure gradients with respect to the reaction plane.
Initial
conditions for hydrodynamic expansions generally assume that matter
has no transverse velocity at the time of equilibration, thus any
anisotropies due to pressure gradients build up with time and are
strongest very late in the collision. In such models, the large
azimuthal momentum anisotropy of thermal photons suggests a late
emission, apparently inconsistent with the large yields and high
inverse slope.  Indeed, models have difficulty describing thermal photon
spectra and elliptic flow consistently~\cite{PhysRevC.84.054906}.

PHENIX has reported excess production of di-electrons at low mass and
low $p_T$~\cite{Adare:2009qk}, which is not consistent with theories that
successfully describe the NA60 data from CERN. The $p_T$ distributions
show two components. One has a slope of approximately 250\,MeV, and
yields the real photon spectrum when extrapolated to zero mass. The
other component is much softer, and has a much higher yield. The
combination of high yield and small slope is not consistent with early
emission, yet these di-leptons are absent in
models of the hadronic medium. The NA60 spectra show hints of a
component with a similar slope, but smaller yield. 
Higher quality data are needed to provide constraints on 
possible explanations of these soft dielectrons.

It will also be extremely interesting to measure $v_2$ and higher
harmonic flow for dileptons.  
Recently it has been suggested that the angular distribution of 
dileptons as a
function of mass can provide information on the degree of local
equilibration~\cite{Shuryak:2012nf}.  The $p_T$-dependent ratio of
dileptons to real photons in a certain mass window is sensitive to
the value of $\eta/s$, particularly at higher dielectron
masses~\cite{Chaudhuri:2012ti}.  It will be crucial, however, to measure
di-leptons in the intermediate mass region, tagging on displaced
vertices to pick out heavy flavor decays.


The PHENIX Hadron Blind Detector improves the signal to background
ratios over that in the published data. 
Analysis of these data is under way. However, even with the 
anticipated improved sensitivity, the precision of the dilepton
data is likely to be insufficient for a good determination of the
flow higher moments.


Further progress in understanding the electromagnetic probes will
require even higher precision data. Significantly reduced systematic
uncertainties are needed; for di-electrons this can be achieved
through Dalitz rejection via identified low momentum electrons. 
Unraveling the intermediate mass spectrum
requires secondary vertex measurement, which will be
provided by the silicon vertex detector at the center of sPHENIX.


Our approach would be to measure both real and virtual
photons in the $e^+e^-$ channel - in one case requiring the collision
vertex as the di-electron source, and in the other utilizing detector
material as the conversion point.
The detector should be sensitive to (at least) 0.5--5\,GeV/$c$
transverse momentum and 0--2.0\,GeV/c$^2$ pair mass, preferably higher.
Qualitatively, both the real photon (via external conversions) and the
dielectron measurements require high resolution ($\delta p/p \approx
1$\% or better) tracking and precise determination of the dielectron
vertex - compatible with sPHENIX requirements for heavy flavor
tagging.  Electron identification (e/$\pi$ rejection) should be in the
range of 1/500 to 1/1000.  This typically requires more than one
detector. We will study how well sPHENIX with the preshower
upgrade can satisfy the electron ID requirements. It may be possible 
that additional electron identification will be needed. R\&D
efforts underway for low mass tracking and for electron ID at
forward angles for ePHENIX study technology of potential 
applicability for dielectron measurements. However, full
specification of the requirements beyond the broad-brush
estimates above will require careful simulation study.

%
%
%


\chapter{Forward Upgrades and Physics}
\chaptermark{Forward upgrades}
\label{chap:fsPHENIX}

The sPHENIX detector described earlier in this proposal replaces the
current PHENIX spectrometer arms at mid-rapidity. The upgrade will
remove the central magnet including the massive iron yoke that
currently provides the hadron absorber located upstream of the PHENIX muon
detectors at forward rapidity.  The sPHENIX open geometry
will allow for the addition of 
spectrometers at forward and backward scattering angles capable of measuring hadrons, electrons, and
photons. Such a forward detector is being designed for the study of
cold nuclear matter effects in proton- and deuteron-nucleus collisions,
precision measurements of single transverse spin asymmetries for
the Drell-Yan process, and measurements of novel observables in jet production in transversely polarized \pp collisions.  
A subsequent upgrade adding an electron detector in the opposite
direction would further evolve sPHENIX into a detector for inclusive,
semi-inclusive and exclusive processes in deep inelastic
electron-proton and electron-nucleus scattering, referred to as ePHENIX, utilizing a
future high intensity electron beam at RHIC, as discussed in
Appendix~\ref{chap:ePHENIX}.

This Appendix highlights selected physics channels that are
presently being explored by the forward upgrade study group in PHENIX
and provides a brief discussion of the detector design used in
these studies.


\section{Transverse Momentum Dependent Phenomena in Nucleon Structure}

Over the past 10 years, RHIC experiments have studied the gluon
helicity distribution, $\Delta g(x)$, in the proton through the
measurement of longitudinal double spin asymmetries in inclusive
hadron and jet production~\cite{Adler:2004ps,Adler:2006bd,Abelev:2009pb,Adare:2007dg,Adare:2008aa,Adare:2010cc}. PHENIX and STAR have completed
instrumentation upgrades that will be used to measure quark and
anti-quark helicity distributions through $W$ production with high
luminosity polarized \pp collisions at $\sqrt{s}=500$\,GeV. The
additional high statistics data samples taken for $W$ measurements
will also be used to significantly improve the precision of the double
spin asymmetries constraining $\Delta g(x)$. RHIC is on course to
complete its measurements of $\Delta q(x)$ and $\Delta \overline{q}(x)$ in
$W$ production over the next few years. RHIC measurements, in
combination with polarized deep inelastic lepton-nucleon
scattering, will provide significant constraints for quark and gluon
helicity distributions for $0.01 < x_{q,g} < 0.3$. On the theoretical
side, the physics of the helicity structure of the nucleon in hard
scattering processes has been described successfully at leading twist
in the Operator Product Expansion (OPE) in collinear approximations of
hard scattering processes ignoring parton momentum components
transverse to the direction of the relativistic probe particles.

In addition to measurements constraining quark and gluon helicity
distributions, the BRAHMS, STAR and PHENIX collaborations have carried
out precise measurements of single transverse spin asymmetries, $A_N$,
at center of mass energies of $\sqrt{s}=62.4$, $200$ and $500$\,
GeV. It was observed that the large single transverse spin asymmetries
found at $\sqrt{s}\sim 20$\,\,GeV in fixed target experiments at FNAL
persist to the higher center of mass energies at RHIC. 
Despite intense effort over the past decade, a quantitative theoretical understanding
of the single transverse spin asymmetries observed in polarized
proton-proton collisions remains elusive.  
However, the theoretical interest in these asymmetries remains high as transverse 
spin observables are related to the orbital angular momentum structure of hadrons~\cite{Burkardt:2012sd}, 
and as such holds promise for understanding these dynamics as well as the complex dynamics of confinement
and chiral symmetry breaking in hadrons~\cite{Sivers:2011ci}. 

In parallel to the measurements of transverse spin asymmetries at
RHIC, deep inelastic scattering experiments at DESY, CERN, and
Jefferson Laboratory have confirmed the existence of two transverse
momentum dependent mechanisms that were proposed by Sivers and
Collins some 20 years ago to explain the large non-zero
transverse single spin asymmetries in polarized proton-proton
scattering. These measurements have been of an exploratory nature with
significant statistical uncertainties and limited coverage in $x$ and
$Q^2$.

A consistent theoretical framework has been developed to describe
transverse momentum dependent observables in hard scattering
processes. At large transverse momenta the Qiu-Sterman mechanism
describes Transverse Momentum Distribution (TMD) observables~\cite{Qiu:1998ia}. Spin dependent transverse
momentum components can be generated through multi-gluon correlations
at higher orders in the OPE. At smaller transverse momentum,
corresponding to the intrinsic transverse momentum of partons in
hadrons, the Sivers and Collins mechanisms are
applicable. Correlations between the transverse spin of the target
proton and intrinsic transverse momentum of quarks in the initial state
(Sivers~\cite{Sivers:1989:cc}) or correlations between quark spin and
the transverse momentum of hadrons in the final state (Collins~\cite{Collins:1992kk}) give rise to anisotropic distributions of
hadrons in the final state with respect to the proton spin.

In an important result, it has been demonstrated by Ji, Qiu, Vogelsang
and Yuan~\cite{Ji:2006ub} that the Qiu-Sterman and Sivers mechanisms lead to identical
results in the region of intermediate transverse momenta. The departure from the collinear approximation and the
description of transverse momentum dependent processes requires the
correct treatment of the exchange of soft (non-collinear) gluons in
the initial and final states of high energy scattering reactions. This
has been accomplished through the inclusion of gauge link integrals
that sum over initial and final state soft gluon exchange. The
inclusion of the gauge link integrals in the hard scattering matrix
elements can give rise to Sivers correlations between transverse
proton spin and quark transverse momentum. Gauge link integrals have
been found to be process dependent. For example, at leading order the final state gauge
link relevant for the Sivers effect in semi-inclusive deep-inelastic
scattering is equal in magnitude but opposite in sign to the initial
state gauge link relevant for the Sivers mechanism in Drell-Yan. The
test of the TMD framework
through observation of the predicted sign change of Sivers asymmetry
observables in Drell-Yan compared to SIDIS is the subject of a NSAC
milestone for hadron physics, HP13.

Significant theoretical progress has been made recently in the
derivation of evolution equations for TMD parton distributions and
fragmentation functions.  The knowledge of TMD evolution equations
will make it possible to carry out a rigorous QCD analysis of TMD
observables measured at collider energies by experiments at RHIC and
future SIDIS experiments using polarized fixed targets at Jefferson
Laboratory and at CERN. Recent theoretical work~\cite{Kang:2012em,Aybat:2011ta} 
suggests that the correct evolution may lead to significantly smaller Sivers
asymmetries at large $Q^2$.  However, the final theoretical analysis
of the evolution for Sivers asymmetries will require additional input
from experiment.

\begin{figure}
\centering
\includegraphics[width=0.75\linewidth]{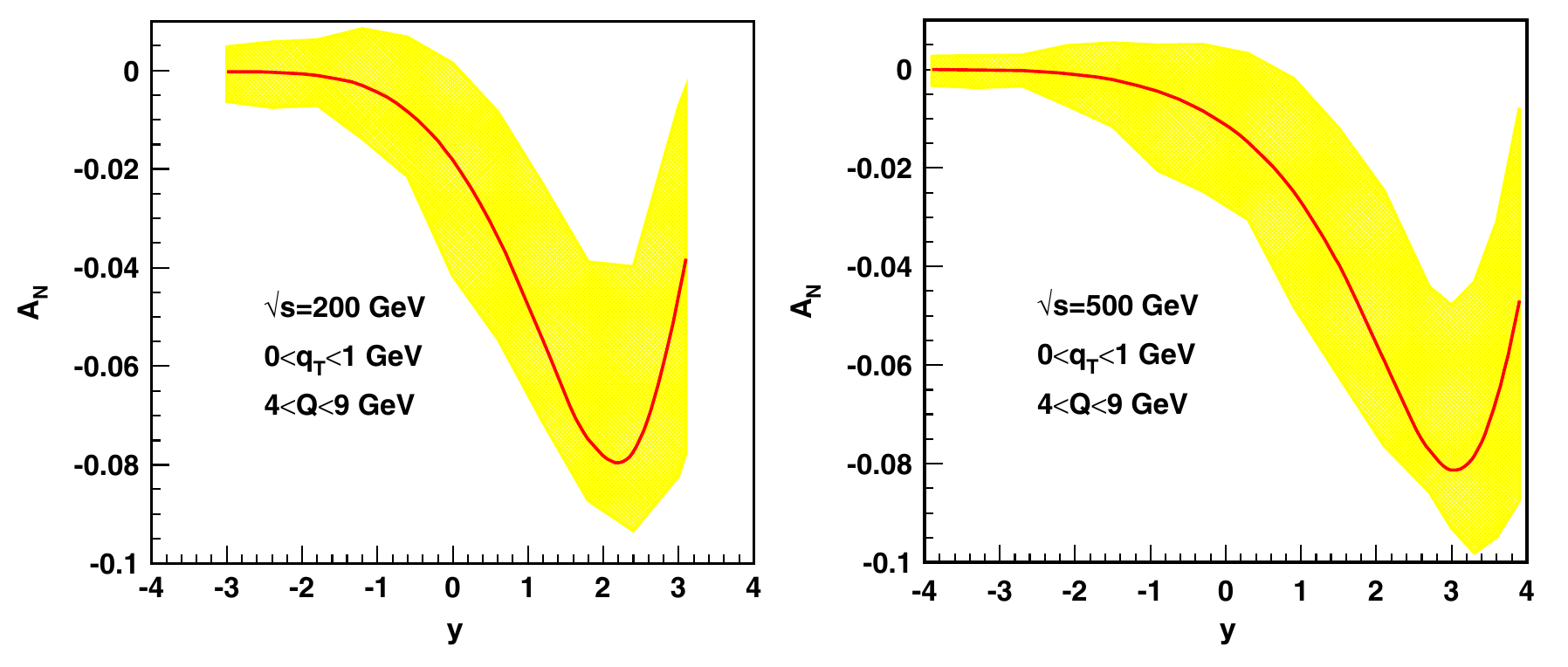}
\caption{The expected $A_N$ for Drell-Yan dilepton production at RHIC for $\sqrt{s}=200$\,GeV and $\sqrt{s}=500$\,GeV~\cite{Kang:2009sm}.  The uncertainties
shown as yellow bands are a result of the uncertainties from the Sivers function extracted from SIDIS data.}
\label{fsPHENIX:DY_AN}
\end{figure}

Single transverse Spin Asymmetries (SSA) in Drell-Yan production of
electron-positron pairs at forward rapidity will uniquely test the
process dependence of the Sivers mechanism (sign change compared to
SIDIS). Similar tests are possible through the observation of SSAs in
jet-photon production in polarized \pp scattering at RHIC. A 
comparison
of the asymmetries measured in proton induced Drell-Yan at RHIC with
SSAs measured in pion induced Drell-Yan processes at fixed target
energies provides a unique test of TMD evolution.  
The predicted Drell-Yan asymmetries at $\sqrt{s} = 200$ and $500$\,GeV
are shown in Figure~\ref{fsPHENIX:DY_AN}~\cite{Kang:2009sm}. 
The precise measurement of SSAs in Drell-Yan in the forward rapidity region probes the Sivers
quark distributions at large $x$ and measurements in the backward
region provide unique access to the Sivers distributions for sea
quarks~\cite{Collins:2005rq}. 

The Collins effect manifests itself through an azimuthal anisotropy in
the distribution of hadrons in final state jets with respect to the
proton spin component normal to the scattering plane. Precision
measurements of single transverse spin asymmetries for single
identified charged pions (Collins effect) and for identified hadron
pairs (Interference Fragmentation Function, IFF) will lead to the
first measurement of quark transversity distributions at large
$x>0.35$. This measurement will make possible a data based
determination of the tensor charge of the nucleon thus testing Lattice
QCD predictions for the tensor charge \cite{ Musch:2010ka}.
A comparison of quark transversity extracted from TMD Collins
observables and from collinear IFF asymmetries will provide a powerful
test of the TMD framework and evolution.

In the intervening time between the current PHENIX and the sPHENIX forward upgrade,
the MPC-EX upgrade~\cite{MPC-EX:web} will add a silicon preshower detector to the current PHENIX forward
calorimeters, the Muon Piston Calorimeter (MPC)~\cite{Chiu:2006zj}.  Two goals of this upgrade are a measurement
of the $A_N$ of direct photons (Sivers) and an exploratory measurement of $\pi^{0}$s in jets to extract
the Collins asymmetry contribution to the forward $A_N$ of $\pi^0$.  Because of the lack of a full jet
reconstruction, the Collins asymmetry measurement will not be able to determine the asymmetry as a function
of the $\pi^0$ fraction of the total jet energy.  This measurement will give important guidance in the design of
the sPHENIX forward upgrade by indicating how much of the forward $A_N$ for $\pi^0$ is derived from
transversity and a Collins fragmentation function, but any attempt to extract transversity
from the MPC-EX data would be model dependent.  Ultimately the full sPHENIX forward upgrade will be 
required for a complete survey of this important physics.

\section{Cold Nuclear Matter Effects}

Our quest to understand QCD processes in Cold Nuclear Matter (CNM)
centers on the following fundamental questions:

\begin{itemize}

\item What are the dynamics of partons at very small and very large
  momentum fraction ($x$) in nuclei, and at high gluon density what
  are the nonlinear evolution effects (i.e., saturation)?

\item What are the pQCD mechanisms that cause energy loss of partons
    in CNM, and is this intimately related to transverse momentum
    broadening?

\item What are the detailed hadronization mechanisms and time scales
  and how are they modified in the nuclear environment?

\end{itemize}

Various aspects of these questions are being attacked by numerous experiments and
facilities around the world.  Deep inelastic scattering on nuclei
addresses many of these questions with results from HERMES at
DESY~\cite{Airapetian:2009jy,Airapetian:2003mi}, CLAS at
JLab~\cite{Brooks:2009xg}, and in the future at the JLab 12\,GeV
upgrade and eventually an Electron-Ion Collider~\cite{Brooks:2010rz}.
This program is complemented with hadron-nucleus reactions in fixed
target $p+$A experiments at Fermilab (E772, E886, and soon
E906)~\cite{Vasilev:1999fa} and at the CERN-SPS.  RHIC has
significantly extended this program to $d+$A reactions at much higher
colliding energies, and also with the key augmentation of being able
to tag impact-parameter categories of the collisions.

The RHIC program has already played a major role in addressing the 
fundamental question of low-$x$ partons in nuclei.  It has been known 
for many years that the population of small momentum fraction (small 
$x$) partons in a nucleon embedded in a nucleus is depleted compared to 
that for a free nucleon. Evidence for this phenomenon has come largely 
from deep-inelastic scattering (DIS) 
measurements~\cite{Geesaman:1995yd,Berger:1987er} and from 
Drell-Yan~\cite{Alde:1990im,Vasilev:1999fa} measurements.  Quarks and 
anti-quarks are both depleted for $x < 10^{-2}$.  For gluons the 
evidence is mostly indirect and relies on the $Q^2$ scaling violations 
observed in lepton DIS measurements.  The state of the art for gluons is 
embodied in the EPS09 gluon nuclear parton distribution functions 
(nPDFs) of Eskola  {\it et al.}~\cite{Eskola:2009uj}.
These modifications are 
extremely uncertain in the $Q^2$ range relevant at RHIC energies, with depletion factors ranging from $\simeq 10\%$ 
to nearly no gluons at $x \simeq 5 \times 10^{-3}$.
Recent gluon saturation models assert that a novel semi-classical
state---the color glass condensate (CGC)---is formed above a critical
saturation scale, $Q^2$, at low enough momentum~\cite{Gelis:2010nm}.
A universal and quantitative description of cold nuclear
matter effects in nucleon structure for different collision systems is
currently not available.

The PHENIX experiment has explored these cold nuclear matter effects with
measurements of \jpsi~and hadron-hadron correlations over a broad range of 
rapidity~\cite{Adare:2007gn,Adare:2011sc,Adare:2012qf} 
which are sensitive to
an extended range in Bjorken $x$.  These results cannot be explained within the
parametrized nuclear modified parton distribution functions~\cite{Eskola:2009uj},
and hint at new physics of gluon saturation and possibly initial-state parton
energy loss.  The PHENIX forward silicon tracker (FVTX) has been installed in 2012 and will enable
detailed measurements of open heavy flavor, multiple quarkonia states, and a first look at 
Drell-Yan at forward rapidity ($1.2 < y < 2.4$).   This provides crucial comparison data to
the \jpsi~nuclear modification with precision open heavy flavor and Drell-Yan, the latter of 
which has no final state interactions to disentangle.  
In the future, the MPC-EX upgrade~\cite{MPC-EX:web} will measure prompt photon production in
p(d)$+$A collisions as another channel to constrain the nuclear gluon distribution at low-$x$.


The sPHENIX forward upgrade discussed here will build upon these current upgrades and substantially
extend their kinematic reach and channels of measurement.  
It is a central goal of the sPHENIX forward
upgrade to systematically survey the cold nuclear matter effects in
high energy nuclear collisions with small statistical and systematic
uncertainties. Important experimental observables will include the
cross sections for quarkonia,  vertex tagged open charm and bottom
quarks, inclusive hadrons and jets, as well jet-jet correlation
measurements, over a broad range of rapidity intervals between the
reconstructed jets. These data sets will serve as input to the
development of a universal and quantitative description of cold
nuclear matter effects in nucleon structure and the initial state of
high energy heavy ion collisions.

The high luminosity in deuteron-ion and proton-ion collisions
achievable at RHIC will make it possible to carry out measurements for
several nuclei at different collision energies, still with small
statistical uncertainties. The high luminosity  will also allow for
the comparison of results from proton- and deuteron-ion collisions
enabling the systematic analysis of the large effects expected from
multiple parton interactions (MPI) in nuclear collisions at forward
rapidity.  These effects are expected to be large and need to be
carefully accounted for in quantitative theoretical analysis~\cite{Strikman:2010bg}. 
Further, in order to keep experimental
systematic uncertainties and model uncertainties in the theoretical
analysis small the sPHENIX forward detector will be able to fully
reconstruct jets and to carry out low background heavy quark and
Drell-Yan measurements.

\section{Detector Considerations}

Precision Drell-Yan measurements require excellent dilepton identification and
the ability to reduce backgrounds from correlated charm and beauty decays.  The current
PHENIX forward silicon vertex tracker (FVTX) combined with the muon spectrometer measure
dimuons over rapidity $1.2 < |y| < 2.4$ and with the ability to tag heavy flavor decays.
Drell-Yan measurements via dimuons will give a first look at this transverse spin physics,
and the forward sPHENIX upgrade presented in this Appendix will substantially advance this program to
precision measurements over a much wider rapidity range.
The forward sPHENIX Drell-Yan measurements will be via dielectrons, and will require an
electromagnetic calorimeter and charged particle tracking as well as heavy
flavor tagging for background rejection. 


The forward sPHENIX detector will greatly extend our ability to 
measure the Collins and Sivers asymmetries at forward rapidities by performing full jet reconstruction. 
Experimentally, the measurement of transverse spin effects within jets
will require electromagnetic and hadronic calorimetry for jet
reconstruction, particle tracking to determine the fractional momentum
of hadrons or hadron pairs in the jet and particle ID to avoid
canceling of transverse spin effects in the fragmentation of different
hadrons. Precise studies of the Sivers and Collins effects in
different channels with the sPHENIX forward upgrade will make it
possible to decompose the large transverse single spin asymmetries
that historically have been observed in polarized proton-proton
collisions and to identify and quantify the contributions from
different spin effects.

The forward sPHENIX study group has been investigating the exact detector performance
design requirements for these physics channels. The currently envisioned sPHENIX 
forward detector will have an acceptance from a pseudorapidity of $1.2 < \eta < 4$. The
acceptances of the mid-rapidity upgrade and the forward upgrade will be matched closely in order to
minimize the missing energy in the event reconstruction.
Currently, a ``straw man'' design is being used for the purpose of sensitivity
studies. This design divides forward sPHENIX into a forward section,
$1.2< \eta < 3 $ and very forward region, $3 < \eta <4$. Currently it
is thought that there will be two sources of magnetic field. For the
forward region an extension or modification of the central solenoid
could provide a sufficiently strong tracking field. For the high
momenta in the very forward region an additional forward coil is
foreseen to be able to reach acceptable momentum resolution. Gas Electron
Multiplier (GEM) detectors will provide charged particle tracking. Particle
identification is based on a Ring Imaging Cherenkov Detector. It
believed that a hadronic calorimeter with a modest energy resolution will
energy smearing for Collins measurements in jets within acceptable
limits. The forward electromagnetic calorimeter may consist of
a re-stack of the current PHENIX electromagnetic calorimeters (EMCal) and the MPC-EX towers. Early simulation results
indicate that the performance of the EMCal will be sufficient.
The ``straw man'' design will form the basis of \geant
studies to better define and demonstrate the capability of an sPHENIX
forward upgrade to address the physics discussed in the previous
sections. It is anticipated that the details of the detector design
and configuration will undergo significant evolution during this
process. 

\begin{figure}
\centering
\includegraphics[width=0.7\linewidth]{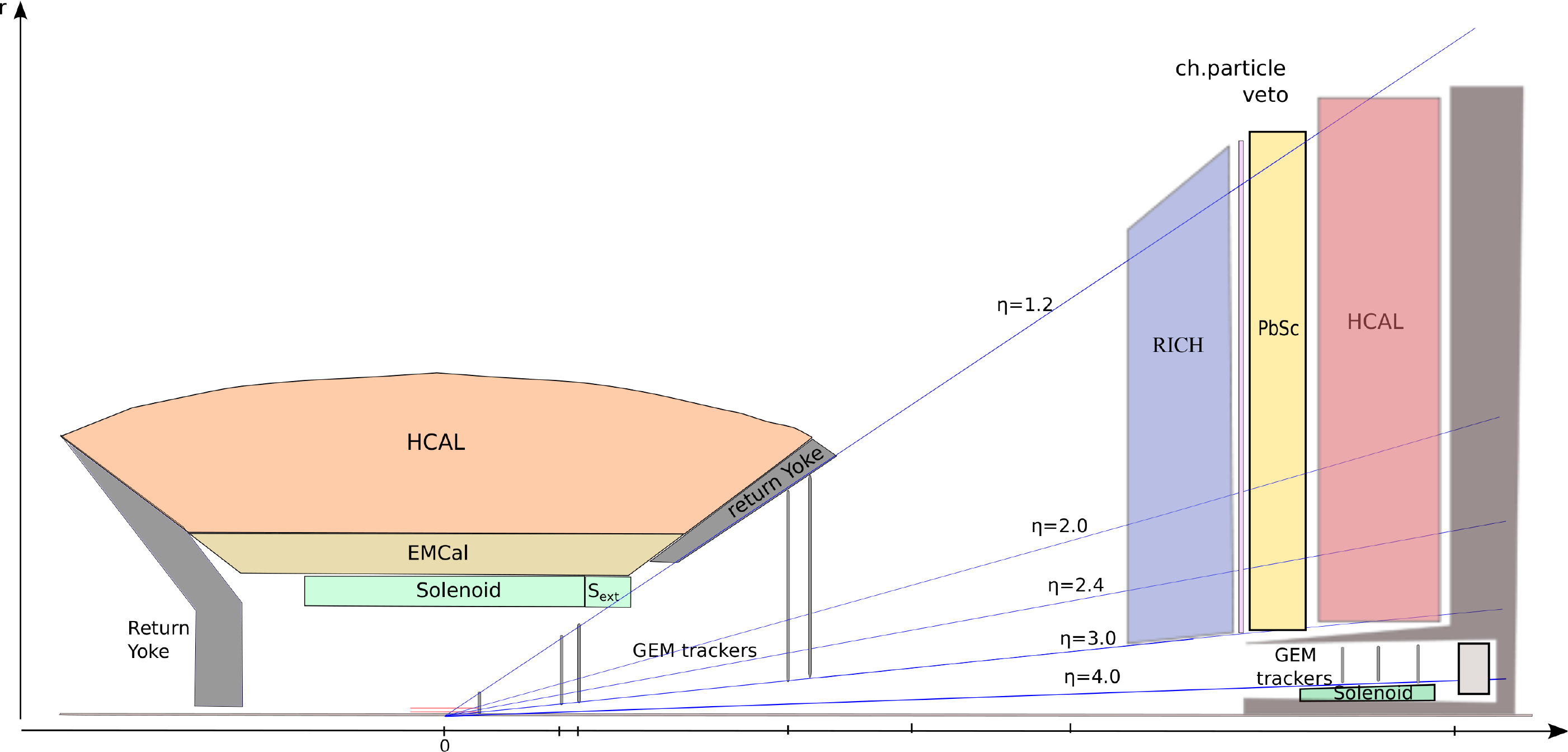}
\caption{A straw man layout of a possible detector for a future forward upgrade to sPHENIX.}
\end{figure}

Finally, PHENIX will host a workshop sponsored by the RIKEN BNL Research Center
with the goal to further develop the physics case for future forward
upgrades at RHIC from July 30th to August 1st.  The experimental
feasibility of the proposed forward physics case will be studied
through \geant based simulations and a final report with detailed
results from the study group will be available by the end of November
2012.
Institutions that are presently participating in the forward upgrades are
Abilene Christian University;
Brookhaven National Laboratory;
University of California, Riverside;
CIAE, Beijing, China;
Georgia State University;
Hanyang University, Seoul, Korea;
University of Illinois, Urbana;
Iowa State University;
KEK, Tsukuba, Japan;
Korea University, Seoul, Korea;
Los Alamos National Laboratory; 
Muhlenberg College;
New Mexico State University;
University of New Mexico;
RIKEN Brookhaven Research Center;
RIKEN Institute;
Rikkyo University, Tokyo, Japan;
Seoul National University, Seoul, Korea;
Stony Brook University;
and 
the University of Michigan.

\chapter{Evolution to ePHENIX}
\label{chap:ePHENIX}

One realization of a future Electron Ion Collider (EIC) consists of
adding a 5--30\,GeV electron beam to the current RHIC hadron and
nuclear beam capabilities.  The proposed initial construction would
consist of a 5--10\,GeV electron beam, referred to as Phase 1 of eRHIC.
Given the large capital investment and expertise in the current PHENIX
experiment, it is logical to determine if the sPHENIX upgrade proposal
presented in this document can serve as the foundation for a future
EIC detector, referred to as ePHENIX.  In designing the sPHENIX 
upgrade (covering $|\eta|<1.0$), we want to make sure that this
is compatible with a world-class EIC program in multiple measurement
channels when combined with future upgrades in the backward ($\eta <
-1.0$) and forward ($\eta > 1.0$) regions (enabled by the open
geometry of the magnetic solenoid).  Thus, in this Appendix, we first
describe the key physics that ePHENIX can access, and then how the
sPHENIX upgrade at midrapidity enables
measurements in terms of the magnetic field strength, tracking
resolution, calorimeter resolutions, and physical radial space for
additional particle identification detectors.  We conclude that the
sPHENIX upgrade serves as an excellent platform for this exciting future ePHENIX project.

\section{ePHENIX at eRHIC}
\label{ePHENIX:physics}


Quantum Chromodyanmics (QCD) is the Standard Model of strong
interactions, and yet our knowledge and understanding of it remains
incomplete. The properties of visible matter in the universe are
greatly influenced by the strong interactions.  The investigations of
the \qgp described in this document explore one such phase of QCD
matter at high temperature. In contrast, quark-gluon interactions of
cold QCD matter are optimally studied using the well established
technique of Deep Inelastic Scattering (DIS).  Many questions remain
unanswered in the regime where the gluons and sea quarks dominate the
landscape of hadrons and nuclei. The US nuclear science community is
considering a high-energy, high-luminosity, polarized electron-proton
and electron-ion collider~\cite{nsac2007} to study and understand the
structure and the properties of cold QCD matter.  In particular, it
aims to understand the role of gluons and sea quarks in QCD through
precision measurements of the structure of protons and nuclei and the
dynamics of the partons inside them~\cite{Boer:2011fh}.

Two possible realizations of the future DIS facility, the Electron Ion
Collider, are under consideration.  One realization, eRHIC, proposes
to add a 5--30\,GeV electron beam facility to the existing RHIC
facility at BNL to collide with one of its hadron (polarized nucleons
and nuclei) beams. The other option proposes colliding electrons from
the 12\,GeV CEBAF with a new hadron beam facility to be built at
Jefferson Laboratory.  This document will focus on the potential
physics program with a future evolved/upgraded PHENIX detector
(ePHENIX) to utilize collisions at eRHIC.

\begin{figure}[!hbt]
  \centering
  \includegraphics[width=0.57\linewidth]{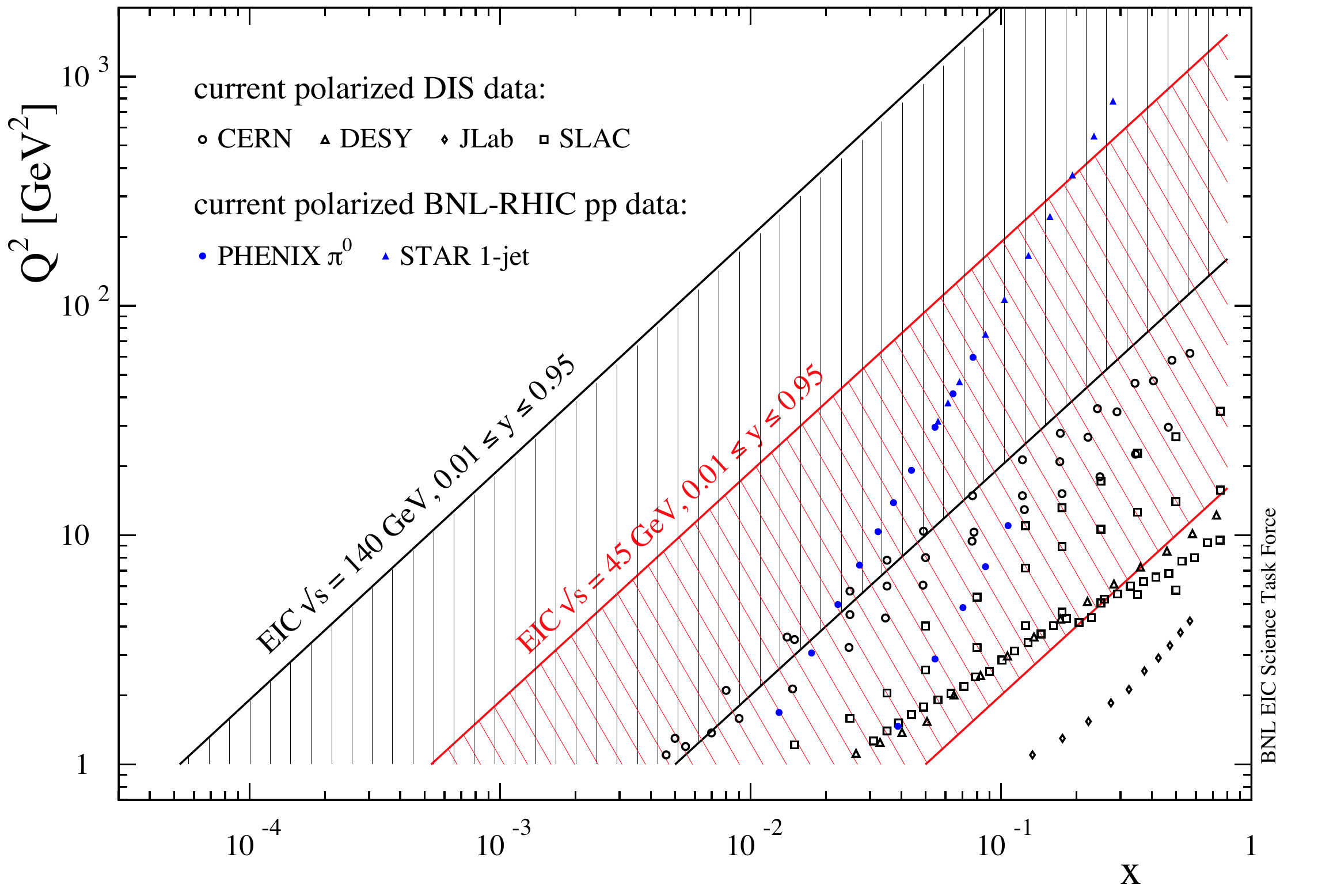}
  \hfill
  \includegraphics[width=0.42\linewidth]{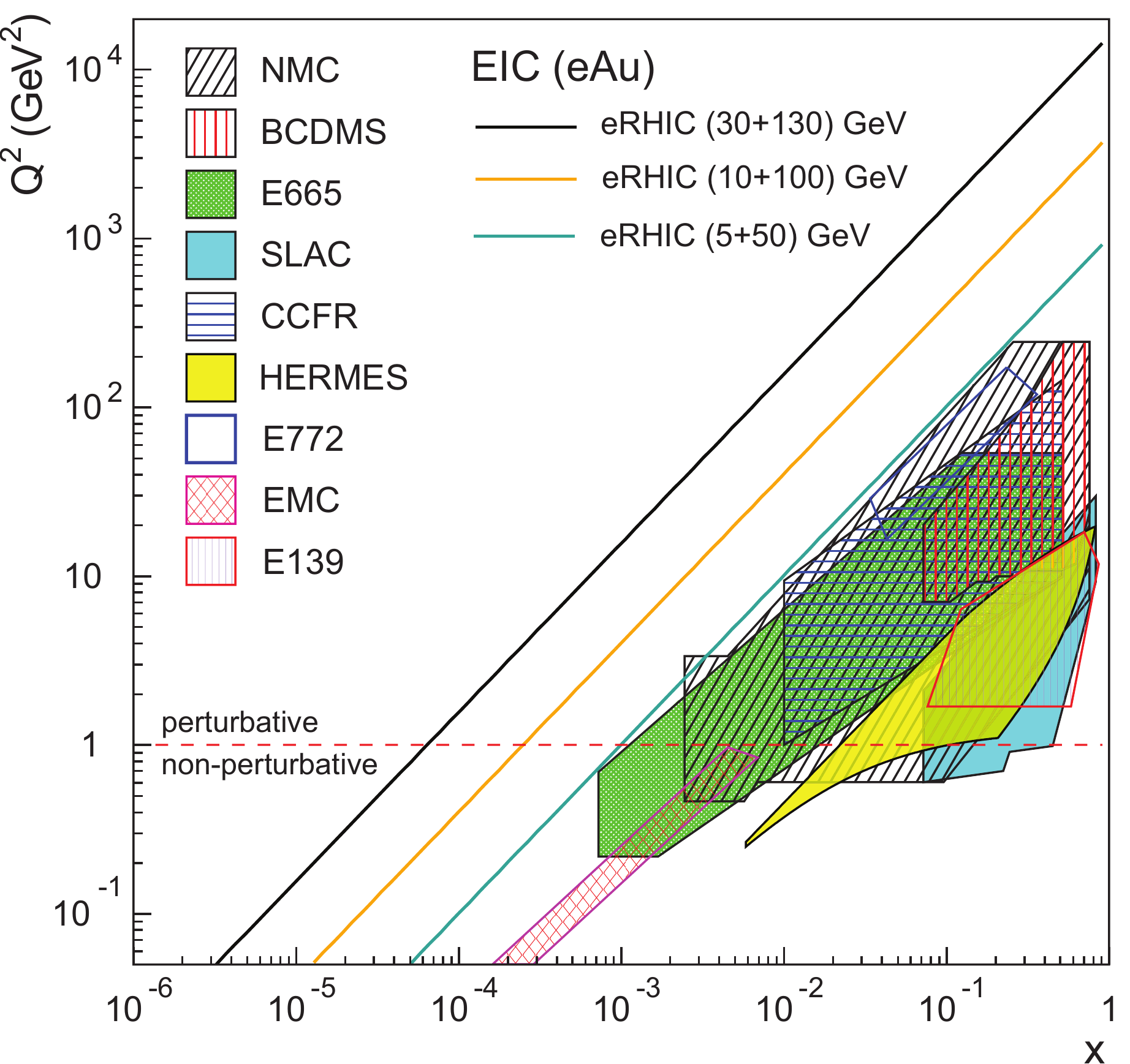}
  \caption{(Left) The $x$-$Q^{2}$ region covered by eRHIC and previous
    experiments for $e$$+$$p$ collisions in (red) Phase 1 and (black) at
    full energy, courtesy of Marco Stratmann.  (Right) Same for $e$$+$A,
    with the Phase 1 maximum $x$-$Q^2$ coverage given by the yellow line,
    courtesy of Thomas Ulrich.}
  \label{fig:xq2}
\end{figure}

The construction proposal for eRHIC includes an initial beam of 5\,GeV
electrons colliding with an existing hadron beam of RHIC with
100--250\,GeV polarized protons, and a wide range in nuclei from
deuterium to uranium with energies from 100\,GeV/nucleon.
Figure~\ref{fig:xq2} shows the $x$ and $Q^{2}$ kinematic region
accessible with 5\,GeV electrons colliding with 100\,GeV protons
($\sqrt{s} = \sqrt{4 E_{e} E_{p}} = 45$\,GeV) in red hatches.  It is
envisioned that the electron beam energy could gradually go up to
10\,GeV ($\sqrt{s}$ up to 100\,GeV) in the following few years after
construction completion.  The luminosities anticipated for Phase 1 of
eRHIC, with 5--10~GeV electrons, range from 0.6--10$\times
10^{33}$~cm$^{-2}$sec$^{-1}$, dependent only on the proton beam energy
for $E_e\leq20$\,GeV~\cite{Boer:2011fh}.  Similarly, for the
nuclei (in particular Au) the luminosity ranges from 0.5--3.9$\times
10^{33}$\,cm$^{-2}$\,sec$^{-1}$, dependent on the energy of the
nucleus, for $E_e\leq20$\,GeV.  This is the kinematic region and
luminosity reach for which ePHENIX is being designed to perform world
class measurements.  We envision full use of the sPHENIX detector
discussed in this document at midrapidity $|\eta|<1.0$, followed by a
sPHENIX-Forward upgrade, and then additional modifications specific to
ePHENIX.

\subsection{Kinematics}

Due to the large imbalance between the electron and hadron beam
energies in eRHIC collisions, the center of mass of such a collision
is moving in the incoming proton direction: this is {\it defined} as
the positive $z$ axis, and all angles will be measured with respect to
it.  For a fixed center of mass energy, inclusive scattering $e+p
\rightarrow eX$ is described by two independent variables $x$ and
$Q^{2}$, which are the Bjorken scaling variable, and the square of the
four momentum transfer in a DIS event. These variables describe the
lowest order process where the electron scatters elastically on a free
constituent of the proton/nucleus.  For neutral current events, which
are the primary concern for this Appendix, these variables can be
determined from the energy and angle of the scattered electron or from
the final state hadron system, or from a mixture of both.
The measurement of the scattered electron, or the hadronic final state when the measurement of scattered electron becomes difficult, yields:
\begin{alignat}{2}
y &= 1 - (E'_{e}/2E_{e}) \cdot (1-\cos{\theta'_{e}})  & &= (E_{j}/2E_{e}) (1 - \cos{\theta_{j}})\\
Q^{2} &= 2 E_{e} E'_{e} (1+ \cos{\theta'_{e}}) &  &= E_{j}^{2} \sin^{2} \theta_{j}/(1-y)\\
x &= E'_{e} (1+\cos{\theta'_{e}})/(2 y E_{p})  &  &= E_{j} (1+\cos{\theta_{j}})/[(1-y)(2E_{p})]
\end{alignat}
where $y$ is a measure of inelasticity in $e$$+$$p$ scattering, $E_{e,p}$ are the initial energies of the electron and proton beam, and 
$E'_{e}$ ($E_j$), $\theta'_{e}$ ($\theta_j$) are the energy and angle of the scattered electron (hadronic final state). 


\section{Physics Goals of the EIC}

There are three areas of nuclear physics where the EIC will greatly
expand our current knowledge: nucleon structure, QCD in nuclei, and
searches beyond the Standard Model.  In Phase 1 of eRHIC, ePHENIX will
be able to make important measurements in the first two, whereas the
third will require the larger electron beam energies available only in
a later phase.  The goals presented in \cite{Boer:2011fh} that ePHENIX
can measure are:
\begin{itemize}
\item Three dimensional structure of the nucleon, including its spin:
\begin{itemize}
\item The gluon and sea quark helicity contributions to the nucleon helicity
\item Quark and gluon transverse momentum distributions
\item The spatial distribution of gluons and sea quarks in the nucleon
\end{itemize}
\item QCD in nuclei
\begin{itemize}
\item Nuclear modification of parton distributions
\item Parton propagation in cold nuclear matter
\end{itemize}
\end{itemize}

Here we briefly describe the physics we will measure with ePHENIX.



\subsection{Polarized and 3d structure of the nucleon}
\label{ePHENIX:physics:spin}


\subsubsection{Helicity structure}
The golden physics channels to study the helicity structure of the
nucleon are detailed in the INT report~\cite{Boer:2011fh}. Two key
channels are within reach during Phase 1 of eRHIC with moderate
electron energies (5--10\,GeV).  The first is the gluon polarization
to low $x$ via the scaling of the structure function $g_1$ obtained
from inclusive DIS.  The second is the quark and antiquark helicity
distributions, especially $\Delta \overline{u}(x) - \Delta
\overline{d}(x)$ and $\Delta s(x) - \Delta \overline{s}(x)$, via the
semi-inclusive measurement of final state pions and kaons.  The last
golden channel, via electroweak measurements, will require the highest
electron energies as well as a more hermetic detector (for charged
current reactions) than what is currently being discussed and is
therefore not considered here.

Inclusive DIS is the simplest $e+p$ process in which only the
scattered electron is reconstructed. In polarized DIS one is able to
access the structure function $g_1(x,Q^2)$ in double spin asymmetries
of the spin orientations of the lepton and nucleon. The structure
function $g_1$ is, at leading order in the strong coupling $\alpha_S$,
the charge-squared-weighted sum of the quark and antiquark helicities:
$g_1(x,Q^2) = \sum_q e_q^2 [ \Delta q(x,Q^2) + \Delta
\overline{q}(x,Q^2)]$.  At next-to-leading order, the gluon helicity
contribution $\Delta g(x,Q^2)$, enters via the splitting functions. It
is therefore possible to extract the gluon helicity via the scaling
violations by measuring $g_1$ over a large $x$-$Q^2$ range. While for
the limited initial electron beam energies the lowest $x$ will not yet
be accessible, a large $Q^2$ range will still be covered for $x$ as
small as $10^{-3}$ and especially at intermediate $x$ (0.01--0.1)
where fixed target measurements exist but only at very low $Q^2$.

In semi-inclusive DIS (SIDIS), one additionally detects at least one
final-state hadron fragmenting from the struck quark.  It is then
possible to perform a flavor decomposition of the semi-inclusive
double spin asymmetries to arrive at the individual quark and
antiquark helicity distribution functions, $\Delta q(x,Q^2)$ and
$\Delta \overline{q}(x,Q^2)$.  This relies on the fragmentation
function, which describes the hadronization of an initial parton into
a final-state hadron, and on the existing knowledge of the unpolarized
parton distribution functions (PDFs).  While the valence quark
helicity distributions at intermediate to high $x$ were already
reasonably well determined, the sea quark helicities and in particular
difference between $\Delta \overline u(x,Q^2)$ and $\Delta \overline
d(x,Q^2)$ as well as $\Delta \overline s(x,Q^2)$ and $\Delta s(x,Q^2)$
are not well determined to date. The knowledge on the former difference
is expected to improve with the RHIC $W$ program at $x\sim0.1-0.4$,
and the data from fixed target SIDIS below that is insufficient
\cite{deFlorian:2009vb}.  
As the fragmentation functions for favored and disfavored
fragmentation become similar at lower fractional energies $z=E_h/E_q$,
it is important to cover the higher $z$ ranges to maintain flavor
discrimination power.  While the strange sea helicity is of interest
from high to low $x$, the $SU(2)_F$ sea asymmetry is expected to be
broken at $x$ of 0.01--0.1 in some
models.  

\subsubsection{3D structure of hadrons: Transverse momentum dependent distributions}   
While the previous section deals only with parton distributions
integrated over transverse momenta, recent theoretical and
experimental developments started to allow us also to study the
transverse momentum dependence (TMD) of partons in hadrons. When the
relevant transverse momentum is sufficiently smaller than the hard
scale, $Q$, the so-called TMD framework can be applied where PDFs
depend not only on the longitudinal momentum fraction $x$, but also on
the transverse momentum, $k_{\perp}$.  At leading twist, eight
functions exist for the nucleon, which originate from the possible
combinations of parton and nucleon spin and transverse momentum
orientations.  From this it becomes evident that orbital motion of
partons is required and that spin-orbit correlations of partons in the
nucleon can be studied via TMDs.

Three of the eight TMDs return to the previously defined collinear
PDFs upon integration of the transverse momentum: the unpolarized
distribution function, the helicity distribution and the quark
transversity distribution.  The quark transversity distribution
function is in itself of interest as it directly connects to the
tensor charge of the nucleon for which lattice QCD and model
predictions exist. Due to its chiral-odd nature, it can only be
accessed via another chiral-odd function; this only became possible
with the measurement of the Collins and the interference fragmentation
functions in recent years. Consequently, the knowledge of this last
leading twist distribution function is rather limited compared to its
unpolarized and helicity counterparts. With additional chiral-odd
fragmentation function measurements from $e^+e^-$ annihilation, it
will become possible to perform a flavor decomposition similar to the
helicity case when several hadronic final states can be detected and
identified in SIDIS. As transversity is expected to be a valence
dominated object, the intermediate $x$ range is of interest. As a
detailed understanding of the evolution of the relevant fragmentation
functions is still lacking, overlap in $x$ with the fixed target
experiments but at higher virtualities is important.

In recent years, increased interest has developed in the Sivers
function and the Boer-Mulders function, which both are spin-orbit
correlations: in the case of the Sivers function between the nucleon
spin and the parton's transverse momentum and in the case of the
Boer-Mulders function between the parton's transverse spin and
momentum.  Accessing these correlations requires a phase interference
which originates in the gauge link structure of QCD.  Again, in Phase
1 of eRHIC, there is overlap in $x$ coverage with previous
measurements but critically at higher scales.
In addition, abundant heavy flavor production will access the unmeasured gluon Sivers distribution function.

\subsubsection{3D structure of hadrons: Spatial imaging}   
Another important goal in Phase 1 of eRHIC is the spatial imaging of the nucleon. Instead of using transverse momentum dependence, the impact parameter dependence is studied via exclusive processes, in particular deeply virtual compton scattering (DVCS) and meson production.  Generalized Parton Distributions (GPDs) describe the correlation between parton momentum and (transverse) position within the nucleon, and in certain limits simplify to the Form Factors and the PDFs.  One important aspect of these measurements is the connection of GPDs to the total quark and gluon angular momentum through the Ji sum rule \cite{Ji:1996ek}.  Luminosities at the $\sqrt{s}$ available in Phase 1 of eRHIC will allow measurements of GPDs with good precision over a wider $x$-$Q^2$ range than at previous high luminosity fixed target experiments.


\subsection{QCD in nuclei}

\begin{figure}
\centering
\includegraphics[width=\onewidth]{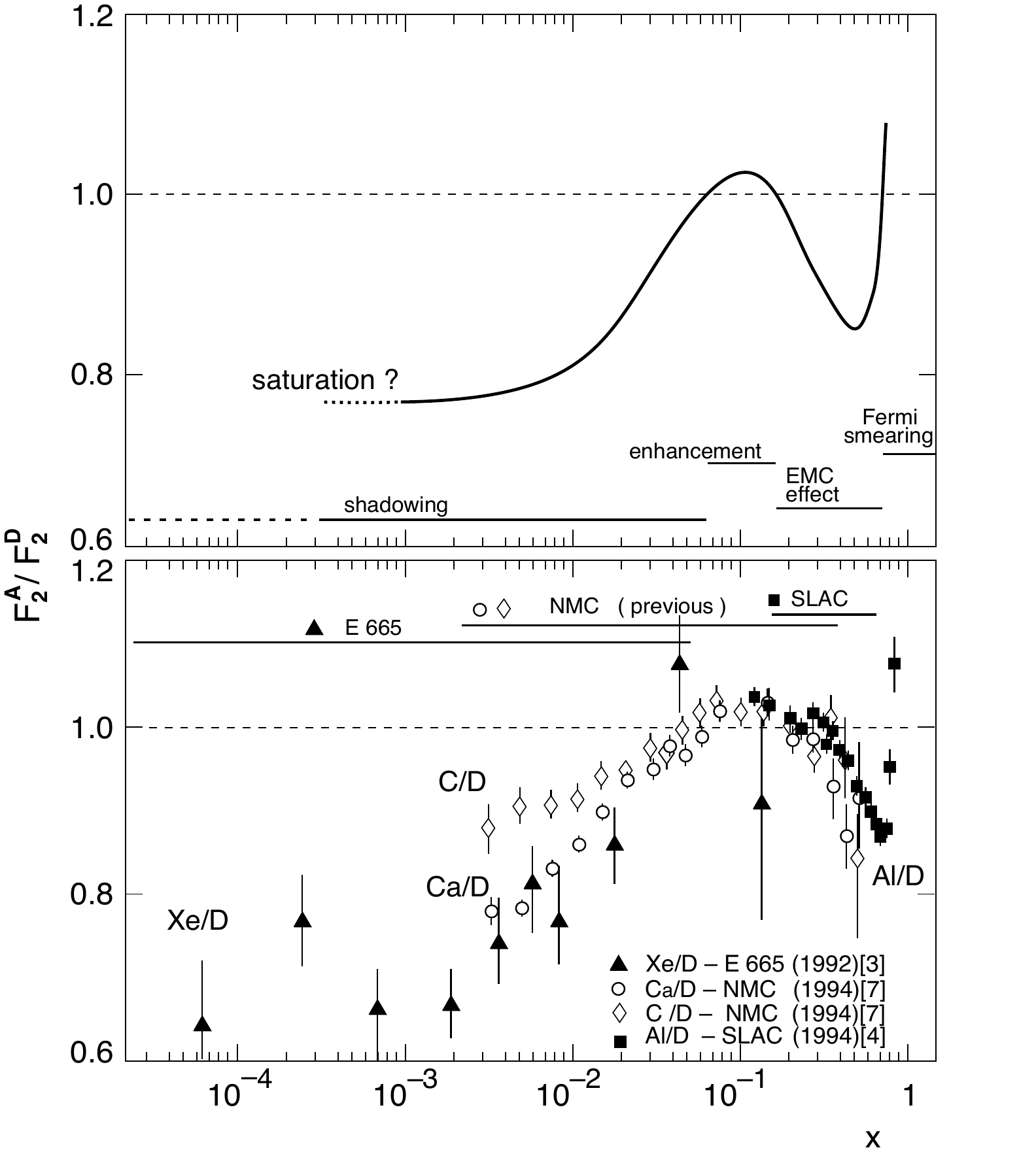}
\caption[]{(Top) Idealized depiction of the ratio of the structure
  function of a nucleus, $F_2^A(x, Q^2)$ per nucleon to $F_2^d (x,
  Q^2)$ of deuterium. Lower: Measured $F_2(x, Q^2)$ structure
  functions for C, Ca, and Xe relative to deuterium. Figure taken
  from~\cite{Arneodo:1995cs}.  
  \label{figure:eAdata}}
\end{figure}

One of the fundamental goals of nuclear physics is the understanding
of the structure of hadrons and nuclei in terms of the degrees of
freedom in the QCD Lagrangian, quarks and gluons.  More than two
decades ago it was discovered that quarks and gluons in bound nucleons
have markedly different distributions from those in the free nucleon,
as illustrated in Figure~\ref{figure:eAdata}.  Such differences can
arise through various mechanisms, for example modification of the free
nucleon structure, the presence of nonnucleonic degrees of freedom,
and quantum mechanical interference of the quark and gluon fields of
multiple nucleons at small $x$ (shadowing).  At even smaller $x$, the
gluon density increases to the point where gluon fields overlap,
leading to a strong field regime of non-linear QCD evolution called
saturation.  This regime is argued to have universal properties for
any hadronic system, but its onset is enhanced in nuclear targets due
to the superposition of the gluon field of many nucleons.
Measurements at ePHENIX will extend the $x$-$Q^2$ range beyond that in
the fixed target data, shown in Figure~\ref{fig:xq2}, and initiate the
first systematic study of non-linear QCD evolution.

$e$$+$A collisions offer a clean probe of nuclear PDFs (nPDFs),
complementing knowledge to be gained by $p$$+$A collisions.  Detailed
studies with an electroweak probe provide precision measurements
without the complications of strong interactions and with full access
to the scattering kinematics at the partonic level.  Precise knowledge
of nPDFs not only provides a deeper understanding of nuclei in terms
of partonic degrees of freedom, but is also crucial input for the
theoretical interpretation of a variety of ongoing and future high
energy physics experiments, such as heavy ion and proton-nucleus
collisions at RHIC and the LHC, or neutrino-nucleus interactions in
long baseline neutrino experiments.  High statistics inclusive
measurements of $F_2^A$ enabled by a 5--10~GeV electron beam would
greatly improve constraints on quark nPDFs, with access to gluons made
possible by examining the $Q^2$ dependence of $F_2^A$ or by measuring
$F_L^A$.  Semi-inclusive DIS measurements with identified hadrons
could provide flavor-separated information on the quark nPDFs.

There is great interest in understanding various features of gluon
distributions in nuclei, such as their $k_{\perp}$ and impact parameter
dependencies.  While the $x$ dependence can be accessed via $F_2^A$
and $F_L^A$ as described above, the simplest process to extract the
gluon TMD distributions is SIDIS.  The impact parameter dependence of
gluon distributions in nuclei can be measured via exclusive $J/\psi$
production~\cite{Caldwell:2010zza}.  Furthermore, measurements of
dihadrons will allow access to multi-gluon correlations in nuclei.

$e$$+$A collisions will also enable clean measurements of transport
coefficients in cold nuclear matter.  Novel observables, namely open
heavy flavors, charmonium, bottomonium, and jets, will be available
due to the high energy reach compared to earlier fixed-target $e$$+$A
experiments, greatly expanding the sensitivity to various nuclear
effects.

\subsection{Hadronization}

Closely linked to the propagation of color charges in nuclear matter is the study of how these color charges neutralize into hadrons.  The unique feature of eRHIC compared to previous fixed target experiments in studying the space-time evolution of hadronization is its large energy range,
allowing one to experimentally boost the hadronization process in and out
of the nuclear medium.  This permits isolation of in-medium parton propagation effects
from color neutralization and hadron formation times.  Subsequent comparison to nuclear Drell-Yan data, which are free from hadronization effects, will further isolate initial-state
parton energy loss from nuclear wave function effects.  The energies at eRHIC will enable
the study of hadronization of charm and bottom quarks in $e + $A collisions for the first time. 

The collider mode will also make it feasible to study in detail for the first time target
fragmentation and its correlation to current fragmentation through multi-particle correlations.  Hadronization data from $e+p$ and $e+$A at eRHIC spanning from the current to the target fragmentation region will offer many opportunities to study the dynamics of color confinement mechanisms.

\section{Detector Considerations}
\label{ePHENIX:detector}

We now turn to the detector requirements to make the above described
physics measurements in terms of momentum and angular resolution for
the scattered electron and particle identification needs for the
electron and hadrons.  The key is understanding the kinematics for the
various measurements and the precision required.  We find that the
current sPHENIX proposal at midrapidity is enabling of these
measurements when augmented with additional hadron identification
fitting into the spatial envelope allowed, in addition to forward and
backward future upgrade spectrometers.

\subsection{Tracking:  Momentum and Angular Resolution}
\label{ePHENIX:detector:tracking}


Measuring $F_L$ is challenging to the physics program as a whole since
it requires a well-devised long term run plan and excellent control of
systematics across a long time period.  Furthermore, as the beam
kinematics are varied, the scattered electrons at any given $(x,Q^2)$
will utilize different parts of the detector.

The desired detector resolution can be solved for analytically.  To
determine a limit on the detector performance we require that
uncertainties on the yield due to bin migration be held below some
acceptable level.  We take as an ansatz that 1\% yield measurements
are possible when the total yield uncertainty due to bin shifts is
held below 20\%.

Using the MRST2002 (NLO) parameterization for the structure functions,
we calculated the resolution requirements necessary to satisfy the
20\% ansatz for a variety of beam kinematics. Shown below in
Figure~\ref{fig:pres} (top) is the required momentum resolution as a
function of lab angle $\theta$ and lab momentum $p$ for different
Phase 1 eRHIC beam energy combinations.  Figure~\ref{fig:pres}
(bottom) shows the required angular resolution in the same manner.

These plots can be used as guidelines for evaluating spectrometer
designs and whether they will provide the necessary performance.  For
the beam energies in eRHIC Phase 1, the detector performance
requirements are not excessively stringent and are satisfied by the
sPHENIX magnetic field strength and next stage additional tracking
update, as detailed in Appendix~\ref{chap:barrel_upgrade}.

\begin{figure}[htbp]
\centering
\includegraphics[width=0.85\linewidth]{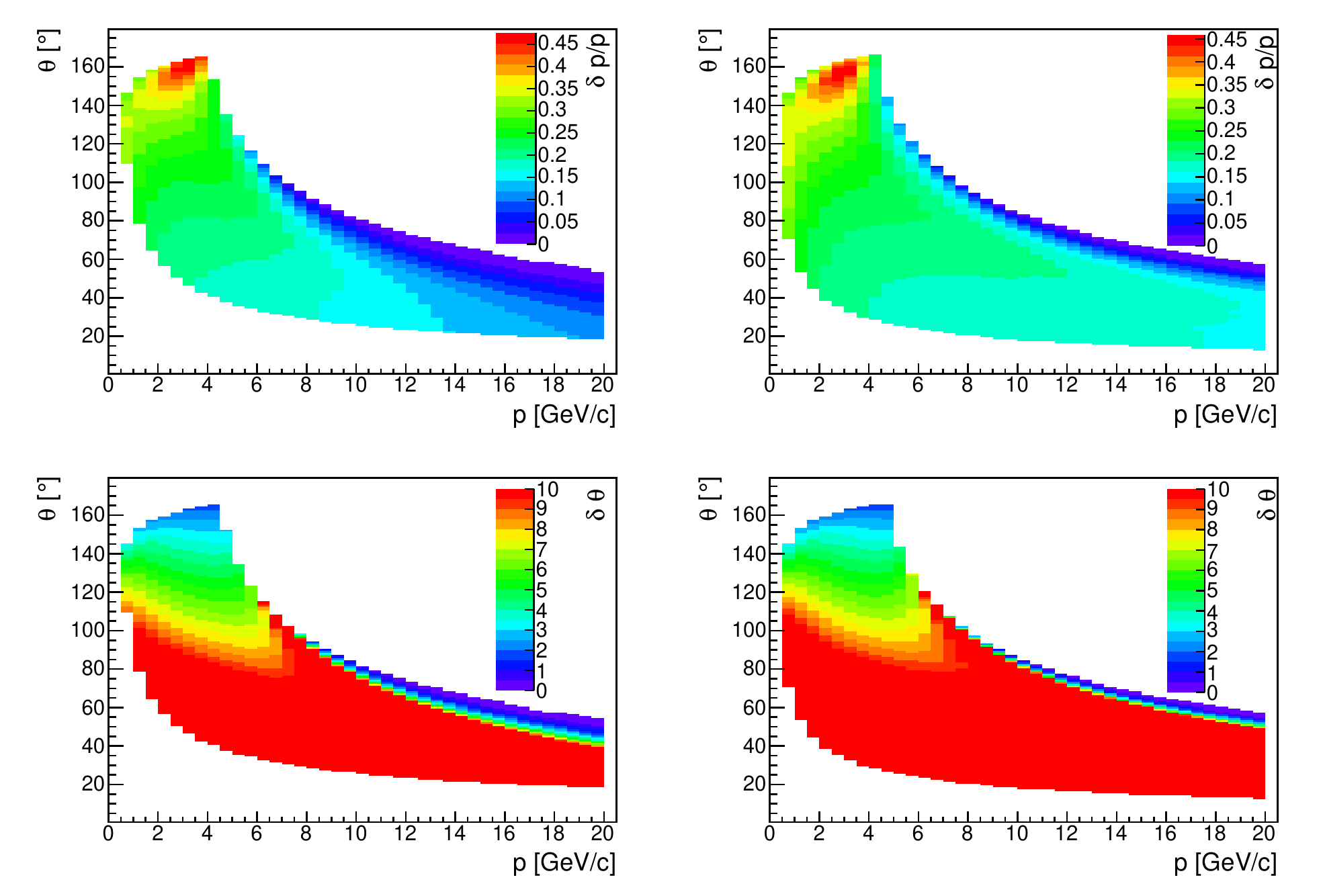}
\caption{ \label{fig:pres} Momentum resolution (top) and angular resolution 
  (bottom) requirements for eRHIC Phase 1 with a 5\,GeV electron beam colliding 
  with a (left) 100\,GeV and (right) 250\,GeV proton beam.  The color scale 
  indicates the required specifications for (top) $\frac{\delta p}{p}$ and 
  (bottom) $\delta \theta$ (in degrees).}
\end{figure}

\subsection{Scattered Electron Energy and Angular Resolution}
\label{ePHENIX:detector:EMCal}

\begin{figure}[!hbt]
\begin{center}
\includegraphics[width=0.95\linewidth]{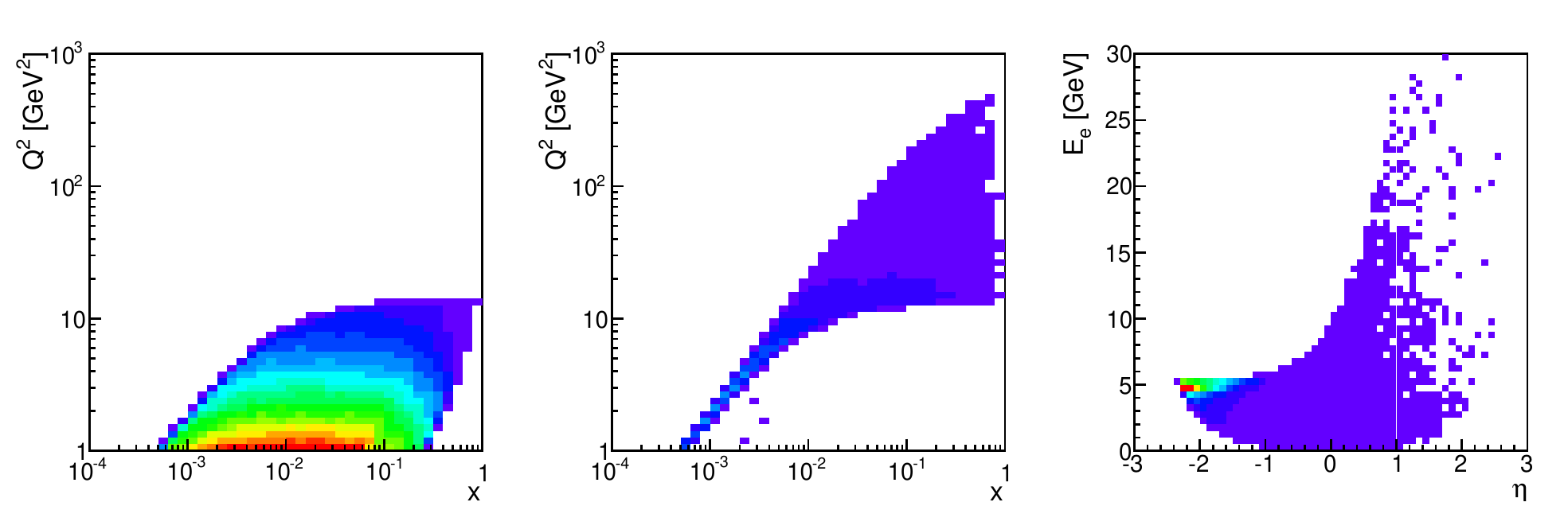}
\end{center}
\caption{For 5\,GeV $\times$ 100\,GeV beam energy configuration: 
$x$-$Q^2$ coverage of DIS for scattered electron detected in backward 
rapidities, $\eta<-1$ (left) and midrapidity, $|\eta|<1$ (middle).  (Right) Scattered electron energy vs pseudorapidity distribution.}
\label{fig:q2x}
\end{figure}

In collider geometry the DIS electrons are scattered mainly in backward 
and central rapidities. Central rapidity selects scattering with higher 
$Q^2$ and higher $x$ (due to its correlation with $Q^2$); 
see Figure~\ref{fig:q2x}.
The higher electron beam energy the more scattering in the backward 
(electron beam) direction. The energy of the scattered electron varies 
in the range from zero to the electron beam energy and even to higher values 
for electrons detected at midrapidity $|\eta|<1.0$, as shown in Figure~\ref{fig:q2x}.

\begin{figure}[!hbt]
\begin{center}
\includegraphics[width=0.9\linewidth]{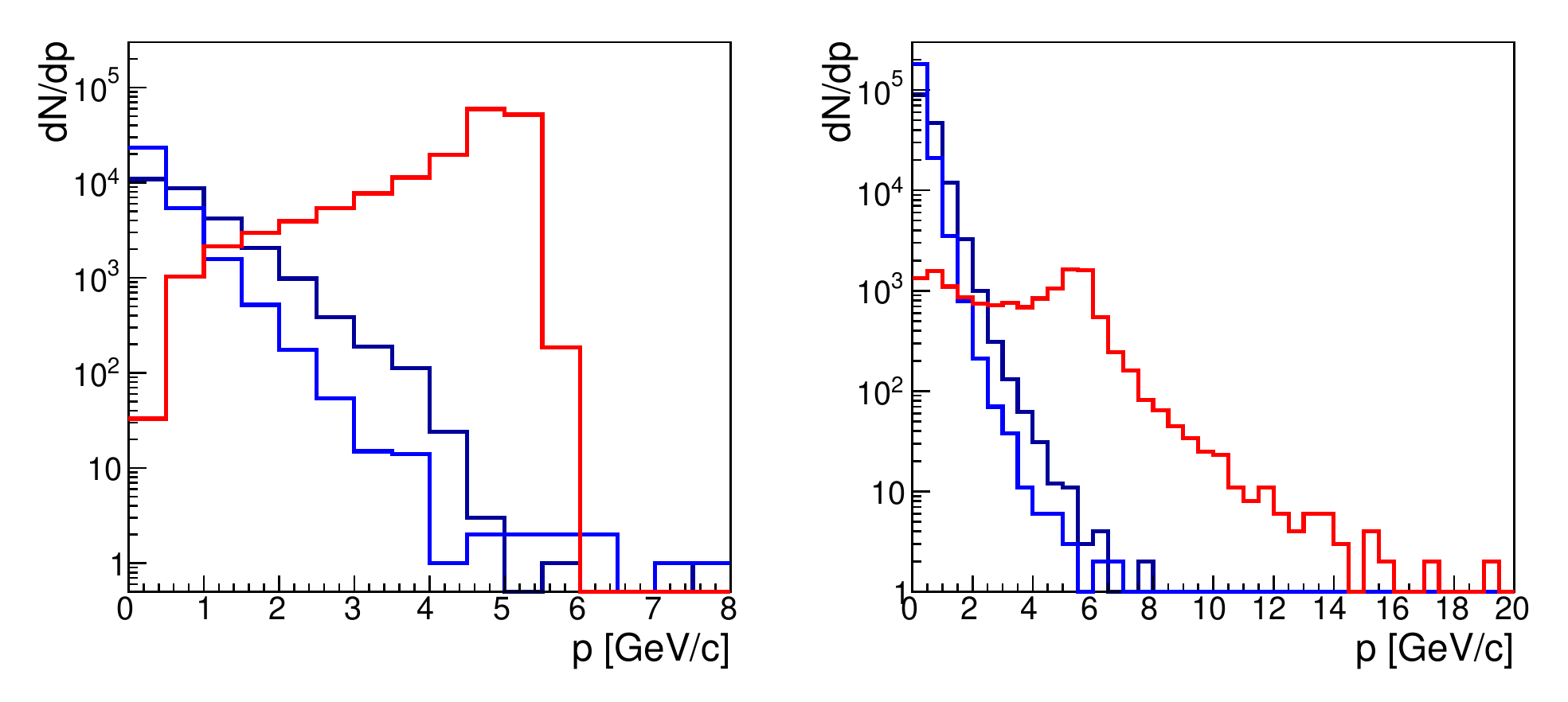}
\end{center}
\caption{For 5\,GeV $\times$ 100\,GeV beam energy configuration: 
Momentum spectra for scattered electron (red), charged pions (black) and 
photons (blue) detected in backward rapidities, $\eta<-1$ (left) and 
in central rapidities, $|\eta|<1$ (right).}
\label{fig:spectra}
\end{figure}

Collider kinematics allows us to clearly separate scattered electrons from 
other DIS fragments - hadrons and their decay products, which are detected 
preferably in the forward region, leaving much softer spectra in central and 
backward rapidities. Figure~\ref{fig:spectra} 
shows the scattered electron momentum spectrum along 
with photon (mainly from hadron decays) and charged pion spectra. 
Reasonable tracking and electromagnetic calorimetry will provide enough 
rejection through $E/p$ matching and shower profile analysis to allow us to 
reliably identify electrons down to momentum at least 1~GeV/$c$. 
Photon conversion in material before the EMCal of up to $10\%$ of a radiation length  
is not expected to contribute sizable background except for very low 
momenta ($<1$\,GeV/c). Lower momentum electrons ($<1$\,GeV/c) only modestly 
extend the $x$-$Q^2$ phase space of DIS kinematics. In addition, these events 
are more contaminated by radiative effects, so other approaches 
(e.g. Jacquet-Blondel method with hadronic final states) should be used for DIS kinematics reconstruction. 

The simplest approach to reconstruct DIS kinematics is from the
scattered electron. While the scattered angle is expected to be
measured with good precision with tracking (see
Section~\ref{ePHENIX:detector:tracking}), the energy resolution will
limit the precision of $x$-$Q^2$ reconstruction.  The energy
resolution $\sigma_E$ is directly propagated to $\sigma_{Q^2}$, so
that $\sigma_{Q^2}/Q^2=\sigma_E/E$, and energy resolution of
$\sim$15--20\% in the range of scattered electron kinematics
(0.04--0.06 in $\log_{10}(Q^2)$ binning) is acceptable.  However, the
$x$ resolution, $\sigma_x$, increases proportionally to $1/y$:
$\sigma_x/x=(1/y) \cdot (\sigma_E/E)$, so the energy resolution
effectively defines the reach of the kinematic region at low $y$. For
the first eRHIC stage with a 5\,GeV electron beam, the electron
momentum measurements will be defined by tracking.  Including an EMCal
with energy resolution of $\sigma_E/E \sim 10\%/\sqrt{E}$ does not
improve measurements if the tracking momentum resolution is
$\sigma_p/p \sim 1\% \cdot p$ or better.  RHIC/eRHIC flexibility to
vary beam energy offers another way to improve the resolutions, where
lowering $\sqrt{s}$ allows access to a given $x$-$Q^2$ bin at higher
$y$ with better $x$-resolution.

QED radiative effects (radiation of real or virtual photons) 
are another source of smearing which is usually 
corrected with an unfolding technique. Unlike energy-momentum resolution 
which introduces gausian-like smearing, radiative corrections introduce a tail toward higher $x$. At higher $y$ they dominate over energy-momentum smearing. 
Jacquet-Blondel method using hadronic 
final states is considered as an alternative approach 
to reconstruct DIS kinematics, which is free of radiative smearing effects.

\begin{figure}[!hbt]
\begin{center}
\includegraphics[width=0.95\linewidth]{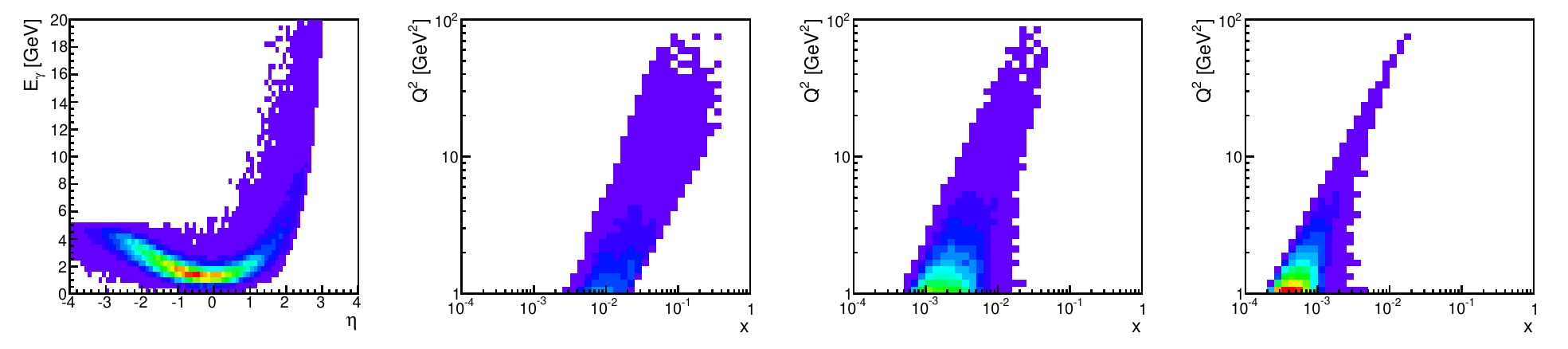}
\end{center}
\caption{For 5\,GeV x 250\,GeV beam energy configuration: 
(Left) DVCS photon energy vs pseudorapidity distribution.  $x$-$Q^2$ 
coverage for DVCS events with photon detected in forward 
rapidities, $\eta>1$ (middle left), central rapidities, $|\eta|<1$ (middle right) 
and backward rapidities, $\eta<-1$  (right).}
\label{fig:dvcs}
\end{figure}

It is also important to measure photons in DVCS. 
The produced DVCS photon energy
versus pseudorapidity distribution is shown in Figure~\ref{fig:dvcs} (left panel). 
For a 5\,GeV electron beam, nearly half of all photons are detected in $|\eta|<1$. For higher electron beam energy
more photons scatter in the backward direction, 
still leaving about a third of photons scattered in $|\eta|<1$ at an electron 
beam energy of 20~GeV. 
The photon momentum in central rapidity varies in the range $\sim$1--4~GeV/c 
nearly independent of beam energy in the range considered for eRHIC. 
Photons in the backward rapidity are more correlated with electron beam and 
have energy from 1~GeV/c to electron beam energy. 
Figure~\ref{fig:dvcs} shows the $x$-$Q^2$ range covered
by DVCS measurements in different rapidity ranges, 
emphasizing the importance of the measurements 
in both backward and central regions. 

\subsection{Particle Identification Needs}
\label{ePHENIX:detector:PID}



Particle Identification (PID) is a requirement of several of the
physics goals of ePHENIX.  Pion and kaon identification are required
for the SIDIS program, both for tagging kaons to extract $\Delta s$,
and for tagging pions and kaons to study the transverse spin structure
of the proton as well as the flavor dependence of nuclear PDFs.  In
addition, PID is essential for a comprehensive program to study
hadronization.  Electron identification is needed to properly
reconstruct event kinematics.  In the case of DVCS, it is important to
tag the minimally scattered proton, which remains in the beam pipe.
In this section, we will briefly discuss how we plan to meet these
requirements, with the main focus on the central region, which is most
relevant to the proposed detector.

\begin{figure}[t]
\centering
\includegraphics[width=\twowidth]{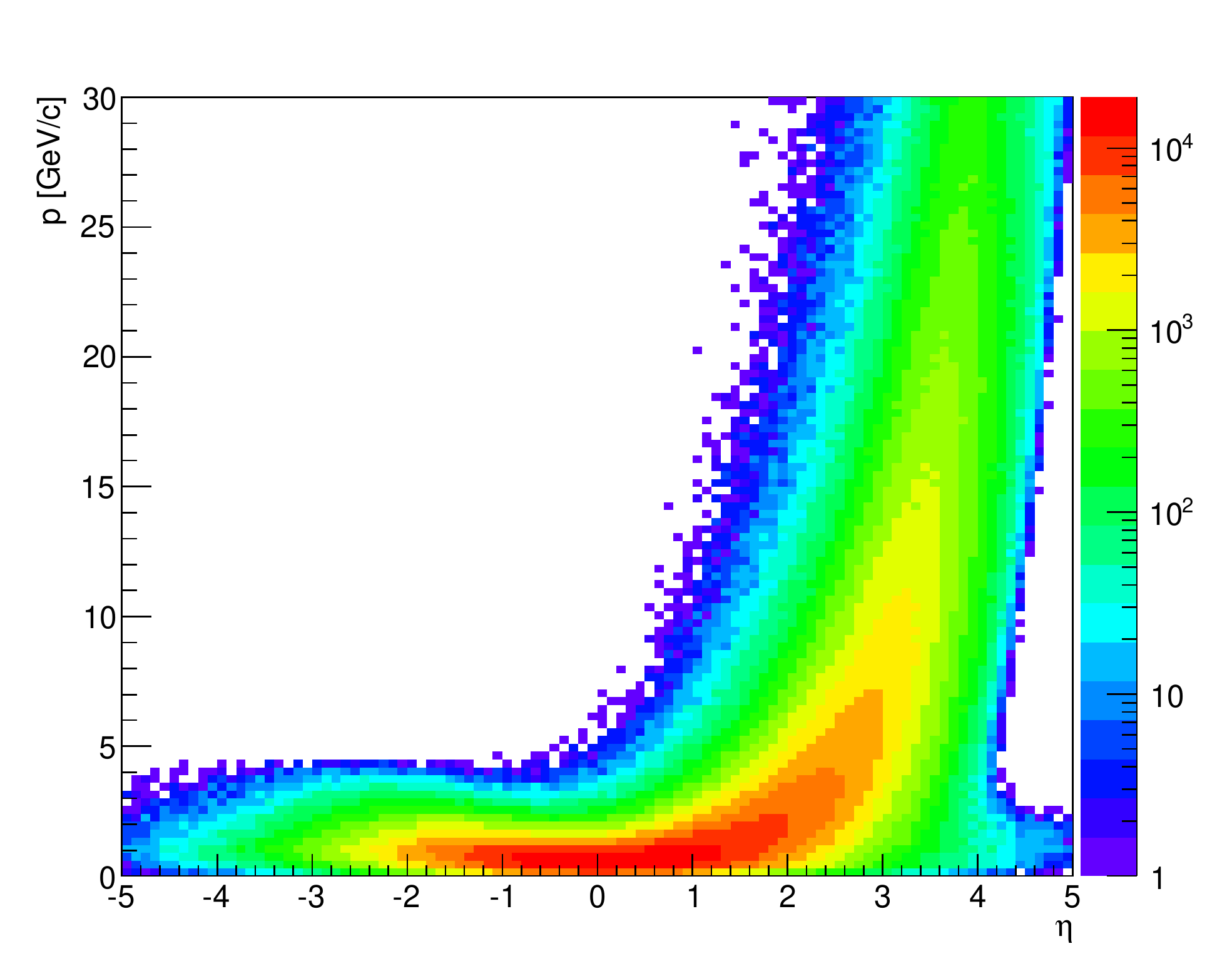}
\hfill
\includegraphics[width=\twowidth]{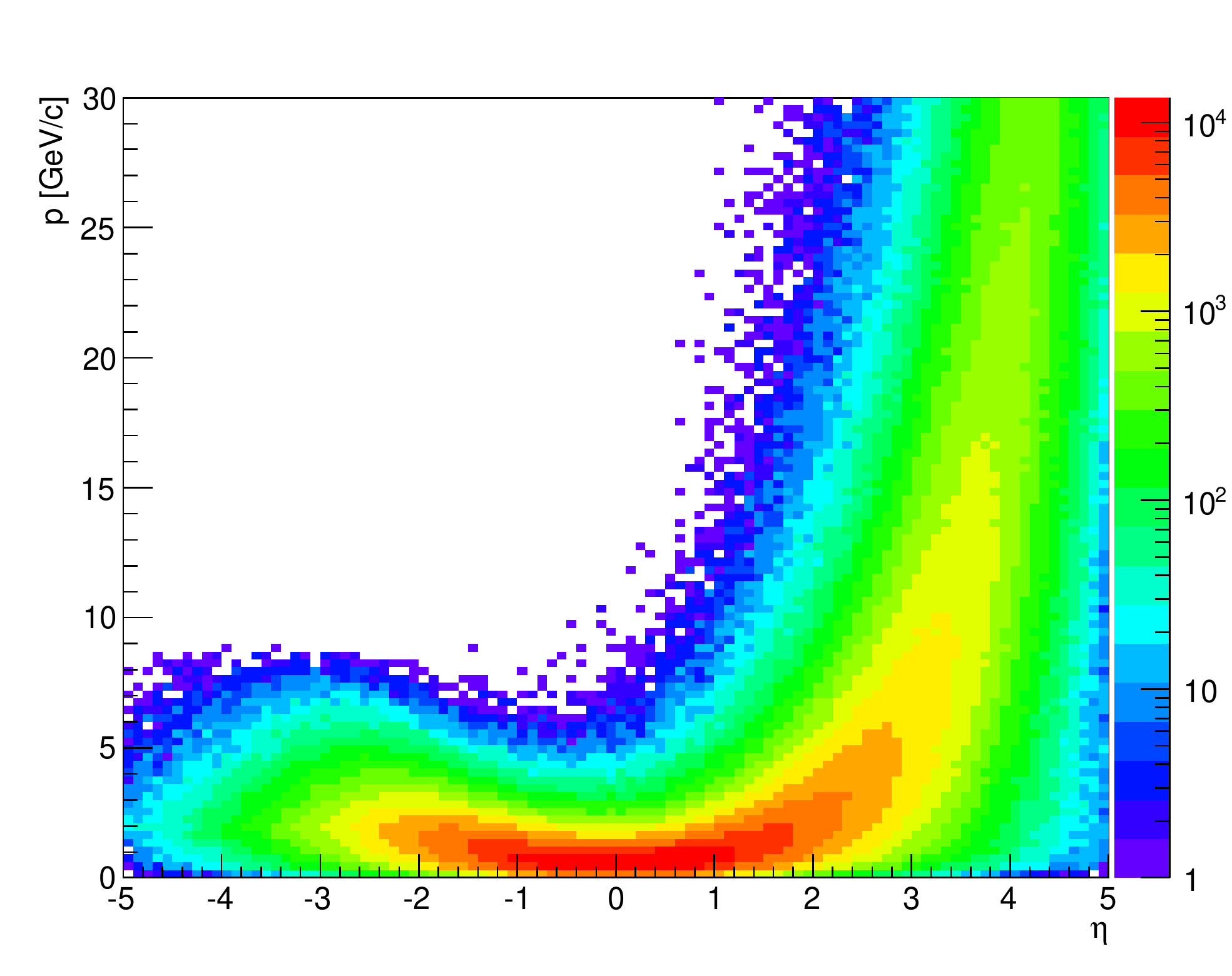}
\caption{Distribution of $\pi^{+}$ momentum vs pseudorapidity  for
  (left) 5\,GeV $\times$ 100\,GeV and (right) 10\,GeV $\times$ 250\,GeV
  beam energy configurations.} 
\label{fig:sidis_pvseta}
\end{figure}

Figure~\ref{fig:sidis_pvseta} shows the $\pi^{+}$ momentum vs
pseudorapidity distribution for two electron beam energy
configurations relevant for the first stage of eRHIC.  Minimal
standard cuts have been applied: $Q^2>1$\,GeV$^2$ and
$W^2>4$~(GeV$/c^2)^2$.  The fractional momentum of the scattered
parton carried by the pion, $z$, is required to be above 0.2 to remove
target fragmentation and below 0.85 to suppress exclusive processes.
For these energies, the majority of hadrons are in the $-1<\eta<4$
range, which will be covered by the central ($-1<\eta<1$) or forward
sPHENIX detector ($1<\eta<4$).  Though hadrons in the forward
direction have the highest momentum due to the boost of the system
center of mass, those in the central and backward region reach higher
$Q^2$ at a given
$x$.  
Hadrons scattered in the forward direction sample events with
$xE_p>E_e$, i.e. events with relatively higher $x$ and lower to
moderate $Q^2$.  In the midrapidity region, $xE_p\simeq E_e$ with a
maximum transverse momentum of $\sqrt{xE_pE_e}$ which for a 5 and
10\,GeV electron beam on a 250\,GeV proton can be as high as
$35.4\sqrt{x}$\,GeV$/c^2$ and $50\sqrt{x}$\,GeV$/c^2$ respectively.
Hadrons in the backwards direction cover a subset of the range covered
by the midrapidity region.

The conceptual design for the forward sPHENIX detector (See
Appendix~\ref{chap:fsPHENIX}) covers $1<\eta<4$ and includes RICH
based PID with $\pi-$K separation up to sufficiently high momentum.
As the forward sPHENIX design moves ahead, all ePHENIX requirements
will be considered.

\begin{figure}[t]
\centering
\includegraphics[width=\twowidth]{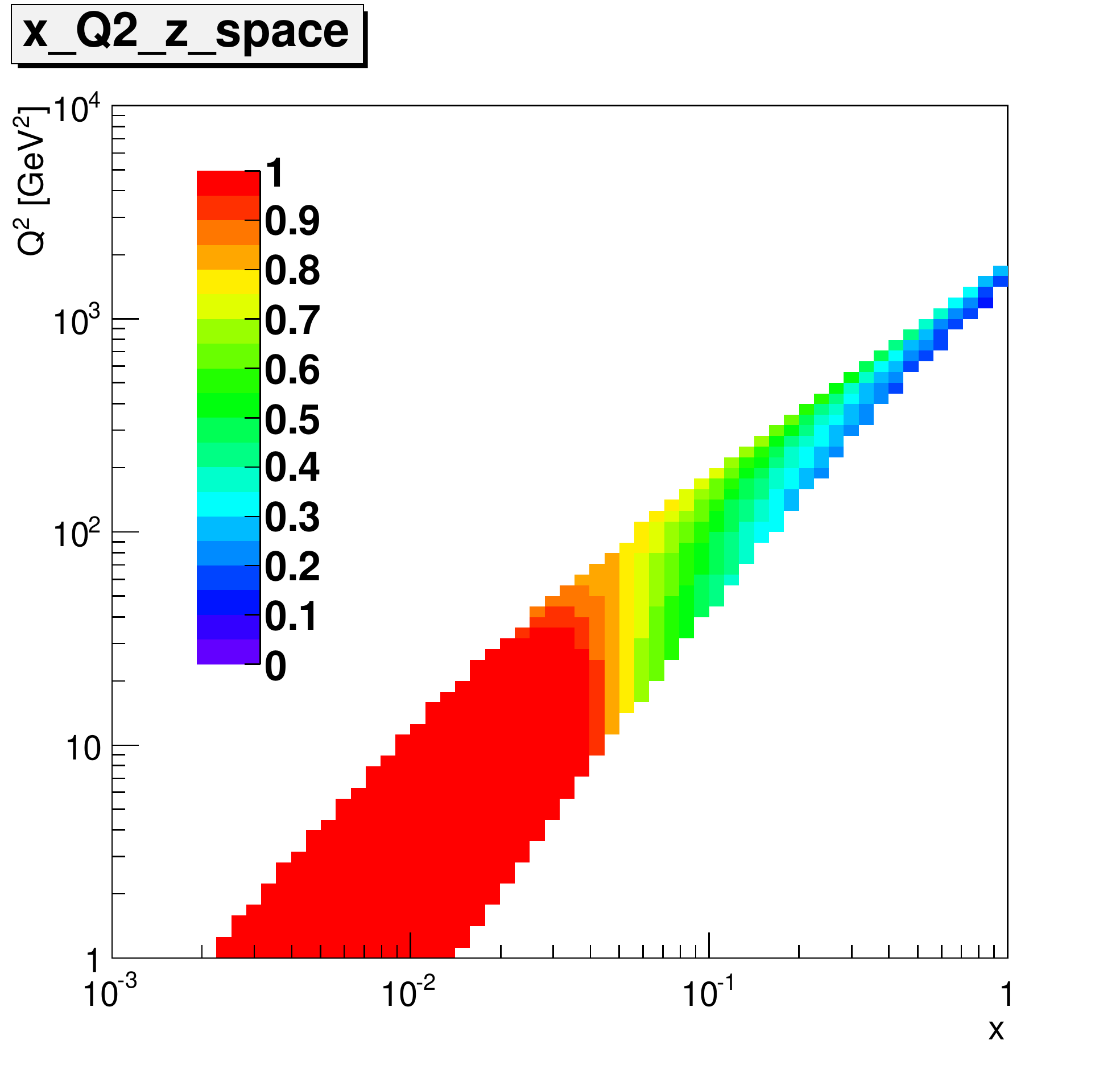}
\hfill
\includegraphics[width=\twowidth]{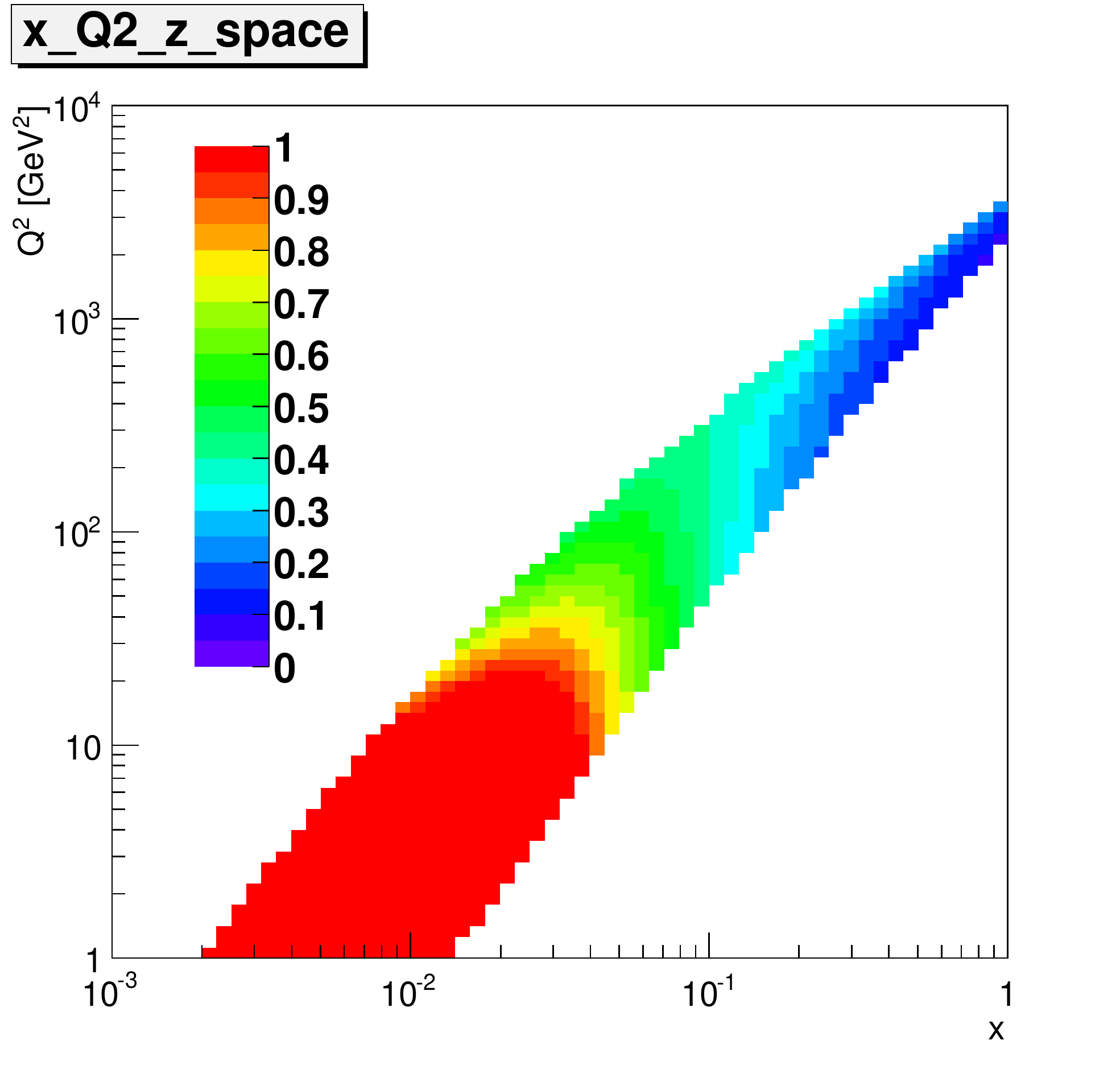}
\caption{(Left) 5\,GeV $\times$ 100\,GeV and (right) 10\,GeV $\times$
  100\,GeV beam energy configurations.  Maximum $z$ accessible (color
  scale) in $|\eta|<1$ at given $x$-$Q^2$ assuming a $p<4$\,GeV/$c$ cut
  is imposed.}
\label{fig:sidis_maxp_4}
\end{figure}

In the central barrel region ($|\eta|<1$), standard RICH-based PID is
not possible due to the large radial space required, and the need for
light collection in the acceptance of the planned forward
spectrometer.  However, we can use other \v{C}erenkov based detectors,
as is discussed in the next section, which have $\pi-$K separation up
to $p\sim 4$\,GeV/$c$.  As stated above, a minimum $z$ cut of 0.2 is
required in SIDIS, therefore limiting the selection to events in which
the scattered parton has $E<20$\,GeV.  The impact of this can be seen
in Figure~\ref{fig:sidis_maxp_4}, where, assuming simple
$2\rightarrow2$ kinematics, the maximum $z$ accessible when imposing a
$p<4$\,GeV/$c$ cut is plotted for the $x$-$Q^2$ ranges sampled in the
midrapidity region.  The dark blue regions coincide with $z<0.2$, and
would be inaccessible.  At the beam energies available in the first
stage of eRHIC, this limitation is minimal.

\subsubsection{Technologies}

Particle identification in ePHENIX will be accomplished using the
existing sPHENIX central and forward detectors, augmented by
additional ePHENIX specific detectors. In the midrapidity region, the
electromagnetic calorimeter will provide the principal means for
electron identification.  It was shown above that with a calorimeter
energy resolution $\sim$15--20$\%/\sqrt{E}$ and a momentum resolution
of $\sigma_p/p\sim 1\%p$ in the central tracking system, good electron
identification can be achieved over the entire kinematic range of the
scattered electron.  It is also assumed that a preshower detector as
detailed in Appendix~\ref{chap:barrel_upgrade} will be installed in
the central sPHENIX detector by the time ePHENIX is running, which
will provide significant electron-hadron separation.

Identification of pions, kaons and protons will require additional detectors both for 
central and forward regions.  In the central region, one is limited by the space available 
inside the solenoid magnet, and therefore requires a compact detector that takes up the 
minimal amount of radial space. In order to achieve PID above 5~GeV/c, we have considered a 
dual radiator gas RICH. This could use a proximity focused configuration 
\cite{Barile:2011zza} and a photosensitive gas detector, such as a CsI photocathode GEM 
detector, similar to what was used in the PHENIX HBD \cite{Anderson:2011jw}, only with a 
more highly segmented readout that would allow complete ring imaging. This technology is 
already being studied by as part of the EIC detector R\&D program.

Two other technologies being explored for PID are a DIRC (Detection of Internally Reflected 
Cherenkov Light)~\cite{Adam:2004fq} and a Time of Propagation (TOP) detector 
\cite{Nishimura:2010gj}, as is being developed for BELLE II. A DIRC would have the 
advantage of a very thin radiator ($\sim$5~cm), but would have limited PID capabilities 
above $\sim3.5$~GeV/c. A TOP detector may extend this somewhat, but would still not cover 
the full momentum range of particles at central rapidity. However, it may be possible to 
extend the capabilities of either a DIRC or TOP detector using silicon photomultipliers as 
photodetectors, which can provide very good position and timing resolution. Currently, 
silicon photomultipliers provide position resolutions $\sim$ 1 mm and time resolutions 
$\sim 200$~ps, but it is in principal possible to achieve $\sim$50~$\mu$m and $\sim$50~ps 
with these devices. This could greatly enhance the capabilities of either a DIRC or TOP 
counter, and also provide the ability to construct such a detector in a much more compact 
arrangement.






\cleardoublepage

\chapter{The PHENIX Collaboration}
\label{chap:collaboration}

\begin{flushleft}
\begin{description}
\item[Abilene Christian University,] 
 {\em  Abilene, Texas 79699, USA} \\
	M.S.~Daugherity,
	K.~Gainey,
	D.~Isenhower,
	H.~Qu,
	A.~Tate,
	R.~Towell,
	R.S.~Towell*,
	T.S.~Watson
 
\item[Department of Physics,] 
 {\em  Banaras Hindu University, Varanasi 221005, India} \\
	P.~Garg,
	P.K.~Khandai,
	B.K.~Singh,
	C.P.~Singh,
	V.~Singh*,
	P.K.~Srivastava,
	S.~Tarafdar
 
\item[Bhabha Atomic Research Centre,] 
 {\em  Bombay 400 085, India} \\
	D.K.~Mishra,
	A.K.~Mohanty*,
	P.K.~Netrakanti,
	P.~Sett,
	P.~Shukla
 
\item[Baruch College,] 
 {\em  City University of New York, New York, New York, 10010 USA} \\
	S.~Bathe*,
	J.~Bryslawskyj
 
\item[Collider-Accelerator Department,] 
 {\em  Brookhaven National Laboratory, Upton, New York 11973-5000, USA} \\
	M.~Bai,
	K.A.~Drees,
	S.~Edwards,
	Y.I.~Makdisi*,
	A.~Zelenski
 
\item[Physics Department,] 
 {\em  Brookhaven National Laboratory, Upton, New York 11973-5000, USA} \\
	B.~Azmoun,
	A.~Bazilevsky,
	S.~Boose,
	M.~Chiu,
	G.~David,
	E.J.~Desmond,
	K.O.~Eyser,
	A.~Franz,
	P.~Giannotti,
	J.S.~Haggerty,
	J.~Jia,
	B.M.~Johnson,
	H.-J.~Kehayias,
	E.~Kistenev,
	D.~Lynch,
	E.~Mannel,
	J.T.~Mitchell,
	D.P.~Morrison,
	R.~Nouicer,
	E.~O'Brien,
	R.~Pak,
	C.~Pinkenburg,
	R.P.~Pisani,
	M.L.~Purschke,
	T.~Sakaguchi,
	A.~Sickles,
	I.V.~Sourikova,
	P.~Steinberg,
	S.P.~Stoll,
	A.~Sukhanov,
	M.J.~Tannenbaum*,
	C.L.~Woody
 
\item[University of California - Riverside,] 
 {\em  Riverside, California 92521, USA} \\
	K.N.~Barish*,
	M.~Beaumier,
	D.~Black,
	K.O.~Eyser,
	T.~Hester,
	R.S.~Hollis,
	A.~Iordanova,
	D.~Kleinjan,
	M.~Mendoza,
	S.D.~Rolnick,
	K.~Sedgwick,
	R.~Seto
 
\item[Charles University,] 
 {\em  Ovocn\'{y} trh 5, Praha 1, 116 36, Prague, Czech Republic} \\
	M.~Finger Jr.,
	M.~Finger*,
	M.~Slune\v{c}ka,
	V.~Slune\v{c}kova
 
\item[Chonbuk National University,] 
 {\em  Jeonju, 561-756, Korea} \\
	J.B.~Choi,
	E.-J.~Kim*,
	K.-B.~Kim,
	G.H.~Lee
 
\item[Science and Technology on Nuclear Data Laboratory,] 
 {\em  China Institute of Atomic Energy, Beijing 102413, P.~R.~China} \\
	X.~Bai,
	X.~Li*,
	S.~Zhou
 
\item[Center for Nuclear Study,] 
 {\em  Graduate School of Science, University of Tokyo, 7-3-1 Hongo, Bunkyo, Tokyo 113-0033, Japan} \\
	R.~Akimoto,
	T.~Gunji,
	H.~Hamagaki*,
	R.~Hayano,
	S.~Hayashi,
	Y.~Hori,
	Y.~Komatsu,
	S.~Masumoto,
	A.~Nukariya,
	H.~Oide,
	K.~Ozawa,
	Y.~Sekiguchi,
	A.~Takahara,
	H.~Torii,
	T.~Tsuji,
	Y.S.~Watanabe,
	Y.L.~Yamaguchi
 
\item[University of Colorado,] 
 {\em  Boulder, Colorado 80309, USA} \\
        A.~Adare,
	T.~Koblesky,
	M.~McCumber,
	J.L.~Nagle*,
	M.R.~Stone,
 
\item[Columbia University,] 
 {\em  New York, New York 10027 and Nevis Laboratories, Irvington, New York 10533, USA} \\
	C.Y.~Chi*,
	B.A.~Cole,
	T.~Engelmore,
	N.~Grau,
	Y.S.~Lai,
	D.~Perepelitsa,
	F.W.~Sippach,
	E.~Vazquez-Zambrano,
	A.~Veicht,
	D.~Winter,
	W.A.~Zajc,
	L.~Zhang
 
\item[Czech Technical University,] 
 {\em  Zikova 4, 166 36 Prague 6, Czech Republic} \\
	T.~Li\v{s}ka,
	M.~Tom\'a\v{s}ek,
	M.~Virius*,
	V.~Vrba
 
\item[Debrecen University,] 
 {\em  H-4010 Debrecen, Egyetem t{\'e}r 1, Hungary} \\
	J.~Imrek, 
	P.~Tarj\'an*
 
\item[ELTE,] 
 {\em  E{\"o}tv{\"o}s Lor{\'a}nd University, H - 1117 Budapest, P{\'a}zm{\'a}ny P.~s.~1/A, Hungary} \\
	M.~Csan\'ad*,
	\'A.~Kiss,
	M.~Kofarago,
	M.I.~Nagy,
	M.~Vargyas
 
\item[Ewha Womans University,] 
 {\em  Seoul 120-750, Korea} \\
	K.I.~Hahn*,
	S.Y.~Han,
	D.H.~Kim,
	J.~Lee,
	I.H.~Park
 
\item[Florida State University,] 
 {\em  Tallahassee, Florida 32306, USA} \\
	A.D.~Frawley*,
	J.R.~Hutchins,
	J.~Klatsky,
	D.~McGlinchey
 
\item[Georgia State University,] 
 {\em  Atlanta, Georgia 30303, USA} \\
	C.~Butler,
	H.~Guragain,
	X.~He*,
	M.~Jezghani,
	L.~Patel,
	M.~Sarsour,
	A.~Sen
 
\item[Hanyang University,] 
 {\em  Seoul 133-792, Korea} \\
	B.H.~Kang,
	J.S.~Kang,
	Y.K.~Kim*,
	M.S.~Ryu
 
\item[Hiroshima University,] 
 {\em  Kagamiyama, Higashi-Hiroshima 739-8526, Japan} \\
	K.~Homma,
	T.~Hoshino,
	K.M.~Kijima,
	Y.~Nakamiya,
	M.~Nihashi,
	K.~Shigaki,
	T.~Sugitate*,
	D.~Watanabe
 
\item[IHEP Protvino,] 
 {\em  State Research Center of Russian Federation, Institute for High Energy Physics, Protvino, 142281, Russia} \\
	V.~Babintsev,
	V.~Bumazhnov,
	S.~Chernichenko,
	A.~Denisov*,
	A.~Durum,
	I.~Shein,
	A.~Soldatov,
	A.~Yanovich
 
\item[University of Illinois at Urbana-Champaign,] 
 {\em  Urbana, Illinois 61801, USA} \\
	J.~Blackburn,
	I.J.~Choi,
	L.~Eberle,
	F.~Giordarno,
	M.~Grosse~Perdekamp*,
	D.S.~Jumper,
	Y.-J.~Kim,
	M.~Leitgab,
	C.~McKinney,
	B.~Meredith,
	D.~Northacker,
	J.-C.~Peng,
	E.~Thorsland,
	S.~Wolin,
	R.~Yang,
	E.~Zarndt
 
\item[Institute for Nuclear Research of the Russian Academy of Sciences,] 
 {\em  prospekt 60-letiya Oktyabrya 7a, Moscow 117312, Russia} \\
	V.~Pantuev*
 
\item[Institute of Physics,] 
 {\em  Academy of Sciences of the Czech Republic, Na Slovance 2, 182 21 Prague 8, Czech Republic} \\
	J.~Popule,
	P.~Sicho,
	L.~Tom\'a\v{s}ek,
	M.~Tom\'a\v{s}ek,
	V.~Vrba*
 
\item[Iowa State University,] 
 {\em  Ames, Iowa 50011, USA} \\
	S.~Campbell,
	L.~Ding,
	J.C.~Hill*,
	J.G.~Lajoie,
	A.~Lebedev,
	R.~McKay,
	C.A.~Ogilvie,
	J.~Perry,
	M.~Rosati,
	A.~Shaver,
	M.~Shimomura,
	A.~Timilsina,
	S.~Whitaker
 
\item[Advanced Science Research Center,] 
 {\em  Japan Atomic Energy Agency, 2-4 Shirakata Shirane, Tokai-mura, Naka-gun, Ibaraki-ken 319-1195, Japan} \\
	K.~Imai*,
	T.~Maruyama,
	H.~Sako,
	S.~Sato
 
\item[Helsinki Institute of Physics and University of Jyv{\"a}skyl{\"a},] 
 {\em  P.O.Box 35, FI-40014 Jyv{\"a}skyl{\"a}, Finland} \\
	D.J.~Kim,
	F.~Krizek,
	N.~Novitzky,
	J.~Rak*
 
\item[KEK,] 
 {\em  High Energy Accelerator Research Organization, Tsukuba, Ibaraki 305-0801, Japan} \\
	Y.~Fukao,
	S.~Kanda,
	M.~Makek,
	T.~Mibe,
	S.~Nagamiya,
	K.~Ozawa,
	N.~Saito,
	S.~Sawada*,
	Y.S.~Watanabe
 
\item[Korea University,] 
 {\em  Seoul, 136-701, Korea} \\
	B.~Hong*,
	C.~Kim,
	K.S.~Lee,
	S.K.~Park,
 
\item[Russian Research Center ``Kurchatov Institute",] 
 {\em  Moscow, 123098 Russia} \\
	D.S.~Blau,
	S.L.~Fokin,
	A.V.~Kazantsev,
	V.I.~Manko*,
	T.V.~Moukhanova,
	A.S.~Nyanin,
	D.Yu.~Peressounko,
	I.E.~Yushmanov
 
\item[Kyoto University,] 
 {\em  Kyoto 606-8502, Japan} \\
	H.~Asano,
	S.~Dairaku,
	K.~Karatsu,
	T.~Murakami*,
	T.~Nagae,
	K.R.~Nakamura
 
\item[Laboratoire Leprince-Ringuet,] 
 {\em  Ecole Polytechnique, CNRS-IN2P3, Route de Saclay, F-91128, Palaiseau, France} \\
	S.~Chollet,
	A.~Debraine,
	O.~Drapier,
	F.~Fleuret*,
	F.~Gastaldi,
	M.~Gonin,
	R.~Granier~de~Cassagnac
 
\item[Lawrence Livermore National Laboratory,] 
 {\em  Livermore, California 94550, USA} \\
	I.~Garishvili,
	A.~Glenn,
	R.A.~Soltz*
 
\item[Los Alamos National Laboratory,] 
 {\em  Los Alamos, New Mexico 87545, USA} \\
	C.~Aidala,
	M.L.~Brooks,
	J.M.~Durham,
	J.~Huang,
	X.~Jiang,
	J.~Kapustinsky,
	K.B.~Lee,
	M.J.~Leitch,
	M.X.~Liu*,
	P.L.~McGaughey,
	C.L.~Silva,
	W.E.~Sondheim,
	H.W.~van~Hecke
 
\item[Department of Physics,] 
 {\em  Lund University, Box 118, SE-221 00 Lund, Sweden} \\
	P.~Christiansen,
	A.~Oskarsson*,
	L.~\"Osterman,
	E.~Stenlund
 
\item[University of Maryland,] 
 {\em  College Park, Maryland 20742, USA} \\
	O.~Baron,
	L.~D'Orazio,
	A.C.~Mignerey*,
	E.~Richardson,
	L.~Stevens 

\item[Department of Physics,] 
 {\em  University of Massachusetts, Amherst, Massachusetts 01003-9337, USA } \\
	N.~Bandara,
	D.~Kawall*,
	M.~Stepanov
 
\item[Muhlenberg College,] 
 {\em  Allentown, Pennsylvania 18104-5586, USA} \\
	G.~Benjamin,
	A.~Carollo,
	N.~Cronin,
	N.~Crossett,
	B.~Fadem*,
	A.~Isinhue,
	M.~Moskowitz,
	S.~Motschwiller,
	A.~Nederlof,
	M.~Skolnik,
	S.~Solano,
	A.~Tullo,
	M.~Young,
	C.~Zumberge
 
\item[Myongji University,] 
 {\em  Yongin, Kyonggido 449-728, Korea} \\
	S.J.~Jeon,
	K.S.~Joo*
 
\item[Nagasaki Institute of Applied Science,] 
 {\em  Nagasaki-shi, Nagasaki 851-0193, Japan} \\
	T.~Fusayasu*,
	Y.~Tanaka
 
\item[University of New Mexico,] 
 {\em  Albuquerque, New Mexico 87131, USA } \\
	B.~Bassalleck,
	J.~Bok,
	S.~Butsyk,
	A.~Datta,
	K.~DeBlasio,
	D.E.~Fields*,
	M.~Hoefferkamp,
	J.A.~Key,
	I.~Younus
 
\item[New Mexico State University,] 
 {\em  Las Cruces, New Mexico 88003, USA} \\
	J.~Bok,
	A.~Meles,
	V.~Papavassiliou,
	S.F.~Pate*,
	G.D.N.~Perera,
	E.~Tennant,
	X.R.~Wang,
	F.~Wei
 
\item[Department of Physics and Astronomy,] 
 {\em  Ohio University, Athens, Ohio 45701, USA} \\
	X.~Bing,
	J.E.~Frantz*,
	D.~Kotchetkov,
	N.~Riveli
 
\item[Oak Ridge National Laboratory,] 
 {\em  Oak Ridge, Tennessee 37831, USA} \\
	T.C.~Awes,
	M.~Bobrek,
	C.L.~Britton,Jr.,
	V.~Cianciolo,
	Y.V.~Efremenko,
	K.F.~Read,
	D.~Silvermyr,
	P.W.~Stankus*,
	M.~Wysocki
 
\item[IPN-Orsay,] 
 {\em  Universite Paris Sud, CNRS-IN2P3, BP1, F-91406, Orsay, France} \\
	D.~Jouan*
 
\item[PNPI,] 
 {\em  Petersburg Nuclear Physics Institute, Gatchina, Leningrad region, 188300, Russia} \\
	V.~Baublis,
	D.~Ivanischev,
	V.~Ivanov,
	A.~Khanzadeev,
	L.~Kochenda,
	B.~Komkov,
	P.~Kravtsov,
	V.~Riabov,
	Y.~Riabov,
	E.~Roschin,
	V.~Samsonov*,
	V.~Trofimov,
	E.~Vznuzdaev
 
\item[RIKEN Nishina Center for Accelerator-Based Science,] 
 {\em  Wako, Saitama 351-0198, Japan} \\
	Y.~Akiba,
	K.~Aoki,
	H.~Asano,
	S.~Baumgart,
	S.~Dairaku,
	A.~Enokizono,
	Y.~Fukao,
	Y.~Goto,
	K.~Hashimoto,
	T.~Ichihara,
	Y.~Ikeda,
	Y.~Imazu,
	K.~Karatsu,
	M.~Kurosawa,
	S.~Miyasaka,
	T.~Murakami,
	J.~Murata,
	I.~Nakagawa,
	K.R.~Nakamura,
	T.~Nakamura,
	K.~Nakano,
	M.~Nihashi,
	K.~Ninomiya,
	R.~Seidl,
	T.-A.~Shibata,
	K.~Shoji,
	A.~Taketani,
	T.~Todoroki,
	K.~Watanabe,
	Y.~Watanabe*,
	S.~Yokkaichi
 
\item[RIKEN BNL Research Center,] 
 {\em  Brookhaven National Laboratory, Upton, New York 11973-5000, USA} \\
	Y.~Akiba*,
	S.~Bathe,
	K.~Boyle,
	C.-H.~Chen,
	A.~Deshpande,
	Y.~Goto,
	T.~Ichihara,
	J.~Koster,
	M.~Kurosawa,
	I.~Nakagawa,
	R.~Nouicer,
	K.~Okada,
	J.~Seele,
	R.~Seidl,
	A.~Taketani,
	K.~Tanida,
	Y.~Watanabe,
	S.~Yokkaichi
 
\item[Physics Department,] 
 {\em  Rikkyo University, 3-34-1 Nishi-Ikebukuro, Toshima, Tokyo 171-8501, Japan} \\
	K.~Hashimoto,
	K.~Kurita*,
	J.~Murata,
	K.~Ninomiya,
	K.~Watanabe
 
\item[Saint Petersburg State Polytechnic University,] 
 {\em  St.~Petersburg, 195251 Russia} \\
	A.~Berdnikov,
	Y.~Berdnikov*,
	D.~Kotov,
	A.~Safonov
 
\item[Universidade de S{\~a}o Paulo,] 
 {\em  Instituto de F\'{\i}sica, Caixa Postal 66318, S{\~a}o Paulo CEP05315-970, Brazil} \\
	O.~Dietzsch*,
	M.~Donadelli,
	M.~Kuriyama,
	M.A.L.~Leite,
	R.~Menegasso,
	E.M.~Takagui
 
\item[Department of Physics and Astronomy,] 
 {\em  Seoul National University, Seoul, Korea} \\
	S.~Choi,
	S.~Park,
	K.~Tanida,
	I.~Yoon
 
\item[Chemistry Department,] 
 {\em  Stony Brook University, SUNY, Stony Brook, New York 11794-3400, USA} \\
	N.N.~Ajitanand,
	J.~Alexander,
	X.~Gong,
	Y.~Gu,
	J.~Jia,
	R.~Lacey*,
	A.~Mwai,
	M.~Soumya,
	S.~Radhakrishnan,
	R.~Reynolds,
	A.~Taranenko,
	R.~Wei
 
\item[Department of Physics and Astronomy,] 
 {\em  Stony Brook University, SUNY, Stony Brook, New York 11794-3400, USA} \\
	N.~Apadula,
	E.T.~Atomssa,
	B.~Bannier,
	K.~Dehmelt,
	A.~Deshpande*,
	A.~Dion,
	A.~Drees,
	H.~Ge,
	C.~Gal,
	J.~Hanks,
	T.K.~Hemmick,
	B.V.~Jacak,
	J.~Kamin,
	V.~Khachatryan,
	P.~Kline,
	S.H.~Lee,
	R.~Lefferts,
	B.~Lewis,
	A.~Lipski,
	M.~Lynch,
	A.~Manion,
	C.~Pancake,
	R.~Petti,
	B.~Sahlmueller,
	E.~Shafto,
	D.~Sharma,
	J.~Sun

\item[Accelerator and Medical Instrumentation Engineering Lab,] 
 {\em  SungKyunKwan University, 53 Myeongnyun-dong, 3-ga, Jongno-gu, Seoul, South Korea} \\
	J.-S.~Chai*
 
\item[University of Tennessee,] 
 {\em  Knoxville, Tennessee 37996, USA} \\
	A.~Garishvili,
	C.~Nattrass,
	K.F.~Read,
	S.P.~Sorensen*,
	E.~Tennant
 
\item[Department of Physics,] 
 {\em  Tokyo Institute of Technology, Oh-okayama, Meguro, Tokyo 152-8551, Japan} \\
	S.~Miyasaka,
	K.~Nakano,
	T.-A.~Shibata*
 
\item[Institute of Physics,] 
 {\em  University of Tsukuba, Tsukuba, Ibaraki 305, Japan} \\
	T.~Chujo,
	S.~Esumi*,
	M.~Inaba,
	Y.~Miake,
	S.~Mizuno,
	T.~Niida,
	M.~Sano,
	T.~Todoroki,
	K.~Watanabe
 
\item[Vanderbilt University,] 
 {\em  Nashville, Tennessee 37235, USA} \\
	R.~Belmont,
	S.V.~Greene*,
	S.~Huang,
	B.~Love,
	C.F.~Maguire,
	D.~Roach,
	B.~Schaefer,
	J.~Velkovska
 
\item[Weizmann Institute,] 
 {\em  Rehovot 76100, Israel} \\
	M.~Makek,
	A.~Milov,
	I.~Ravinovich,
	I.~Tserruya*
 
\item[Institute for Particle and Nuclear Physics,] 
 {\em  Wigner Research Centre for Physics, Hungarian Academy of Sciences (Wigner RCP, RMKI) H-1525 Budapest 114, POBox 49, Budapest, Hungary} \\
	T.~Cs\"org\H{o}*,
	M.I.~Nagy,
	T.~Novak,
	A.~Ster,
	J.~Sziklai,
	R.~V\'ertesi
 
\item[Yonsei University,] 
 {\em  IPAP, Seoul 120-749, Korea} \\
	J.H.~Do,
	J.H.~Kang*,
	H.J.~Kim,
	Y.~Kwon,
	S.H.~Lim
 
\end{description}
 
\end{flushleft}

\vspace{0.3cm}
\noindent *PHENIX Institutional Board member

\begin{table}[hb]
\begin{flushleft}
\begin{tabbing}
\hspace*{0.2in}Spokesperson\hspace{2.7in}\=Barbara Jacak \\
                    \>  \hspace{0.25in}\=\emph{Stony Brook University} \\
\hspace*{0.2in}Deputy Spokesperson   \>Jamie Nagle \\
                    \> \hspace{0.25in}\={\em University of Colorado} \\
\hspace*{0.2in}Deputy Spokesperson   \>Yasuyuki Akiba \\
                    \> \hspace{0.25in}\={\em RIKEN Nishina Center for} \\
                    \> \hspace{0.30in}\={\em Accelerator-Based Science} \\
\hspace*{0.2in}Deputy Spokesperson   \>David Morrison \\
                    \> \hspace{0.25in}\={\em Brookhaven National Laboratory} \\
\hspace*{0.2in}Operations Director   \>Ed O'Brien \\
                    \> \hspace{0.25in}\={\em Brookhaven National Laboratory} \\
\hspace*{0.2in}Deputy Operations Director for Upgrades   \>Mike Leitch \\
                    \> \hspace{0.25in}\={\em Los Alamos National Laboratory} \\
\hspace*{0.2in}Deputy Operations Director for Operations \>John Haggerty \\
                    \> \hspace{0.25in}\={\em Brookhaven National Laboratory} \\
\end{tabbing}
\end{flushleft}
\end{table}

\backmatter

\cleardoublepage
\phantomsection
\addcontentsline{toc}{chapter}{References}

\bibliographystyle{unsrturl}

\end{document}